\documentclass[a4paper,11pt]{article}
\pdfoutput=1
\usepackage{jheppub}

\usepackage{amsmath}
\usepackage{float}
\usepackage{slashed}
\usepackage{graphicx}
\usepackage{dsfont}
\usepackage{epstopdf}

\usepackage{amssymb}
\usepackage{siunitx}
\PassOptionsToPackage{hyphens}{url}\usepackage{hyperref}
\usepackage{cleveref}
\usepackage[utf8]{inputenc}
\usepackage[right]{lineno}
\usepackage{csquotes}
\usepackage{booktabs}
\usepackage{longtable}
\usepackage{adjustbox}
\usepackage{array}
\usepackage{url}
\usepackage{titlesec}
\usepackage[numbers,sort&compress]{natbib}
\usepackage{enumitem}

\numberwithin{equation}{section}

\newcommand{\Tr}{\mathrm{Tr}}
\newcommand{\der}{\mathrm{d}}
\newcommand{\N}{N_{c}}
\newcommand{\Perp}[1]{\boldsymbol{#1}_{\perp}}
\newcommand{\Transv}[2]{\boldsymbol{#1}_{#2\perp}}
\newcommand{\WL}[1]{V(\Perp{#1})}
\newcommand{\WLadj}[1]{V^{\dagger}(\Perp{#1})}
\newcommand{\PolVect}[1]{\boldsymbol{\epsilon}^{#1}_{\perp}}
\newcommand{\gtoqqbar}[2]{\Gamma^{#1,#2}_{g\rightarrow q \bar{q}} }
\newcommand{\qtoqg}[2]{\Gamma^{#1,#2}_{q\rightarrow qg} }

\newcommand{ \dipvec}[1]{ \boldsymbol{r}_{#1}}
\newcommand{ \Dcal}{\mathcal{D}_{Y}}
\newcommand{ \Qcal}{\mathcal{Q}_{Y}}
\newcommand{ \Ncal}{\mathcal{N}}
\newcommand{\zbar}{z_{\bar{q}}}
\newcommand{\avg}[2]{\left\langle #1 \right\rangle_{#2}}
\newcommand{\et}{\boldsymbol{\epsilon}_\perp}
\newcommand{\Gt}{\boldsymbol{G}_\perp}

\usepackage{xcolor} 
\renewcommand{\thetable}{\Roman{table}}

\title{Forward trijet production in proton-nucleus collisions:\\ gluon-initiated channel}

\author[a]{Paul Caucal}
\author[b]{, Marcos Guerrero Morales}
\author[b,c,d]{, Farid Salazar}

\affiliation[a]{SUBATECH UMR 6457 (IMT Atlantique, Universite de Nantes, IN2P3/CNRS), 4 rue Alfred Kastler, 44307 Nantes, France}
\affiliation[b]{Department of Physics, Temple University, Philadelphia, Pennsylvania 19122, USA}
 \affiliation[c]{RIKEN-BNL Research Center, Brookhaven National Laboratory, Upton, New York 11973, USA}
 \affiliation[d]{Physics Department, Brookhaven National Laboratory, Upton, New York 11973, USA}

\emailAdd{caucal@subatech.in2p3.fr}
\emailAdd{marcos.guerrero.morales@temple.edu}
\emailAdd{farid.salazar@temple.edu}

\abstract{In this paper, we present analytical results for the forward three-parton production differential cross section in the gluon-initiated channel in proton-nucleus collisions. This result is the leading-order contribution to forward trijet production and provides the real-emission building block required for the NLO dijet/dihadron cross section. The calculations are carried out within the Color Glass Condensate (CGC) effective theory, and in the dilute-dense approximation, using effective vertices for the quark and gluon propagators interacting with the small-$x$ background gluon field. We employ the covariant perturbation theory approach and disentangle the amplitudes into regular and instantaneous contributions. Our results are expressed as convolutions of multiparton color correlators of light-like Wilson lines and perturbative impact factors, organized in compact expressions in terms of the ``bare" topologies of the contributing diagrams. The gluon-initiated channel receives contributions from a $q\bar{q}g$ and a $ggg$ final state. Interestingly, when considering the $ggg$ final state, we observe, for the first time, that the four-gluon vertex topology follows a structure similar to the instantaneous contributions. This observation suggests a simplification in the organization of multigluon CGC amplitudes and may prove useful for future one-loop calculations. Furthermore, when integrating (one of) the real gluon(s) in the final state, we identify that the rapidity divergence is absorbed into the real part of the JIMWLK evolution of the leading-order Wilson-line correlator. In addition, we isolate the divergences arising when the unobserved parton becomes collinear to one of the observed hadrons in the final state. These divergences are absorbed by renormalizing the initial-state parton distribution functions and final-state fragmentation functions, which are shown to obey DGLAP evolution. These results provide nontrivial one-loop consistency checks of the dilute-dense hybrid formalism, and are key ingredients toward the complete next-to-leading order calculation of dijet/dihadron production in proton--nucleus collisions. Our results can also be used to compute the ``double-real" contribution to the two-loop calculation for gluon production in proton--nucleus collisions.}

\begin{document}
\maketitle

\flushbottom

\section{Introduction}
\label{sec:introduction}

HERA measurements of deep inelastic electron--proton scattering \cite{H1:2009pze, H1:2015ubc} have established that, at small Bjorken-$x$, the gluon distribution inside hadrons grows rapidly. Due to non-linear QCD effects such as gluon recombination and screening, this growth is expected to be tamed, eventually leading to a saturated state of nuclear matter \cite{Gribov:1983iv, Mueller:1985wy}. The characterization of this state has been one of the central challenges in high-energy QCD over the last decades.

The Color Glass Condensate (CGC) is an effective field theory (EFT) that describes the dynamics of partons in this high-density regime \cite{McLerran:1994vd, McLerran:1994Gf, McLerran:1994ka, Iancu:2003ad, Morreale:2021}. The dense gluonic environment inside hadrons gives rise to an emergent momentum scale that characterizes the typical transverse momentum of small-$x$ gluons. This scale, known as the saturation scale $Q_s(A,x)$, depends parametrically on the nuclear size $A$ and on the longitudinal momentum fraction $x$ of the probed partons. Within the CGC EFT, high-energy (eikonal) multiple scattering off a dense target is described by light-like Wilson lines, which rotate the color of the partons. Physical observables, such as cross sections, are expressed as convolutions of correlators of these Wilson lines --- encoding the effects of multiple eikonal scattering with the classical gluon background field --- and process-dependent impact factors, which can be calculated perturbatively to the desired accuracy.

The energy dependence of the $n$-point Wilson-line correlators is governed by a set of renormalization-group equations known as the BK--JIMWLK (Balitsky, Kovchegov, Jalilian-Marian, Iancu, McLerran, Weigert, Leonidov, Kovner) equations \cite{Balitsky:1995ub, Kovchegov:1999yj, Iancu:2001ad, JalilianMarian:1997jx, Jalilian-Marian:1997dw, Kovner:2000pt, Iancu:2001md, Ferreiro:2001qy}. At leading-logarithmic (LL) accuracy, these equations resum terms of the form $\alpha_s^n \ln^n(1/x)$. In order to obtain reliable predictions for comparison with experimental data, significant efforts have been devoted in recent years to extending the CGC EFT to next-to-leading order (NLO) accuracy. One such effort has been the derivation of the BK--JIMWLK evolution equations at NLO \cite{Balitsky:2013fea, Kovner:2013ona, Kovner:2014lca, Lublinsky:2016meo, Lappi:2020srm, Dai:2022wmb} (see also \cite{Brunello:2025rhh} for progress toward NNLO). Another essential ingredient required to achieve full NLO accuracy is the computation of impact factors, which have by now been calculated at NLO for a variety of processes \cite{Balitsky:2012bs, Beuf:2011xd, Beuf:2016wdz, Beuf:2017bpd, Hanninen:2017ddy, Ducloue:2017ftk, Boussarie:2016bkq, Boussarie:2016ogo, Boussarie:2019ero, Mantysaari:2021ryb, Chirilli:2011km, Chirilli:2012jd, Altinoluk:2014eka, Stasto:2013cha, Roy:2019hwr, Roy:2019ieb, Beuf:2021qqa, Caucal:2021lgf, Fucilla:2022wfz, Beuf:2024dqe}.

Among the observables sensitive to the physics of saturation, dijet and dihadron production in proton--nucleus (pA) collisions are particularly promising. Multiple scattering off a dense target is expected to decorrelate the back-to-back peak of azimuthal correlations. This suppression is predicted to increase for larger nuclear targets, smaller values of $x$, and more forward rapidities \cite{Kharzeev:2004bw, Albacete:2010pg,Lappi:2012nh, Albacete:2018ruq, vanHameren:2016ftu, vanHameren:2019ysa, Al-Mashad:2022zbq, Kutak:2020ika,Stasto:2011ru,Zheng:2014vka,Stasto:2018rci, vanHameren:2020rqt, Benic:2022ixp,Caucal:2023fsf, Caucal:2025ncp}. So far this observable has been calculated to leading order (LO) accuracy, with only partial progress toward NLO precision.

In the hybrid factorization approach, a collinear parton from the proton interacts with the background field of the nucleus, subsequently producing the two jets in the final state. At LO in $\alpha_s$, the dijet amplitude receives contributions from both quark- and gluon-initiated channels. The contribution from the quark channel is described by the radiation of a gluon from the quark, while the contribution from the gluon-initiated channel arises either from the splitting of the incoming gluon into a quark--antiquark pair or from the splitting of the gluon into a pair of gluons. These contributions have been calculated in the literature \cite{Baier:2005qf, Marquet:2007vb, Dominguez:2011wm, Iancu:2013dta}. Moreover, in Ref.~\cite{Dominguez:2011wm} it was shown that, in the back-to-back limit relevant for away-side azimuthal correlations, the leading order dijet cross sections admit a TMD-like factorization. This result significantly simplifies CGC calculations and provides a more tractable framework for confronting theoretical predictions with experimental data \cite{Petreska:2018cbf}.

Additional mechanisms also contribute to the decorrelation of the back-to-back peak and must be carefully disentangled from genuine saturation effects. In particular, Sudakov logarithms arising from initial- and final-state radiation have been extensively studied and shown to induce significant broadening of the away-side peak \cite{Stasto:2011ru,Zheng:2014vka,Stasto:2018rci, vanHameren:2020rqt, Benic:2022ixp,Caucal:2023fsf, Caucal:2025ncp}. The emergence of these contributions in the context of small-$x$ was first studied in Refs.~\cite{Mueller:2012uf,Mueller:2013wwa}.
Subsequently, Refs.~\cite{Xiao:2017yya, Zhou:2018lfq} proposed the transverse-momentum–dependent (TMD) evolution of gluon and quark distributions by incorporating both Collins–Soper–Sterman (CSS) \cite{Collins:1981uk,Collins:1981uw,Collins:1984kg,Collins:2011zzd} and small-$x$ evolution. Sudakov effects have also been studied in the context of azimuthal anisotropies in jet and dijet production \cite{Hatta:2020bgy, Hatta:2021jcd}, as well as in deep inelastic scattering at small-$x$ \cite{Taels:2022tza, Caucal:2022ulg, Altinoluk:2024vgg, Caucal:2025mth} (see also \cite{Marquet:2025jdr} in the context of heavy flavor). Despite their importance as part of the NLO corrections to azimuthal correlations, these contributions alone do not provide a complete description of the observable. A consistent separation between saturation effects, perturbative radiation, and hard scattering dynamics requires the inclusion of the full next-to-leading order contributions to the cross section.

So far in the program toward a complete NLO calculation of dijet production in pA collisions\footnote{The only two-particle correlation process in proton-nucleus collision at NLO accuracy is Drell-Yan \cite{Taels:2023czt}. }, only the real corrections in the quark-initiated channel --- corresponding to three-parton final states with one unmeasured parton --- have been addressed~\cite{Iancu:2018hwa, Iancu:2020mos} (see also \cite{Altinoluk:2018byz,Altinoluk:2020qet,Kang:2023doo} for three-particle production processes involving photons in the initial or final state). These computations, however, were performed in the full kinematic regime and did not focus on the back-to-back (correlation) limit. Recently, a study of some NLO contributions to dijets and $\gamma $-jet in eA and pA collisions in the back-to-back configuration was carried out \cite{Caucal:2025sea}. In this study, we analyzed the contribution of dijet production coming from an integrated quark (or antiquark) and demonstrated factorization in terms of sea-quark TMD distributions which are sensitive to saturation. In order to obtain the full NLO dijet (and dihadron) cross section, two ingredients remain missing: the real corrections in the gluon-initiated channel and the virtual corrections in both channels. The present work addresses the calculation of real contributions to gluon-initiated channels.

The computation of trijets in the gluon-initiated channel in pA collisions has been studied under an extended Improved-TMD (ITMD) factorization framework \cite{Bury:2020ndc}. Within this framework, the cross section is expressed in terms of gauge-invariant hard factors with off-shell small-$x$ gluons \cite{vanHameren:2012if}, and gluon TMD distributions. This formalism encapsulates both the TMD limit, where $Q_s^2\ll \Transv{K}{}^2 \ll\Transv{P}{}^2$ (with $\Transv{P}{}$ being the hard momentum of the jet and $\Transv{K}{}$ the momentum imbalance of the trijet), and the High Energy Factorization (HEF) limit, where $Q_s^2\ll \Transv{K}{}^2 \sim \Transv{P}{}^2$. To compute the NLO correction to dijet production, one has to integrate over the phase space of one of the final-state partons. A sizable contribution to this integral comes from regions outside the domain of validity of the ITMD formalism.

The main purpose of this paper is to compute analytical expressions for forward trijet production in the gluon-initiated channel without imposing any constraints on the transverse momenta of the particles in the final state, so long as the jets are forward to justify small-$x$ methods. These results constitute a crucial step toward the completion of the real NLO corrections to dijet and dihadron production in pA collisions. Our computation combines techniques developed in previous NLO studies: in particular, we employ the covariant perturbation theory approach used in Refs.~\cite{Roy:2019ieb, Caucal:2021lgf}, together with spinor-helicity methods developed in Ref.~\cite{Ayala:2017rmh}. This combination provides an efficient framework for computing trijet cross sections in the CGC. From a broader perspective, this approach may facilitate the development of automated tools for performing CGC amplitude calculations at one-loop order and beyond.

The computation of trijet production in the gluon-initiated channel involves two distinct final states: $q\bar q g$ and $g g g$. The $q\bar q g$ channel receives contributions from three ``bare'' topologies, corresponding to gluon emission from the quark, the antiquark, or the parent gluon. The first two topologies lead to impact factors closely related to those obtained for $q\bar q g$ production in DIS \cite{Ayala:2017rmh, Caucal:2021lgf}, while the third topology represents a genuinely new feature of this process. The $g g g$ final state receives contributions from both a double-splitting diagram and a four-gluon-vertex diagram, the latter of which has not, to our knowledge, been previously considered in CGC calculations. 

Our paper is organized as follows. In section~\ref{sec: Dijet prod LO}, we introduce the basic ingredients required for the computation of CGC amplitudes and review, as a warm-up, the calculation of dijet production in pA at leading order. In section~\ref{sec: Amplitudes trijet}, we begin reviewing the general strategy to tackle the computation of amplitudes for trijet production, identifying regular and instantaneous contributions, as well as anticipating unitarity constraints of the perturbative factors for a given topology.  We then present the analytical results for the amplitudes for both channels $g\to q\bar{q}g$ and $g \to ggg$, organizing the results for each case according to the topologies of the Feynman diagrams. The summary of these results can be found in Section \,\ref{sec:summary-amplitudes}. In section\,\ref{sec:Slow gluon lim}, we study the logarithmic rapidity divergence arising by the integration of (one of) the gluon(s) in the final state, and verify the consistency of our results with the expected real part of the JIMWLK evolution of the LO cross section, providing a non-trivial cross-check of our calculation. Another important region is the DGLAP regime, associated with initial- and final-state collinear divergences when the unobserved parton is integrated out. These DGLAP limits are studied in section~\ref{sec:DGLAP}; they constitute an essential step in establishing the hybrid ``collinear+CGC" framework for inclusive dijet/dihadron production in pA collisions, and also provide a complementary validation of our analytical results. Section~\ref{sec: Cross section trijets} collects the complete results for the differential cross sections. Finally, section~\ref{sec: Conclusions} is devoted to our conclusions and perspectives for future work.

In addition, we supplement this manuscript with six appendices which serve as a guide for the reader who is interested in following our calculations in more detail. Appendix \ref{app:Conventions and general identities} summarizes our conventions for our notation, Feynman rules, and some useful identities with color matrices. Appendix \ref{app: Spinor contractions} is devoted to our conventions for Dirac spinors, along with some identities useful to simplify the Dirac algebra in the perturbative factors, followed by some identities used to simplify the tensor structures involved in our calculations of gluon propagators. Appendix \ref{app: Useful integrals} includes the derivation of the ``plus" integrals via complex contours, as well as the transverse integrals (Fourier transforms) needed to simplify the integration over the loop momenta in the perturbative factors. Appendix \ref{app: Detailed of calculation of diagram R7} contains the detailed calculation of diagram R7, corresponding to the $g\to ggg$ subprocess with a double splitting topology and where all final state partons interact with the shockwave. In Appendix \ref{app: Kernels_and_Smatrices} we list our definitions for the kernels and $S$ matrices featured in our results for the trijet cross section presented in section \ref{sec: Cross section trijets}. Finally, in Appendix \ref{app: JIMWLK_cross_section} we present the explicit expressions for the high-energy evolution of the LO cross section for both the $gg$ and the $q\bar{q}$ channels, which are intended to complement the results presented in section \ref{sec:Slow gluon lim}.

\section{Dijet production at LO}
\label{sec: Dijet prod LO}

Within the CGC formalism, the interactions of color-charged partons with the dense target are described by effective vertices that rotate the color of the partons while leaving the rest of their quantum numbers unchanged at leading eikonal accuracy.  For a quark propagator, the expression of the effective vertex in the $A^{-}=0$ gauge is \cite{McLerran:1994Gf, Ayala:1995kg}
\begin{align} \label{quark eff vertex}
    T^q_{ij}(l,l')= 2\pi \delta(l^- -l'^-) \gamma^- \mathrm{sgn}(l^-) \int \der^2 \Transv{x}{} e^{-i(\Transv{l}{}-\Transv{l}{}')\cdot \Transv{x}{}} V^{\mathrm{sgn}(l-)}_{ij}(\Transv{x}{})\,.
\end{align}
Here, $l$($l$') and $i$($j$) denote the momentum and color of the quark after (before) the shockwave, respectively. The object $V_{ij}$ is a light-like Wilson line in the fundamental representation of SU(3) which resums the multiple eikonal interactions of the parton with the target. It is defined as \cite{McLerran:1994vd,Balitsky:1995ub}
\begin{align}
    V_{ij}(\Transv{x}{}) = \mathcal{P}\, \left[\mathrm{exp}\left( -ig\int \der z^{-} A^{+,c}(z^-, \Transv{x}{}) t^{c}_{ij} \right) \right]\,,
\end{align}
with $t^{a}_{ij}$ being $SU(3)$ generators in the fundamental representation. The notation $\mathcal{P}$ denotes path ordering in $z^-$. The field $A^{+,c}(z^-, \Transv{x}{})$ corresponds to the ``plus" component of the classical gluon background gauge field. In light-cone gauge $ A^-=0$, the field $A^\mu= \delta^{\mu+} A^+$ is the solution to the classical Yang-Mills equations in the presence of large current $J^{\mu} = \delta^{\mu+} \rho$, provided it satisfies the Poisson equation $\nabla_\perp^2 \,A^{+,a}(z^-, \Transv{x}{}) = - \rho^a(z^-, \Transv{x}{})$, where $\rho^a(z^-, \Transv{x}{})$ represents the density of the classical sources of color charge, generated by the large-$x$ partons. The transverse components of the gauge field are suppressed in the high-energy limit. 

The gluon propagator in $A^-=0$ has a similar expression \cite{Gelis:2005pt}:
\begin{align} \label{gluon eff vertex}
    T^{g,\mu \nu}_{ab}(l,l')= -2\pi \delta(l^- -l'^-) (2l^-) g^{\mu \nu} \mathrm{sgn}(l^-) \int \der^2\Transv{x}{} e^{-i(\Transv{l}{}-\Transv{l}{}')\cdot \Transv{x}{}} U^{\mathrm{sgn}(l-)}_{ab}(\Transv{x}{}) \,.
\end{align}
Compared to the quark effective vertex, the gluon vertex features a Wilson line in the adjoint representation, $U_{ab}(\Transv{x}{})$, defined as
\begin{align}
    U_{ab}(\Transv{x}{}) = \mathcal{P}\, \left[\mathrm{exp}\left( -ig\int \der z^{-} A^{+,c}(z^-, \Transv{x}{}) T^{c}_{ab} \right) \right] \,,
\end{align}
where $T^{c}_{ab}$ are the generators in the adjoint representation of $SU(3)$. The effective vertices in Eq.\,\eqref{quark eff vertex} and Eq.\,\eqref{gluon eff vertex} are the building blocks to calculate amplitudes within the CGC. 

Within the dilute-dense (commonly referred to as hybrid) approach, the projectile is treated as being dilute, and can therefore be described in terms of Parton Distribution Functions (PDFs). The target, on the other hand, is treated as dense and is described within the CGC framework as a (semi-) classical gluon background field \cite{Dumitru:2005gt}. Thus, in this picture and at leading order, in order to obtain a dijet cross section in  pA collisions, it is sufficient to compute the partonic cross section of the different contributing channels, and convolve them with the appropriate PDF. The differential cross section for dijets in pA is of the form
\begin{align}\label{dijet_cross_section_LO}
    \frac{\der\sigma^{pA\rightarrow \textrm{dijet} +X}}{\der^2\boldsymbol{k}_{1\perp}\der\eta_1 \der^2\boldsymbol{k}_{2\perp} \der\eta_{2}} = \sum_{a} \int \der x_p\,  f_{a}(x_p,\mu^2) \sum_{b,c} \frac{\der\sigma^{aA\rightarrow bc +X}} {\der^2\boldsymbol{k}_{1\perp}\der\eta_1 \der^2\boldsymbol{k}_{2\perp} \der\eta_{2} } \Bigg|_{p^-=x_p q^-} \,,
\end{align}
where $f_{a}(x, \mu^2)$ is the PDF of the parton $a$, which could be a quark or a gluon, evaluated at the factorization scale $\mu^2$. The letters $b,c$ denote the possible final state partons that contribute to the partonic cross section for a given channel. The variable $x_p= p^-/q^-$ corresponds to the longitudinal momentum of the incoming parton relative to the proton, where $q$ is the proton's momentum. Beyond leading order, there is no longer a one-to-one correspondence between jets and partons. The real NLO correction to the dijet cross section arises from integrating the three-parton production cross section over all channels, subject to appropriate jet-measure constraints on the phase space. Alternatively, one may consider the dihadron production cross section, as we do in section~\ref{sec:DGLAP}; in that case, Eq.\,\eqref{dijet_cross_section_LO} must be extended to include a convolution with two collinear fragmentation functions into hadrons.

The amplitude of a given CGC diagram can then be calculated using standard pQCD Feynman rules together with the expressions given for the effective vertices. In a CGC diagram, we describe the shockwave with a red rectangle. The interaction of a quark with the shockwave is described by a crossed dot, whereas the interaction of a gluon with the shockwave is described by a filled circle. It is sufficient to consider diagrams in which the shockwave interacts instantaneously and simultaneously with all partons at a given time \footnote{Observe that in the effective vertices in Eqs.\,\eqref{quark eff vertex}-\eqref{gluon eff vertex}, the Wilson lines include the non-interacting contribution; hence, the shockwave diagrams include all combinations of scattering or no scattering of the partons crossing the shockwave. For an explicit mathematical exposition, we refer the reader to Sec. II.C. in \cite{Roy:2018jxq}}. Furthermore, due to the instantaneous nature of the shockwave, QCD splittings cannot occur within the shockwave (see Appendix\,C in \cite{Roy:2018jxq}).

Before introducing the calculations of interest in this paper and for the sake of completeness of the material, let us start with the simpler example of the LO cross section of dijet production in pA collisions. These results have been computed several times under different approaches (see for example \cite{Baier:2005qf, Marquet:2007vb, Dominguez:2011wm, Iancu:2013dta, vanHameren:2016ftu}). As mentioned in the introduction, at leading order, there are two channels contributing to the dijet production in the gluon-initiated channel. The first one corresponds to a gluon splitting into a quark and an antiquark pair $g \to q\bar{q}$. The second one corresponds to a gluon splitting into a pair of gluons $g \to gg$. In the following calculations, we will work in the frame where the proton is left moving, i.e. it propagates in the ``minus" direction. The gluon in the hybrid approach is treated as collinear to the proton. Hence it is left moving as well with large minus component $p^- > 0$ and with no transverse momentum $\Transv{p}{} = 0$, as appropriate for the collinear treatment of the projectile\footnote{One can relax this approximation and work with a dilute unintegrated gluon distribution, see e.g. \cite{Gelis:2003vh,Blaizot:2004wu,Blaizot:2004wv}}. In addition, we assume that quarks and antiquarks are massless. The fraction of longitudinal momentum relative to the incoming parton of a parton with momentum $k_{i}$ is denoted $z_{i} = k_{i}^-/p^-$. The rapidity is then given by
\begin{align}
    \eta_{i} = \ln\left(\sqrt{2} z_i p^{-}/k_{i\perp}   \right)\,,
\end{align}
where positive rapidity is in the proton-going direction.

\subsection{The $g \to q\bar{q}$ channel}

\begin{figure}
    \centering
    \includegraphics[width=0.8\linewidth]{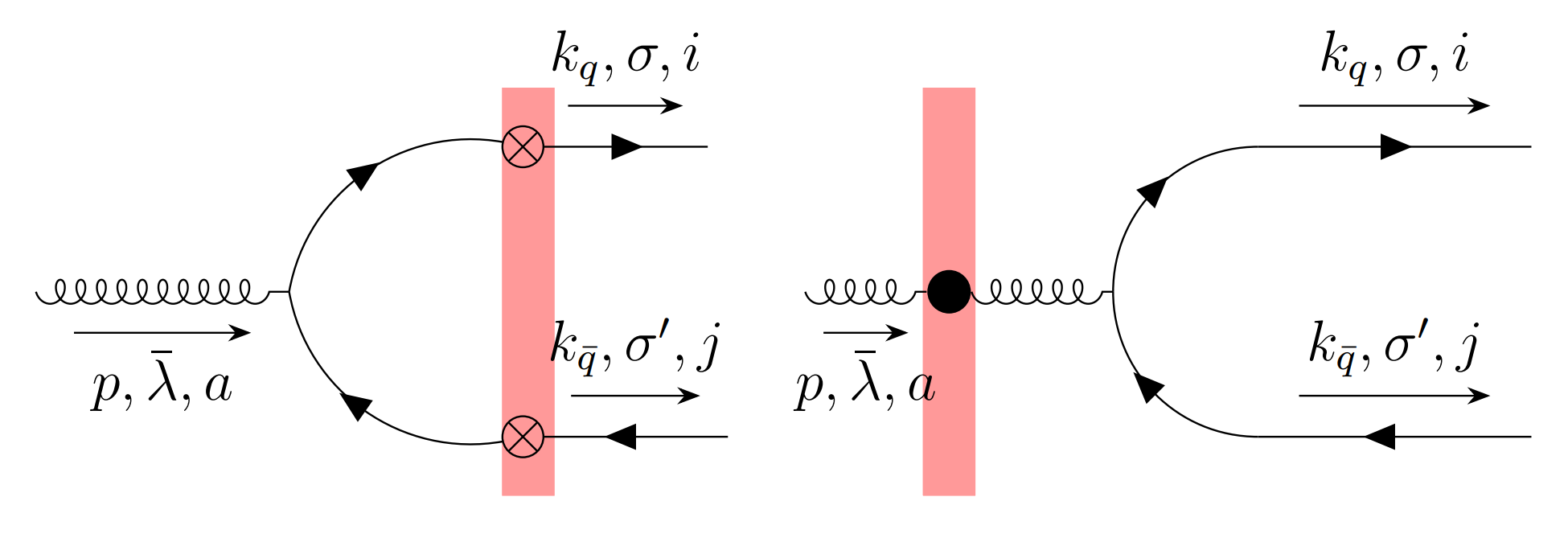}
    \caption{Diagrams contributing to the $g\rightarrow q\bar{q}$ channel at LO for dijet production in pA collisions. The red rectangle represents the interaction with the ``shockwave". The crossed dots represent quark effective vertices, while the black dot represents a gluon effective vertex.}
    \label{fig:LOg-qbarqDiagrams}
\end{figure}

In the $g \to q\bar{q}$ channel there are two diagrams contributing to leading  order. These diagrams share the same topology, and differ by whether the shockwave interaction occurs before or after the splitting. In Figure \ref{fig:LOg-qbarqDiagrams}, we depict the two Feynman diagrams contributing to the total amplitude of the process. The first diagram corresponds to the interaction of the $q\bar{q}$ pair with the shockwave, while the second diagram represents the interaction of the shockwave with the incoming gluon before fluctuating into the final $q\bar{q}$ state. The difference between these two amplitudes lies in the color operator of the Wilson lines. The first diagram features a product of a Wilson line and a conjugate Wilson line, both in their fundamental representation, which describe the scattering of a $q\bar{q}$ dipole, the quark being at $\Transv{x}{}$ and the antiquark at $\Transv{y}{}$ at the time of the scattering. The second diagram features a single Wilson line in the adjoint representation. This accounts for the scattering of the gluon with the shockwave evaluated at transverse coordinate $\Transv{w}{} = (z_1 \Transv{x}{} + z_2 \Transv{y}{})/(z_1 + z_2)=z_1\Perp{x}+(1-z_1)\Perp{y}$. 

The total amplitude of the process can be expressed as a convolution of the difference of these color operators with the perturbative factor of the $g \to q\bar{q}$ production as

\begin{align}
    \mathcal{M}^{\bar{\lambda}\sigma\sigma'}_{q\bar{q},LO,ij} =  \int \der^2\Transv{x}{} \der^2\Transv{y}{} e^{-i\Transv{k}{q}\cdot \Transv{x}{}}e^{-i\Transv{k}{\bar{q}}\cdot \Transv{y}{}}C^{a}_{q\bar{q},LO,ij}\,  \Ncal^{\bar{\lambda} \sigma \sigma'}_{q\bar{q},LO}(z_q,z_{\bar q},\boldsymbol{r}_{xy})\,,
\end{align}
with perturbative factor 
\begin{align}
    \Ncal^{\bar{\lambda} \sigma \sigma'}_{q\bar{q},LO}(z_q,z_{\bar q},\boldsymbol{r}_{xy}) = -\frac{2igp^-}{\pi} \sqrt{z_qz_{\bar{q}}}\, \Gamma^{\sigma \bar{\lambda}}_{g\xrightarrow{}q\bar{q}}(z_q) \delta^{\sigma,-\sigma'} \frac{\et^{\bar{\lambda}}\cdot \dipvec{xy} }{r_{xy}^2}\,,
\end{align}
and color operator
\begin{equation}
    C^{a}_{q\bar{q},LO,ij}(\Transv{x}{}, \Transv{y}{}) = [ V(\Transv{x}{})t^a V^\dagger(\Transv{y}{}) - t^b U_{ba}(\boldsymbol{w}_{\perp})]_{ij}\,.
\end{equation}
In the perturbative factor we have introduced the notation $\dipvec{xy} = \Transv{x}{} - \Transv{y}{}$ for the $q\bar{q}$ dipole vector.  We will be using this notation throughout this paper. The variables $k_q$, $k_{\bar{q}}$ are the momenta of the quark and the antiquark, with polarizations $\sigma$ and $\sigma'$, and colors $i$ and $j$, respectively. The polarization and color of the gluon are denoted by $\lambda$ and $a$, respectively. In the perturbative factor, we have also introduced the $g\to q\bar{q}$ ``square root" of the splitting function defined as
\begin{align}\label{g_to_qqbar_splitting_function}
    \gtoqqbar{\sigma}{\lambda} (\xi) \equiv (1-\xi)\delta^{\sigma,\lambda} - \xi\delta^{\sigma,-\lambda}\,.
\end{align}
This function when squared, and summed over the polarization and helicities of the partons involves the $g\to q\bar{q}$ QCD splitting function. Indeed, we have
\begin{equation}
    \frac{1}{4}\sum_{\lambda,\sigma}\,\left[\gtoqqbar{\sigma}{\lambda} (\xi)\right]^2 = P^{real}_{qg}(\xi)\,,
\end{equation}
where
\begin{equation}
    P^{real}_{qg}(\xi) \equiv \frac{\xi^2 + (1-\xi)^2}{2}\,.
\end{equation}
The differential cross section is then obtained by squaring the amplitude and summing (averaging) over final (initial) state quantum numbers,
\begin{multline}\label{crosssection LO general}
    \frac{\der\sigma^{gA\rightarrow q\bar{q} X}}{\der^2\Transv{k}{q}\der^2\Transv{k}{\bar q}\der\eta_{q}\der\eta_{\bar{q}}}\\
    = \frac{1}{4(2\pi)^6}\frac{1}{2(p^-)^2}(2\pi)\delta(1-z_q-z_{\bar{q}})\frac{1}{2(\N^2-1)} \sum_{\sigma\sigma', \rm colors} \left\langle  \mathcal{M}^{\bar{\lambda}\sigma\sigma'\dag}[\rho_A]\mathcal{M}^{\bar{\lambda}\sigma\sigma'}[\rho_A]\right\rangle_Y\,.
\end{multline}
The notation $\langle\dots\rangle_{Y}$ denotes the average over the CGC charge configurations for a fixed rapidity $Y=\ln(1/x)$. The factor of $1/(2(\N^2 -1))$ comes from the average over initial polarization and color states of the gluon. When squaring the amplitudes, we obtain four terms coming from the four diagrams contributing at this level. The resulting cross section then reads \cite{Dominguez:2011wm, Iancu:2018hwa}
\begin{multline}\label{LO cross section qqbar}
    \frac{\der\sigma^{gA\rightarrow q\bar{q} X}}{\der^2\Transv{k}{q}\der^2\Transv{k}{\bar{q}}\der\eta_{q}\der\eta_{\bar{q}}} = \frac{\alpha_s}{(2\pi)^6} \delta(1-z_q-z_{\bar{q}}) \int \der^2\boldsymbol{x}_{\perp} \der^2\boldsymbol{y}_{\perp}\der^2\boldsymbol{x'}_{\perp}\der^2\boldsymbol{y'}_{\perp} e^{-i\boldsymbol{k}_{q_{\perp}}\cdot (\boldsymbol{x}_{\perp}-\boldsymbol{x}'_{\perp})} e^{-i\boldsymbol{k}_{\bar{q}_{\perp}}\cdot (\boldsymbol{y}_{\perp}-\boldsymbol{y}'_{\perp})}\\
    \times \mathcal{K}_{q\bar{q},LO}(z_q,z_{\bar q},\boldsymbol{r}_{xy}, \boldsymbol{r}_{x'y'}) \left\langle \Xi_{q\bar q,LO}(\Perp{x},\Perp{y},\Perp{x'},\Perp{y'}) \right\rangle_Y\,,
\end{multline}
with the LO kernel
\begin{equation}
     \mathcal{K}_{q\bar{q},LO}(z_q,z_{\bar q},\boldsymbol{r}_{xy}, \boldsymbol{r}_{x'y'}) =  z_qz_{\bar{q}} (z_q^2 + z_{\bar{q}}^2) \frac{\boldsymbol{r}_{xy}\cdot \boldsymbol{r}_{x'y'}}{r^2_{xy}r^2_{x'y'}}\,.
\end{equation}
For later convenience, we also define the LO color structure $\Xi_{q\bar q,LO}$ for the $g\to  q\bar q$ channel as
\begin{align}
    \left\langle \Xi_{q\bar q,LO}(\Perp{x},\Perp{y},\Perp{x'},\Perp{y'}) \right\rangle_Y = \ &S^{q\bar{q}}_{q\bar{q}}(\boldsymbol{x}'_{\perp}, \boldsymbol{y}'_{\perp}, \boldsymbol{x}_{\perp}, \boldsymbol{y}_{\perp}) - S^{q\bar{q}}_{g}(\boldsymbol{x}'_{\perp}, \boldsymbol{y}'_{\perp},\boldsymbol{w}_{\perp}) \nonumber\\
    &- S^{q\bar{q}}_{g}(\boldsymbol{x}_{\perp}, \boldsymbol{y}_{\perp},\boldsymbol{w}'_{\perp}) + S^{g}_{g}(\boldsymbol{w}'_{\perp},\boldsymbol{w}_{\perp})\,.
\end{align}
Although not manifest in our notation, the color structure implicitly depends on the longitudinal momentum fractions $z_q$ and $z_{\bar q}$ of the outgoing quark and antiquark through the transverse coordinates $\Perp{w}$ and $\Perp{w'}$.

The scattering of the different partons with the gluon background field is encoded in the S-matrices. Following a notation similar to that of \cite{Iancu:2018hwa}, our notation intends to represent the partons involved in the scattering, with the partons in the direct amplitude being in the lower index with unprimed coordinates, and the partons from the c.c. amplitude in the upper index, with primed coordinates. The simplest of the matrices appearing in the expression corresponds to the gluon dipole correlator
\begin{align}\label{S-ggmatrix qbarq scattering}
    \begin{split}
        S^{g}_{g}(\boldsymbol{w}'_{\perp},\boldsymbol{w}_{\perp}) &= \frac{1}{C_F\N} \left\langle\Tr(V(\Transv{w}{}')t^aV^{\dagger}(\Transv{w}{}')V(\Transv{w}{})t^aV^\dagger\Transv{w}{})) \right\rangle_{Y} \\
        &= \frac{1}{2C_F\N}\left\langle N_c^2 D_{ww'}D_{w'w} - 1 \right\rangle_{Y}\,,
    \end{split}
\end{align}
where we have used the Fierz identities to arrive to the second equality. The S-matrix corresponding to quark-antiquark-gluon triplet is defined as
\begin{align}\label{S-qbarqgmatrix qbarq scattering}
    \begin{split}
        S^{q\bar{q}}_{g}(\boldsymbol{x}'_{\perp}, \boldsymbol{y}'_{\perp},\boldsymbol{w}_{\perp}) &= \frac{1}{C_F\N} \left\langle \Tr(V(\Transv{y}{}')t^aV^{\dagger}(\Transv{x}{}')V(\Transv{w}{})t^aV^\dagger(\Transv{w}{})) \right\rangle_{Y} \\
        &= \frac{1}{2C_F\N} \left\langle N_c^2 D_{y'w}D_{wx'} - D_{y'x'} \right \rangle_{Y}\,.
    \end{split}
\end{align} 
And finally, the one corresponding to two quarks and two antiquarks 
\begin{align}\label{S-qbarq qbarq matrix qbarq scattering}
    \begin{split}
        S^{q\bar{q}}_{q\bar{q}}(\boldsymbol{x}'_{\perp}, \boldsymbol{y}'_{\perp};\Transv{x}{}, \Transv{y}{}) &= \frac{1}{C_F\N} \left\langle \Tr(V(\Transv{y}{}')t^aV^{\dagger}(\Transv{x}{}')V(\Transv{x}{})t^aV^{\dagger}(\Transv{y}{}) )\right\rangle_{Y} \\
        &= \frac{1}{2C_F\N} \left\langle N_c^2 D_{xx'}D_{y'y} - Q_{xyy'x'}\right\rangle_{Y}\,.
    \end{split}
\end{align} 
In the second equalities of Eqs.\,\eqref{S-ggmatrix qbarq scattering}, \eqref{S-qbarqgmatrix qbarq scattering} and \eqref{S-qbarq qbarq matrix qbarq scattering} we have introduced the dipole and quadrupole operators. The dipole is defined as
\begin{align}\label{dipole_op}
    D_{x_1x_2} = \frac{1}{\N} \Tr(V(\boldsymbol{x}_{1\perp})V^{\dag}(\boldsymbol{x}_{2\perp}) )\,,
\end{align}
and the quadrupole as
\begin{align}\label{quad_op}
    Q_{x_1x_2x_3x_4} = \frac{1}{\N} \Tr(V(\boldsymbol{x}_{1\perp})V^{\dagger}(\boldsymbol{x}_{2\perp})V(\boldsymbol{x}_{3\perp})V^{\dagger}(\boldsymbol{x}_{4\perp}) )\,.
\end{align}
By taking the large $\N$ and the mean field approximation, we can write the S-matrices in terms of the dipole distribution
\begin{align}
    S^{q\bar{q}}_{q\bar{q}}(\boldsymbol{x}'_{\perp}, \boldsymbol{y}'_{\perp};\Transv{x}{}, \Transv{y}{}) &\approx \mathcal{D}_{Y}(\Transv{x}{}, \Transv{x}{}' ) \mathcal{D}_{Y}(\Transv{y}{}, \Transv{y}{}' )\\
    S^{q\bar{q}}_{g}(\boldsymbol{x}'_{\perp}, \boldsymbol{y}'_{\perp},\boldsymbol{w}_{\perp}) &\approx \mathcal{D}_{Y}(\Transv{w}{}, \Transv{x}{}' ) \mathcal{D}_{Y}(\Transv{w}{}, \Transv{y}{}' )\\
    S^{g}_{g}(\boldsymbol{w}'_{\perp},\boldsymbol{w}_{\perp}) & \approx \mathcal{D}_{Y}(\Transv{w}{}, \Transv{w}{}' ) \mathcal{D}_{Y}(\Transv{w}{}, \Transv{w}{}' )\,,
\end{align}
with $\mathcal{D}_{Y}(\Transv{x}{}, \Transv{x}{}' ) \equiv\left \langle D_{xx'} \right\rangle_{Y}$ being the dipole correlator defined in Eq.\,\eqref{dipole_op} averaged over the charge configurations.

\subsection{The $g\rightarrow gg$ channel}
\label{sub:gggLO}

Analogous to the $g\rightarrow q\bar{q}$ channel, there are two diagrams contributing to the leading order amplitude of this process, which can be combined into the following amplitude:
\begin{figure}
    \centering
    \includegraphics[width=0.8\linewidth]{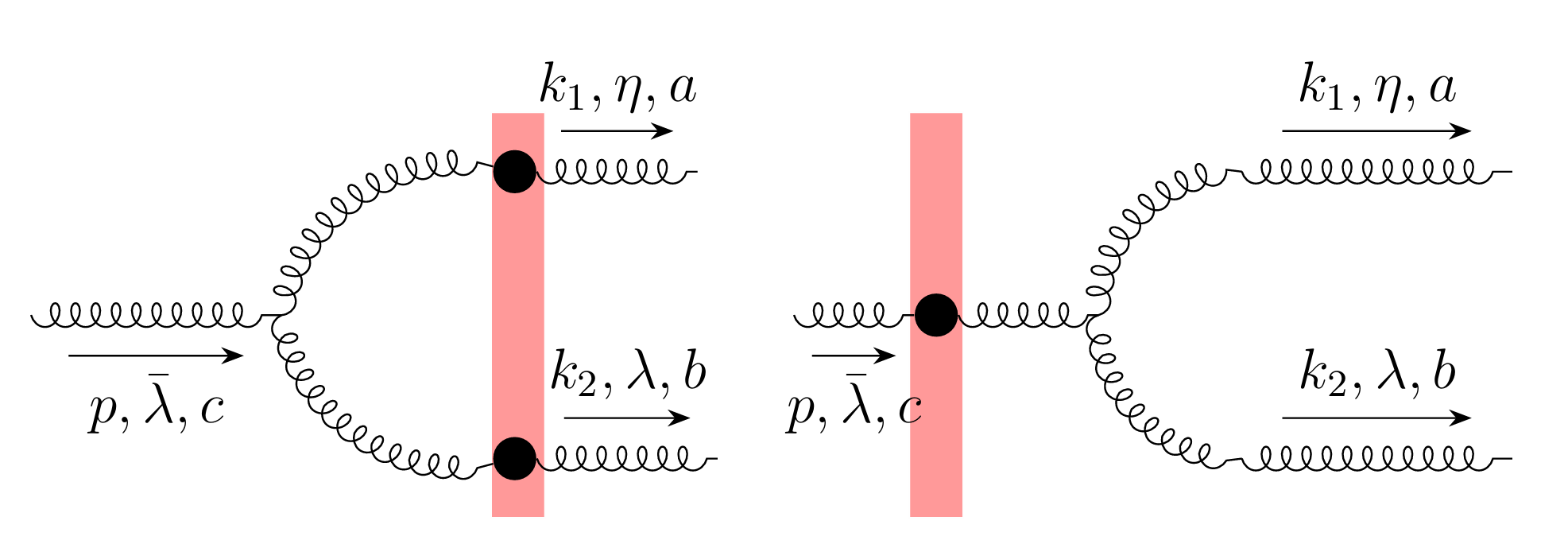}
    \caption{Diagrams contributing to the $g\rightarrow gg$ channel at LO for dijet production in pA collisions.}
    \label{fig:LOg-ggDiagrams}
\end{figure}
\begin{align}
    \mathcal{M}^{ \bar{\lambda} \lambda \eta,abc}_{gg,LO}=\int \der^2\boldsymbol{x}_{\perp} \der^2\boldsymbol{y}_{\perp} e^{-i \boldsymbol{k}_{1\perp}\cdot \boldsymbol{x}_{\perp}}e^{-i \boldsymbol{k}_{2\perp}\cdot \boldsymbol{y}_{\perp}} C^{abc}_{gg,LO}(\Transv{x}{}, \Transv{y}{})  \Ncal^{\bar{\lambda} \lambda \eta}_{gg,LO}(\boldsymbol{x}_{\perp}, \boldsymbol{y}_{\perp})\,,
\end{align}
with perturbative factor
\begin{align}
    \Ncal^{\bar{\lambda} \lambda \eta}_{gg,LO}(z_1,z_2,\boldsymbol{r}_{xy}) = -\frac{2igp^-}{\pi} \frac{\Gt^{\bar{\lambda} \lambda \eta} (z_1) \cdot \boldsymbol{r}_{xy}}{r^2_{xy}}\,.
\end{align}
and color operator
\begin{equation}
    C^{abc}_{gg,LO}(\Transv{x}{}, \Transv{y}{}) =\left[ if^{dce} U_{ad}(\boldsymbol{x}_{\perp})U_{be}(\boldsymbol{y}_{\perp}) - if^{bae} U_{ec}(\Transv{w}{}) \right] \,.
\end{equation}
Here, the momenta of the outgoing gluons are denoted by $k_{i},\, i=1,2$. Their polarizations and colors are denoted by $\eta$, $\lambda$ and $a, b$ respectively. The polarization and color of the incoming gluon are given by $\lambda$ and $c$. The first term in the color operator describes the scattering of the two final state gluons with the shockwave with two Wilson lines in the adjoint representation at the position of the scattering. The second term represents the scattering of the incoming gluon off the shockwave. The triple gluon vertex introduces a new square root of the splitting function defined as
\begin{align}\label{g_to_gg_splitting_function}
    \Gt^{\bar{\lambda}\lambda\eta} (\xi) = (1-\xi)\delta^{\lambda\bar{\lambda}} \et^{\eta*} +\xi \delta^{\bar{\lambda}\eta} \et^{\lambda*} - \xi(1-\xi) \delta^{\lambda,-\eta} \et^{\bar{\lambda}}\,.
\end{align}
As we shall see in section~\ref{sec:DGLAP}, when squared, this vertex features the $g\to gg$ DGLAP splitting function.

Using the formula \eqref{crosssection LO general}, we obtain the following expression for the cross section \cite{Dominguez:2011wm, Iancu:2013dta, Iancu:2018hwa}
\begin{multline}\label{LO gg cross section}
    \frac{\der\sigma^{gA\rightarrow gg X}}{\der^2\Transv{k}{1}\der^2\Transv{k}{2}\der\eta_{1}\der\eta_{2}} = \frac{\alpha_s \N}{(2\pi)^6} \delta(1-z_1-z_2) \int \der^2\boldsymbol{x}_{\perp} \der^2\boldsymbol{y}_{\perp}\der^2\boldsymbol{x'}_{\perp}\der^2\boldsymbol{y'}_{\perp} e^{-i\boldsymbol{k}_{1_{\perp}}\cdot (\boldsymbol{x}_{\perp}-\boldsymbol{x}'_{\perp})} e^{-i\boldsymbol{k}_{2_{\perp}}\cdot (\boldsymbol{y}_{\perp}-\boldsymbol{y}'_{\perp})}\\
    \times \mathcal{K}_{gg,LO}(z_1,z_2,\boldsymbol{r}_{xy}, \boldsymbol{r}_{x'y'})\left\langle \Xi_{gg,LO}(\Perp{x},\Perp{y},\Perp{x'},\Perp{y'}) \right\rangle_Y\,,
\end{multline}
where the kernel of the cross section reads
\begin{align}\label{LO gg kernel}
    \mathcal{K}_{gg,LO}(z_1,z_2,\boldsymbol{r}_{xy}, \boldsymbol{r}_{x'y'}) &=4(z_1^2+z_2^2+z_1z_2)^2\frac{\boldsymbol{r}_{xy} \cdot \boldsymbol{r}_{x'y'}}{r^2_{xy}r^2_{x'y'}}\,,
\end{align}
and the color structure of the $g\to gg$ channel is given by
\begin{align}
    \left\langle \Xi_{gg,LO}(\Perp{x},\Perp{y},\Perp{x'},\Perp{y'}) \right\rangle_Y = \ &S^{gg}_{gg}(\boldsymbol{x}'_{\perp}, \boldsymbol{y}'_{\perp}, \boldsymbol{x}_{\perp}, \boldsymbol{y}_{\perp}) - S^{gg}_{g}(\boldsymbol{x}'_{\perp}, \boldsymbol{y}'_{\perp},\boldsymbol{w}_{\perp}) \nonumber\\
    &- S^{gg}_{g}(\boldsymbol{x}_{\perp}, \boldsymbol{y}_{\perp},\boldsymbol{w}'_{\perp}) + S^{g}_{g}(\boldsymbol{w}'_{\perp},\boldsymbol{w}_{\perp})\,.
\end{align}
At LO, since $z_1$ and $z_2$ sum to one, the polynomial function of $z_1$ and $z_2$ in the parentheses of Eq.\,\eqref{LO gg kernel} can equivalently be written as $z_1^2+z_2^2+z_1z_2=1 - z_1 z_2$.

In this process, we have the appearance of two new S-matrices corresponding to the interaction of four and three gluons, respectively, defined as
\begin{align} \label{LO Sgggg and Sggg matrices}
    S^{gg}_{gg}(\boldsymbol{x}'_{\perp}, \boldsymbol{y}'_{\perp}, \boldsymbol{x}_{\perp}, \boldsymbol{y}_{\perp}) &=\frac{1}{2C_F \N} \left\langle N_c^2 D_{xx'}D_{yy'}Q_{yxx'y'} -  O_{x'y'yx'xyy'x} \right \rangle_{Y}\,, \\
    S^{gg}_{g}(\boldsymbol{x}'_{\perp}, \boldsymbol{y}'_{\perp},\boldsymbol{w}_{\perp}) &= \frac{1}{2C_F \N^2} \left\langle N_c^2 D_{wx'}D_{x'y'}D_{y'w} - S_{y'x'wy'x'w} \right \rangle_{Y}\,.
\end{align}
Notice that already in this process we have the emergence of a six-point (hexapole) and even eight-point (octupole) correlators
\begin{align}
    S_{x_1x_2x_3x_4x_5x_6}&=\frac{1}{N_c}\textrm{Tr}(V(\Perp{x_1})V^\dagger(\Perp{x_2})V(\Perp{x_3})V^\dagger(\Perp{x_4})V(\Perp{x_5})V^\dagger(\Perp{x_6}))\,,\\
    O_{x_1x_2x_3x_4x_5x_6x_7x_8}&=\frac{1}{N_c}\textrm{Tr}(V(\Perp{x_1})V^\dagger(\Perp{x_2})V(\Perp{x_3})V^\dagger(\Perp{x_4})V(\Perp{x_5})V^\dagger(\Perp{x_6})V(\Perp{x_7})V^\dagger(\Perp{x_8}))\,.
\end{align}
In the large $\N$ limit however, all S-matrices can be expressed in terms of the dipole and quadrupole correlators~\cite{Dominguez:2013}:
\begin{align}
    S^{gg}_{gg}(\boldsymbol{x}'_{\perp}, \boldsymbol{y}'_{\perp}, \boldsymbol{x}_{\perp}, \boldsymbol{y}_{\perp}) & \approx \mathcal{D}_{Y}(\Transv{x}{}, \Transv{x}{}' ) \mathcal{D}_{Y}(\Transv{y}{}, \Transv{y}{}' )\mathcal{Q}_{Y}(\Transv{y}{}, \Transv{x}{}, \Transv{x}{}', \Transv{y}{}') \,, \\
    S^{gg}_{g}(\boldsymbol{x}'_{\perp}, \boldsymbol{y}'_{\perp},\boldsymbol{w}_{\perp}) &\approx \mathcal{D}_{Y}(\Transv{x}{}, \Transv{x}{}' )\mathcal{D}_{Y}(\Transv{y}{}, \Transv{y}{}' ) \mathcal{D}_{Y}(\Transv{y}{}, \Transv{y}{}' )\,,
\end{align}
where $\mathcal{Q}_{Y}(\Transv{y}{}, \Transv{x}{}, \Transv{x}{}', \Transv{y}{}') \equiv \left\langle Q_{yxx'y'} \right \rangle_{Y}$. Using the Gaussian approximation one can obtain explicit expressions for the quadrupole (and higher $n$-point) correlators in terms of the dipole. These correlators and their $x$ dependence have been studied extensively in the literature \cite{Iancu:2011nj, Iancu:2011ns,Dumitru:2011vk}.

\section{Amplitudes for trijet production}\label{sec: Amplitudes trijet}
Let us now move on to the main goal of this paper, the trijet production in pA collisions in the gluon-initiated channel. Two final states contribute to the trijet production in this channel: the $q\bar{q}g$ and the $ggg$ final state. We will start this section by discussing the general strategy of the computation of these amplitudes, followed by the presentation of our main results for the amplitudes of these processes.

\subsection{General strategy for the computation}
The calculation of the amplitudes for both the $q\bar{q}g$ and the $ggg$ final states involve the computation of several diagrams. To be more concrete, the $q\bar{q}g$ final state has contributions from nine different diagrams at the level of the amplitude, which can be classified in terms of three different topologies. These topologies differ by the emitter of the final state gluon, this being the quark, the antiquark or the gluon. The $ggg$ final state, due to the symmetries of the process, can only have two topologies in principle: the first one being a gluon splitting into two gluons followed by a subsequent splitting of one of the daughter gluons into another pair of gluons, or a gluon splitting into a gluon triplet. However, for the double splitting topology, we must account for three permutations corresponding to the exchange of identical particles. As we will see, it is convenient to organize the amplitudes according to these topologies, as it allows us to express their amplitudes in a relatively compact way.

We follow a strategy very similar to that developed in \cite{Caucal:2021lgf}. The general structure of the amplitudes for each diagram is of the form
\begin{align}
    \int \der^2 \boldsymbol{x}_{1\perp}\der^2 \boldsymbol{x}_{2\perp}\dots C(\boldsymbol{x}_{1\perp}, \boldsymbol{x}_{2\perp},\dots) \int \frac{\der^4 l_1}{(2\pi)^4} \frac{\der^4 l_2}{(2\pi)^4} \dots N(l_1,l_2,\dots) \,, 
\end{align}
where $C(\boldsymbol{x}_{1\perp}, \boldsymbol{x}_{2\perp},\dots)$ is a color correlator of Wilson lines, and $N(l_1,l_2,\dots)$ is a perturbatively calculable function encapsulating QCD propagators and vertices. 

The latter can be simplified using spinor helicity methods \cite{Ayala:2017rmh}. The idea is to rewrite the internal propagators in terms of the sum of a regular part and an instantaneous part. For a fermionic propagator with momentum $k$, the numerator can be rewritten as
\begin{align}\label{fermion propagator decomposition}
    \slashed{k} = \slashed{\bar{k}} + \frac{k^2}{2k^-}\slashed{n}\,,
\end{align}
where $\bar{k}^\mu = (\frac{k_{\perp}^2}{2k^-}, k^-, \Transv{k}{})$ is on shell, and $n^\mu = (1,0,\boldsymbol{0}_{\perp})$ in light-cone basis. The regular term can then be rewritten in terms of Dirac spinors by making use of the completeness relation
\begin{align}
    \slashed{\Bar{k}} = \sum_{\sigma}u(\bar{k},\sigma)\bar{u}(\bar{k},\sigma)\,,
\end{align}
which allows us to rewrite the Dirac structure in terms of simple structures of splitting vertices. For a gluon propagator a similar decomposition is applied:
\begin{align}\label{gluon propagator decomposition}
    \Pi_{\alpha\beta}(l) = \sum_{\eta} \epsilon^{*}_{\alpha}(l,\lambda)\epsilon_{\beta}(l,\lambda) + \frac{l^2}{(l^-)^2}n_{\alpha}n_{\beta}\,.
\end{align}

In principle, each propagator contains a regular and an instantaneous piece. However, due to the instantaneous nature of the shockwave, the instantaneous pieces of the propagators next to the shockwave will vanish. Physically, this can be understood from the time ordering imposed by the shockwave interaction at $x^{+}=0$. Any fluctuation of the partonic wavefunction must either happen long before or after the scattering. Mathematically, this is justified as follows. For a fermion, the instantaneous term of the propagator and the fermionic effective vertex both contain a Dirac matrix $\gamma^-$, and since $(\gamma^-)^2 = 0$, the term vanishes. For a gluon propagator, the effective vertex is proportional to $g^{\mu\nu}$, so the instantaneous term of Eq.\,\eqref{gluon propagator decomposition} will either contain a factor of $n^2=0$ or $\epsilon(k,\lambda)\cdot n= 0$, in our choice of light-cone gauge $A^-=0$.

Once the Dirac structure is simplified, we can address the $l$ integrals. The $l^-$ integrals can be performed trivially by using a delta function that preserves longitudinal momentum, present in every calculation. The $l^+$ dependence is fully contained in the denominator (scalar part of propagator) and the integration can then be performed using complex contour and the residue theorem. Finally, we are left with the transverse integrations, which can be performed using some two-dimensional Fourier transform identities. 

In principle, the number of transverse coordinate integrals on a given diagram depends on the number of partons participating in the scattering. However, using a strategy similar to that of Section 4 of Ref.~\cite{Caucal:2021lgf}, we can express the amplitude of each diagram as an integral over the three coordinates of the final state partons. The final structure of the amplitude of every diagram will be
\begin{align}
   \int \der^2\boldsymbol{x_{\perp}} \der^2\Perp{y} \der^2 \Perp{z} e^{-i\boldsymbol{k_{1\perp}}\cdot \Perp{x}} e^{-i\boldsymbol{k_{2\perp}}\cdot \Perp{y}} e^{-i\boldsymbol{k_{3\perp}}\cdot \Perp{z}} C_{Rn}(\Perp{x},\Perp{y},\Perp{z}) \Ncal_{Rn}(\Perp{x},\Perp{y},\Perp{z}) \,,
\end{align}
where
\begin{align}
    \Ncal_{Rn} = \Ncal_{Rn,reg} + \Ncal_{Rn,inst}\,.
\end{align}
At the order of interest in this calculation there is at most one instantaneous contribution per diagram, corresponding to the cancellation of the scalar part of the propagator of the intermediate quark (or gluon) which is not adjacent to the shockwave.

\begin{figure}
    \centering
    \includegraphics[width=1\linewidth]{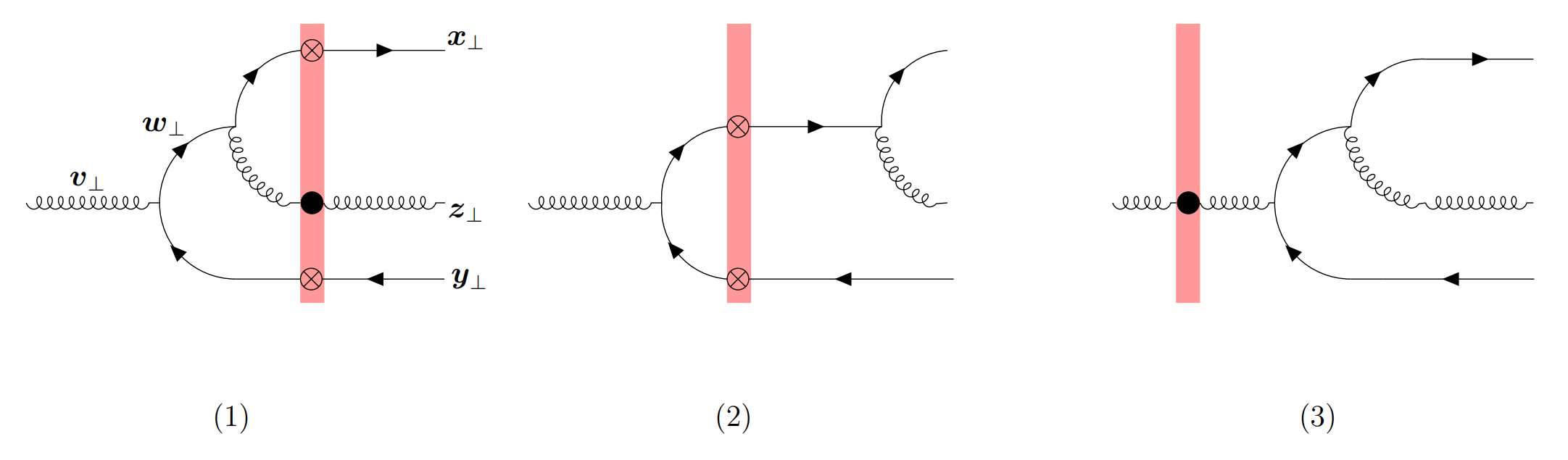}
    \caption{An example of a set of diagrams with the same topology for the $g \to q\bar{q}g$ channel. The diagrams represent the gluon emission by a quark.}
    \label{fig:Topology agroupation example}
\end{figure}

While this general strategy is shared with~\cite{Caucal:2021lgf}, 
an interesting novelty of the present study is the use of 
unitarity constraints to relate the various perturbative 
factors of each diagram, thereby allowing for a more compact  expression of the final amplitude. As mentioned at the beginning of this section, it is convenient to group these amplitudes according to the different topologies of the diagrams\footnote{The gluon four-vertex diagrams have the topology of the instantaneous diagrams, so instead of combining them together we can reabsorb them as part of the instantaneous pieces of the triple vertex topology}. For the diagrams computed here, each topology has three different possible shockwave insertions, which result in three different diagrams. The amplitude of these diagrams can be written in general as
\begin{align}
    &\int \der^2\boldsymbol{x_{\perp}} \der^2\Perp{y} \der^2 \Perp{z} e^{-i\boldsymbol{k_{1\perp}}\cdot \Perp{x}} e^{-i\boldsymbol{k_{2\perp}}\cdot \Perp{y}} e^{-i\boldsymbol{k_{3\perp}}\cdot \Perp{z}} \nonumber \\
    & \times \left\{(\Ncal_{reg,1} + \Ncal_{inst,1} )C_{1} + (\Ncal_{reg,2} + \Ncal_{inst,2} ) C_{2} + (\Ncal_{reg,3} + \Ncal_{inst,3} ) C_{3}  \right\} \,.
\end{align}
The indices $1,2,3$  in the color operators represent final, intermediate and initial state interactions of the shockwave, respectively, for a given topology (see for example Figure \ref{fig:Topology agroupation example}).

When the shockwave acts between the splittings, there is no instantaneous contribution as all internal propagators are adjacent to the shockwave; hence, $\Ncal_{inst,2} = 0$. Furthermore, in the limit of no interactions, i.e. $V(\Transv{x}{}) \to \mathds{1}$, the amplitude must vanish (for a formal demonstration of this fact using LCPT see section 2 of Ref. \cite{Iancu:2018hwa}); hence, this imposes the condition:
\begin{align}
    \Ncal_{reg,1} + \Ncal_{reg,2} + \Ncal_{reg,3} = 0 \,, \\
    \Ncal_{inst,1}  + \Ncal_{inst,3} = 0 \,.
\end{align}
This allows us to write the amplitude in the form:
\begin{align} \label{general amplitude trijets}
    &\int \der^2\boldsymbol{x_{\perp}} \der^2\Perp{y} \der^2 \Perp{z} e^{-i\boldsymbol{k_{1\perp}}\cdot \Perp{x}} e^{-i\boldsymbol{k_{2\perp}}\cdot \Perp{y}} e^{-i\boldsymbol{k_{3\perp}}\cdot \Perp{z}} \nonumber \\
    & \times \left\{\Ncal_{reg,1} \left( C_{1} - C_3\right) - \Ncal_{reg,2} \left( C_{2} - C_3 \right) + (C_{1} - C_{3})\Ncal_{inst,1} \right\} \,,
\end{align}
which no longer depends anymore on $\Ncal_{reg,3}$ nor  $\Ncal_{inst,3}$. Another advantage of this rewriting is that the difference 
between color structures leads to simplifications at the 
cross section level when taking the modulus squared, 
particularly in the slow gluon or DGLAP limits.

\subsection{The $g \to q\bar{q}g$ channel}

\begin{figure}
    \centering
    \includegraphics[width=1\linewidth]{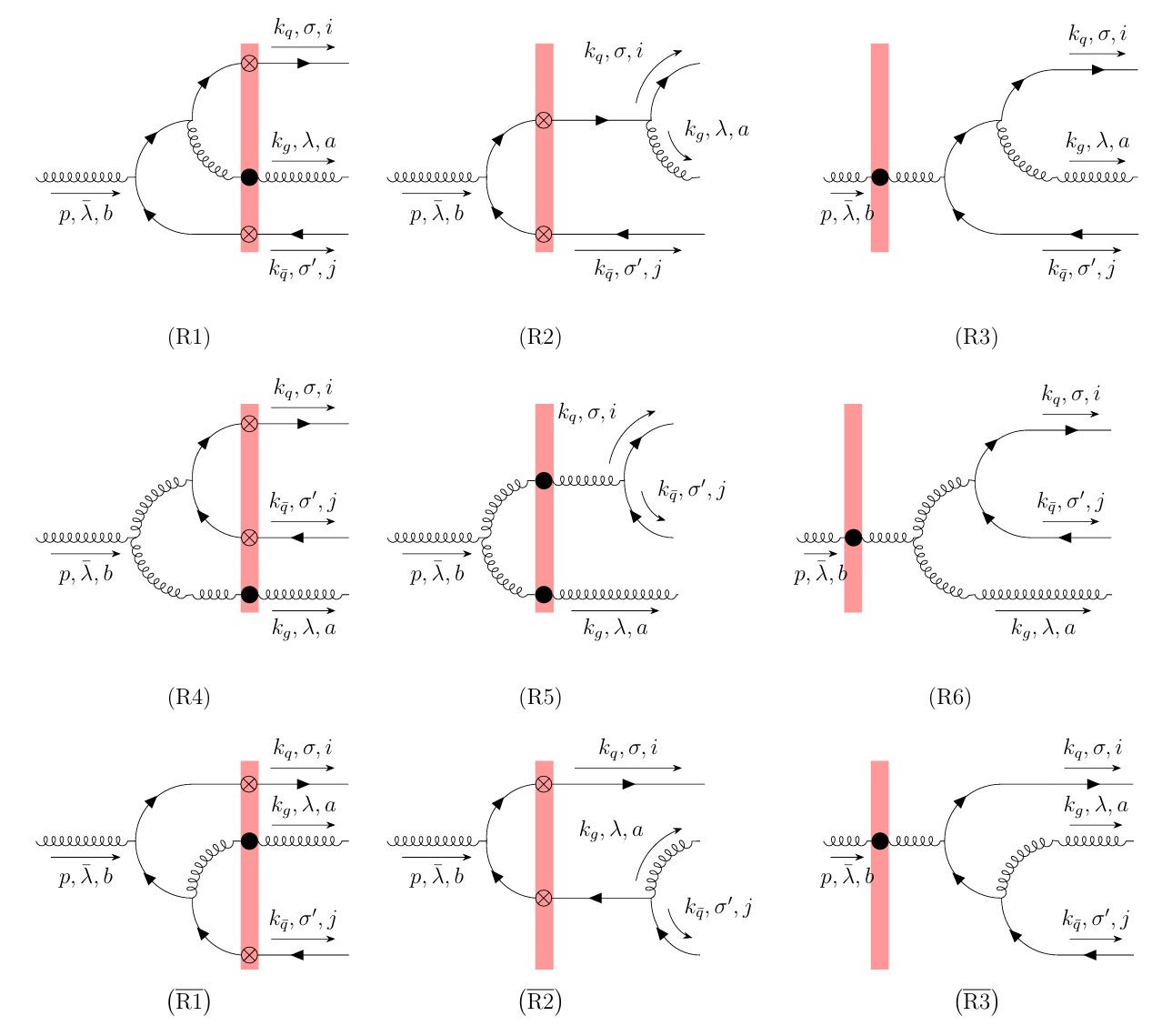}
    \caption{Diagrams contributing to trijet production in the gluon-initiated channel with $q\bar{q}g$ final state particles. }
    \label{fig:Trijet qbarqg}
\end{figure}

Let us start with the calculation of the amplitude for the $g \to q\bar{q}g$. The diagrams involved at the level of the amplitude for this process are shown in Figure \ref{fig:Trijet qbarqg}. Each row in the figure corresponds to different ``bare" topologies of this channel which can be classified by which parent parton emits the final-state gluon. The first, second and third row correspond to the  gluon emission by the quark, gluon and the antiquark, respectively. Due to the symmetries of the diagrams with gluon emission by a quark and an antiquark, the amplitude for the gluon emission by antiquark can be inferred using charge conjugation. We must therefore only compute six of the nine possible diagrams of the process. We will show an example of the calculation of the gluon emission by a quark with the shockwave interacting with all final state partons, diagram R1. The calculations of the rest of the diagrams follow a similar strategy, so we will only present our final expressions for them in the form of the amplitude introduced in Eq.\,\eqref{general amplitude trijets}.

Before starting with the calculation, let us list the relevant variables for the process at hand. In table \ref{momentum labels} we display the momentum, polarization, and color labels employed for this process.
\begin{table}
\centering
\begin{tabular}{l|c}
\hline
\textbf{Definition} & \textbf{Symbol} \\
\hline
Momentum of the incoming gluon & $p$ \\
Momentum of the outgoing partons & $k_i$, $i=q,\, \bar{q},\, g$\\
Longitudinal fraction of momentum of the final state partons & $z_i \equiv k_{i}^{-}/p^-$, $i=q,\, \bar{q},\, g$\\ 
Polarization of the incoming/outgoing gluon & $\bar{\lambda}, \lambda$ \\
Color indices of the incoming/outgoing gluon (adjoint rep.) & $b,a$ \\
Spin index of the outgoing quark/antiquark & $\sigma, \sigma'$  \\
Color indices of the quark/antiquark (fundamental rep.) & $i,j$ \\
\hline
\end{tabular}
\caption{Definitions of variables and indices appearing in the diagrams of the $g\to q\bar{q}g$ channel.}
\label{momentum labels}
\end{table}

\subsubsection{Gluon emission by quark}

\begin{center}
    \textbf{Calculation of diagram R1}
\end{center}
\vspace{3mm}

\begin{figure}
    \centering
    \includegraphics[width=0.5\linewidth]{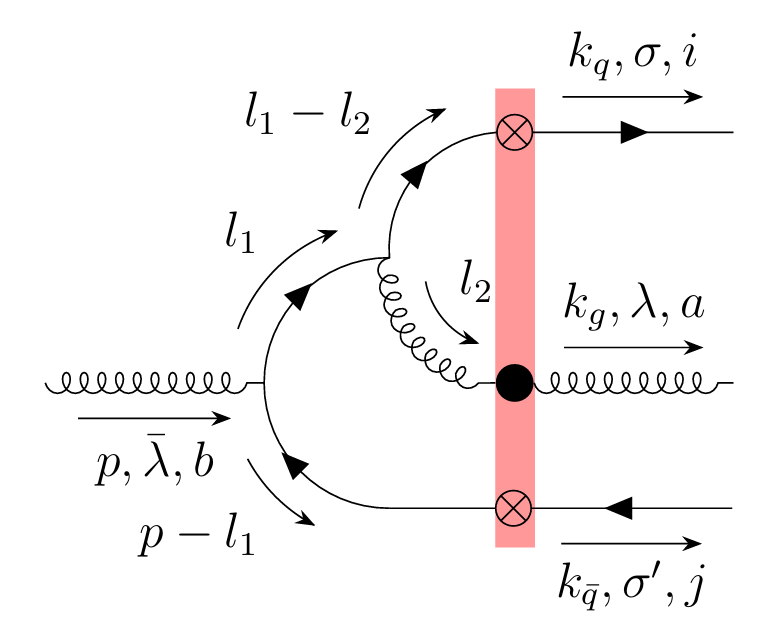}
    \caption{One of the three diagrams contributing to the gluon emission by a quark topology. In this diagram, the classical field interacts with all final state partons.}
    \label{fig:R1-detail}
\end{figure}
Diagram R1 with the definition of momenta, polarization, helicity and color indices is shown in Fig.\,\ref{fig:R1-detail}.
Using the standard pQCD rules, defined in Appendix \ref{app: Feynman rules}, along with the definition of the effective vertices for a quark and a gluon propagator, Eqs.\,\eqref{quark eff vertex} and \eqref{gluon eff vertex}, we can write the scattering amplitude of this diagram as 
\begin{multline}
    S = \int \frac{\der^4l_1}{(2\pi)^4} \frac{\der^4l_2}{(2\pi)^4} \bar{u}(k_q,\sigma)T^q(k_q,l_1-l_2)S^0(l_1 - l_2) (igt^c\gamma^{\mu}) S^0(l_1) (igt^b \slashed{\epsilon}(p,\bar{\lambda})) S^0(l_1-p)\\
    \times T^q(l_1-p,-k_{\bar{q}})v(k_{\bar{q}},\sigma') \epsilon^*_{\rho}(k_g,\lambda) T^{g,\rho\nu}_{ac}(k_g,l_2) G^0_{\nu\mu}(l_2)\,.
\end{multline}
Factoring out a delta function $2\pi \delta(p^- - k_q^- - k_{\bar{q}}^- - k_{g}^-)$, we obtain the physical amplitude
\begin{align}
     \mathcal{M}_{R1,ij}^{\bar{\lambda}\sigma\lambda\sigma',ab} =  \int \der^2\boldsymbol{x}_{\perp} \der^2\boldsymbol{y}_{\perp} \der^2\boldsymbol{z}_{\perp} e^{-i\boldsymbol{k}_{q\perp}\cdot \boldsymbol{x}_{\perp}} e^{-i\boldsymbol{k}_{\Bar{q}\perp}\cdot \boldsymbol{y}_{\perp}} e^{-i\boldsymbol{k}_{g\perp}\cdot \boldsymbol{z}_{\perp}} C_{R1,ij}^{ab}(\boldsymbol{x}_{\perp}, \boldsymbol{y}_{\perp}, \boldsymbol{z}_{\perp}) \Ncal_{R1}^{\bar{\lambda}\sigma\lambda\sigma'}(\boldsymbol{r}_{xz},\boldsymbol{r}_{xy})\,,
\end{align}
with color factor
\begin{align}\label{color factor R1}
    C_{R1, ij}^{ab}(\boldsymbol{x}_{\perp}, \boldsymbol{y}_{\perp}, \boldsymbol{z}_{\perp}) = V(\boldsymbol{x}_{\perp})t^ct^bV^{\dagger}(\boldsymbol{y}_\perp) U_{ac}(\boldsymbol{z}_\perp)\,,
\end{align}
and a perturbative factor
\begin{align}
    \Ncal_{R1}^{\bar{\lambda}\sigma\lambda\sigma'}(\boldsymbol{r}_{xz},\boldsymbol{r}_{xy}) = \frac{g^2}{4\pi^2}\int \frac{\der^4l_1}{(2\pi)^2} \frac{\der^4l_2}{(2\pi)^2}  \frac{e^{i\boldsymbol{l}_{1\perp}\cdot \boldsymbol{r}_{xy}} e^{i\boldsymbol{l}_{2\perp}\cdot \boldsymbol{r}_{zx}} \delta(l_1^- -(p^- - k_{\bar{q}}^-)) \delta(l_2^- - k_{g}^-) N_{R1}^{\bar{\lambda}\sigma\lambda\sigma'}(l_1,l_2)}{((l_1- l_2)^2 + i\varepsilon)(l_2^2+ i\varepsilon)(l_1^2 + i\varepsilon)((p-l_1)^2+i\varepsilon)}\,.
\end{align}
The Dirac structure appearing in the perturbative factor reads
\begin{align}
    N_{R1}^{\Bar{\lambda}\sigma\lambda\sigma'}(l_1,l_2) = -2l_2^- \Bar{u}(k_q,\sigma)\gamma^- (\slashed{l_1} - \slashed{l_2})\slashed{\epsilon}^*(l_2,\lambda) \slashed{l_1} \slashed{\epsilon}(p,\Bar{\lambda})(\slashed{p} - \slashed{l_1})\gamma^- v(k_{\Bar{q}},\sigma')\,.
\end{align}
In this expression we have made use of 
identity Eq.\,\eqref{external_to_internal_gluon_id} to convert the external gluon polarization vector into an internal one. Expressing $\slashed{l_1}$ as $\slashed{\bar{l_1}} + \frac{l_1^2}{2l_1^-}\slashed{\bar{n}}$ and expanding in terms of Dirac spinors yields
\begin{align}
    N_{R1}^{\Bar{\lambda}\sigma\lambda\sigma'}(l_1,l_2) = N_{R1,reg}^{\Bar{\lambda}\sigma\lambda\sigma'}(l_1,l_2) + l_1^2 N_{R1,inst}^{\Bar{\lambda}\sigma\lambda\sigma'}(l_1,l_2) \,,
\end{align}
where the regular part and the instantaneous parts of the Dirac structure are, respectively,
\begin{align}
    \begin{split}
        N_{R1,reg}^{\Bar{\lambda}\sigma\lambda\sigma'}(l_1,l_2) &= -2l_2^- \Bar{u}(k_q,\sigma)\gamma^- (\slashed{l_1} - \slashed{l_2})\slashed{\epsilon}^*(l_2,\lambda) u(\Bar{l_1}, \sigma_1)\Bar{u}(\Bar{l_1},\sigma_1) \slashed{\epsilon}(p,\Bar{\lambda})(\slashed{p} - \slashed{l_1})\gamma^- v(k_{\Bar{q}},\sigma'),\\
        N_{R1,inst}^{\Bar{\lambda}\sigma\lambda\sigma'}(l_1,l_2) &= \frac{-l_2^-}{l_1^-} \Bar{u}(k_q,\sigma)\gamma^- (\slashed{l_1} - \slashed{l_2})\slashed{\epsilon}^*(l_2,\lambda) u(n, \sigma_1)\Bar{u}(n,\sigma_1) \slashed{\epsilon}(p,\Bar{\lambda})(\slashed{p} - \slashed{l_1})\gamma^- v(k_{\Bar{q}},\sigma')\,.
    \end{split}
\end{align}
Here, the sum over the intermediate polarization $\sigma_1$ is implied. As mentioned before, these structures can be simplified by expressing the rest of the propagators in terms of Dirac spinors and using the identities listed in Appendix \ref{app: Spinor contractions}.\\ 

\begin{center}
    \textit{Regular contribution}
\end{center}
\vspace{3mm}
Making use of such identities, and noticing that all the propagators left in the regular term do not have an instantaneous term since they are interacting with the shockwave, we can simplify the structure to
\begin{align}
    N_{R1,reg}^{\bar{\lambda}\sigma\lambda\sigma'}(l_1,l_2) = 32 p^-l_1^- \sqrt{k_q^- k_{\bar{q}}^-} \delta^{\sigma, -\sigma'}  \Gamma^{\sigma \lambda}_{q\xrightarrow{}qg}\left(\frac{l_1^- -l_2^-}{l_1^-}\right) \Gamma^{\sigma \bar{\lambda}}_{g\xrightarrow{}q\bar{q}}\left(\frac{l_1^-}{p^-}\right) \et^{\lambda*}\cdot \boldsymbol{L}_{\perp} \et^{\bar{\lambda}}\cdot \boldsymbol{l}_{1\perp}\,,
\end{align}
where we have introduced the vector $\Transv{L}{} \equiv \Transv{l}{2} - \left(l_2^-/l_{1}^- \right) \Transv{l}{1}$. The new splitting function appearing here corresponds to the $q\to qg$ splitting function and is defined as
\begin{equation}\label{q_to_qg_splpitting_function}
    \Gamma^{\sigma \lambda}_{q\xrightarrow{}qg}(\xi) = \xi \delta^{\sigma, \lambda} + \delta^{\sigma, -\lambda}\,.
\end{equation}

Analogous to the other splitting functions, this function is related to the $q\to qg$ QCD splitting function when squared and summed over the polarization and helicity:
\begin{equation}
    \frac{C_F}{2(1-\xi)}\sum_{\lambda,\sigma} \left[ \qtoqg{\sigma}{\lambda}(\xi) \right]^2 = P_{qq}^{real}(\xi)\,,
\end{equation}
where the real part of $q\to qg$ splitting function is given by
\begin{equation}
    P^{real}_{qq}(\xi) \equiv C_F \frac{1+\xi^2}{1-\xi}\,.
\end{equation}
We can now perform the inner momenta integrals. The $l^-$ integrals can be done trivially using the two Dirac delta functions. The perturbative factor of the regular contribution is then
\begin{align}
    \Ncal_{R1,reg}^{\Bar{\lambda}\sigma\lambda\sigma'}(\boldsymbol{r}_{xz},\boldsymbol{r}_{xy}) = 
    \frac{g^2}{4\pi^2}\int \frac{\der^2\boldsymbol{l}_{1\perp}}{2\pi} \frac{\der^2\boldsymbol{l}_{2\perp}}{2\pi} e^{i\boldsymbol{l}_{1\perp}\cdot \boldsymbol{r}_{xy}} e^{i\boldsymbol{l}_{2\perp}\cdot \boldsymbol{r}_{zx}} N_{R1,reg}^{\Bar{\lambda}\sigma\lambda\sigma'}(l_1,l_2) I_{R1,reg}\,,
\end{align}
where the $l^+$ integrals read
\begin{align}
    I_{R1,reg} = \int \frac{dl_1^+}{2\pi} \frac{dl_2^+}{2\pi} \frac{1}{((l_1- l_2)^2 + i\varepsilon)(l_2^2+ i\varepsilon)(l_1^2 + i\varepsilon)((p-l_1)^2+i\varepsilon)}\,.
\end{align}
To solve this integral, we use contour integration and close the contour in the upper plane. The integral simplifies to (see Appendix \ref{app: Useful integrals}), 
\begin{align}\label{reg integral 4 poles}
   I_{R1,reg} =  -\frac{1}{2l_1^- \left[L_{\perp}^2 + \frac{p^-l_2^-(l_1^- - l_2^-)}{(l_1^-)^{2}(p^- - l_1^-)}l_{1\perp}^2 \right] 2p^- l^2_{1\perp}}\,.
\end{align}
Substituting this expression in the perturbative factor we obtain, after some simplifications, 
\begin{multline}
    \Ncal_{R1,reg}^{\Bar{\lambda}\sigma\lambda\sigma'}(\boldsymbol{r}_{xz},\boldsymbol{r}_{xy}) = -2g^2p^- \frac{\sqrt{z_q z_{\bar{q}}}}{\pi^2 }  \delta^{\sigma, -\sigma'} \Gamma^{\sigma \lambda}_{q\xrightarrow{}qg}\left(\frac{z_q}{z_q+z_g}\right) \Gamma^{\sigma \bar{\lambda}}_{g\xrightarrow{}q\bar{q}}(z_q + z_g)\\
    \times \int \frac{\der^2\boldsymbol{l}_{1\perp}}{2\pi} \frac{\der^2\boldsymbol{l}_{2\perp}}{2\pi} e^{i\boldsymbol{l}_{1\perp}\cdot \boldsymbol{r}_{xy}} e^{i\boldsymbol{l}_{2\perp}\cdot \boldsymbol{r}_{zx}} \frac{\et^{\lambda*}\cdot \boldsymbol{L}_{\perp} \et^{\bar{\lambda}}\cdot \boldsymbol{l}_{1\perp}}{l^2_{1\perp} \left[L_{\perp}^2 + \frac{z_q z_g}{z_{\bar{q}}(z_q + z_g)^2}l_{1\perp}^2 \right]}\,.
\end{multline}
The transverse integration can be performed using the identity Eq.\,\eqref{Trijet_Transverse_integral_final_state_reg}, giving the result
\begin{align}\label{transverse integral diag R1}
    \int \frac{\der^2\boldsymbol{l}_{1\perp}}{2\pi} \frac{\der^2\boldsymbol{l}_{2\perp}}{2\pi} e^{i\boldsymbol{l}_{1\perp}\cdot \boldsymbol{r}_{xy}} e^{i\boldsymbol{l}_{2\perp}\cdot \boldsymbol{r}_{zx}} \frac{\et^{\lambda*}\cdot \boldsymbol{L}_{\perp} \et^{\bar{\lambda}}\cdot \boldsymbol{l}_{1\perp}}{l^2_{1\perp} \left[L_{\perp}^2 + \frac{z_q z_g}{z_{\bar{q}}(z_q + z_g)^2}l_{1\perp}^2 \right]} = -\frac{\et^{\lambda*}\cdot \boldsymbol{r}_{zx}}{r_{zx}^2} \frac{\et^{\bar{\lambda}}\cdot \boldsymbol{r}_{w_1y}}{r_{w_1y}^2 + \frac{z_q z_g}{z_{\bar{q}}(z_q + z_g)^2} r_{zx}^2}\,,
\end{align}
where we have introduced the variable $\boldsymbol{w}_{1\perp} = (z_q\boldsymbol{x}_{\perp} + z_g\boldsymbol{z}_{\perp})/(z_q + z_g)$, corresponding to the transverse position of the quark before gluon emission. Defining the variable $\Theta_{q,1} = r_{w_1y}^2/(r_{w_1y}^2 + \frac{z_q z_g}{z_{\bar{q}}(z_q + z_g)^2} r_{zx}^2)$, we can write the final expression of the regular part of the perturbative impact factor as
\begin{multline}
    \Ncal_{R1,reg}^{\Bar{\lambda}\sigma\lambda\sigma'}(\Transv{x}{}, \Transv{y}{}, \Transv{z}{}) = \frac{2g^2p^-}{\pi^2} \sqrt{z_q z_{\Bar{q}}} \delta^{\sigma,-\sigma'}  \Gamma^{\sigma \lambda}_{q\xrightarrow{}qg}\left(\frac{z_q}{z_q+z_g}\right) \Gamma^{\sigma \bar{\lambda}}_{g\xrightarrow{}q\bar{q}}(z_q + z_g)\\
    \times \frac{\et^{\lambda*}\cdot \boldsymbol{r}_{zx}}{r_{zx}^2} \frac{\et^{\bar{\lambda}}\cdot \boldsymbol{r}_{w_1y}}{r_{w_1y}^2}\Theta_{q,1}\,.
\end{multline}\\

\begin{center}
    \textit{Instantaneous contribution}
\end{center}
\vspace{3mm}
Using identities Eq.\,\eqref{gluon->quark-antiquark wavefunction inst} and Eq.\,\eqref{quark->quark-gluon wavefunction inst}, we can simplify the instantaneous Dirac structure to
\begin{align}
    N_{R1,inst}^{\Bar{\lambda}\sigma\lambda\sigma'}(l_1,l_2) = -\frac{16(p^-)^3z_gz_qz_{\bar{q}}}{z_q + z_g} \sqrt{z_q z_{\bar{q}}} \delta^{\sigma,-\sigma'} \delta^{\sigma,\lambda} \delta^{\lambda,\bar{\lambda}} \,.
\end{align}
Then the instantaneous contribution to the perturbative factor reads
\begin{align}
    \Ncal_{R1,inst}^{\Bar{\lambda}\sigma\lambda\sigma'}(\boldsymbol{r}_{xz},\boldsymbol{r}_{xy}) = 
    \frac{g^2}{4\pi^2}\int \frac{\der^2\boldsymbol{l}_{1\perp}}{2\pi} \frac{\der^2\boldsymbol{l}_{2\perp}}{2\pi} e^{i\boldsymbol{l}_{1\perp}\cdot \boldsymbol{r}_{xy}} e^{i\boldsymbol{l}_{2\perp}\cdot \boldsymbol{r}_{zx}} N_{R1,inst}^{\Bar{\lambda}\sigma\lambda\sigma'}(l_1,l_2) I_{R1,inst} \,,
\end{align}
with
\begin{align}
    I_{R1,inst} = \int \frac{dl_1^+}{2\pi} \frac{dl_2^+}{2\pi} \frac{1}{((l_1- l_2)^2 + i\varepsilon)(l_2^2+ i\varepsilon)((p-l_1)^2+i\varepsilon)}\,.
\end{align}
Notice that the pole structure of the $l^+$ integral differs from the regular one, since the pole from the intermediate quark propagator canceled out. Using the same method as for the regular piece the integral simplifies to (see Appendix \ref{app: Useful integrals} for more details of the calculation)
\begin{align}\label{inst integral diagram R1}
    I_{R1,inst} = \frac{1}{4l_1^- (p^- - l_1^-) \left[L_{\perp}^2 + \frac{z_q z_g}{z_{\bar{q}}(z_q + z_g)^2} l_{1\perp}^2 \right]} \,.
\end{align}
The transverse integration left can be solved using identity Eq.\,\eqref{Trijet_Transverse_integral_final_state_inst}, giving
\begin{align}\label{instantaneous transverse integral diagram R1}
    \int \frac{\der^2\boldsymbol{l}_{1\perp}}{2\pi} \frac{\der^2\boldsymbol{l}_{2\perp}}{2\pi} e^{i\boldsymbol{l}_{1\perp}\cdot \boldsymbol{r}_{xy}} e^{i\boldsymbol{l}_{2\perp}\cdot \boldsymbol{r}_{zx}} \frac{1}{\left[L_{\perp}^2 + \frac{z_q z_g}{z_{\bar{q}}(z_q + z_g)^2} l_{1\perp}^2 \right]} = \frac{(z_q + z_g)z_{\bar{q}}}{X_R^2} \,,
\end{align}
where we have introduced the square of the size of the tripole $q\bar{q}g$, 
\begin{equation}
    X_R^2 = z_qz_{\bar{q}}r_{xy}^2  + z_gz_q r_{zx}^2  + z_gz_{\bar{q}} r_{yz}^2\,.
\end{equation}
Putting all the factors together, and after some simplifications, we arrive to the following expression for the instantaneous perturbative factor
\begin{align}
    \Ncal_{R1,inst}^{\Bar{\lambda}\sigma\lambda\sigma'}(\boldsymbol{r}_{xz},\boldsymbol{r}_{xy}) = -\frac{ g^2p^- z_g (z_qz_{\bar{q}})^{3/2}}{\pi^2(z_q + z_g)} \frac{ \delta^{\sigma,-\sigma'} \delta^{\sigma,\lambda} \delta^{\lambda,\bar{\lambda}}  }{X_R^2}\,.
\end{align}

\begin{center}
    \textit{Total perturbative factor}
\end{center}

\vspace{3mm}
Combining the regular and instantaneous pieces of the perturbative factor, we get the perturbative factor for diagram R1,
\begin{multline} \label{eq: impact factor R1}
    \Ncal_{R1}^{\bar{\lambda}\sigma\lambda\sigma'}(\boldsymbol{x}_{\perp},\boldsymbol{y}_{\perp},\boldsymbol{z}_{\perp}) = \frac{2g^2p^-}{\pi^2} \sqrt{z_q z_{\bar{q}}} \delta^{\sigma,-\sigma'}  \Gamma^{\sigma \lambda}_{q\xrightarrow{}qg}\left(\frac{z_q}{z_q+z_g}\right) \Gamma^{\sigma \bar{\lambda}}_{g\xrightarrow{}q\bar{q}}(z_q + z_g)\\
    \times \frac{\et^{\lambda*}\cdot \boldsymbol{r}_{zx}}{r_{zx}^2} \frac{\et^{\bar{\lambda}}\cdot \boldsymbol{r}_{w_1y}}{r_{w_1y}^2}\Theta_{q,1} -\frac{ g^2p^- z_g (z_qz_{\bar{q}})^{3/2}}{\pi^2(z_q + z_g)} \frac{ \delta^{\sigma,-\sigma'} \delta^{\lambda,\bar{\lambda}}  }{X_R^2}\,.
\end{multline}
The impact factor displayed in Eq.\,\eqref{eq: impact factor R1} provides an intuitive pictorial representation of the physical process at hand. The regular contribution can be viewed as the product of two factors, which represent the two splittings of the process. The first factor arises from the splitting of the incoming gluon into a quark-antiquark pair, providing the kernel $\boldsymbol{r}^{i}_{w_1y}/r^2_{w_1y}$ and the $g\to q\bar{q}$ splitting function. Here, the vector $\dipvec{wy}$ corresponds to the dipole vector of the intermediate quark-antiquark pair. The second factor corresponds to the radiation of a gluon from the produced quark, with kernel $\boldsymbol{r}^{i}_{zx}/r^2_{zx}$ and the corresponding $q \to qg$ splitting function. Similarly, the vector $\dipvec{zx}$ represents the dipole vector  $qg$ dipole in the final state. In the instantaneous term, the gluon emitted will preserve the polarization of the initial gluon. The kernel of the amplitude is $1/X_R^2$, where $X_{R}^2$ can be interpreted as the effective size of the three-parton $q\bar{q}g$ system.

The color operator of the diagram, Eq.\,\eqref{color factor R1}, has a straightforward interpretation as well. Because of the interaction of the shockwave with all three final state partons, the structure features three Wilson lines. The first one is a Wilson line in the fundamental representation at $\Transv{x}{}$ coming from the scattering of the quark. The scattering of the antiquark off the nucleus generates a similar Wilson line at $\Transv{y}{}$. Finally, the Wilson line in the adjoint representation accounts for the interaction of the radiated gluon from the quark at position $\Transv{z}{}$. The color matrix $t^a$ comes from the radiation of the gluon from the quark.

The calculation of diagrams R2 and R3 follows a very similar strategy, with analogous impact factors. The color operators of these diagrams naturally differ from that of diagram R1, since in diagram R2, only two partons participate in the scattering off the shockwave, while in diagram R3 only the incoming gluon participates in the scattering. Diagram R2 will thus only feature two light-like Wilson lines, coming from the scattering of a $q\bar{q}$ dipole. Diagram R3 features a single Wilson line in the adjoint representation from the interaction of the incoming gluon with the field. By defining the regular and instantaneous common perturbative factors as
\begin{align}
    \begin{split}
        \Ncal^{\bar{\lambda}\lambda\sigma\sigma'}_{q,reg} &= \frac{2g^2p^-}{\pi^2} \sqrt{z_q z_{\bar{q}}} \delta^{\sigma, -\sigma'}\Gamma^{\sigma \lambda}_{q\xrightarrow{}qg}\left(\frac{z_q}{z_q+z_g}\right) \Gamma^{\sigma \bar{\lambda}}_{g\xrightarrow{}q\bar{q}}(z_q + z_g) \frac{\et^{\lambda*}\cdot \boldsymbol{r}_{zx}}{r_{zx}^2} \frac{\et^{\bar{\lambda}}\cdot \boldsymbol{r}_{w_1y}}{r_{w_1y}^2},\label{eq:Nqreg-def_first}\\
        \Ncal^{\bar{\lambda}\lambda\sigma\sigma'}_{q,inst} &= - \frac{g^2p^- z_{g} (z_q z_{\bar{q}})^{3/2}}{\pi^2(z_q + z_g)}   \frac{\delta^{\sigma,-\sigma'} \delta^{\sigma,\lambda} \delta^{\lambda,\bar{\lambda}}}{X_R^2},
    \end{split}
\end{align}
we can write the amplitude of the sum of the diagrams contributing to the gluon emission by a quark (R1,R2 and R3) as 
\begin{multline}\label{quark parent qbarqg amplitude_first}
    \mathcal{M}^{\bar{\lambda}\lambda\sigma\sigma',ab}_{q,ij} = \int \der^6 \boldsymbol{\Pi} \left\{\Ncal^{\bar{\lambda}\lambda\sigma\sigma'}_{q,reg} \left[\Theta_{q,1}( C_{R1,ij}^{ab} - C_{R3,ij}^{ab}) - (C_{R2,ij}^{ab} - C_{R3,ij}^{ab}) \right]\right.\\
    \left.+ (C_{R1,ij}^{ab} - C_{R3,ij}^{ab})\Ncal^{\bar{\lambda}\lambda\sigma\sigma'}_{q,inst} \right\},
\end{multline}
where the phase space of the integral, including their Fourier phases, is
\begin{align}
    \der^6\boldsymbol{\Pi} = \der^2\boldsymbol{x}_{\perp} \der^2\boldsymbol{y}_{\perp} \der^2\boldsymbol{z}_{\perp} e^{-i\boldsymbol{k}_{q\perp}\cdot \boldsymbol{x}_{\perp}} e^{-i\boldsymbol{k}_{\bar{q}\perp}\cdot \boldsymbol{y}_{\perp}} e^{-i\boldsymbol{k}_{g\perp}\cdot \boldsymbol{z}_{\perp}}\,.
\end{align}
The corresponding color factors are
\begin{align}
    C_{R1,ij}^{ab}(\Perp{x}, \Perp{y}, \Perp{z}) &= \left[\WL{x}t^ct^b\WLadj{y} U_{ac}(\Perp{z}) \right]_{ij},\\
    C_{R2,ij}^{ab}(\Transv{w}{1}, \Perp{y})  &= \left[t^a V(\Transv{w}{1})t^b\WLadj{y}\right]_{ij},\label{eq:CR2-def}\\
    C_{R3,ij}^{ab}(\Perp{v}) &= \left[t^at^c U_{cb}(\Perp{v})\right]_{ij}.\label{eq:CR3-def}
\end{align}
In the color operator of diagram R3, we have introduced the coordinate of the incoming gluon at the point of interaction with the shockwave:
\begin{equation}
    \Transv{v}{} = \frac{z_q \Transv{x}{} + z_{\bar{q}} \Transv{y}{} + z_g \Transv{z}{}}{z_q + z_{\bar{q}} + z_g} \,.
\end{equation}
Equation Eq.\,\eqref{quark parent qbarqg amplitude_first} is our final expression for the amplitude of the gluon emission by a quark topology.

\subsubsection{Gluon emission by a gluon}
The amplitude for diagrams with gluon emission by a gluon can be cast in an analogous way. Due to the triple gluon vertex, the regular perturbative factor features the $g \to gg$ splitting function, Eq.\,\eqref{g_to_gg_splitting_function}, introduced in section \ref{sec: Dijet prod LO}. The Wilson lines featured in the color operators of the diagrams with final and initial state interactions are analogous to the ones from the color operators of diagrams R1 and R3 in $q\bar{q}g$ production, as the partons involved in the scattering are the same. The structure of diagram R5, with intermediate interactions will feature a Wilson line in the adjoint representation, instead of the fundamental representation like diagram R2, from the interaction of the intermediate gluon with the nucleus.

The amplitude of the sum of the three diagrams contributing to the gluon emission by the gluon topology reads:
\begin{multline}\label{gluon parent qbarqg amplitude_first}
    \mathcal{M}^{\bar{\lambda}\lambda\sigma\sigma',ab}_{g,ij} = \int \der^6 \boldsymbol{\Pi} \big\{\Ncal^{\bar{\lambda}\lambda\sigma\sigma'}_{g,reg} \left[\Theta_{g,1} (C_{R4,ij}^{ab} - C_{R6,ij}^{ab}) - (C_{R5,ij}^{ab} - C_{R6,ij}^{ab}) \right] \\
    + (C_{R4,ij}^{ab} - C_{R6,ij}^{ab})\Ncal^{\bar{\lambda}\lambda\sigma\sigma'}_{g,inst} \big\}\,,
\end{multline}
where the regular and instantaneous perturbative factors are
\begin{align}
    \Ncal^{\bar{\lambda}\lambda\sigma\sigma'}_{g,reg} &= -\frac{2g^2p^-}{\pi^2} \sqrt{\frac{z_qz_{\bar{q}}}{(z_q + z_{\bar q})^2}}\delta^{\sigma,-\sigma'} \Gamma^{\sigma \eta}_{g\xrightarrow{}q\bar{q}}\left(\frac{z_q}{z_q+\zbar}\right) \frac{\et^{\eta} \cdot \boldsymbol{r}_{yx}}{r_{yx}^2} \frac{\Gt^{\bar{\lambda}\lambda \eta}(z_q + z_{\bar{q}}) \cdot \boldsymbol{r}_{w_2z}}{r_{w_2z}^2},\,\label{eq:Ngreg-def_first}\\
    \Ncal^{\bar{\lambda}\lambda\sigma\sigma'}_{g,\mathrm{inst}}  &= - \frac{g^2p^- z_g(1+z_g) (z_qz_{\bar{q}})^{3/2}}{\pi^2(z_q + z_{\bar{q}})^2} \frac{\delta^{\sigma,-\sigma'} \delta^{\lambda,\bar{\lambda}}}{X_R^2}\,,
\end{align}
and the color operators are
\begin{align}
    C_{R4,ij}^{ab}(\Perp{x}, \Perp{y}, \Perp{z}) &=\left[iV(\boldsymbol{x}_{\perp})t^d V^{\dagger}(\boldsymbol{y}_{\perp})U_{ac}(\boldsymbol{z}_{\perp}) f^{bcd} \right]_{ij},\\
    C_{R5,ij}^{ab}(\Transv{w}{2},\Perp{z}) &= \left[ i f^{cdb} U_{ed}(\Transv{w}{2})t^e U_{ac}(\Perp{z})\right]_{ij},\\
    C_{R6,ij}^{ab}(\Perp{v}) &= \left[ i t^cf^{dac} U_{db}(\Perp{v})\right]_{ij}.
\end{align}
Here, the variable $\Transv{w}{2}$ corresponds to the transverse coordinate of the intermediate gluon before splitting defined as
\begin{align}
    \Transv{w}{2} = \frac{z_q \Perp{x} + z_{\bar{q}}\Perp{y} }{z_q + z_{\bar{q}}}\,.
\end{align}
Finally, the function $\Theta_{g,1}$ is defined analogous to the previous case
\begin{equation}
    \Theta_{g,1} = \frac{r_{w_2z}^2}{r_{w_2z}^2 + \frac{z_qz_{\bar{q}}}{z_g(z_q + z_{\bar{q}})^2}r_{yx}^2}\,.
\end{equation}

\subsubsection{Gluon emission by an antiquark}
The amplitudes for the diagrams where the gluon is emitted by an antiquark can be obtained from the gluon emission by quarks using charge conjugation. The amplitude is obtained by performing the following transformations to the quark amplitudes:
\begin{itemize}
    \item Interchange quark and antiquark momenta $k_{q}\leftrightarrow k_{\bar{q}} $. This implies changing $z_q \leftrightarrow z_{\bar{q}}$ and $\Transv{k}{q}\leftrightarrow \Transv{k}{\bar{q}} $.
    \item Interchange transverse coordinates $\Transv{x}{}\leftrightarrow \Transv{y}{} $
    \item Flip signs of helicities $\sigma\rightarrow -\sigma $ and $\sigma' \rightarrow -\sigma' $
    \item Hermitian conjugate of color operator $C(\Transv{x}{},\Transv{y}{},\Transv{z}{}) \rightarrow C^{\dagger}(\Transv{y}{},\Transv{x}{},\Transv{z}{})$
    \item Include an additional minus sign.
\end{itemize}

The total amplitude of the gluon emission by an antiquark is then
\begin{multline}\label{antiquark parent qbarqg amplitude_first}
    \mathcal{M}^{\bar{\lambda}\lambda\sigma\sigma',ab}_{\bar q,ij} = \int \der^6 \boldsymbol{\Pi} \left\{\Ncal^{\bar{\lambda}\lambda\sigma\sigma'}_{\bar{q},reg} \left[\Theta_{\bar{q},1}( C_{\overline{R1},ij}^{ab} - C_{\overline{R3},ij}^{ab}) - (C_{\overline{R2},ij}^{ab} - C_{\overline{R3},ij}^{ab}) \right]\right.\\
    \left.+ (C_{\overline{R1},ij}^{ab} - C_{\overline{R3},ij}^{ab})\Ncal^{\bar{\lambda}\lambda\sigma\sigma'}_{\bar{q},inst} \right\},
\end{multline}
with perturbative factors
\begin{align}
    \begin{split}
        \Ncal^{\bar{\lambda}\lambda\sigma\sigma'}_{\bar{q},reg} &= -\frac{2g^2p^-}{\pi^2} \sqrt{z_q z_{\bar{q}}} \delta^{\sigma, -\sigma'}\Gamma^{\sigma' \lambda}_{q\xrightarrow{}qg}\left(\frac{\zbar}{\zbar+z_g}\right) \Gamma^{\sigma' \bar{\lambda}}_{g\xrightarrow{}q\bar{q}}(z_{\bar{q}} + z_g) \frac{\et^{\lambda*}\cdot \boldsymbol{r}_{zy}}{r_{zy}^2} \frac{\et^{\bar{\lambda}}\cdot \boldsymbol{r}_{w_3x}}{r_{w_3x}^2},\\
        \Ncal^{\bar{\lambda}\lambda\sigma\sigma'}_{\bar{q},inst} &= \frac{g^2p^- z_{g} (z_q z_{\bar{q}})^{3/2}}{\pi^2(z_{\bar{q}} + z_g)}   \frac{\delta^{\sigma,-\sigma'} \delta^{\sigma',\lambda} \delta^{\lambda,\bar{\lambda}}}{X_R^2},
    \end{split}
\end{align}
The corresponding color operators are
\begin{align}
    C_{\overline{R1},ij}^{ab}(\Perp{x}, \Perp{y}, \Perp{z}) &= \left[\WL{x}t^bt^c\WLadj{y} U_{ac}(\Perp{z}) \right]_{ij},\\
    C_{\overline{R2},ij}^{ab}(\Transv{w}{3}, \Perp{x})  &= \left[\WL{x} t^b V^{\dagger}(\Transv{w}{3})t^a\right]_{ij},\\
    C_{\overline{R3},ij}^{ab}(\Perp{v}) &= \left[t^ct^a U_{cb}(\Perp{v})\right]_{ij}.
\end{align}
The transverse coordinate of the intermediate antiquark $\Transv{w}{3}$ is defined as
\begin{align}
    \Transv{w}{3} = \frac{z_{\bar{q}}\Perp{y} + z_g \Transv{z}{} }{z_{\bar{q}} + z_g}\,,
\end{align}
and the corresponding theta function
\begin{equation}
    \Theta_{\bar{q},1} = \frac{r_{w_3x}^2}{r_{w_3x}^2 + \frac{z_{\bar{q}} z_{g}}{z_q(z_{\bar{q}} + z_g)^2} r_{yz}^2}\,.
\end{equation}

\subsection{The $g\rightarrow ggg$ channel}
\begin{figure}
    \centering
    \includegraphics[width=1\linewidth]{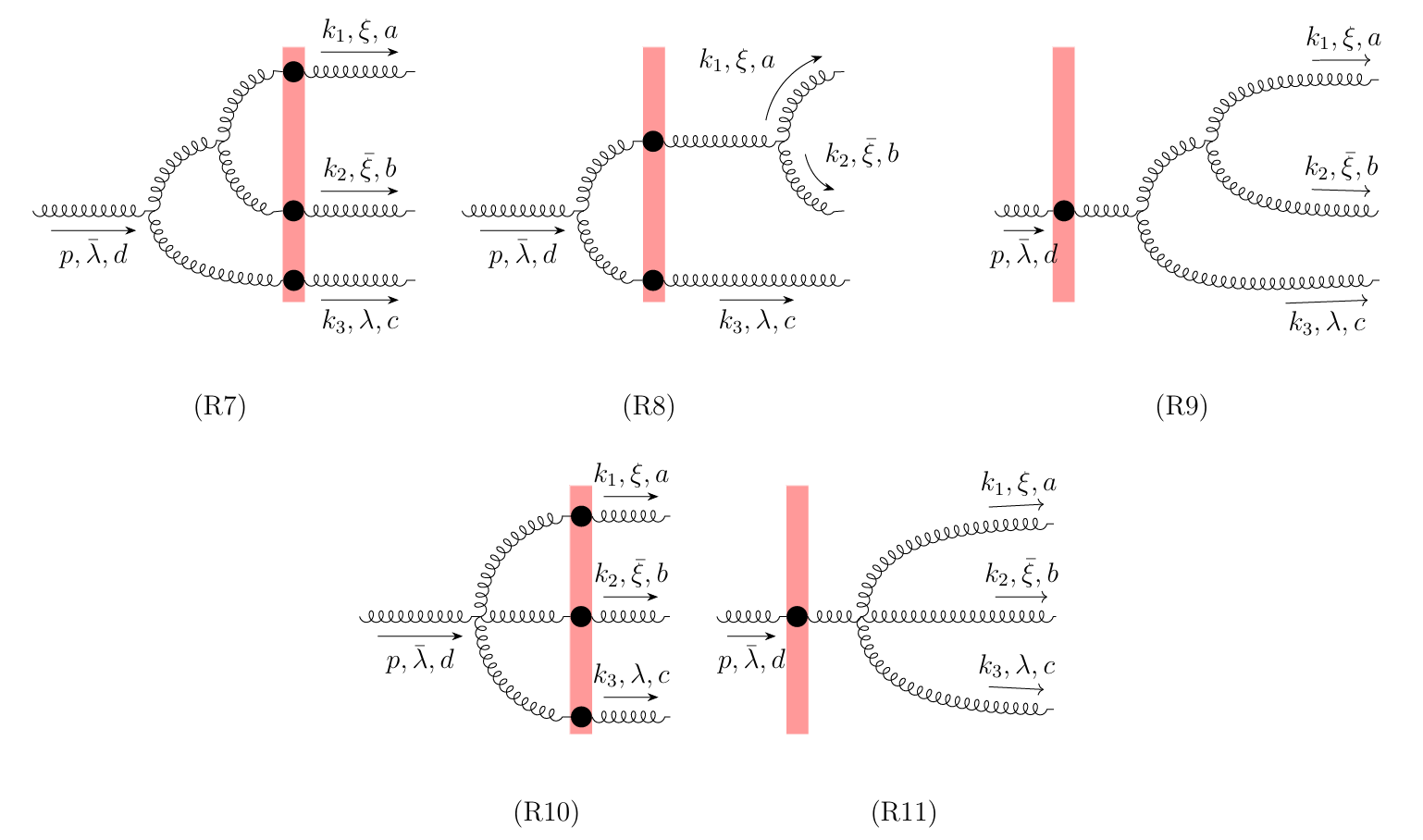}
    \caption{Diagrams involved in the production of three gluons in the final state. The six diagrams obtained by permutations of the final state gluons in R7, R8 and R9 are not shown.}
    \label{fig:Diag contributing to ggg channel}
\end{figure}

In the $ggg$ channel, five diagrams contribute at the amplitude level (see Fig.\,\ref{fig:Diag contributing to ggg channel}). The first three correspond to a double gluon-gluon splitting, where the incoming gluon splits into a pair of gluons, followed by the splitting of one of the daughter gluons into another pair of gluons. The two diagrams on the second row correspond to a 4-gluon vertex, where the incoming gluon splits into three gluons. The four-gluon vertex diagrams will naturally combine with the instantaneous contributions to the double splitting topology diagrams, and their corresponding permutations. Similarly to the $q\bar{q}g$ case, we list the definition of the pertinent variables for the process in table~\ref{momentum labels ggg}.

\begin{table}[H]
\centering
\begin{tabular}{l|c}
\hline
\textbf{Definition} & \textbf{Symbol} \\
\hline
Momentum of the incoming gluon & $p$ \\
Momentum of the outgoing partons & $k_i$, $i=1,\, 2,\, 3$\\
Longitudinal fraction of momentum of the final state partons & $z_i \equiv k_{i}^{-}/p^-$, $i=1,\, 2,\, 3$\\ 
Polarization of the incoming gluon & $\bar{\lambda}$ \\
Polarization of the outgoing gluons & $\xi, \bar{\xi}, \lambda$ \\
Color index of the incoming gluon (adjoint rep.) & $d$ \\
Color indices of the outgoing gluons (adjoint rep.) & $a,b,c$\\
\hline
\end{tabular}
\caption{Definitions of variables and indices appearing in the diagrams of the $g\to ggg$ process.}\label{momentum labels ggg}
\end{table}

\subsubsection{Double gluon splitting}

To streamline the main text, we have relegated to 
Appendix~\ref{app: Detailed of calculation of diagram R7} 
the details of the calculation of diagram R7 as an example.
The sum of the amplitudes of the diagrams corresponding to the gluon double splitting can be cast in the same way as the amplitudes of the $q\bar{q}g$ channel. Due to the presence of two triple gluon vertices, the algebra involved in the calculation becomes more cumbersome, but follows the same structure as the previous calculations. After some algebra and simplifications, the final expression for the sum of the amplitudes for the diagrams R7, R8, and R9, reads
\begin{multline}\label{ggg_double_splitting_amplitude}
    \mathcal{M}^{\bar{\lambda}\lambda\bar{\xi}\xi,abcd}_{R7+R8+R9} = \int \der^6 \boldsymbol{\Pi} \big\{\Ncal^{\bar{\lambda}\lambda\bar{\xi}\xi}_{ggg,reg\,A} \left[\Theta_{ggg,1A} (C_{R7}^{abcd} - C_{R9}^{abcd}) - (C_{R8}^{abcd} - C_{R9}^{abcd}) \right]\\
        + (C_{R7}^{abcd} - C_{R9}^{abcd})\Ncal^{\bar{\lambda}\lambda\bar{\xi}\xi}_{ggg,inst} \big\}\,,
\end{multline}
where the perturbative factors are:
\begin{align}\label{ggg_reg_and_inst_perturbative_factors}
    \begin{split}
        \Ncal^{\bar{\lambda}\lambda\bar{\xi}\xi}_{ggg,reg\,A}(\boldsymbol{x}_{\perp}, \boldsymbol{y}_{\perp}, \boldsymbol{z}_{\perp}) &= - \frac{2g^2p^-}{\pi^2} \frac{\Gt^{\eta\bar{\xi}\xi}\left(\frac{z_1}{z_1+z_2}\right) \cdot \boldsymbol{r}_{yx}}{\boldsymbol{r}_{yx}^2} \frac{\Gt^{\bar{\lambda} \lambda\eta}(z_1+z_2)\cdot \boldsymbol{r}_{w_A z}}{r_{w_A z}^2}\,,\\
        \Ncal^{\bar{\lambda}\lambda\bar{\xi}\xi}_{ggg,inst}(\boldsymbol{x}_{\perp}, \boldsymbol{y}_{\perp}, \boldsymbol{z}_{\perp}) &= -\frac{g^2p^- z_1z_2z_3(z_2-z_1)(1 + z_3)}{2\pi^2(z_1 + z_2)^2} \frac{\delta^{\xi,-\bar{\xi}} \delta^{\lambda \bar{\lambda}}}{X_R^2} \,,
    \end{split}
\end{align}
and the color operators:
\begin{align}
    \begin{split}
        C_{R7}^{abcd}(\boldsymbol{x}_{\perp}, \boldsymbol{y}_{\perp}, \boldsymbol{z}_{\perp}) &= f^{efg}f^{dhf} U_{ae}(\boldsymbol{x}_{\perp})U_{bg}(\boldsymbol{y}_{\perp})U_{ch}(\boldsymbol{z}_{\perp})\,,\\
        C_{R8}^{abcd}(\boldsymbol{w}_{A\perp}, \boldsymbol{z}_{\perp}) &= f^{eba}f^{gfd} U_{ef}(\boldsymbol{w}_{A\perp}) U_{cg}(\boldsymbol{z}_{\perp})\,,\\
        C_{R9}^{abcd}(\boldsymbol{v}_{\perp}) &= f^{eba}f^{fce} U_{fd}(\boldsymbol{v}_{\perp}) \,.
    \end{split}
\end{align}
The following variables were used in the expressions above:
\begin{align}\label{ggg_variable_def_first}
    \begin{split}
        \boldsymbol{w}_{A\perp} = \frac{z_1 \boldsymbol{x}_{\perp} + z_2\boldsymbol{y}_{\perp}}{z_1 + z_2}, &\qquad\boldsymbol{v}_{\perp} = \frac{z_1 \boldsymbol{x}_{\perp} + z_2\boldsymbol{y}_{\perp} + z_3\boldsymbol{z}_{\perp}}{z_1 + z_2 + z_3} \,, \\
        \Theta_{ggg,1A} &= \frac{r_{w_A z}^2}{r_{w_A z}^2 + \frac{z_1z_2}{z_3(z_1 + z_2)^2}r_{yx}^2} \,.
    \end{split}
\end{align}

\subsubsection{Four-gluon vertex diagrams}

As mentioned above, these diagrams will have an amplitude similar to that of the instantaneous contributions of the double-gluon splitting diagrams, which will allow us to combine them together. The 4-gluon vertex, according to the Feynman rules, is defined as 
\begin{align}
    \mathcal{V}^{\mu\nu\rho\sigma}_{abcd} = -ig^2 \bigg[f^{abe}f^{cde}(g^{\mu\rho}g^{\nu\sigma}-g^{\mu\sigma}g^{\nu\rho}) + f^{ace}f^{bde}(g^{\mu\nu}g^{\rho\sigma} - g^{\mu\sigma}g^{\nu\rho}) + f^{ade}f^{bce}(g^{\mu\nu}g^{\rho\sigma}-g^{\mu\rho}g^{\nu\sigma}) \bigg] \,.
\end{align}
Using this definition of the 4-gluon vertex, and after writing the physical amplitude, simplifying the gluon structures, and doing the momentum integrals we obtain the amplitude, 
\begin{multline}
    \mathcal{M}_{4-gluon}^{\bar{\lambda}\lambda\bar{\xi}\xi} = \int \der^6 \boldsymbol{\Pi}  \bigg\{\left( C^{abcd}_{R10,A} - C^{abcd}_{R11,A} \right)\Ncal_{4-gluon,A}^{\bar{\lambda}\lambda\bar{\xi}\xi}(\Perp{x}, \Perp{y}, \Perp{z})\\
    + \left( C^{abcd}_{R10,B} - C^{abcd}_{R11,B} \right)\Ncal_{4-gluon,B}^{\bar{\lambda}\lambda\bar{\xi}\xi}(\Perp{x}, \Perp{y}, \Perp{z})\\
    + \left(C^{abcd}_{R10,C} - C^{abcd}_{R11,C} \right)\Ncal_{4-gluon,C}^{\bar{\lambda}\lambda\bar{\xi}\xi}(\Perp{x}, \Perp{y}, \Perp{z})  \bigg\} \,,
\end{multline}
where the perturbative factors are
\begin{align}
    \Ncal_{4-gluon,A}^{\bar{\lambda}\lambda\bar{\xi}\xi}(\Transv{x}{}, \Transv{y}{}, \Transv{z}{}) &= -\frac{g^2p^- z_1 z_2 z_3 ( \delta^{\bar{\lambda}\xi}\delta^{\lambda,-\bar{\xi}} - \delta^{\bar{\lambda}\bar{\xi}}\delta^{\xi,-\lambda} )}{2\pi^2X_R^2} \,, \\
    \Ncal_{4-gluon,B}^{\bar{\lambda}\lambda\bar{\xi}\xi}(\Transv{x}{}, \Transv{y}{}, \Transv{z}{}) &= -\frac{g^2p^- z_1 z_2 z_3 ( \delta^{\bar{\lambda}\lambda}\delta^{\xi,-\bar{\xi}}  - \delta^{\bar{\lambda}\bar{\xi}}\delta^{\xi,-\lambda} )}{2\pi^2X_R^2} \,, \\
    \Ncal_{4-gluon,C}^{\bar{\lambda}\lambda\bar{\xi}\xi}(\Transv{x}{}, \Transv{y}{}, \Transv{z}{}) &= -\frac{g^2p^- z_1 z_2 z_3 (  \delta^{\bar{\lambda}\xi}\delta^{\lambda,-\bar{\xi}} - \delta^{\bar{\lambda}\lambda}\delta^{\xi,-\bar{\xi}} )}{2\pi^2X_R^2} \,.
\end{align}
and the color operators
\begin{align}\
    \begin{split}
        C^{abcd}_{R10,A}(\Transv{x}{}, \Transv{y}{}, \Transv{z}{}) &= U_{ae}(\Transv{x}{})U_{bf}(\Transv{y}{})U_{cg}(\Transv{z}{}) f^{dgn}f^{efn}\,, \\
        C^{abcd}_{R10,B}(\Transv{x}{}, \Transv{y}{}, \Transv{z}{}) &= U_{ae}(\Transv{x}{})U_{bf}(\Transv{y}{})U_{cg}(\Transv{z}{}) f^{den}f^{gfn} \,, \\
        C^{abcd}_{R10,C}(\Transv{x}{}, \Transv{y}{}, \Transv{z}{}) &= U_{ae}(\Transv{x}{})U_{bf}(\Transv{y}{})U_{cg}(\Transv{z}{}) f^{dfn}f^{egn}\,, \\
        C^{abcd}_{R11,A}(\Transv{v}{}) &= U_{ed}(\Transv{v}{}) f^{ecn}f^{abn}\,, \\
        C^{abcd}_{R11,B}(\Transv{v}{}) &= U_{ed}(\Transv{v}{})f^{ean}f^{cbn}\,, \\
        C^{abcd}_{R11,C}(\Transv{v}{}) &= U_{ed}(\Transv{v}{})f^{ebn}f^{acn} \,.
    \end{split}
\end{align}
Using the identity $f^{efg}U_{ae}(\Transv{x}{})U_{bf}(\Transv{x}{})=f^{abn}U_{ng}(\Transv{x}{})$ twice, one can show that $C_{R11,\alpha}(\Transv{v}{}) = C_{R10,\alpha}(\Transv{v}{}, \Transv{v}{}, \Transv{v}{})$ for $\alpha=A,B,C$. This relation allows us to show that, in the non-interacting limit, the amplitudes of the four gluon vertex vanish, as we should expect from our previous discussion.

\subsubsection{Exchange of identical particles}
\label{sub:ggg-permutations}

\begin{figure}[H]
    \centering
    \includegraphics[width=0.9\linewidth]{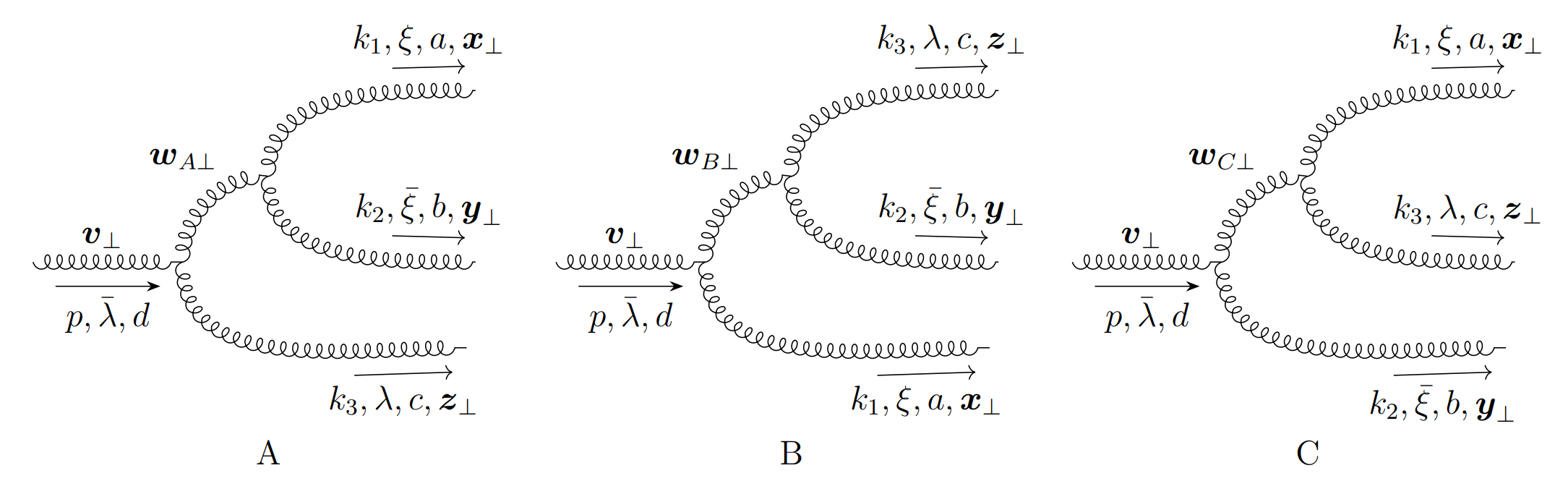}
    \caption{The three non-trivial permutations of identical gluons in the double splitting topology, labeled A, B, and C for the original configuration, the exchange of gluons $1\leftrightarrow 3$ and the exchange of gluons $2\leftrightarrow 3$, respectively. For simplicity, we do not depict the shockwave, but it should be understood that this permutation is applied to all diagrams.}
    \label{fig:ggg topology permutations}
\end{figure}

Since there are three identical gluons in the final state of this process, we must account for the exchange of these particles in the final expression of the total amplitude. Due to the symmetries of the gluon vertices, we only need to consider three non-trivial permutations for diagrams R7-R9 (the four-gluon vertex diagrams are identical under exchange of any final state partons so we do not need to account for them). The three permutations we need to account for correspond to the exchange of either particles $k_1$ with $k_3$, or $k_2$ with $k_3$ (see Figure \ref{fig:ggg topology permutations}).

Let us relabel the color operators in a more convenient way to account for these permutations. We introduce
\begin{align}\label{colo structures R7}
    C_{R7,A}^{abcd} (\Perp{x}, \Perp{y}, \Perp{z}) &= U_{ae}(\Transv{x}{})U_{bg}(\Transv{y}{})U_{ch}(\Transv{z}{}) f^{efg}f^{dhf}\,,\\
    C_{R7,B}^{abcd} (\Perp{x}, \Perp{y}, \Perp{z}) & = C_{R7,A}^{cbad} (\Perp{z}, \Perp{y}, \Perp{x})  = U_{ce}(\Perp{z}) U_{bg}(\Perp{y}) U_{ah}(\Perp{x}) f^{efg}f^{dhf}\,,\\
    C_{R7,C}^{abcd} (\Perp{x}, \Perp{y}, \Perp{z}) &= C_{R7,A}^{acbd} (\Perp{x}, \Perp{z}, \Perp{y})  = U_{ae}(\Perp{x}) U_{cg}(\Perp{z}) U_{bh}(\Perp{y}) f^{efg}f^{dhf} \,.
\end{align}
Notice that the operator $C_{R7,A}^{abcd}$ is just the color operator $C_{R7}^{abcd}$, while the structures B and C are color permutations of the transverse coordinates and color indices of the gluons. With these definitions we can rewrite the color operators of diagram R10 as follows
\begin{align}
    C_{R10,\alpha}^{abcd} (\Perp{x}, \Perp{y}, \Perp{z}) &= -C_{R7,\alpha}^{abcd} (\Perp{x}, \Perp{y}, \Perp{z})\,.
\end{align}
We define analogous color operators for diagrams R8 and R9. They can be obtained by taking appropriate limits of the color operators of diagram R7. For R8, we take $\Perp{x}, \Perp{y} \to \Transv{w}{A}$, and for R9 we take the limit $\Perp{x}, \Perp{y}, \Perp{z} \to \Perp{v}$. The color operators of diagram R11 can then be related to the structures of R9 in the same way as R7 to R10:
\begin{align}
    C_{R11,\alpha}^{abcd} (\Transv{v}{}) &= -C_{R9,\alpha}^{abcd} (\Transv{v}{}).
\end{align}
Using the Jacobi identity, we can establish a relation between the structures A,B, and C:
\begin{align}
    C_{Ri,A}^{abcd} = C_{Ri,B}^{abcd} + C_{Ri,C}^{abcd}\,,
    \label{eq:Jacobi-color}
\end{align}
with $i=7,8,9$.

As our notation suggests, and from the identities found above relating the color operators of the four-gluon vertex diagrams to the double splitting diagrams, we should be able to combine them in a compact way. Because the regular amplitudes will have different kernels under the exchange of partons, we will only combine the instantaneous amplitudes with the amplitudes of the four-gluon vertices. To that aim, let us write the explicit expressions for such instantaneous amplitudes of the double splitting diagrams under the exchange of the final partons. Under the exchange of gluons with momentum $k_1$ and $k_3$ we have

\begin{align}
    \int \der^6 \boldsymbol{\Pi}\, (C_{R7}^{abcd} - C_{R9}^{abcd})\Ncal^{\bar{\lambda}\sigma\lambda\sigma'}_{ggg,inst}\bigg|_{1\leftrightarrow 3} =  \int \der^6 \boldsymbol{\Pi}  \left(C^{abcd}_{R7,B} - C^{abcd}_{R9,B}\right) \Ncal^{\bar{\lambda}\lambda\bar{\xi}\xi}_{ggg, inst,B}\,.
\end{align}
The instantaneous perturbative factor reads
\begin{align}
   \Ncal^{\bar{\lambda}\lambda\bar{\xi}\xi}_{ggg, inst,B} &= -\frac{g^2p^- z_1 z_2 z_3 }{2\pi^2 X_R^2} \bigg(\frac{(z_2 - z_3)(1+z_1)}{(z_3+z_2)^2} \delta^{\bar{\lambda}\xi}\delta^{\lambda,-\bar{\xi}} \bigg)\,.
\end{align}
Similarly, under the exchange of gluons $k_2$ and $k_3$ we have
\begin{align}
     \int \der^6 \boldsymbol{\Pi}\, (C_{R7}^{abcd} - C_{R9}^{abcd})\Ncal^{\bar{\lambda}\sigma\lambda\sigma'}_{ggg,inst}\bigg|_{2\leftrightarrow 3} =  \int \der^6 \boldsymbol{\Pi}  \left(C^{abcd}_{R7,C} - C^{abcd}_{R9,C}\right) \Ncal^{\bar{\lambda}\lambda\bar{\xi}\xi}_{ggg, inst,C}\,,
\end{align}
with perturbative factor
\begin{align}
    \Ncal^{\bar{\lambda}\lambda\bar{\xi}\xi}_{ggg, inst,C} &= -\frac{g^2p^- z_1 z_2 z_3 }{2\pi^2 X_R^2} \bigg(\frac{(z_3 - z_1)(1+z_2)}{(z_1+z_3)^2} \delta^{\xi, -\lambda}\delta^{\bar{\xi}, \bar{\lambda}} \bigg)\,.
\end{align}
Combining the amplitudes of these permutations with the amplitudes of the four-gluon vertex, and using Eq.\,\eqref{eq:Jacobi-color} to express the operators of topology C in terms of A and B yields
\begin{align}\label{ggg_total_inst_amplitude}
    \int \der^6\boldsymbol{\Pi}\,\left\{ (C^{abcd}_{R7,A} - C^{abcd}_{R9,A}) \overline{\Ncal}^{\bar{\lambda}\lambda\bar{\xi}\xi}_{ggg,inst,A}+ (C^{abcd}_{R7,B} - C^{abcd}_{R9,B}) \overline{\Ncal}^{\bar{\lambda}\lambda\bar{\xi}\xi}_{ggg,inst,B} \right\} \,,
\end{align}
where the perturbative factors of the combined amplitudes are
\begin{multline}\label{ggg_eff_inst_perturbative_factor_A}
    \overline {\Ncal}^{\bar{\lambda}\lambda\bar{\xi}\xi}_{ggg,inst,A} = \Ncal^{\bar{\lambda}\lambda\bar{\xi}\xi}_{ggg, inst,A} - \Ncal^{\bar{\lambda}\lambda\bar{\xi}\xi}_{4-gluon,A} + \Ncal^{\bar{\lambda}\lambda\bar{\xi}\xi}_{ggg, inst,C} - \Ncal^{\bar{\lambda}\lambda\bar{\xi}\xi}_{4-gluon,C}\\
    =-\frac{g^2p^- z_1 z_2 z_3 }{2\pi^2X_R^2} \left\{\left( \frac{(z_2 - z_1)(1+z_3)}{(z_1+z_2)^2} +1 \right)\delta^{\bar{\lambda}\lambda}\delta^{\xi,-\bar{\xi}}  + \left( \frac{(z_3 - z_1)(1+z_2)}{(z_1+z_3)^2} + 1 \right) \delta^{\bar{\lambda}\bar{\xi}}\delta^{\xi,-\lambda} -2\delta^{\bar{\lambda}\xi}\delta^{\lambda, -\bar{\xi}}  \right\} \,,
\end{multline}
and
\begin{multline}\label{ggg_eff_inst_perturbative_factor_B}
    \overline {\Ncal}^{\bar{\lambda}\lambda\bar{\xi}\xi}_{ggg,inst,B} = \Ncal^{\bar{\lambda}\lambda\bar{\xi}\xi}_{ggg, inst,B} - \Ncal^{\bar{\lambda}\lambda\bar{\xi}\xi}_{4-gluon,B} - \Ncal^{\bar{\lambda}\lambda\bar{\xi}\xi}_{ggg, inst,C} + \Ncal^{\bar{\lambda}\lambda\bar{\xi}\xi}_{4-gluon,C}\\
    =-\frac{g^2p^- z_1 z_2 z_3 }{2\pi^2X_R^2} \left\{\left( \frac{(z_2 - z_3)(1+z_1)}{(z_2 + z_3)^2} +1 \right)\delta^{\bar{\lambda}\xi}\delta^{\lambda, -\bar{\xi}}  + \left( \frac{(z_1 - z_3)(1+z_2)}{(z_1+z_3)^2} + 1 \right) \delta^{\bar{\lambda}\bar{\xi}}\delta^{\xi,-\lambda} -2 \delta^{\bar{\lambda}\lambda}\delta^{\xi,-\bar{\xi}} \right\}.
\end{multline}
Combining the regular part of amplitude Eq.\,\eqref{ggg_double_splitting_amplitude} along with its corresponding permutations, and the ``total" instantaneous amplitude, Eq.\,\eqref{ggg_total_inst_amplitude}, we can write the total amplitude for $ggg$ production as
\begin{multline} \label{gluon_to_ggg_total_amplitude_first}
    \mathcal{M}_{ggg}^{\bar{\lambda} \lambda \bar{\xi} \xi} =  \int \der^6 \boldsymbol{\Pi} \bigg[\big\{\Theta_{ggg,1A}(C^{abcd}_{R7,A} - C^{abcd}_{R9,A}) - (C^{abcd}_{R8,A} - C^{abcd}_{R9,A}) \big\} \Ncal^{\bar{\lambda}\lambda\bar{\xi}\xi}_{ggg,reg\,A} + (k_1,\xi,a, \Perp{x}) \leftrightarrow (k_3,\lambda, c, \Perp{z})\\
    + (k_2,\bar{\xi},b, \Perp{y}) \leftrightarrow (k_3,\lambda, c, \Perp{z}) \big\} + (C^{abcd}_{R7,A} - C^{abcd}_{R9,A}) \overline {\Ncal}^{\bar{\lambda}\lambda\bar{\xi}\xi}_{ggg,inst,A} + (C^{abcd}_{R7,B} - C^{abcd}_{R9,B}) \overline {\Ncal}^{\bar{\lambda}\lambda\bar{\xi}\xi}_{ggg,inst,B} \bigg].
\end{multline}
with the theta function and coordinates defined in Eq.\,\eqref{ggg_variable_def_first}. This is our final expression for the total amplitude of this channel. Since by construction this amplitude is completely symmetric under the exchange of any pair of final state gluons, there is no need for additional symmetry factors in the expression.

\subsection{Summary of results}
\label{sec:summary-amplitudes}
In this section we provide a summary of the results obtained for the amplitudes of the two contributing channels, including the perturbative factors, the color operators and the variables used for each topology. This section is intended to be self-contained, so that the main analytical results of this paper can be readily extracted.

\subsubsection{The $g \to q \bar{q} g$ channel}

The nine diagrams for the amplitudes are illustrated in Fig.\,\ref{fig:Trijet qbarqg}, the total amplitude can be written as a sum of three contributions
\begin{align}
    \mathcal{M}^{\bar{\lambda}\lambda\sigma\sigma',ab}_{q\bar{q}g,ij} = \mathcal{M}^{\bar{\lambda}\lambda\sigma\sigma',ab}_{q,ij} + \mathcal{M}^{\bar{\lambda}\lambda\sigma\sigma',ab}_{\bar{q},ij} + \mathcal{M}^{\bar{\lambda}\lambda\sigma\sigma',ab}_{g,ij} \,,
\end{align}
corresponding to the topologies where the final state gluon is emitted by the quark, antiquark and gluon respectively. Each contribution combines the three different insertions of the shockwave for a given topology. The amplitudes of each contribution are found to be

\begin{multline}\label{quark parent qbarqg amplitude}
    \mathcal{M}^{\bar{\lambda}\lambda\sigma\sigma',ab}_{q,ij} = \int \der^6 \boldsymbol{\Pi} \left\{\Ncal^{\bar{\lambda}\lambda\sigma\sigma'}_{q,reg} \left[\Theta_{q,1}( C_{R1,ij}^{ab} - C_{R3,ij}^{ab}) - (C_{R2,ij}^{ab} - C_{R3,ij}^{ab}) \right]\right.\\
    \left.+ (C_{R1,ij}^{ab} - C_{R3,ij}^{ab})\Ncal^{\bar{\lambda}\lambda\sigma\sigma'}_{q,inst} \right\},
\end{multline}
\begin{multline}\label{gluon parent qbarqg amplitude}
    \mathcal{M}^{\bar{\lambda}\lambda\sigma\sigma',ab}_{g,ij} = \int \der^6 \boldsymbol{\Pi} \big\{\Ncal^{\bar{\lambda}\lambda\sigma\sigma'}_{g,reg} \left[\Theta_{g,1} (C_{R4,ij}^{ab} - C_{R6,ij}^{ab}) - (C_{R5,ij}^{ab} - C_{R6,ij}^{ab}) \right] \\
    + (C_{R4,ij}^{ab} - C_{R6,ij}^{ab})\Ncal^{\bar{\lambda}\lambda\sigma\sigma'}_{g,inst} \big\}\,,
\end{multline}
\begin{multline}\label{antiquark parent qbarqg amplitude}
    \mathcal{M}^{\bar{\lambda}\lambda\sigma\sigma',ab}_{\bar q,ij} = \int \der^6 \boldsymbol{\Pi} \left\{\Ncal^{\bar{\lambda}\lambda\sigma\sigma'}_{\bar{q},reg} \left[\Theta_{\bar{q},1}( C_{\overline{R1},ij}^{ab} - C_{\overline{R3},ij}^{ab}) - (C_{\overline{R2},ij}^{ab} - C_{\overline{R3},ij}^{ab}) \right]\right.\\
    \left.+ (C_{\overline{R1},ij}^{ab} - C_{\overline{R3},ij}^{ab})\Ncal^{\bar{\lambda}\lambda\sigma\sigma'}_{\bar{q},inst} \right\},
\end{multline}
where the regular perturbative factors are
\begin{align} 
\Ncal^{\bar{\lambda}\lambda\sigma\sigma'}_{q,reg} &= \frac{2g^2p^-}{\pi^2} \sqrt{z_q z_{\bar{q}}} \delta^{\sigma, -\sigma'}\Gamma^{\sigma \lambda}_{q\xrightarrow{}qg}\left(\frac{z_q}{z_q+z_g}\right) \Gamma^{\sigma \bar{\lambda}}_{g\xrightarrow{}q\bar{q}}(z_q + z_g) \frac{\et^{\lambda*}\cdot \boldsymbol{r}_{zx}}{r_{zx}^2} \frac{\et^{\bar{\lambda}}\cdot \boldsymbol{r}_{w_1y}}{r_{w_1y}^2},\,\\
\Ncal^{\bar{\lambda}\lambda\sigma\sigma'}_{g,reg} &= -\frac{2g^2 p^-}{\pi^2} \sqrt{\frac{z_qz_{\bar{q}}}{(z_q + z_{\bar q})^2}} \delta^{\sigma,-\sigma'} \Gamma^{\sigma \eta}_{g\xrightarrow{}q\bar{q}}\left(\frac{z_q}{z_q+\zbar}\right) \frac{\et^{\eta} \cdot \boldsymbol{r}_{yx}}{r_{yx}^2} \frac{\Gt^{\bar{\lambda}\lambda \eta}(z_q + z_{\bar{q}}) \cdot \boldsymbol{r}_{w_2z}}{r_{w_2z}^2},\,\\    \Ncal^{\bar{\lambda}\lambda\sigma\sigma'}_{\bar{q},reg} &= -\frac{2g^2p^-}{\pi^2} \sqrt{z_q z_{\bar{q}}} \delta^{\sigma, -\sigma'}\Gamma^{\sigma' \lambda}_{q\xrightarrow{}qg}\left(\frac{\zbar}{\zbar + z_g}\right) \Gamma^{\sigma' \bar{\lambda}}_{g\xrightarrow{}q\bar{q}}(z_{\bar{q}} + z_g) \frac{\et^{\lambda*}\cdot \boldsymbol{r}_{zy}}{r_{zy}^2} \frac{\et^{\bar{\lambda}}\cdot \boldsymbol{r}_{w_3x}}{r_{w_3x}^2}\,.
\end{align}
For convenience, let us also express the perturbative factors in terms of the LO $g\to q\bar{q}$ perturbative factors
\begin{align} 
    \Ncal^{\bar{\lambda}\lambda\sigma\sigma'}_{q,reg} &=\frac{i g}{\pi} \sqrt{\frac{z_q}{(z_q + z_g)}} \Gamma^{\sigma \lambda}_{q\xrightarrow{}qg}\left(\frac{z_q}{z_q+z_g}\right) \frac{\et^{\lambda*}\cdot \boldsymbol{r}_{zx}}{r_{zx}^2} \mathcal{N}^{\bar{\lambda} \sigma \sigma'}_{q\bar{q},LO}(z_q+z_g,z_{\bar q}, \boldsymbol{r}_{w_{1}y}) \,,\label{N_q_LO_factorization}\\
    \Ncal^{\bar{\lambda}\lambda\sigma\sigma'}_{g,reg} &= \frac{i g}{\pi}  \frac{\Gt^{\bar{\lambda}\lambda \eta}(z_q + z_{\bar{q}}) \cdot \boldsymbol{r}_{w_2z}}{r_{w_2z}^2} \mathcal{N}^{\eta \sigma \sigma'}_{q\bar{q},LO}\left( \frac{z_q}{z_q + \zbar},\frac{\zbar}{z_q + \zbar}, \boldsymbol{r}_{xy} \right) \,, \label{N_g_LO_factorization}\\    \Ncal^{\bar{\lambda}\lambda\sigma\sigma'}_{\bar{q},reg} &= -\frac{i g}{\pi} \sqrt{\frac{z_{\bar q}}{(z_{\bar{q}} + z_g)}} \Gamma^{\sigma' \lambda}_{q\xrightarrow{}qg}\left(\frac{\zbar}{\zbar+z_g}\right) \frac{\et^{\lambda*}\cdot \boldsymbol{r}_{zy}}{r_{zy}^2} \mathcal{N}^{\bar{\lambda} \sigma' \sigma}_{q\bar{q},LO}(z_{\bar q}+z_g,z_q, \boldsymbol{r}_{w_3x})\,, \label{N_qbar_LO_factorization}
\end{align}
where the LO perturbative factor is
\begin{align}
    \Ncal^{\bar{\lambda} \sigma \sigma'}_{q\bar{q},LO}(z_q,z_{\bar q},\boldsymbol{r}_{xy}) = -\frac{2igp^-}{\pi} \sqrt{z_qz_{\bar{q}}}\, \Gamma^{\sigma \bar{\lambda}}_{g\xrightarrow{}q\bar{q}}(z_q) \delta^{\sigma,-\sigma'} \frac{\et^{\bar{\lambda}}\cdot \dipvec{xy} }{r_{xy}^2}\,.
\end{align}
These expressions will prove useful in obtaining the JIMWLK and DGLAP limits in sections \ref{sec:Slow gluon lim} and \ref{sec:DGLAP}.

\begin{figure}
    \centering
    \includegraphics[width=0.9\linewidth]{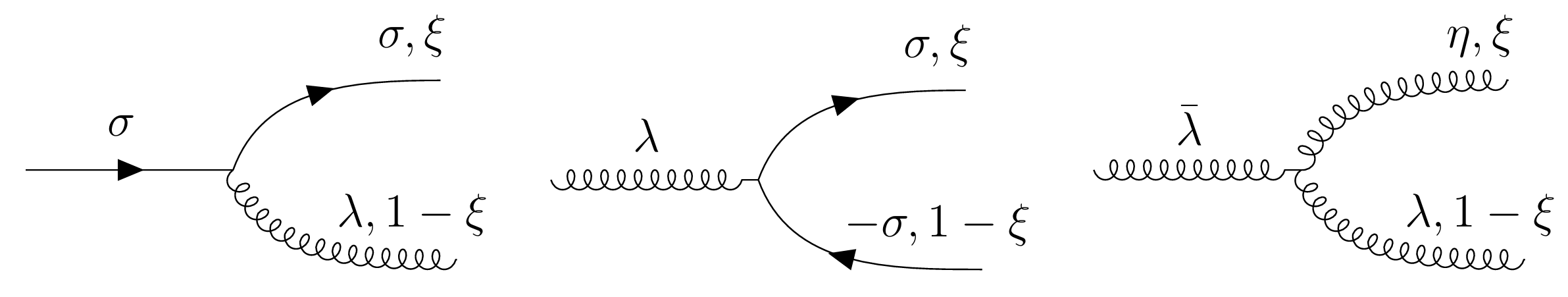}
    \caption{The diagrammatic representation of the ``square root" of the splitting functions appearing in the perturbative factors. The variable $\xi$ denotes the longitudinal momentum fraction of a daughter parton relative to the parent parton.}
    \label{fig:splitting_functions}
\end{figure}

The instantaneous perturbative factors are
\begin{align}
    \Ncal^{\bar{\lambda}\lambda\sigma\sigma'}_{q,inst} &= - \frac{g^2p^- z_{g} (z_q z_{\bar{q}})^{3/2}}{\pi^2(z_q + z_g)}   \frac{\delta^{\sigma,-\sigma'} \delta^{\sigma,\lambda} \delta^{\lambda,\bar{\lambda}}}{X_R^2},\\
    \Ncal^{\bar{\lambda}\lambda\sigma\sigma'}_{g,\mathrm{inst}} &= - \frac{g^2 p^- z_g(1+z_g) (z_qz_{\bar{q}})^{3/2}}{\pi^2(z_q + z_{\bar{q}})^2} \frac{\delta^{\sigma,-\sigma'} \delta^{\lambda,\bar{\lambda}}}{X_R^2}\,, \\
    \Ncal^{\bar{\lambda}\lambda\sigma\sigma'}_{\bar{q},inst} &= \frac{g^2p^- z_{g} (z_q z_{\bar{q}})^{3/2}}{\pi^2(z_{\bar{q}} + z_g)}   \frac{\delta^{\sigma,-\sigma'} \delta^{\sigma',\lambda} \delta^{\lambda,\bar{\lambda}}}{X_R^2}\,.
\end{align}
These perturbative factors feature the square of the $q\bar{q}g$ dipole size:
\begin{equation}
    X_R^2 = z_qz_{\bar{q}}r_{xy}^2  + z_gz_q r_{zx}^2 +z_gz_{\bar{q}} r_{yz}^2\,.
\end{equation}
The ``square root" of the splitting functions appearing in the regular perturbative factors are defined as
\begin{align}
    \Gamma^{\sigma \lambda}_{q\xrightarrow{}qg}(\xi) & \equiv \xi \delta^{\sigma, \lambda} + \delta^{\sigma, -\lambda}\,,\\
    \gtoqqbar{\sigma}{\lambda} (\xi) &\equiv (1-\xi)\delta^{\sigma,\lambda} - \xi\delta^{\sigma,-\lambda}\,,\\
    \Gt^{\bar{\lambda}\lambda\eta} (\xi) & \equiv (1-\xi)\delta^{\lambda\bar{\lambda}} \et^{\eta*} +\xi \delta^{\bar{\lambda}\eta} \et^{\lambda*} - \xi (1-\xi) \delta^{\lambda,-\eta} \et^{\bar{\lambda}}\,.
\end{align}
In Figure \ref{fig:splitting_functions} we illustrate the splitting vertices associated to each of the defined splitting functions. In the splitting functions involving a daughter quark, the variable $\xi$ corresponds to the longitudinal momentum of the quark relative to the parent parton. In the $g\to gg$ splitting function, it represents the longitudinal momentum of the daughter gluon with polarization $\eta$ relative to the parent gluon.

The color operators appearing in the amplitudes of the quark, gluon and antiquark topologies are, respectively,
\begin{align}
    C_{R1,ij}^{ab}(\Perp{x}, \Perp{y}, \Perp{z}) &= \left[\WL{x}t^ct^b\WLadj{y} U_{ac}(\Perp{z}) \right]_{ij},\\
    C_{R2,ij}^{ab}(\Transv{w}{1}, \Perp{y})  &= \left[t^a V(\Transv{w}{1})t^b\WLadj{y}\right]_{ij},\\
    C_{R3,ij}^{ab}(\Perp{v}) &= \left[t^at^c U_{cb}(\Perp{v})\right]_{ij}.
\end{align}
\begin{align}
    C_{R4,ij}^{ab}(\Perp{x}, \Perp{y}, \Perp{z}) &=\left[iV(\boldsymbol{x}_{\perp})t^d V^{\dagger}(\boldsymbol{y}_{\perp})U_{ac}(\boldsymbol{z}_{\perp}) f^{bcd} \right]_{ij},\\
    C_{R5,ij}^{ab}(\Transv{w}{2},\Perp{z}) &= \left[ i f^{cdb} U_{ed}(\Transv{w}{2})t^e U_{ac}(\Perp{z})\right]_{ij},\\
    C_{R6,ij}^{ab}(\Perp{v}) &= \left[ i t^cf^{dac} U_{db}(\Perp{v})\right]_{ij}.
\end{align}
\begin{align}
    C_{\overline{R1},ij}^{ab}(\Perp{x}, \Perp{y}, \Perp{z}) &= \left[\WL{x}t^bt^c\WLadj{y} U_{ac}(\Perp{z}) \right]_{ij},\\
    C_{\overline{R2},ij}^{ab}(\Transv{w}{3}, \Perp{x})  &= \left[\WL{x} t^b V^{\dagger}(\Transv{w}{3})t^a\right]_{ij},\\
    C_{\overline{R3},ij}^{ab}(\Perp{v}) &= \left[t^ct^a U_{cb}(\Perp{v})\right]_{ij}.
\end{align}
The theta functions of the three topologies are
\begin{align}
    \Theta_{q,1} = \frac{r_{w_1y}^2}{r_{w_1y}^2 + \frac{z_q z_g}{z_{\bar{q}}(z_q + z_g)^2} r_{zx}^2}\,, \quad \Theta_{g,1} = \frac{r_{w_2z}^2}{r_{w_2z}^2 + \frac{z_qz_{\bar{q}}}{z_g(z_q + z_{\bar{q}})^2}r_{yx}^2}\,,\quad
    \Theta_{\bar{q},1} = \frac{r_{w_3x}^2}{r_{w_3x}^2 + \frac{z_{\bar{q}} z_{g}}{z_q(z_{\bar{q}} + z_g)^2} r_{yz}^2}\,,
\end{align}
the transverse coordinates of the intermediate partons read
\begin{align}
    \boldsymbol{w}_{1\perp} = \frac{z_q\boldsymbol{x}_{\perp} + z_g\boldsymbol{z}_{\perp}}{z_q + z_g}\,,\quad \Transv{w}{2} = \frac{z_q \Perp{x} + z_{\bar{q}}\Perp{y} }{z_q + z_{\bar{q}}}\,,\quad
    \Transv{w}{3} = \frac{z_{\bar{q}}\Perp{y} + z_g \Transv{z}{} }{z_{\bar{q}} + z_g}\,,
\end{align}
and, finally, the transverse coordinate of the incoming gluon is given by
\begin{equation}
    \Transv{v}{} = \frac{z_q \Transv{x}{} + z_{\bar{q}} \Transv{y}{} + z_g \Transv{z}{}}{z_q + z_{\bar{q}} + z_g}\,.
\end{equation}

\subsubsection{The $g \to ggg$ channel}

Eleven diagrams contribute to the amplitude, five are shown in Fig.\,\ref{fig:Diag contributing to ggg channel}, and the rest are obtained by permutations of the final state gluons. The sum of the amplitudes is organized as a sum of four terms. The first term, labeled $A$, corresponds to the sum of the regular contribution to diagrams R7, R8 and R9; while the second and third term, labeled $B$ and $C$ correspond to the two distinct permutations of $A$. The fourth term gathers the sum of all instantaneous contributions and the contribution of the four-gluon vertex diagrams.

\begin{align}
    \mathcal{M}^{\bar{\lambda} \lambda \bar{\xi} \xi, abcd}_{ggg} =  \mathcal{M}^{\bar{\lambda} \lambda \bar{\xi} \xi,abcd}_{ggg,A} + \mathcal{M}^{\bar{\lambda} \lambda \bar{\xi} \xi,abcd}_{ggg,B} + \mathcal{M}^{\bar{\lambda} \lambda \bar{\xi} \xi,abcd}_{ggg,C} + \mathcal{M}_{ggg,inst}^{\bar{\lambda} \lambda \bar{\xi} \xi, abcd} \,.
\end{align}
The regular amplitude of topology A is given by
\begin{align} \label{gluon_to_ggg_reg_amp_A}
    \mathcal{M}_{ggg,A}^{\bar{\lambda} \lambda \bar{\xi} \xi} =  \int \der^6 \boldsymbol{\Pi} \bigg[\big\{\Theta_{ggg,1A}(C^{abcd}_{R7,A} - C^{abcd}_{R9,A}) - (C^{abcd}_{R8,A} - C^{abcd}_{R9,A}) \big\} \Ncal^{\bar{\lambda}\lambda\bar{\xi}\xi}_{ggg,reg\,A} \bigg].
\end{align}
The amplitudes of topologies B and C can be obtained by interchanging the quantum numbers of the final state gluons in the amplitude of topology A. For topology B, we exchange $(k_1,\xi,a, \Perp{x}) \leftrightarrow (k_3,\lambda, c, \Perp{z})$. For topology C we interchange $(k_2,\bar{\xi},b, \Perp{y}) \leftrightarrow (k_3,\lambda, c, \Perp{z})$.

The amplitude of the ``total" instantaneous contribution reads\footnote{The contribution from topology $C$ has been absorbed by employing the Jacobi identity (c.f. Eq.\,\eqref{eq:Jacobi-color}), and redefining the perturbative factors.}
\begin{align} \label{gluon_to_ggg_total_amplitude}
    \mathcal{M}_{ggg,inst}^{\bar{\lambda} \lambda \bar{\xi} \xi} =  \int \der^6 \boldsymbol{\Pi} \bigg[ (C^{abcd}_{R7,A} - C^{abcd}_{R9,A}) \overline {\Ncal}^{\bar{\lambda}\lambda\bar{\xi}\xi}_{ggg,inst,A} + (C^{abcd}_{R7,B} - C^{abcd}_{R9,B}) \overline {\Ncal}^{\bar{\lambda}\lambda\bar{\xi}\xi}_{ggg,inst,B} \bigg]\,.
\end{align}
The perturbative factor of amplitude A is given by
\begin{align}
    \Ncal^{\bar{\lambda}\lambda\bar{\xi}\xi}_{ggg,reg\,A}(\boldsymbol{x}_{\perp}, \boldsymbol{y}_{\perp}, \boldsymbol{z}_{\perp}) &= - \frac{2g^2p^-}{\pi^2} \frac{\Gt^{\eta\bar{\xi}\xi}\left(\frac{z_1}{z_1 + z_2} \right) \cdot \boldsymbol{r}_{yx}}{\boldsymbol{r}_{yx}^2} \frac{\Gt^{\bar{\lambda} \lambda\eta}(z_1+z_2)\cdot \boldsymbol{r}_{w_A z}}{r_{w_A z}^2}\,.
\end{align}
Similarly to the $q\bar{q}g$ case, let us express this perturbative factor in terms of the LO $g\to gg$ perturbative factor for later convenience:
\begin{align}
    \Ncal^{\bar{\lambda}\lambda\bar{\xi}\xi}_{ggg,reg\,A}(\boldsymbol{x}_{\perp}, \boldsymbol{y}_{\perp}, \boldsymbol{z}_{\perp}) &= \frac{ig}{\pi} \frac{\Gt^{\bar{\lambda} \lambda\eta}(z_1+z_2)\cdot \boldsymbol{r}_{w_A z}}{r_{w_A z}^2} \Ncal^{\eta\bar{\xi}\xi}_{gg,LO}\left(\frac{z_1}{z_1 + z_2} \frac{z_2}{z_1 + z_2}, \dipvec{xy}\right)\,,
    \label{ggg-reg-perturbative-factor}
\end{align}
where the LO perturbative factor is
\begin{align}
    \Ncal^{\bar{\lambda} \lambda \eta}_{gg,LO}(z_1,z_2,\boldsymbol{r}_{xy}) = -\frac{2igp^-}{\pi} \frac{\Gt^{\bar{\lambda} \lambda \eta} (z_1) \cdot \boldsymbol{r}_{xy}}{r^2_{xy}}\,.
\end{align}
The instantaneous perturbative factors are given by
\begin{multline}
      \overline {\Ncal}^{\bar{\lambda}\lambda\bar{\xi}\xi}_{ggg,inst,A} \\
      =-\frac{g^2p^- z_1 z_2 z_3 }{2\pi^2X_R^2} \left\{\left( \frac{(z_2 - z_1)(1+z_3)}{(z_1+z_2)^2} +1 \right)\delta^{\bar{\lambda}\lambda}\delta^{\xi,-\bar{\xi}}  + \left( \frac{(z_3 - z_1)(1+z_2)}{(z_1+z_3)^2} + 1 \right) \delta^{\bar{\lambda}\bar{\xi}}\delta^{\xi,-\lambda} -2\delta^{\bar{\lambda}\xi}\delta^{\lambda, -\bar{\xi}}  \right\}\,,
\end{multline}
\begin{multline}
    \overline {\Ncal}^{\bar{\lambda}\lambda\bar{\xi}\xi}_{ggg,inst,B} \\
    =-\frac{g^2p^- z_1 z_2 z_3 }{2\pi^2X_R^2} \left\{\left( \frac{(z_2 - z_3)(1+z_1)}{(z_2 + z_3)^2} +1 \right)\delta^{\bar{\lambda}\xi}\delta^{\lambda, -\bar{\xi}}  + \left( \frac{(z_1 - z_3)(1+z_2)}{(z_1+z_3)^2} + 1 \right) \delta^{\bar{\lambda}\bar{\xi}}\delta^{\xi,-\lambda} -2 \delta^{\bar{\lambda}\lambda}\delta^{\xi,-\bar{\xi}} \right\}\,.
\end{multline}
The color operators appearing in the amplitudes are defined as
\begin{align}
    \begin{split}
        C_{R7,A}^{abcd}(\boldsymbol{x}_{\perp}, \boldsymbol{y}_{\perp}, \boldsymbol{z}_{\perp}) &= f^{efg}f^{dhf} U_{ae}(\boldsymbol{x}_{\perp})U_{bg}(\boldsymbol{y}_{\perp})U_{ch}(\boldsymbol{z}_{\perp})\,,\\
        C_{R8,A}^{abcd}(\boldsymbol{w}_{A\perp}, \boldsymbol{z}_{\perp}) &= f^{eba}f^{gfd} U_{ef}(\boldsymbol{w}_{A\perp}) U_{cg}(\boldsymbol{z}_{\perp})\,,\\
        C_{R9,A}^{abcd}(\boldsymbol{v}_{\perp}) &= f^{eba}f^{fce} U_{fd}(\boldsymbol{v}_{\perp}) \,.
    \end{split}
\end{align}
Finally, the transverse coordinates and theta function of topology A are
\begin{align}\label{ggg_variable_def}
    \begin{split}
        \boldsymbol{w}_{A\perp} = \frac{z_1 \boldsymbol{x}_{\perp} + z_2\boldsymbol{y}_{\perp}}{z_1 + z_2}, &\quad\boldsymbol{v}_{\perp} = \frac{z_1 \boldsymbol{x}_{\perp} + z_2\boldsymbol{y}_{\perp} + z_3\boldsymbol{z}_{\perp}}{z_1 + z_2 + z_3} \,, \quad\Theta_{ggg,1A} = \frac{r_{w_A z}^2}{r_{w_A z}^2 + \frac{z_1z_2}{z_3(z_1 + z_2)^2}r_{yx}^2} \,.
    \end{split}
\end{align}

\section{Rapidity divergence: recovering the real part of the JIMWLK}\label{sec:Slow gluon lim}

We now focus on the regime where (one of) the emitted gluon(s) carries a very small longitudinal momentum fraction relative to its parent but with no constraint on its transverse momentum. We shall refer to this regime as the ``slow gluon" limit. After integrating over the phase space of this emitted gluon, one encounters a logarithmic rapidity divergence arising from the slow gluon regime. We will demonstrate that for each trijet channel, this rapidity divergent term is proportional to the action of the real part of the JIMWLK Hamiltonian on the corresponding dijet (leading order) cross section. 

To this end, we will first show that in the slow gluon limit, the trijet amplitudes factorize into the product of the dijet  perturbative factor, a gluon emission kernel, and the action of left and right $SU(3)$ generators on the color operator corresponding to the dijet amplitude; schematically,
\begin{align}
    \mathcal{M}^{trijet} \sim \sum_{k}  \Ncal^{dijet} \frac{\et^{\lambda*}\cdot \boldsymbol{r}_{x_k z}}{r_{x_k z}^2} [T^{a}_{x_k,L}-U_{ac}( \Transv{z}{})T^{c}_{x_k,R}] \mathcal{C}^{dijet}(\Transv{x}{1}, \Transv{x}{2}, \dots) \,.
\end{align}
This will be sufficient to demonstrate that the rapidity divergent part after integration over the gluon phase space satisfies
\begin{equation}
    \frac{\der\sigma^{gA\rightarrow dijet (g) +X}}{\der^2\boldsymbol{k}_{q\perp}\der\eta_q \der^2\boldsymbol{k}_{{\bar{q}}\perp} \der\eta_{\bar{q}} } = \int \frac{\der z_g}{z_g}\, H^{(\rm real)}_{\rm JIMWLK} \frac{\der\sigma^{gA\rightarrow dijet+ X}}{\der^2\Transv{k}{q}\der^2\Transv{k}{\bar q}\der\eta_{q}\der\eta_{\bar{q}}} \,,
\end{equation}
where the overall $1/z_g$ factor, which gives rise to the logarithmic rapidity divergence, comes from the Lorentz-invariant phase space measure of the radiated gluon, and $H^{(\rm real)}_{\rm JIMWLK}$ is the real part of the JIMWLK Hamiltonian.

Before carrying out the calculation, let us anticipate which contributions possess rapidity divergence. Only diagrams where the slow gluon is emitted adjacent to the shockwave contribute to the rapidity divergence; and hence, to the JIMWLK evolution of the color operators. This was noted in the calculation of trijets for the quark-initiated channel in Ref.~\cite{Iancu:2018hwa}. The instantaneous amplitudes do not contribute in the slow gluon limit, as they are power suppressed by the longitudinal momentum of the gluon. 

\subsection{Slow gluon limit for $g \to q\bar{q}g$ channel}

This channel features a single gluon in the final state. Hence, we can unambiguously set $z_g \ll z_q, z_{\bar q}$. This includes ignoring the recoil of the parent parton emitting the slow gluon; i.e., the transverse position before and after the gluon emission is unchanged.

Let us start with the amplitude for gluon emission by the quark. In the slow gluon limit $z_g\ll z_q, z_{\bar q}$, the ratio of transverse coordinates $\Theta_{q,1}$ can be approximated to unity. Hence the amplitude in Eq.\,\eqref{quark parent qbarqg amplitude} collapses to
\begin{equation}
    \mathcal{M}^{\bar{\lambda}\lambda\sigma\sigma',ab}_{q,ij, \rm{slow}} = \int \der^6 \boldsymbol{\Pi} ( C_{R1,ij}^{ab} - C_{R2,ij}^{ab}) \Ncal^{\bar{\lambda}\lambda\sigma\sigma'}_{q,reg}\,.
\end{equation}
Notice that the only diagrams that contribute in this limit are precisely those where the gluon radiation is adjacent to the shockwave. In this limit, the coordinate of the quark before the gluon radiation can be approximated to $\Transv{w}{1} \approx \Transv{x}{}$. Using  $\Gamma^{\sigma \lambda}_{q\xrightarrow{}qg}(1)=1$, we can approximate the regular perturbative factor as
\begin{align}
    \Ncal^{\bar{\lambda}\lambda\sigma\sigma'}_{q,reg} \approx \frac{-ig}{\pi} \Ncal^{\bar{\lambda}\sigma \sigma'}_{q\bar{q},LO} \frac{\et^{\lambda*}\cdot \boldsymbol{r}_{xz}}{r_{xz}^2}\,.
\end{align}
To deal with the color operators, let us introduce the left and right $SU(3)$ generators $T^{a}_{i,L}$ and $T^{a}_{i,R}$ which can be defined by their action on Wilson lines:
\begin{equation}
    T^{a}_{x_k,L}[V(\Transv{x}{j})]\equiv\delta_{kj}t^aV(\Transv{x}{j}), \qquad T^{a}_{x_k,R}[V(\Transv{x}{j})]\equiv\delta_{kj}V(\Transv{x}{j})t^a\,.
\end{equation}
Unitarity demands that the actions of these generators on the Hermitian conjugate of Wilson lines are
\begin{equation}
    T^{a}_{x_k,L}[V^{\dagger}(\Transv{x}{j})]\equiv-\delta_{kj}V^{\dagger}(\Transv{x}{j})t^a, \qquad T^{a}_{x_k,R}[V^{\dagger}(\Transv{x}{i})]\equiv -\delta_{kj}t^a V^{\dagger}(\Transv{x}{j}) \,.
\end{equation}
The actions on a Wilson line in the adjoint representation are:
\begin{equation}
    T^{a}_{x_k,L}[U_{bc}(\Transv{x}{j})]\equiv\delta_{kj}if^{ban}U_{nc}(\Transv{x}{j}), \qquad T^{a}_{x_k,R}[U_{bc}(\Transv{x}{j})]\equiv \delta_{kj}if^{acn}U_{bn}(\Transv{x}{j})\,.
\end{equation}
Making use of these identities, we can write the color operator $C_{R1,ij}^{ab} - C_{R2,ij}^{ab}$ as
\begin{equation}
    C_{R1,ij}^{ab} - C_{R2,ij}^{ab} = [U_{ac}(\Transv{z}{})T^{c}_{x,R} - T^{a}_{x,L}] V(\Transv{x}{})t^bV^{\dagger}(\Transv{y}{}) \,,
\end{equation}
where we recognize $V(\Transv{x}{})t^bV^{\dagger}(\Transv{y}{})$ as the part of the color operator $g \to q \bar{q}$ associated to the scattering of the shockwave with the quark-antiquark pair in the final state.

The action $[U_{ac}(\Transv{z}{})T^{c}_{x,R} - T^{a}_{x,L}]$ has a clear physical interpretation. The first term generates a gluon emission from the quark before the shockwave. As a consequence, the gluon will scatter off the shockwave, rotating its color from $c$ to $a$. This rotation is encoded in the adjoint Wilson line $U_{ac}(\Transv{z}{})$. On the other hand, the second term generates a gluon emission from the quark after the shockwave, so the gluon cannot participate in the scattering. 

The amplitude for the gluon emission from the quark in the slow gluon limit can then be written as
\begin{align}
    \mathcal{M}^{\bar{\lambda}\lambda\sigma\sigma',ab}_{q,ij, \rm{slow}} = \frac{ig}{\pi}\int \der^2\Transv{z}{} e^{-i\Transv{z}{}\cdot \Transv{k}{g}} \der^4 \boldsymbol{\Pi}_{LO} [ T^{a}_{x,L}-U_{ac}(\Transv{z}{})T^{c}_{x,R}]V(\Transv{x}{})t^bV^{\dagger}(\Transv{y}{}) \Ncal^{\bar{\lambda}\sigma \sigma'}_{q\bar{q},LO} \frac{\et^{\lambda*}\cdot \boldsymbol{r}_{xz}}{r_{xz}^2}\,,
\end{align}
where the phase space of the LO amplitude is
\begin{equation}
    \der ^4\boldsymbol{\Pi}_{LO} \equiv \der^2\Transv{x}{} \der^2\Transv{y}{} e^{-i\Transv{k}{q}\cdot \Transv{x}{}}e^{-i\Transv{k}{\bar{q}}\cdot \Transv{y}{}}\,.
\end{equation}
Let us proceed in a similar way with amplitude for gluon emission from initial state gluon. Starting from Eq.\,\eqref{gluon parent qbarqg amplitude}, and recognizing that in the slow gluon limit the fraction $\Theta_{g,1}\to 0$, we have 
\begin{equation}
    \mathcal{M}^{\bar{\lambda}\lambda\sigma\sigma',ab}_{g,ij, \rm{slow}} = \int \der^6 \boldsymbol{\Pi} ( -C_{R5,ij}^{ab} + C_{R6,ij}^{ab}) \Ncal^{\bar{\lambda}\lambda\sigma\sigma'}_{g,reg}\,,
\end{equation}
where only diagrams with the gluon adjacent to the shockwave contribute. Furthermore, $\Gt^{\bar{\lambda}\lambda \eta}(z_q + z_{\bar{q}})$ simplifies to $\delta^{\bar{\lambda}\eta}\PolVect{\lambda*}$ and the transverse coordinate of the parent gluon can be approximated to $\Transv{v}{}\approx \Transv{w}{2} \approx \Transv{w}{}$, where $\Transv{w}{}$ is the transverse coordinate of the gluon in the LO cross section. The perturbative factor can be simplified to
\begin{align}
    \Ncal^{\bar{\lambda}\lambda\sigma\sigma'}_{g,reg} \approx \frac{ig}{\pi} \Ncal^{\bar{\lambda}\sigma \sigma'}_{q\bar{q},LO} \frac{\et^{\lambda*}\cdot \boldsymbol{r}_{wz}}{r_{wz}^2}\,.
\end{align}
The color operators can also be expressed in terms of the left and right generators as
\begin{equation}
    -C_{R5,ij}^{ab} + C_{R6,ij}^{ab} = -[ T^{a}_{w,L}-U_{ac}(\Transv{z}{})T^{c}_{w,R}] t^eU_{eb}(\Transv{w}{})\,.
\end{equation}
Putting all the expressions together this amplitude in the slow gluon limit reads:
\begin{align}
    \mathcal{M}^{\bar{\lambda}\lambda\sigma\sigma',ab}_{g,ij, \rm{slow}} = -\frac{ig}{\pi}\int \der^2\Transv{z}{} e^{-i\Transv{z}{}\cdot \Transv{k}{g}} \der^4 \boldsymbol{\Pi}_{LO} [ T^{a}_{w,L}-U_{ac}(\Transv{z}{})T^{c}_{w,R}]t^eU_{eb}(\Transv{w}{}) \Ncal^{\bar{\lambda}\sigma \sigma'}_{q\bar{q},LO} \frac{\et^{\lambda*}\cdot \boldsymbol{r}_{wz}}{r_{wz}^2}\,.
\end{align}
Finally, let us address the antiquark contribution. This contribution is analogous to the quark case. In fact, the resulting expression only differs in what variable the generators acts on --- in this case, on the position of the antiquark $\Transv{y}{}$ --- and the kernel of the radiated gluon. We thus have
\begin{align}
    &\mathcal{M}^{\bar{\lambda}\lambda\sigma\sigma',ab}_{\bar{q},ij, \rm{slow}}\nonumber\\
    &= \frac{ig}{\pi}\int \der^2\Transv{z}{} e^{-i\Transv{z}{}\cdot \Transv{k}{g}} \der^4 \boldsymbol{\Pi}_{LO} [ T^{a}_{y,L}-U_{ac}(\Transv{z}{})T^{c}_{y,R}]V(\Transv{x}{})t^bV^{\dagger}(\Transv{y}{}) \Ncal^{\bar{\lambda}\sigma \sigma'}_{q\bar{q},LO} \frac{\et^{\lambda*}\cdot \boldsymbol{r}_{yz}}{r_{yz}^2}\,.
\end{align}
In the slow gluon limit, the sum of the amplitude is given by 
\begin{align}\label{amplitude_qqbarg_slow_limit}
    &\mathcal{M}^{\bar{\lambda}\lambda\sigma\sigma',ab}_{q\bar{q}g,ij, \rm{slow}} \nonumber \\
    &= \frac{ig}{\pi}\int \der^2\Transv{z}{} e^{-i\Transv{z}{}\cdot \Transv{k}{g}} \der^4 \boldsymbol{\Pi}_{LO} \Ncal^{\bar{\lambda}\sigma \sigma'}_{q\bar{q},LO}  \sum_{x_k=x,y,w} \frac{\et^{\lambda*}\cdot \boldsymbol{r}_{x_kz}}{r_{x_kz}^2} [ T^{a}_{x_k,L}-U_{ac}(\Transv{z}{})T^{c}_{x_k,R}]C^{b}_{q\bar{q},LO,ij}(\Transv{x}{}, \Transv{y}{}) \,,
\end{align}
where the variable $\Transv{w}{}$ should be treated as independent of $\Transv{x}{}$ and $\Transv{y}{}$ with respect to the action of the generators on the Wilson lines.

The rapidity divergent part of the differential cross section can be readily obtained using Eq.\,\eqref{trijet cross section general} from section~\ref{sec: Cross section trijets} using the slow gluon amplitude, integrating over the phase space of the gluon, and neglecting the longitudinal momentum of the gluon $z_g$ inside the delta function:
\begin{equation}
    \frac{\der\sigma^{gA\rightarrow q\bar{q} (g) +X}}{\der^2\boldsymbol{k}_{q\perp}\der\eta_q \der^2\boldsymbol{k}_{{\bar{q}}\perp} \der\eta_{\bar{q}} } = \int \frac{\der z_g}{z_g} \int \der^2 \Transv{k}{g} \frac{1}{8(2\pi)^8} \frac{\delta(1-z_q - z_{\bar{q}})}{2(p^-)^2} \frac{1}{2(\N^2 -1)} \sum_{\bar{}}\left\langle  \big|\mathcal{M}_{\rm slow}\big|^2\right\rangle_Y\,.
\end{equation}
The square of the amplitude Eq.\,\eqref{amplitude_qqbarg_slow_limit} gives
\begin{align}
    &\sum_{\begin{smallmatrix} \bar{\lambda}\lambda\sigma\sigma',  \\ ab,ij \end{smallmatrix}}\left\langle  \big|\mathcal{M}_{\rm slow}\big|^2\right\rangle_Y = -\frac{g^2}{\pi^2}\int \der^2 \Transv{z}{} \der^2 \Transv{z}{}' e^{-i\Transv{k}{g} \cdot (\Transv{z}{} - \Transv{z}{}')} \der^4 \boldsymbol{\Pi}_{LO}\der^4 \boldsymbol{\Pi}'_{LO} \sum_{\bar{\lambda}\sigma\sigma'} \Ncal^{\bar{\lambda} \sigma \sigma'}_{q\bar{q},LO} \left(\Ncal^{\bar{\lambda} \sigma \sigma'}_{q\bar{q},LO} \right)^{\dagger}\nonumber\\
    &\times \sum_{\begin{smallmatrix} x_k=x,y,w\\ x_j=x',y',w'\end{smallmatrix}} \frac{\dipvec{x_jz'}\cdot \boldsymbol{r}_{x_kz}}{r_{x_jz'}^2r_{x_kz}^2} [ T^{a}_{x_j,L}-U_{ac}(\Transv{z}{}')T^{c}_{x_j,R}] [ T^{a}_{x_k,L}-U_{ac}(\Transv{z}{})T^{c}_{x_k,R}]\Tr\left[C^{b,\dagger}_{q\bar{q},LO} C^{b}_{q\bar{q},LO}  \right]\,.
\end{align}
The appearance of the extra $-1$ comes from the action of the generators on the Hermitian conjugate of the color operator, which results in the inverse of the color operators of the c.c.~trijet amplitudes. Finally, doing the integration over the transverse momentum of the gluon $\Transv{k}{g}$ which identifies the coordinates $\Transv{z}{}=\Transv{z}{}'$ gives the result
\begin{equation}\label{qqbar_cross_section_JIMWLK_evol}
    \frac{\der\sigma^{gA\rightarrow q\bar{q} (g) +X}}{\der^2\boldsymbol{k}_{q\perp}\der\eta_q \der^2\boldsymbol{k}_{{\bar{q}}\perp} \der\eta_{\bar{q}} } = \int \frac{\der z_g}{z_g}\, H^{(\rm real)}_{\rm JIMWLK} \frac{\der\sigma^{gA\rightarrow q\bar{q} +X}}{\der^2\Transv{k}{q}\der^2\Transv{k}{\bar q}\der\eta_{q}\der\eta_{\bar{q}}}\,,
\end{equation}
where we have introduced the real part of the JIMWLK Hamiltonian
\begin{equation}
    H^{(\rm real)}_{\rm JIMWLK} \equiv \frac{\alpha_s}{2\pi^2} \int \der^2 \Transv{z}{}  \sum_{\begin{smallmatrix} x_k=x,y,w\\ x_j=x',y',w'\end{smallmatrix}} \left(-\frac{2 \dipvec{x_jz}\cdot \boldsymbol{r}_{x_kz}} {r_{x_jz}^2r_{x_kz}^2} \right) [ T^{a}_{x_j,L}-U_{ac}(\Transv{z}{})T^{c}_{x_j,R}] [T^{a}_{x_k,L}-U_{ac}(\Transv{z}{})T^{c}_{x_k,R}]\,.
    \label{eq:JIMWLK-real}
\end{equation}
In order to connect this result to the full JIMWLK Hamiltonian, let us now discuss the expected structure of the virtual corrections. They come in two types: 1) vertex corrections (gluon is emitted and absorbed by different partons) and 2) self-energy diagrams (gluon is emitted and absorbed by same parton). Summing real and vertex contributions, we have
\begin{equation}
    H^{(\rm real+vertex)}_{\rm JIMWLK} = \frac{\alpha_s}{4\pi^2} \int \der^2 \Transv{z}{}  \sum_{\begin{smallmatrix} x_k,x_j\\ x_j \neq x_k\end{smallmatrix}} \left(-\frac{2 \dipvec{x_jz}\cdot \boldsymbol{r}_{x_kz}} {r_{x_jz}^2r_{x_kz}^2} \right) [ T^{a}_{x_j,L}-U_{ac}(\Transv{z}{})T^{c}_{x_j,R}] [T^{a}_{x_k,L}-U_{ac}(\Transv{z}{})T^{c}_{x_k,R}]\,,
    \label{eq:JIMWLK-real-vertex}
\end{equation}
where the effect of the vertex corrections is that the sum now includes the possibility that $x_j$ and $x_k$ are coordinates both belonging to the direct amplitude or both to the conjugate amplitude. The sum now runs over all coordinates $x_j,x_k \in \{x,y,w,x',y',w'\}$. The factor of $1/2$ in front Eq.\,\eqref{eq:JIMWLK-real-vertex} compared to Eq.\,\eqref{eq:JIMWLK-real} avoid for double counting.

Eq.\,\eqref{eq:JIMWLK-real-vertex} has a divergence coming from the transverse
coordinate of the integrated gluon, where the kernel presents an infrared divergence for large values of $\Transv{z}{}$. This divergence is cured by including self-energy contributions. Indeed, these corrections bring terms proportional to the kernel $1/r_{x_j z}^2$. Using the identity
\begin{align}
\frac{1}{r_{x_j z}^2} + \frac{1}{r_{x_k z}^2} -\frac{2 \dipvec{x_jz}\cdot \boldsymbol{r}_{x_kz}} {r_{x_jz}^2r_{x_kz}^2} =\frac{r_{x_jx_k}^2}{r_{x_jz}^2r_{x_kz}^2}\,,
\end{align}
the complete JIMWLK Hamiltonian reads \cite{Caron-Huot:2013fea,Hentschinski:2018rrf}
\begin{equation}
    H_{\rm JIMWLK} \equiv \frac{\alpha_s}{4\pi^2} \int \der^2 \Transv{z}{}  \sum_{\begin{smallmatrix} j,k\end{smallmatrix}} \frac{r_{x_jx_k}^2}{r_{x_jz}^2r_{x_kz}^2} [ T^{a}_{x_j,L}-U_{ac}(\Transv{z}{})T^{c}_{x_j,R}] [T^{a}_{x_k,L}-U_{ac}(\Transv{z}{})T^{c}_{x_k,R}]\,,
    \label{eq:JIMWLK-hamiltonian-full}
\end{equation}
which is infrared safe.
Then, by writing the integration over the rapidity variable in terms of the longitudinal momentum of the gluon $\int \der z_g/z_g$, we see that the lower bound of the integral provides one step in the JIMWLK evolution of the leading order  $g\to q\bar{q}$ cross section. The evolution equation, for a given operator $O$ is given by
\begin{equation}
    \frac{d}{d\eta}\langle O\rangle_{\eta} = H_{\rm JIMWLK} \langle O\rangle_{\eta}\,,\label{eq:JIMWLK-full}
\end{equation}
with $\der z_g/z_g=\der\eta$ and $\eta$ the rapidity of the gluon.

In the physical cross section, the $1/z_g$ divergence would naturally be regulated by imposing conservation of the plus longitudinal momentum in the $qg \to q\bar q g$ process where the gluon on the right hand side is part of the shockwave in our formalism.


\subsection{Slow gluon limit for $g \to ggg$ channel}

\begin{figure}
    \centering
    \includegraphics[width=1\linewidth]{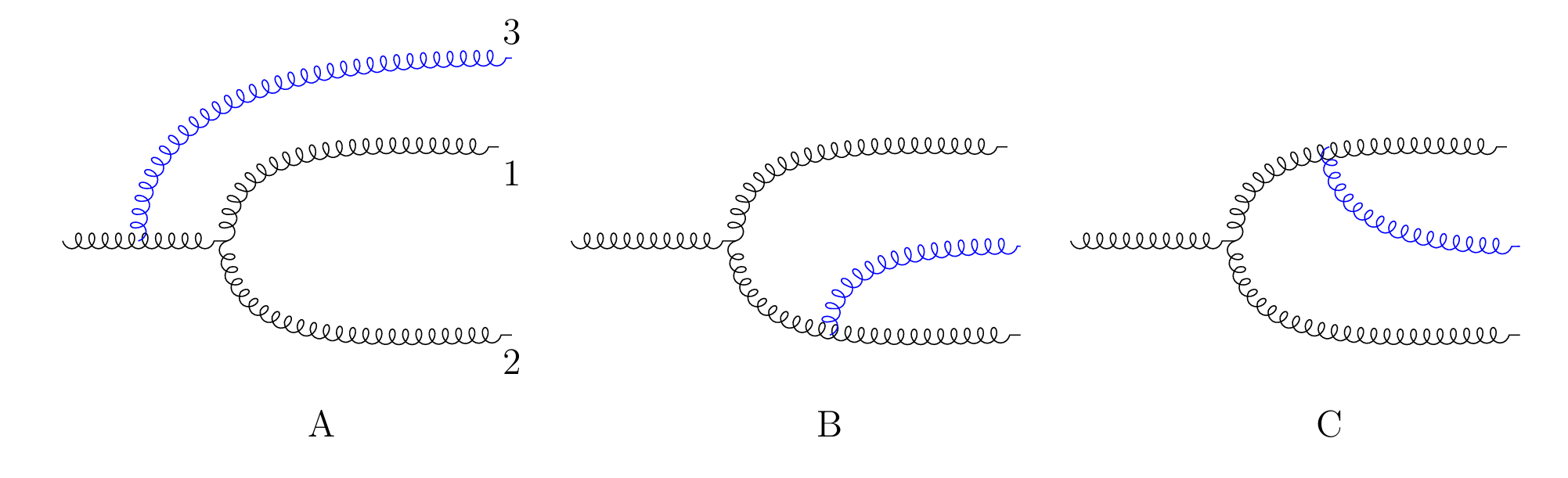}
    \caption{The three different topologies A, B, and C contributing to the slow gluon limit amplitude for $ggg$ production}
    \label{fig:ggg_slow_gluon_topologies}
\end{figure}

Since the amplitudes are symmetric with respect to the exchange of any pair of momenta, it is sufficient to focus on the limit where either of the gluons is slow. Let us work in the limit where the gluon with momenta $k_3$ is slow, i.e., $z_3\ll z_1,z_2$. As for the $q\bar{q}g$ case, in this limit, we can set the longitudinal momentum of the gluon to $z_3$ to 0 in the perturbative factor. The recoil of the parent gluon which emits the slow gluon can be neglected, so its transverse coordinates before and after the emission are the same. In Figure \ref{fig:ggg_slow_gluon_topologies}, we depict the diagrams of the three contributing topologies in the case where gluon $3$ is slow. As the picture suggests, topology A corresponds to the case where the slow gluon is emitted by the incoming gluon. Topologies B and C on the other hand describe a slow gluon emission by gluons $3$ and $2$, respectively.

Let us start with topology A. In the $z_3\to 0$ limit, we can make the approximations $\Theta_{ggg,1A}\to 0$ and $\Transv{w}{A} \approx \Transv{v}{} \approx z_1\Transv{x}{} + z_2\Transv{y}{} =\Transv{w}{}$. The amplitude in Eq.\,\eqref{gluon_to_ggg_reg_amp_A} then simplifies to
\begin{equation}
    \mathcal{M}_{ggg,A, \mathrm{slow}}^{\bar{\lambda} \lambda \bar{\xi} \xi, abcd} =  \int \der^6 \boldsymbol{\Pi} (-C_{R8,A}^{abcd} + C^{abcd}_{R9,A}) \Ncal^{\bar{\lambda}\lambda\bar{\xi}\xi}_{ggg,regA} \,.
\end{equation}
Using the fact that $z_3 G_{\perp}^{\bar{\lambda} \lambda \eta}(z_1+z_2) \to \delta^{\bar{\lambda} \eta} \PolVect{\lambda*}$ in this limit, the perturbative factor reduces to 
\begin{equation}
\Ncal^{\bar{\lambda}\lambda\bar{\xi}\xi}_{ggg,regA} \approx \frac{ig}{\pi} \frac{\PolVect{\lambda*}\cdot \dipvec{wz}}{r_{wz}^2} \Ncal^{\bar{\lambda}\xi\bar{\xi}}_{gg,LO}\,.
\end{equation}
The color operator can be expressed in terms of the action of the generators on the initial state color operator of the LO $g\to gg$ amplitude:
\begin{equation}
    -C_{R8,A}^{abcd} + C^{abcd}_{R9,A} = [ T^{c}_{w,L} - U_{ce}(\Transv{z}{})T^{e}_{w,R}][if^{gba}U_{gd}(\Transv{w}{})]\,.
\end{equation}
Substituting these expressions in the amplitude gives
\begin{equation}
    \mathcal{M}_{ggg,A,\mathrm{slow}}^{\bar{\lambda} \lambda \bar{\xi} \xi, abcd} = \frac{ig}{\pi} \int \der^2\Transv{z}{} \der^4 \boldsymbol{\Pi}_{LO} \Ncal^{\bar{\lambda}\xi\bar{\xi}}_{gg,LO} [T^{c}_{w,L}-U_{ce}(\Transv{z}{})T^{e}_{w,R}][if^{gba}U_{gd}(\Transv{w}{})] \frac{\PolVect{\lambda*}\cdot \dipvec{wz}}{r_{wz}^2}\,.
\end{equation}
We can carry a similar exercise for topology B. We remind the reader that the amplitude for this topology can be obtained from exchanging the quantum numbers of gluons $1$ and $3$ from Eq.\,\eqref{gluon_to_ggg_reg_amp_A}. Then, if $z_3\to 0$ we have 
\begin{align}\label{theta_ggg_B}
    \Theta_{ggg,1B}&=\frac{r_{w_Bx}^2}{r_{w_Bx}^2+\frac{z_3z_2}{z_1(z_3+z_2)^2}r_{yz}^2}\to 1\,,
\end{align}
with $\boldsymbol{w}_{B\perp}=(z_3\Perp{z}+z_2\Perp{y})/(z_3+z_2) \approx \Transv{y}{}$. The amplitude thus simplifies to 
\begin{equation}
    \mathcal{M}_{ggg,B,\mathrm{slow}}^{\bar{\lambda} \lambda \bar{\xi} \xi, abcd} =  \int \der^6 \boldsymbol{\Pi} (C_{R7,B}^{abcd} - C^{abcd}_{R8,B}) \Ncal^{\bar{\lambda}\lambda\bar{\xi}\xi}_{ggg,regB} \,.
\end{equation}
Analogous to the previous cases, the perturbative factor in this limit becomes
\begin{equation}
    \Ncal^{\bar{\lambda}\lambda\bar{\xi}\xi}_{ggg,regB} \approx \frac{ig}{\pi} \frac{\PolVect{\lambda*}\cdot \dipvec{yz}}{r_{yz}^2} \Ncal^{\bar{\lambda}\xi\bar{\xi}}_{gg,LO}\,.
\end{equation}
The color operator can be written in terms of the left and right operators acting on the initial state color operator of the LO $g\to gg$ amplitude:
\begin{equation}
    C_{R7,B}^{abcd} - C^{abcd}_{R8,B} = - [ T^{c}_{y,L}-U_{ce}(\Transv{z}{})T^{e}_{y,R}][if^{hdf}U_{ah}(\Transv{x}{})U_{bf}(\Transv{y}{})]\,.
\end{equation}
Then, the amplitude of this topology can be written as
\begin{equation}
    \mathcal{M}_{ggg,B, \mathrm{slow}}^{\bar{\lambda} \lambda \bar{\xi} \xi, abcd} = -\frac{ig}{\pi} \int \der^2\Transv{z}{} \der^4 \boldsymbol{\Pi}_{LO} \Ncal^{\bar{\lambda}\xi\bar{\xi}}_{gg,LO} [T^{c}_{y,L}-U_{ce}(\Transv{z}{})T^{e}_{y,R}][if^{hdf}U_{ah}(\Transv{x}{})U_{bf}(\Transv{y}{})]\frac{\PolVect{\lambda*}\cdot \dipvec{yz}}{r_{yz}^2}\,.
\end{equation}
The remaining task is to obtain the slow gluon approximation for topology C. As a reminder, this topology interchanges gluons $2$ and $3$ from topology A. In this topology, the slow gluon $3$ is emitted from the gluon $1$. The resulting amplitude in this approximation will thus be closely related to the amplitude of topology B, in the sense that the color operator can be expressed as the generators acting on the same LO operator, but acting on $\Transv{x}{}$ instead of $\Transv{y}{}$. This is analogous to the quark and the antiquark amplitudes in the $q\bar{q}g$ case. We will then skip the intermediate steps of the derivation and present the final expression for the amplitude. The expression reads
\begin{equation}
    \mathcal{M}_{ggg,C,\mathrm{slow}}^{\bar{\lambda} \lambda \bar{\xi} \xi, abcd} =- \frac{ig}{\pi} \int \der^2\Transv{z}{} \der^4 \boldsymbol{\Pi}_{LO} \Ncal^{\bar{\lambda}\xi\bar{\xi}}_{gg,LO} [T^{c}_{x,L} - U_{ce}(\Transv{z}{})T^{e}_{x,R}][if^{hdf}U_{ah}(\Transv{x}{})U_{bf}(\Transv{y}{})]\frac{\PolVect{\lambda*}\cdot \dipvec{xz}}{r_{xz}^2}\,.
\end{equation}
Combining the amplitudes of the three contributing topologies, we have the following expression for the $ggg$ amplitude in the slow gluon limit:
\begin{equation}\label{amplitude_ggg_slow_limit}
    \mathcal{M}^{\bar{\lambda}\lambda\xi \bar{\xi},abcd}_{ggg,\mathrm{slow}} = -\frac{ig}{\pi}\int \der^2\Transv{z}{} e^{-i\Transv{z}{}\cdot \Transv{k}{3}} \der^4 \boldsymbol{\Pi}_{LO} \Ncal^{\bar{\lambda} \xi \bar{\xi}}_{gg,LO}  \sum_{x_k=x,y,w} \frac{\et^{\lambda*}\cdot \boldsymbol{r}_{x_kz}}{r_{x_kz}^2} [ T^{c}_{x_k,L}-U_{ce}(\Transv{z}{})T^{e}_{x_k,R}]C^{abd}_{gg,LO}(\Transv{x}{}, \Transv{y}{})\,. 
\end{equation}
From here, the remaining steps to obtain the JIMWLK evolution of the LO $g\to gg$ cross section are exactly the same as for the $q\bar{q}g$ case. This concludes our derivation of the (real) JIMWLK evolution of the dijet cross section in forward pA collisions in the gluon-initiated channel.

\section{Collinear divergences: recovering the real part of DGLAP}
\label{sec:DGLAP}

In this section, we isolate the collinear divergences arising from the real corrections to the dijet production cross section in gluon-initiated channels. Our goal is twofold: (i) the collinear limits of the trijet cross section provide a nontrivial cross-check of the results obtained in this work; and (ii) this analysis constitutes an essential step toward deriving the complete NLO cross section for dihadron or dijet production in the CGC framework. In particular, the factorization of collinear divergences into the evolution of the initial-state PDF and the final-state fragmentation functions allows for a clean separation of the finite contributions, commonly referred to as the NLO ``impact factor".

Before outlining the calculation of the collinear limits, let us first state a few general observations that apply to all channels, independently of whether the divergences arise from the initial or final state:
\begin{itemize}
\item As in the previous section on gluon rapidity divergences, the color structures of the relevant 
topologies can be combined together, since the perturbative factor also factorizes in the collinear limit. However, unlike in the $z_g \to 0$ limit, it is crucial to retain the exact kinematic dependence of the transverse coordinates at which the color structures are evaluated, as it encodes the recoil associated with the collinear emission, which cannot be neglected in the DGLAP regime.
\item Instantaneous terms (as well as four-gluon vertex contributions) do not contribute to the DGLAP limits.
\end{itemize}

\subsection{DGLAP evolution of the initial state gluon in the $g\to q\bar q$ channel}\label{sec: DGLAP-initial-pA-to-qqb}

We study the collinear limit at the amplitude level for simplicity. In coordinate space, the collinear divergence manifests itself as an infrared divergence when the transverse distance between the emitted gluon and the quark-antiquark pair, of the order of $r_{w_2z}$, is much larger than the transverse separation $r_{xy}$ between the quark and the antiquark. Considering Eq.\,\eqref{gluon parent qbarqg amplitude}, one realizes that in this limit, the function $\Theta_{g,1}$ reduces to unity such that the color structures nicely combine together to give
\begin{align}
    \mathcal{M}^{\bar{\lambda}\lambda\sigma\sigma',ab}_{g,ij} \simeq \int \der^6 \boldsymbol{\Pi} \ \Ncal^{\bar{\lambda}\lambda\sigma\sigma'}_{g,reg} \left[C_{R4,ij}^{ab}- C_{R5,ij}^{ab} \right],
\end{align}
where the instantaneous term is negligible in the collinear limit. The integral over the gluon transverse momentum $\boldsymbol{k}_{g\perp}$ fixes $\Perp{z}=\Perp{z'}$ since $\boldsymbol{k}_{g\perp}$ only appears in the overall phase. Therefore, at the cross section level (and averaging over the colors of the incoming gluon), the color structure in the limit where the emitted gluon is collinear to the incoming one reads
\begin{align}
    &\frac{1}{N_c^2-1}\textrm{Tr}\left\{\left[V(\boldsymbol{x}_{\perp})t^d V^{\dagger}(\boldsymbol{y}_{\perp})U_{ac}(\boldsymbol{z}_{\perp}) f^{bcd} - f^{bcd} U^{\dagger}_{de}(\Perp{w_2})t^e U_{ac}(\Perp{z})\right]\right.\nonumber\\
    &\left.\times\left[f^{bc'd'}U_{c'a}^\dagger(\Perp{z})V(\Perp{y}')t^{d'}V^\dagger(\Perp{x}')-f^{bc'd'}U_{e'd'}(\Perp{w_2}')t^{e'}U^\dagger_{c'a}(\Perp{z})\right]\right\}\,,
\end{align}
where we recall that the transverse coordinate $\Perp{w_2}=(z_q\Perp{x}+z_{\bar q}\Perp{y})/(z_q+z_{\bar q})$ is the center-of-mass coordinate of the $q\bar q$ pair. A straightforward calculation of this trace, using the unitarity of Wilson lines and the color algebra identities  $f^{acd}f^{acd'}=N_c\delta^{dd'}$, $\textrm{Tr}(t^{a}t^b)=\frac{1}{2}\delta^{ab}$ yields the correlator 
\begin{align}
   \frac{N_c}{2}\Xi_{q\bar q,LO}(\Perp{x},\Perp{y},\Perp{x'},\Perp{y'}) =\frac{N_c}{2(N_c^2-1)}&\left\{N_c^2 D_{xx'}D_{y'y}-Q_{xyy'x'}+N_c^2D_{w_2w_2'}D_{w_2'w_2}-1\right.\nonumber\\
&\left.-N_c^2D_{xw_2'}D_{w_2'y}+D_{xy}-N_c^2D_{w_2x'}D_{y'w_2}+D_{y'x'}\right\}\,,
\end{align}
at the cross section level, which is almost identical to the LO color correlator up to an overall $N_c/2$ factor. The only difference lies in the definition of the transverse coordinate $\Perp{w_2}$ in terms of $\Perp{x}$ and $\Perp{y}$ (and likewise for $\Perp{w_2'}$), which is apparently distinct from the LO case because $z_q+z_{\bar q}=1-z_g\neq 1$. As we shall see, this difference is actually essential to obtain factorization.

In order to make appear the $g\to gg$ DGLAP splitting function, it is simpler to express the perturbative factor $\Ncal^{\bar{\lambda}\lambda\sigma\sigma'}_{g,reg}$ computed in Eq.\,\eqref{N_g_LO_factorization} in the following form
\begin{align}
    \Ncal^{\bar{\lambda}\lambda\sigma\sigma'}_{g,reg}&=\frac{ig}{\pi} \sum_{\eta} \frac{\Gt^{\bar{\lambda}\lambda \eta}(\xi) \cdot \boldsymbol{r}_{w_2 z}}{r_{w_2 z}^2}
\mathcal{N}_{q\bar{q},LO}^{\eta \sigma \sigma'}\left(\frac{z_q}{\xi},\frac{z_{\bar q}}{\xi},\boldsymbol{r}_{xy} \right)\,,
\end{align}
where $\xi\equiv 1-z_g=z_q+z_{\bar q}$. As mentioned in section~\ref{sub:gggLO}, the ``square" of the triple gluon vertex $\Gt$ can be related to the $g\to gg$ splitting function thanks the key identity
\begin{align}
    \sum_{\lambda \bar{\lambda}} \Gt^{\bar{\lambda}\lambda \eta,i}(\xi) \Gt^{\bar{\lambda}\lambda \eta'*,j}(\xi)&= \frac{\xi (1-\xi)}{2N_c} \left[   (2\et^{\eta*,i} \et^{\eta',j} -\delta^{ij} \delta^{\eta \eta'}) P_{g g_L}^{real}(\xi) +  \delta^{\eta \eta'} \delta^{ij} P^{real}_{gg}(\xi) \right]\nonumber\\
    &+ 2 \xi^2 (1-\xi) \left( \et^{\eta*,i} \et^{\eta',j}-\et^{\eta',i} \et^{\eta*,j} \right)\,.\label{eq:GG-tensor}
\end{align}
In this expression, $P_{gg}^{real}$ is the real part of the unpolarized $g\to gg$ DGLAP splitting function:
\begin{align}
    P_{gg}^{real}(\xi)&=\frac{2N_c[1-\xi(1-\xi)]^2}{\xi(1-\xi)}\,,
\end{align}
while $P_{gg_L}^{real}$ describes the splitting of an unpolarized gluon into a linearly polarized one and reads
\begin{align}
    P_{gg_L}^{real}(\xi)&=\frac{2N_c(1-\xi)}{\xi}\,.
\end{align}
The first line of Eq.\,\eqref{eq:GG-tensor} features the unpolarized DGLAP splitting function which associated to the $\delta^{ij}$ tensor, whereas the splitting function into linearly polarized gluons multiplies the structure $2\epsilon^{\eta*,i}_{\perp} \epsilon^{\eta',j}_{\perp} -\delta^{ij} \delta^{\eta \eta'}$, a tensor orthogonal to $\delta^{ij}$. On the other hand, the second line is fully antisymmetric under $i\leftrightarrow j$ exchange.
When squaring the perturbative factor $\Ncal^{\bar{\lambda}\lambda\sigma\sigma'}_{g,reg}$, the $(i,j)$ tensor in Eq.\,\eqref{eq:GG-tensor} is ultimately contracted with the tensor
\begin{align}
\int\der^2\Perp{z}\frac{ \dipvec{w_2z}^i\dipvec{w_2'z}^{j}}{r_{w_2'z}^2 r_{w_2z}^2}\,,
\end{align}
which is symmetric under the exchange $(i,j)$. Moreover, the leading IR divergence of this integral (for instance, when computed using dimensional regularization as below) arises from the trace part of the $(i,j)$ tensor:
\begin{align}
\int\der^2\Perp{z}\frac{ \dipvec{w_2z}^i\dipvec{w_2'z}^{j}}{r_{w_2'z}^2 r_{w_2z}^2}
=\frac{\delta^{ij}}{2}\int\der^2\Perp{z}\frac{\dipvec{w_2'z}\cdot \dipvec{w_2z}}{r_{w_2'z}^2 r_{w_2z}^2}+\textrm{IR finite}\,.
\end{align}
Therefore, the contraction with Eq.\,\eqref{eq:GG-tensor} ultimately selects the $\delta^{ij}$ term in the first line of Eq.\,\eqref{eq:GG-tensor}, which is precisely the term proportional to $P_{gg}^{real}(\xi)$. It also depends on $\delta^{\eta\eta'}$, implying that the polarization of the intermediate gluon producing the $q\bar q$ pair is the same in both the amplitude and its complex conjugate, as in the LO case.

Finally, after integrating over $z_g$ with the Lorentz-invariant measure $\der z_g/z_g$, we obtain
\begin{align}
    &\int\frac{\der z_g}{z_g}\int\der^2\Perp{z}\sum_{\bar\lambda\lambda\sigma\sigma'}\Ncal^{\bar{\lambda}\lambda\sigma\sigma'}_{g,reg}\Ncal^{\bar{\lambda}\lambda\sigma\sigma'*}_{g,reg}=\frac{g^2}{2\pi^2N_c}\int_0^1 \der \xi\ \xi P_{gg}^{real}(\xi) \nonumber \\
    & \times \sum_{\eta\sigma\sigma'}\mathcal{N}_{q\bar{q},LO}^{\eta \sigma \sigma'}\left(\frac{z_q}{\xi},\frac{z_{\bar q}}{\xi},\boldsymbol{r}_{xy} \right) \mathcal{N}_{q\bar{q},LO}^{\eta \sigma \sigma',*}\left(\frac{z_q}{\xi},\frac{z_{\bar q}}{\xi},\boldsymbol{r}_{x'y'} \right)\int\der^2\Perp{z}\frac{\dipvec{w_2'z}\cdot \dipvec{w_2z}}{r_{w_2'z}^2 r_{w_2z}^2}+\textrm{IR finite}\,.\label{eq:DGLAP-init-qqg-step}
\end{align}
We have checked that the same result can be obtained by performing the contraction of the kernel $\mathcal{K}_4$ defined in in Eq.\,\eqref{eq:K4-def}  with $\delta^{ll'}/2$. This result explicitly shows the factorization of the LO perturbative factor for the $g\to q\bar q$ process.

Gathering all prefactors (see e.g.~Eq.\,\eqref{trijet cross section general} in the next section) needed to relate the amplitude squared to the actual cross section, we obtain the following simple result for the limit of the cross section when the radiated gluon is collinear to the incoming one:
\begin{align}
    \frac{\der\sigma^{gA\to q\bar q(g)+X}}{\der^2\boldsymbol{k}_{q\perp}\der\eta_q \der^2\boldsymbol{k}_{{\bar{q}}\perp} \der\eta_{\bar{q}}}&\simeq\frac{\alpha_s}{(2\pi)^6}\int_0^1 \der \xi \frac{\alpha_s}{2\pi}P_{gg}^{real}(\xi)\delta\left(1-\frac{z_q}{\xi}-\frac{z_{\bar q}}{\xi}\right)\nonumber\\
&\times\int\der^2\Perp{x}\der^2\Perp{y}\der^2\Perp{x'}\der^2\Perp{y'} e^{-i\boldsymbol{k}_{q\perp}\cdot(\Perp{x}-\Perp{x'})-i\boldsymbol{k}_{\bar q\perp}\cdot(\Perp{y}-\Perp{y'})}\nonumber\\
    &\times \mathcal{K}_{q\bar q,LO}\left(\frac{z_q}{\xi},\frac{z_{\bar q}}{\xi},\boldsymbol{r}_{xy},\boldsymbol{r}_{x'y'}\right)\left\langle\Xi_{q\bar q,LO}(\Perp{x},\Perp{y},\Perp{x'},\Perp{y'})\right\rangle_Y\int\frac{\der^2\Perp{z}}{\pi}\frac{\boldsymbol{r}_{w_2z}\cdot\boldsymbol{r}_{w_2'z}}{r_{w_2z}^2r_{w_2'z}^2}\,,\label{eq:NLO-collinear-limit-is}
\end{align}
where we used
\begin{align}
    \delta(\xi-z_q-z_{\bar q})&=\frac{1}{\xi} \left(1-\frac{z_q}{\xi}-\frac{z_{\bar q}}{\xi}\right)\,.
\end{align}
The integral over $\Perp{z}$ is IR divergent. In dimensional regularization, 
\begin{align}
    \mu^{-\varepsilon}\int\frac{\der^{2-\varepsilon}\Perp{z}}{\pi}\frac{\boldsymbol{r}_{w_2z}\cdot\boldsymbol{r}_{w_2'z}}{r_{w_2z}^2r_{w_2'z}^2}&=\frac{2}{\varepsilon}-\ln(e^{\gamma_E}\pi\mu^2r_{w_2w_2'}^2)+\mathcal{O}(\varepsilon)\,.\label{eq:WWkernel-dimreg}
\end{align}
Recall that the LO cross-section in Eq.\,\eqref{dijet_cross_section_LO} is evaluated at $p^- =  x _p q^-$. On the other hand, Eq.\,\eqref{eq:NLO-collinear-limit-is} features the LO cross-section but with $z_i \to z_i/\xi$. This rescaling can be effectively absorbed by evaluating the cross-section with $p^- =\xi x_p q^-\,$ \footnote{Note that the transverse coordinate scale $\Perp{w_2}$ in the LO color structure is invariant under the rescaling $x_p'=x_p\xi$.}:
\begin{align}
    \frac{\der\sigma^{gA\to q\bar q(g)+X}}{\der^2\boldsymbol{k}_{q\perp}\der\eta_q \der^2\boldsymbol{k}_{{\bar{q}}\perp} \der\eta_{\bar{q}}}&\simeq \int_0^1 \der \xi \frac{\alpha_s}{2\pi}P_{gg}^{real}(\xi)  \left.\frac{\der\sigma^{gA\to q\bar q+X}}{\der^2\boldsymbol{k}_{q\perp}\der\eta_q \der^2\boldsymbol{k}_{{\bar{q}}\perp} \der\eta_{\bar{q}}}\right|_{LO,p^-= \xi x_p q^-} \nonumber \\
    & \times \left[ \frac{2}{\varepsilon}-\ln(e^{\gamma_E}\pi\mu^2r_{w_2w_2'}^2) \right]\,.
\end{align}
This expression has a factorized form where the LO cross-section is expressed in terms of the intermediate gluon which splits into the $q\bar q$ pair. This intermediate gluon has longitudinal momentum fraction $(1-z_g)p^-=\xi x_pq^-$ with $\xi\equiv 1-z_g$. The radiated gluon which is integrated out has longitudinal momentum fraction $1-\xi$ with respect to the parent gluon.

The divergence is absorbed into the renormalization of the gluon PDF. We recall indeed that the LO cross section is
\begin{align}
    \frac{\der\sigma^{pA\to q\bar q+X}}{\der^2\boldsymbol{k}_{q\perp}\der\eta_q \der^2\boldsymbol{k}_{{\bar{q}}\perp} \der\eta_{\bar{q}}}&=\int_0^1\der x_p \ \left.\frac{\der\sigma^{gA\to q\bar q+X}}{\der^2\boldsymbol{k}_{q\perp}\der\eta_q \der^2\boldsymbol{k}_{{\bar{q}}\perp} \der\eta_{\bar{q}}}\right|_{p^-= x_pq^-}\times g^{(0)}(x_p)\,.
\end{align}
In this expression $g^{(0)}(x_p)$ is the bare gluon distribution function of the proton.

The NLO contribution is
\begin{align}
    \frac{\der\sigma^{pA\to q\bar q+X}}{\der^2\boldsymbol{k}_{q\perp}\der\eta_q \der^2\boldsymbol{k}_{{\bar{q}}\perp} \der\eta_{\bar{q}}}&=\int_0^1\der x_p g^{(0)}(x_p) \int_0^1 \der \xi \frac{\alpha_s}{2\pi}P_{gg}^{real}(\xi)  \left.\frac{\der\sigma^{gA\to q\bar q+X}}{\der^2\boldsymbol{k}_{q\perp}\der\eta_q \der^2\boldsymbol{k}_{{\bar{q}}\perp} \der\eta_{\bar{q}}}\right|_{LO,p^-= \xi x_p q^-} \nonumber \\
    & \times \left[ \frac{2}{\varepsilon}-\ln(e^{\gamma_E}\pi\mu^2r_{w_2w_2'}^2) \right]\,.
\end{align}
Performing the change of variable $x_p'=\xi x_p$ in the integral over $x_p$ imposes the constraint $x_p'\le \xi$ (as $x_p\le 1$), which is then taken into account by using $x_p'$ as the lower limit of the $\xi$ integral. These manipulations yield
\begin{align}
    \frac{\der\sigma^{pA\to q\bar q(g)+X}}{\der^2\boldsymbol{k}_{q\perp}\der\eta_q \der^2\boldsymbol{k}_{{\bar{q}}\perp} \der\eta_{\bar{q}}}&=\int_0^1\der x_p' \ \left.\frac{\der\sigma^{gA\to q\bar q(g)+X}}{\der^2\boldsymbol{k}_{q\perp}\der\eta_q \der^2\boldsymbol{k}_{{\bar{q}}\perp} \der\eta_{\bar{q}}}\right|_{LO,p^-= x_p'q^-}\times \int_{x_p'}^1\frac{\der \xi}{\xi}\frac{\alpha_s}{2\pi}P_{gg}^{real}(\xi) g^{(0)}\left(\frac{x_p'}{\xi}\right)\nonumber\\
    &\times\left[\frac{2}{\varepsilon}-\ln(e^{\gamma_E}\pi\mu^2/\mu_F^2)+...\right]\,,\label{eq:pAtoqqb-dglap-init}
\end{align}
where we have introduced a factorization scale $\mu_F$ to isolate the universal part of the collinear divergence. The dotted terms inside the square brackets represents the NLO finite pieces --- depending on $\ln(\mu_F^2r_{w_2w_2'}^2)$ --- to be eventually combined with the finite part of the virtual cross section. The introduction of the factorization scale $\mu_F$ is also 
important when combining real and virtual corrections. In 
particular, the collinear limit of the virtual diagrams, which completes the structure of the $g \to gg$ splitting function, is expected to generate a logarithm similar to that in Eq.\,\eqref{eq:WWkernel-dimreg}, but involving a different transverse coordinate, such as $r_{xy}^2$ or $r_{x'y'}^2$. As shown in~\cite{Caucal:2024nsb}, this ``mismatch'' between real and virtual contributions gives rise to Sudakov (single) logarithms when one further considers the back-to-back limit of the inclusive dijet cross section.

One now recognizes in Eq.\,\eqref{eq:pAtoqqb-dglap-init} the real part of the gluon to gluon splitting function
\begin{align}
    P_{gg}(\xi)&=\frac{2N_c[1-\xi(1-\xi)]^2}{\xi(1-\xi)_+}+\beta_0\delta(1-\xi)\,.
\end{align}
Here $\beta_0=(11N_c-4n_f T_R)/6$.
The virtual part of the splitting function, corresponding to the plus prescription and the $\delta$ function, comes from the virtual corrections to the dijet process which are not computed in this paper.
The collinear divergence is then removed by promoting the bare gluon PDF into a renormalized one with
\begin{align}
    g(x_p,\mu_F^2)&=g^{(0)}(x_p)+\frac{\alpha_s}{2\pi}\left[\frac{2}{\varepsilon}-\ln(e^{\gamma_E}\pi\mu^2/\mu_F^2)\right]\int_{x_p}^1\frac{\der\xi}{\xi}P_{gg}(\xi)g\left(\frac{x_p}{\xi}\right)\,.\label{eq:renorm_gluon-PDF}
\end{align}
As usual, the statement that $g^{(0)}(x_p)$ is independent of $\mu_F$ yields the standard DGLAP equation for the gluon distribution function of the dilute proton. The quark-to-gluon contribution to the DGLAP evolution of the initial state gluon PDF will arise from the channel $q \to q q \bar{q}$ in \cite{Iancu:2020mos}. In particular, this contribution comes from the configuration where a final state quark is collinear to incoming quark.

\subsection{DGLAP evolution of the final state quark in the $g\to q\bar q$ channel}
\label{sub:dglap-final-qqb}

We now turn to the collinear divergence in the final state, namely when the radiated gluon is collinear to either the final quark or antiquark. Since the calculation is identical between the antiquark and quark case, we shall focus on the latter kinematic configuration. As discussed in the introduction to this section, the final state collinear divergence comes from diagrams where the gluon is emitted by the quark and like for the initial state case, it manifests itself in coordinate space as an infrared divergence when the transverse separation $r_{zx}$ between the quark and the gluon is much larger than the distance $r_{xy}$ between the quark and the antiquark. In this limit, the function $\Theta_{q,1}$ in Eq.\,\eqref{quark parent qbarqg amplitude} goes to zero such that the total amplitude simplifies into
\begin{align}\label{quark parent qbarqg amplitude dglap}
    \mathcal{M}^{\bar{\lambda}\lambda\sigma\sigma',ab}_{q,ij} \simeq -\int \der^6 \boldsymbol{\Pi} \ \Ncal^{\bar{\lambda}\lambda\sigma\sigma'}_{q,reg} \left[C_{R2,ij}^{ab}- C_{R3,ij}^{ab}\right]\,.
\end{align}
Because the instantaneous perturbative factor has no $\Perp{\epsilon}^{\lambda*}\cdot\boldsymbol{r}_{zx}$ enhancement, it does not contribute to the collinear limit. Let us pursue with the same steps as in the previous subsection. Knowing that, at the cross section level, the integration over the transverse momentum of the emitted gluon enforces $\Perp{z}=\Perp{z'}$, we first compute the color structure in the DGLAP limit, averaged over the number of color states of the incoming gluon. Using the expressions Eq.\,\eqref{eq:CR2-def} and Eq.\,\eqref{eq:CR3-def} for $C_{R2}$ and $C_{R3}$, we thus consider
\begin{align}
    \frac{1}{N_c^2-1}\textrm{Tr}&\left\{\left[t^a \WL{w}t^b\WLadj{y} - t^at^c U_{cb}(\Perp{v}) \right]\left[\WL{y'}t^b \WLadj{w'}t^a-t^{c'}t^aU^\dagger_{bc'}(\Perp{v'})\right]\right\}\,,\label{eq:color_square_DGLAP_final_pA-qqb}
\end{align}
where $\Perp{w}=(z_q\Perp{x}+z_g\Perp{z})/(z_q+z_g)$, $\Perp{w'}=(z_q\Perp{x'}+z_g\Perp{z})/(z_q+z_g)$, $\Perp{v}=z_q\Perp{x}+z_{\bar q}\Perp{y}+z_g\Perp{z}$ and $\Perp{v}'=z_q\Perp{x'}+z_{\bar q}\Perp{y'}+z_g\Perp{z}$ as a consequence of the identification $\Perp{z}=\Perp{z'}$. Note that $\Perp{v}=(1-z_{\bar q})\Perp{w}+z_{\bar q}\Perp{y}$ and likewise for $\Perp{v'}$. The calculation of the trace and color algebra gives
\begin{align}
    \frac{C_F}{2}\Xi_{q\bar q,LO}(\Perp{w},\Perp{y},\Perp{w'},\Perp{y'})\,.
\end{align}
Here also, while the color structure involves the same operators as in the LO cross section, one should keep in mind that it is evaluated at different transverse coordinates since $\Perp{w}\neq \Perp{x}$ and $\Perp{w'}\neq \Perp{x'}$. These differences will be necessary to factorize the final state collinear divergence.

As in the previous subsection, we first express the perturbative factor $\Ncal^{\bar{\lambda}\lambda\sigma\sigma'*}_{q,reg}$ in terms of the LO one as
\begin{align}
\Ncal^{\bar{\lambda}\lambda\sigma\sigma'}_{q,reg}&=\frac{i g}{\pi} \sqrt{\xi}\, \Gamma^{\sigma \lambda}_{q\xrightarrow{}qg}\left(\xi\right) \frac{\et^{\lambda*}\cdot \boldsymbol{r}_{zx}}{r_{zx}^2} \mathcal{N}^{\bar{\lambda} \sigma \sigma'}_{q\bar{q},LO}(z_q/\xi,z_{\bar q}, \boldsymbol{r}_{wy}) \,,
\end{align}
where $\xi=z_q/(z_q+z_g)$ now represents the longitudinal momentum fraction of the final quark with respect to its parent.
Then,
\begin{align}
\sum_{\lambda}\left[\Gamma^{\sigma \lambda}_{q\xrightarrow{}qg}\left(\xi\right)\right]^2\epsilon^{\lambda*,i}_{\perp}\epsilon^{\lambda,j}_{\perp}&=\frac{(1-\xi)}{2C_F}P_{qq}^{real}(\xi)\delta^{ij}-i\sigma \frac{(1-\xi^2)}{2}\epsilon^{ij} \,.
\end{align}
The spin independent term, which is also the symmetric trace part of the $(i,j)$ tensor displays the real $q\to qg$ DGLAP splitting function defined as
\begin{align}
    P^{real}_{qq}(\xi)=\frac{C_F(1+\xi^2)}{1-\xi}\,.
\end{align}
As for the collinear divergence in the initial state, the contraction with the integral over $\Perp{z}$ selects the symmetric component of the above tensor, such that
\begin{align}
    \int\frac{\der z_g}{z_g}\int\der^2\Perp{z}\sum_{\lambda\bar\lambda\sigma\sigma'}\Ncal^{\bar{\lambda}\lambda\sigma\sigma'}_{q,reg}\Ncal^{\bar{\lambda}\lambda\sigma\sigma'*}_{q,reg}&=\frac{g^2}{2\pi^2C_F}\int_0^1\der\xi \ P_{qq}^{real}(\xi)\int\der^2\Perp{z}\frac{\dipvec{zx'}\cdot\dipvec{zx}}{r_{zx'}^2 r_{zx}^2} \nonumber\\
&\hspace{-0.5cm}\times\sum_{\sigma\sigma'\bar\lambda} \mathcal{N}^{\bar{\lambda} \sigma \sigma'}_{q\bar{q},LO}(z_q/\xi,z_{\bar q}, \boldsymbol{r}_{wy}) \mathcal{N}^{\bar{\lambda} \sigma \sigma'*}_{q\bar{q},LO}(z_q/\xi,z_{\bar q}, \boldsymbol{r}_{w'y})+\textrm{IR finite} \,.\label{eq:dglap-step-qqbar}
\end{align}
The final step before discussing the factorization of the final state collinear divergence consists in performing the change of variable $\Perp{x}\to \Perp{w}$ and $\Perp{x'}\to\Perp{w'}$. After doing so and renaming $\Perp{w}$ into $\Perp{x}$ for simplicity, we get
\begin{align}
    \frac{\der\sigma^{gA\to q\bar q(g)+X}}{\der^2\boldsymbol{k}_{q\perp}\der\eta_q \der^2\boldsymbol{k}_{{\bar{q}}\perp} \der\eta_{\bar{q}}}&\simeq\frac{\alpha_s}{(2\pi)^6}\int_0^1\frac{\der \xi}{\xi^2}\frac{\alpha_s}{2\pi}P_{qq}^{real}(\xi) \ \delta\left(1-\frac{z_q}{\xi}-z_{\bar q}\right)\nonumber\\
&\times\int\der^2\Perp{x}\der^2\Perp{y}\der^2\Perp{x'}\der^2\Perp{y'} e^{-i\frac{\boldsymbol{k}_{q\perp}}{\xi}\cdot (\Perp{x}-\Perp{x'})-i\boldsymbol{k}_{\bar q\perp}\cdot (\Perp{y}-\Perp{y'})}\nonumber\\
    &\times \mathcal{K}_{q\bar q,LO}\left(\frac{z_q}{\xi},z_{\bar q},\boldsymbol{r}_{xy},\boldsymbol{r}_{x'y'}\right)\left\langle\Xi_{q\bar q,LO}(\Perp{x},\Perp{y},\Perp{x'},\Perp{y'})\right\rangle_Y\int\frac{\der^2\Perp{z}}{\pi}\frac{\boldsymbol{r}_{zx}\cdot\boldsymbol{r}_{zx'}}{r_{zx}^2r_{zx'}^2}\,.\label{eq:NLO-collinear-limit-fs-2}
\end{align}
Note that the phase depending on the transverse momentum of the final quark has changed after these manipulations and that the change of variable has brought an overall factor $1/\xi^2$ inside the $\xi$-integral.

To absorb the collinear final state divergence one introduces in the LO cross section the fragmentation function of the quark into hadrons
\begin{align}
    \frac{\der\sigma^{gA\to h\bar q+X}}{\der^2\boldsymbol{k}_{h\perp}\der\eta_h \der^2\boldsymbol{k}_{{\bar{q}}\perp} \der\eta_{\bar{q}}}&=\int_0^1\frac{\der\zeta}{\zeta^2} \ \left.\frac{\der\sigma^{gA\to q\bar q(g)+X}}{\der^2\boldsymbol{k}_{q\perp}\der\eta_q \der^2\boldsymbol{k}_{{\bar{q}}\perp} \der\eta_{\bar{q}}}\right|_{k_q^\mu= k_h^\mu/\zeta}\times D^{(0)}_{h/q}(\zeta)\,.
\end{align}
The following steps are identical to those performed in appendix~A of~\cite{Caucal:2024nsb}. After convolving Eq.\,\eqref{eq:NLO-collinear-limit-fs-2} with the quark fragmentation function into hadron, one performs the change of variable  $\zeta\to \zeta'=\xi\zeta$ to factorize the LO cross section at the scale $\zeta'$:  
\begin{align}
     \frac{\der\sigma^{gA\to h\bar q(g)+X}}{\der^2\boldsymbol{k}_{h\perp}\der\eta_h \der^2\boldsymbol{k}_{{\bar{q}}\perp} \der\eta_{\bar{q}}}&=\int_0^1\frac{\der\zeta'}{\zeta'^2} \ \left.\frac{\der\sigma^{gA\to q\bar q+X}}{\der^2\boldsymbol{k}_{q\perp}\der\eta_q \der^2\boldsymbol{k}_{{\bar{q}}\perp} \der\eta_{\bar{q}}}\right|_{LO,k_q^\mu= k_h^\mu/\zeta'}\times\int_{\zeta'}^1\frac{\der\xi}{\xi}\frac{\alpha_s}{2\pi}P_{qq}^{real}(\xi)D^{(0)}_{h/q}\left(\frac{\zeta'}{\xi}\right)\nonumber\\
    &\times\left[\frac{2}{\varepsilon}-\ln(e^{\gamma_E}\pi\mu^2/\mu_F^2)+...\right]\,.\label{eq:pAtoqqb-dglap-final}
\end{align}
One recognizes in the convolution over $\xi$ the real part of the $P_{qq}$ DGLAP splitting function defined as
\begin{align}
    P_{qq}(\xi)&=\frac{C_F(1+\xi^2)}{(1-\xi)_+}+\frac{3C_F}{2}\delta(1-\xi) \,.
\end{align}
Let us only emphasize the role of the $1/\xi$ factor in the phase of Eq.\,\eqref{eq:NLO-collinear-limit-fs-2} as the change in the phase is a specific feature of the final state collinear singularity: the convolution with the collinear fragmentation function implies that the measured transverse momentum of the quark is related to that of the hadron by $\boldsymbol{k}_{q\perp}=\boldsymbol{k}_{h\perp}/\zeta$. The presence of the $1/\xi$ factor in the phase thus ensures that after the change of variable $\zeta\to \zeta'$, the LO cross section remains evaluated under the same constraint, with $\zeta$ replaced by $\zeta'$, namely $\boldsymbol{k}_{q\perp}=\boldsymbol{k}_{h\perp}/\zeta'$. Another remark is that one can also decide to integrate out the final quark and measure the gluon contribution to the fragmentation. In that case, as shown in~\cite{Bergabo:2024ivx,Caucal:2024nsb}, the collinear divergence contributes to the mixing between the evolution of $D_{h/q}$ and that of $D_{h/g}$ through the $P_{gq}(\xi)$ splitting function. 

Integrating out the quark (collinear to gluon), then contributes to the mixing between $D_{h/q}$ and $D_{h/g}$ in the DGLAP evolution:
\begin{equation}
D_{h/q}(\zeta,\mu_F^2)
=
D_{h/q}^{(0)}(\zeta)
+
\frac{\alpha_s}{2\pi}
\left[
\frac{2}{\varepsilon}
-\ln\!\left(e^{\gamma_E}\pi\mu^2/\mu_F^2\right)
\right]
\int_{\zeta}^{1}\frac{d\xi}{\xi}\,
\left[ P_{qq}(\xi)\,
D_{h/q}^{(0)}\!\left(\frac{\zeta}{\xi}\right) + P_{gq}(\xi)\,
D_{h/g}^{(0)}\!\left(\frac{\zeta}{\xi}\right)
\right]\,.
\end{equation}
To obtain this result we convolve the three-parton differential cross-section with the collinear gluon-to-hadron fragmentation function. We observe that the color correlator obtained by identifying $\Transv{x}{} = \Transv{x}{}'$---from the integration of the quark kinematics--- instead of $\Transv{z}{} = \Transv{z}{}'$ in Eq.\,\eqref{eq:color_square_DGLAP_final_pA-qqb} yields an identical result. Then one can perform the change of variable $\xi \to 1 - \xi$, which corresponds to the longitudinal momentum of the gluon relative to the intermediate quark, in Eq.\,\eqref{eq:dglap-step-qqbar} and integrate over the quark kinematics instead of the gluon ones. Finally, by rewriting $z_q/(1-\xi) = z_g/\xi$ and making the change of variables $\Transv{x}{} \to \Transv{w}{}$ (as opposed to $\Transv{z}{}\to \Transv{w}{}$) and similarly for the primed variables in Eq.\,\eqref{eq:NLO-collinear-limit-fs-2}, one obtains an analogous expression for the integrated quark case. 

The same strategy would apply for the collinear divergence coming from the phase space where the gluon is collinear to the antiquark. Combining these results with those of section \ref{sec: DGLAP-initial-pA-to-qqb}, we thus demonstrate that the collinear divergences can be absorbed into the evolution of non-perturbative objects in the case of forward dihadron production in proton-nucleus collisions:
\begin{align}
         \frac{\der\sigma^{pA\to h_1h_2+X}}{\der^2\boldsymbol{k}_{h_1\perp}\der\eta_{h_1} \der^2\boldsymbol{k}_{h_2\perp} \der\eta_{h_2}}=\int_0^1\der x_p &\int_0^1\frac{\der \zeta_1}{\zeta_1^2}\int_0^1\frac{\der \zeta_2}{\zeta_2^2} \ g(x_p,\mu_F^2)D_{h_1/q}(\zeta_1,\mu_F^2)D_{h_2/\bar q}(\zeta_2,\mu_F^2)\nonumber\\
         &\times \left.\frac{\der\sigma^{gA\to q\bar q+X}}{\der^2\boldsymbol{k}_{q\perp}\der\eta_q \der^2\boldsymbol{k}_{{\bar{q}}\perp} \der\eta_{\bar{q}}}\right|_{k_q^\mu= k_{h_1}^\mu/\zeta_1,k_{\bar q}^\mu=k_{h_2}^\mu/\zeta_2,p^-=x_pq^-}\,.
\end{align}
This formula constitutes the core of the ``hybrid" factorization approach for dilute dense collisions; it is demonstrated here to
to be consistent to NLO accuracy for the inclusive dihadron production process.

\subsection{DGLAP evolution of the initial state gluon in the $g\to gg$ channel}

We consider here the $g\to ggg$ channel and focus on the collinear limit where one of the emitted gluon is collinear to the incoming one. The initial state collinear singularity comes from the diagrams labeled R7 and R8. As we shall see, the final state collinear singularity comes instead from the other two permutations discussed in section~\ref{sub:ggg-permutations}. The four-gluon vertex topology does not contribute to the DGLAP limit as their structure is similar to instantaneous terms. In the sum of amplitudes R7, R8 and R9, the function $\Theta_{ggg,1A}$ can be approximated by one in the limit $r_{w_Az}\gg r_{xy}$ such that the leading collinear divergence in the initial state arises from the amplitude
\begin{align}
 \mathcal{M}^{\bar{\lambda}\lambda\bar{\xi}\xi}_{ggg} \simeq \int \der^6 \boldsymbol{\Pi} \ \Ncal^{\bar{\lambda}\lambda\bar{\xi}\xi}_{ggg,reg} \left[C_{R7,A}^{abcd} - C_{R8,A}^{abcd} \right] \,.
\end{align}
After integrating out the gluon with transverse momentum conjugate to $\Perp{z}-\Perp{z'}$, the colour structure at the cross section level simplifies thanks to the identification $\Perp{z}=\Perp{z'}$:
\begin{align}
    \frac{1}{N_c^2-1}&\left[f^{efg}f^{dhf}U_{ae}(\Perp{x})U_{bg}(\Perp{y})U_{ch}(\Perp{z})-f^{eba}f^{gfd}U_{ef}(\Perp{w_A})U_{cg}(\Perp{z})\right]\nonumber\\
    &\times\left[f^{e'f'g'}f^{dh'f'}U^\dagger_{e'a}(\Perp{x'})U^\dagger_{g'b}(\Perp{y'})U_{h'c}(\Perp{z})-f^{e'ba}f^{g'f'd}U^\dagger_{f'e'}(\Perp{w_A'})U^\dagger_{g'c}(\Perp{z})\right]\\
    &=N_c^2\Xi_{gg,LO}(\Perp{x},\Perp{y},\Perp{x'},\Perp{y'}) \,,
\end{align}
where we have used $f^{dhf}f^{dhf'}=N_c\delta^{ff'}$.
Following the same strategy as for the $g\to q\bar q$ channel, we express the perturbative factor (see Eq.\,\eqref{ggg-reg-perturbative-factor}) of the two diagrams R7 and R8 in terms of the LO one as
\begin{align}
\Ncal^{\bar{\lambda}\lambda\bar{\xi}\xi}_{ggg,reg}&=-\frac{ig}{\pi} \sum_\eta\frac{\Gt^{\bar\lambda\lambda\eta}(\xi)\cdot\boldsymbol{r}_{w_Az}}{r_{w_Az}^2}\Ncal_{gg,LO}^{\eta\bar\xi\xi}\left(\frac{z_1}{\xi}, \frac{z_2}{\xi},\boldsymbol{r}_{xy} \right)\,,
\end{align}
with $\xi\equiv 1-z_3$ in this case. In the square of the gluon vertex $\Gt$ given by Eq.\,\eqref{eq:GG-tensor}, one keeps only the term proportional to $\delta^{ij}$ in the collinear limit, which also identifies the polarization index of the intermediate gluon before splitting into the gluon-gluon pair both in the amplitude and in the complex conjugate amplitude. This identification enables one to recognize the square of the LO perturbative factor. In the end,
\begin{align}
    & \int\frac{\der z_3}{z_3}\int\der^2\Perp{z}\sum_{\bar\lambda\lambda\bar\xi\xi}  \Ncal^{\bar{\lambda}\lambda\bar{\xi}\xi}_{ggg,reg}  \Ncal^{\bar{\lambda}\lambda\bar{\xi}\xi*}_{ggg,reg}=\frac{g^2}{2\pi^2N_c}\int_0^1 \der\xi\ \xi  P_{gg}^{real}(\xi)\nonumber\\
    &\times \sum_{\eta\bar\xi\xi}\Ncal_{gg,LO}^{\eta\bar\xi\xi}\left(\frac{z_1}{\xi}, \frac{z_2}{\xi},\boldsymbol{r}_{xy} \right)\Ncal_{gg,LO}^{\eta\bar\xi\xi*}\left(\frac{z_1}{\xi}, \frac{z_2}{\xi},\boldsymbol{r}_{x'y'} \right) \int\der^2\Perp{z}\frac{\boldsymbol{r}_{w_Az}\cdot\boldsymbol{r}_{w_A'z}}{r_{w_Az}^2r_{w_A'z}^2}+\textrm{IR finite\,,}
\end{align}
which is identical to Eq.\,\eqref{eq:NLO-collinear-limit-is} up to the change in the LO partonic channel. At the cross section level in the collinear limit, we get
\begin{align}
    \frac{\der\sigma^{gA\to gg(g)+X}}{\der^2\boldsymbol{k}_{1\perp}\der\eta_1 \der^2\boldsymbol{k}_{2\perp} \der\eta_{2}}&\simeq\frac{\alpha_sN_c}{(2\pi)^6}\int_0^1 \der \xi \frac{\alpha_s}{2\pi}P_{gg}^{real}(\xi)\delta\left(1-\frac{z_1}{\xi}-\frac{z_{2}}{\xi} \right)\nonumber\\
    &\times\int\der^2\Perp{x}\der^2\Perp{y}\der^2\Perp{x'}\der^2\Perp{y'} e^{-i\boldsymbol{k}_{1}\cdot (\Perp{x}-\Perp{x'})-i\boldsymbol{k}_{2} \cdot (\Perp{y}-\Perp{y'})}\nonumber\\
    &\times \mathcal{K}_{gg,LO}\left( \frac{z_1}{\xi}, \frac{z_2}{\xi},\boldsymbol{r}_{xy},\boldsymbol{r}_{x'y'} \right)\left\langle\Xi_{gg,LO}(\Perp{x},\Perp{y},\Perp{x'},\Perp{y'})\right\rangle_Y\int\frac{\der^2\Perp{z}}{\pi}\frac{\boldsymbol{r}_{w_Az}\cdot\boldsymbol{r}_{w_A'z}}{r_{w_Az}^2r_{w_A'z}^2}\,,\label{eq:NLO-collinear-limit-is-ggg}
\end{align}
such that, following the same steps as in between Eq.\,\eqref{eq:NLO-collinear-limit-is} and \eqref{eq:pAtoqqb-dglap-init}, the initial-state collinear divergence factorizes in terms of the real DGLAP evolution of the gluon PDF:
\begin{align}
    \frac{\der\sigma^{pA\to gg(g)+X}}{\der^2\boldsymbol{k}_{1\perp}\der\eta_1 \der^2\boldsymbol{k}_{2\perp} \der\eta_{2}}&=\int_0^1\der x_p' \ \left.\frac{\der\sigma^{gA\to gg(g)+X}}{\der^2\boldsymbol{k}_{1}\der\eta_1 \der^2\boldsymbol{k}_{2\perp} \der\eta_{2}}\right|_{LO,p^-= x_p'q^-}\times \int_{x_p'}^1\frac{\der \xi}{\xi}\frac{\alpha_s}{2\pi}P_{gg}^{real}(\xi) g^{(0)}\left(\frac{x_p'}{\xi}\right)\nonumber\\
    &\times\left[\frac{2}{\varepsilon}-\ln(e^{\gamma_E}\pi\mu^2/\mu_F^2)+...\right]\,.\label{eq:pAtogg-dglap-init}
\end{align}
Similarly to the previous channel, the quark-to-gluon contribution to the DGLAP evolution of the initial state gluon PDF will arise from the channel $q \to q gg$ in \cite{Iancu:2020mos}, where the final state quark is collinear to incoming quark.

\subsection{DGLAP evolution of the final state gluon in the $g\to gg$ channel}

Finally, the final state collinear divergence when two outgoing gluons are collinear comes from the diagrams obtained from R7, R8 and R9 after permutation of the gluons. The two topologies $B$ and $C$ contribute to the final state collinear divergences, and each one can be attributed to one real DGLAP step of each outgoing gluon in the $gA\to gg$ LO dijet process. Focusing on topology $B$ as an example, the collinear divergence arises when the gluons $2$ and $3$ are collinear, or equivalently in coordinate space when $r_{zy}^2\gg r_{xy}^2$. In this regime, 
\begin{align}
    \Theta_{ggg,1B}&=\frac{r_{w_Bx}^2}{r_{w_Bx}^2+\frac{z_3z_2}{z_1(z_3+z_2)^2}r_{yz}^2}\to 0\,,
\end{align}
with $\boldsymbol{w}_{B\perp}=(z_3\Perp{z}+z_2\Perp{y})/(z_3+z_2)$.  The amplitude in the limit where gluons $2$ and $3$ are collinear thus reads
\begin{align}
    \mathcal{M}_{ggg}^{\bar\lambda\lambda\bar\xi\xi}\simeq  -\int \der^6 \boldsymbol{\Pi} \ \Ncal^{\bar{\lambda}\lambda\bar{\xi}\xi}_{ggg,reg,B} \left[C_{R8,B}^{abcd} - C_{R9,B}^{abcd} \right]\,,
\end{align}
where $C_{R8,B}$ and $C_{R9,B}$ are obtained from $C_{R8,A}$ and $C_{R9,A}$ by exchange $\Perp{x}\leftrightarrow\Perp{z}$ of transverse coordinates and exchange $a\leftrightarrow c$ of color indices as explained in section~\ref{sub:ggg-permutations}. Likewise, the perturbative factor $\Ncal^{\bar{\lambda}\lambda\bar{\xi}\xi}_{ggg,reg,B}$ follows from $\Ncal^{\bar{\lambda}\lambda\bar{\xi}\xi}_{ggg,reg}$ after exchanges of gluon kinematics $1$ and $3$.

Considering first the color structure at the cross section level with $\Perp{z}=\Perp{z}'$, we have the result
\begin{align}
    \frac{1}{N_c^2-1}&\left[f^{ebc}f^{gfd}U_{ef}(\boldsymbol{w}_{B\perp})U_{ag}(\Perp{x})-f^{abc}f^{efd}U_{eh}(\Perp{v})\right]\nonumber\\
    &\times\left[f^{e'bc}f^{g'f'd}U^\dagger_{f'e'}(\boldsymbol{w}_{B\perp}')U^\dagger_{g'a}(\Perp{x}')-f^{abc}f^{e'f'd}U^\dagger_{h'e'}(\Perp{v}')\right]\\
    &=N_c^2\Xi_{LO,gg}(\Perp{x},\boldsymbol{w}_{B\perp},\Perp{x}',\boldsymbol{w}_{B\perp}')\,,
\end{align}
which features the LO color structure modulo the replacement $\Perp{y}\to \boldsymbol{w}_{B\perp}$. As noticed in subsection~\ref{sub:dglap-final-qqb}, the change of variable $\Perp{y}\to\boldsymbol{w}_{B\perp}$ brings the additional $1/\xi$ factor in the phase $e^{i\boldsymbol{k}_{2\perp}\cdot(\Perp{y}-\Perp{y'})}$. 

After the permutation of $1$ and $3$, the perturbative factor $\Ncal^{\bar{\lambda}\lambda\bar{\xi}\xi}_{ggg,reg,B}$ is expressed in terms of the LO one as 
\begin{align}
\Ncal^{\bar{\lambda}\lambda\bar{\xi}\xi}_{ggg,reg,B}=-\frac{ig}{\pi}\sum_\eta \frac{\Gt^{\eta\bar\xi\lambda}\left(1-\xi \right)\cdot\boldsymbol{r}_{yz}}{r_{yz}^2}\Ncal_{gg,LO}^{\bar\lambda\xi\eta}\left(\frac{z_2}{\xi},z_1,\boldsymbol{r}_{w_Bx}\right)\,,
\end{align}
where $\xi=z_2/(z_2+z_3)$ represents the longitudinal momentum fraction of the outgoing gluon labeled ``2" with respect to its parent. Thanks to the symmetry property $G_\perp^{\bar\lambda\eta\lambda}(1-\xi)=G_\perp^{\bar\lambda\lambda\eta}(\xi)$, we can reformulate the previous expression as 
\begin{align}
\Ncal^{\bar{\lambda}\lambda\bar{\xi}\xi}_{ggg,reg,B}=-\frac{ig}{\pi}\sum_\eta \frac{\Gt^{\eta\lambda\bar\xi}\left(\xi \right)\cdot\boldsymbol{r}_{zy}}{r_{zy}^2}\Ncal_{gg,LO}^{\bar\lambda\eta\xi}\left(z_1,\frac{z_2}{\xi},\boldsymbol{r}_{xw_B}\right)\,.
\end{align}
In the case of a final state $g\to gg$ gluon splitting, the relevant tensor contracted with the IR divergent $\Perp{z}$ integral is no  longer $\sum_{\bar\lambda\lambda} \Gt^{\bar\lambda\lambda\eta,i} \Gt^{\bar\lambda\lambda\eta',j*}$ given by Eq.\,\eqref{eq:GG-tensor} but instead $\sum_{\lambda\bar\xi} \Gt^{\eta\lambda\bar\xi,i} \Gt^{\eta'\lambda\bar\xi,j*}$ (in Eq.\,\eqref{eq:GG-tensor}, the last polarization index is not summed over, while now, the first polarization index is kept fixed). Nevertheless, this tensor satisfies the same key property which tremendously simplifies the derivation of the DGLAP limit: its component proportional to $\delta^{ij}$ is still proportional to the $P_{gg}^{real}(\xi)$ splitting function. Indeed,
\begin{align}
    \sum_{\lambda\bar\xi} \Gt^{\eta\lambda\bar\xi,i}(\xi) \Gt^{\eta'\lambda\bar\xi,j*}(\xi)&=\frac{\xi (1-\xi)}{2N_c}\left[(2\et^{\eta,i}\et^{\eta'*,j}-\delta^{ij}\delta^{\eta\eta'})2N_c\xi(1-\xi)+P_{gg}^{real}(\xi)\delta^{ij}\delta^{\eta\eta'}\right]\nonumber\\
    &+2\xi^2 (1-\xi)(\et^{\eta'*,i}\et^{\eta,j}-\et^{\eta,i}\et^{\eta'*,j})\,.
\end{align}
However, one should note that the splitting function in front of the $2\epsilon_\perp^{\eta,i}\epsilon_\perp^{\eta'*,j}-\delta^{ij}\delta^{\eta\eta'}$ tensor structure is not $P_{gg_L}(\xi)=2N_c(1-\xi)/\xi$ as in Eq.\,\eqref{eq:GG-tensor}, but rather $2N_c\xi(1-\xi)$.
Using this identity, the rest of the calculation is identical to the $g\to q\bar q$ channel. We have first
\begin{align}
    \int\frac{\der z_3}{z_3}\int\der^2\Perp{z}\Ncal^{\bar{\lambda}\lambda\bar{\xi}\xi}_{ggg,reg,B}\Ncal^{\bar{\lambda}\lambda\bar{\xi}\xi*}_{ggg,reg,B}&=\frac{g^2}{2\pi^2N_c}\int_0^1\der\xi \ P_{gg}^{real}(\xi)\int\der^2\Perp{z}\frac{\boldsymbol{r}_{zy}\cdot\boldsymbol{r}_{zy'}}{r_{zy}^2r_{zy'}^2}\nonumber\\
    &\hspace{-1cm}\times\sum_{\bar\lambda\eta\xi}\Ncal_{gg,LO}^{\bar\lambda\eta\xi}\left(z_1,\frac{z_2}{\xi},\boldsymbol{r}_{xw_B}\right)\Ncal_{gg,LO}^{\bar\lambda\eta\xi*}\left(z_1,\frac{z_2}{\xi},\boldsymbol{r}_{x'w_B'}\right)+\textrm{IR finite} \,,
\end{align}
such that the divergence of the cross section when the outgoing gluons ``2" and ``3" become collinear reads, after the change of variable $\Perp{y}\to \boldsymbol{w}_B,\Perp{y'}\to\boldsymbol{w}_B'$ (followed by the relabeling of $\boldsymbol{w}_B,\boldsymbol{w}_B'$ into $\Perp{y},\Perp{y}'$ in order to use the same notations as in the LO cross section):
\begin{align}
    \frac{\der\sigma^{gA\to gg(g)+X}}{\der^2\boldsymbol{k}_{1}\der\eta_1 \der^2\boldsymbol{k}_{2\perp} \der\eta_{2}}&\simeq\frac{\alpha_sN_c}{(2\pi)^6}\int_0^1\frac{\der \xi}{\xi^2}\frac{\alpha_s}{2\pi}P_{gg}^{real}(\xi) \ \delta\left(1-\frac{z_2}{\xi}-z_{1}\right)\nonumber\\
&\times\int\der^2\Perp{x}\der^2\Perp{y}\der^2\Perp{x'}\der^2\Perp{y'} e^{-i\boldsymbol{k}_{1\perp}\cdot (\Perp{x}-\Perp{x'})-i\frac{\boldsymbol{k}_{2\perp}}{\xi}\cdot (\Perp{y}-\Perp{y'})}\nonumber\\
    &\times \mathcal{K}_{gg,LO}\left(z_1,\frac{z_2}{\xi},\boldsymbol{r}_{xy},\boldsymbol{r}_{x'y'}\right)\left\langle\Xi_{gg,LO}(\Perp{x},\Perp{y},\Perp{x'},\Perp{y'})\right\rangle_Y\int\frac{\der^2\Perp{z}}{\pi}\frac{\boldsymbol{r}_{zy}\cdot\boldsymbol{r}_{zy'}}{r_{zy}^2r_{zy'}^2}\,.\label{eq:NLO-collinear-limit-fs-ggg}
\end{align}
This result immediately implies the following factorization of final state divergence in terms of the real DGLAP evolution of the gluon ``2" fragmentation function into hadrons:
\begin{align}
     \frac{\der\sigma^{gA\to gh(g)+X}}{\der^2\boldsymbol{k}_{1\perp}\der\eta_1 \der^2\boldsymbol{k}_{h\perp} \der\eta_{h}}&=\int_0^1\frac{\der\zeta'}{\zeta'^2} \ \left.\frac{\der\sigma^{gA\to gg+X}}{\der^2\boldsymbol{k}_{1}\der\eta_1 \der^2\boldsymbol{k}_{2\perp} \der\eta_{2}}\right|_{LO,k_2^\mu= k_h^\mu/\zeta'}\times\int_{\zeta'}^1\frac{\der\xi}{\xi}\frac{\alpha_s}{2\pi}P_{gg}^{real}(\xi)D^{(0)}_{h/g}\left(\frac{\zeta'}{\xi}\right)\nonumber\\
    &\times\left[\frac{2}{\varepsilon}-\ln(e^{\gamma_E}\pi\mu^2/\mu_F^2)+...\right]\,.\label{eq:pAtogg-dglap-final}
\end{align}
There is another divergence which contributes to the gluon-quark mixing term of the evolution of the fragmentation function. This contribution comes from the $q\to q\bar{q}g$ channel; specifically when the quark and antiquark are collinear. In this case the leading contribution to the amplitude is
\begin{align}
    \mathcal{M}^{\bar{\lambda}\lambda\sigma\sigma',ab}_{g,ij} = -\int \der^6 \boldsymbol{\Pi} \ \Ncal^{\bar{\lambda}\lambda\sigma\sigma'}_{g,reg}  (C_{R5,ij}^{ab} - C_{R6,ij}^{ab}) \,.
\end{align}
Let us consider the case where one integrates out the antiquark. The integration over the antiquark kinematics identifies the transverse coordinates $\Transv{y}{}=\Transv{y}{}'$. The color operator obtained from de modulus squared of the color operator $C_{R5} - C_{R6}$ summed over the color of the partons reads
\begin{align}
    &\Tr\bigg\{ \left[  f^{cdb} U_{ed}(\Transv{w}{2})t^e U_{ac}(\Perp{z}) -  t^cf^{dac} U_{db}(\Perp{v})\right] \nonumber\\
    &\times \left[  f^{c'd'b} U_{e'd'}(\Transv{w}{2}')t^{e'} U_{ac'}(\Perp{z}') -  t^{c'}f^{d'ac'} U_{d'b}(\Perp{v}')\right]\bigg\} = \N^2 C_F \Xi_{gg,LO}(\Transv{w}{2}', \Transv{z}{}'; \Transv{w}{2}, \Transv{z}{})\,,
\end{align}
where the transverse coordinates are $\Transv{w}{2}=\xi\Transv{x}{} + (1-\xi)\Transv{y}{}$ and $\Transv{w}{2}'=\xi\Transv{x}{}' + (1-\xi)\Transv{y}{}$ thanks to the identification of the transverse coordinates of the antiquark. The variable $\xi$ in this context represents the longitudinal momentum of the quark relative to the intermediate gluon, $z_q/(z_q+\zbar)$. Analogous to the previous cases, we factorize the amplitude as
\begin{equation}
    \Ncal^{\bar{\lambda}\lambda\sigma\sigma'}_{g,reg} = \frac{ig}{\pi} \sqrt{\xi(1-\xi)} \delta^{\sigma,-\sigma'} \Gamma^{\sigma \eta}_{g\xrightarrow{}q\bar{q}}\left(\xi\right) \Ncal^{\bar{\lambda}\lambda\eta}_{gg,LO}\left( \frac{z_q}{\xi}, z_g, \dipvec{w_2z}\right)\,.
\end{equation}
The square of the splitting function multiplied by the polarization vectors of the intermediate gluon and summed over the helicity of the quark can be cast as
\begin{equation}
    \sum_{\sigma} \left[\Gamma^{\sigma \eta}_{g\xrightarrow{}q\bar{q}}\left(\xi\right) \right]^2 \et^{\eta,i} \et^{\eta*,j} = P_{qg}^{real}(\xi)\left( \delta^{ij} - i2\eta\varepsilon^{ij} \right)\,.
\end{equation}
As in the previous cases, the divergent piece of the transverse integral over $\Transv{y}{}$ will choose the diagonal component of the splitting function. Making the change of variables $\Transv{x}{}\to \Transv{w}{2}$ and similarly for the primed variables, and relabeling $\Transv{w}{2}, \Transv{w}{2}'$ as $\Transv{x}{}, \Transv{x}{}
'$ and $\Transv{z}{},\Transv{z}{}'$ to $\Transv{y}{},\Transv{y}{}'$, we find the expression of the cross section for an integrated antiquark to be
\begin{align}
    \frac{\der\sigma^{gA\to qg(\bar{q})+X}}{\der^2\boldsymbol{k}_{1}\der\eta_1 \der^2\boldsymbol{k}_{2\perp} \der\eta_{2}}&\simeq\frac{\alpha_sN_c}{(2\pi)^6}\int_0^1\frac{\der \xi}{\xi^2}\frac{\alpha_s}{2\pi}P_{qg}^{real}(\xi) \ \delta\left(1-\frac{z_1}{\xi}-z_{2}\right)\nonumber\\
&\times\int\der^2\Perp{x}\der^2\Perp{y}\der^2\Perp{x'}\der^2\Perp{y'} e^{-i\frac{\boldsymbol{k}_{1\perp}}{\xi}\cdot (\Perp{x}-\Perp{x'})-i\boldsymbol{k}_{2\perp}\cdot (\Perp{y}-\Perp{y'})}\nonumber\\
    &\times \mathcal{K}_{gg,LO}\left(\frac{z_1}{\xi}, z_2,\boldsymbol{r}_{xy},\boldsymbol{r}_{x'y'}\right)\left\langle\Xi_{gg,LO}(\Perp{x},\Perp{y},\Perp{x'},\Perp{y'})\right\rangle_Y\left[\frac{2}{\varepsilon}-\ln(e^{\gamma_E}\pi\mu^2/\mu_F^2)+...\right]\,,\label{eq:NLO-collinear-limit-fs-ggg-mixing}
\end{align}
where we have also relabeled the momenta of the quark and gluon to $k_1$ and $k_2$ respectively to facilitate the comparison with the LO expression. From here, one can convolve the expression with the $D_{h/q}$ fragmentation function and factorize the divergence in the same way as for the gluon-gluon contribution. In the end we find
\begin{align}
     \frac{\der\sigma^{gA\to gh(\bar{q})+X}}{\der^2\boldsymbol{k}_{1\perp}\der\eta_1 \der^2\boldsymbol{k}_{h\perp} \der\eta_{h}}&=\int_0^1\frac{\der\zeta'}{\zeta'^2} \ \left.\frac{\der\sigma^{gA\to gg+X}}{\der^2\boldsymbol{k}_{1}\der\eta_1 \der^2\boldsymbol{k}_{2\perp} \der\eta_{2}}\right|_{LO,k_2^\mu= k_h^\mu/\zeta'}\times\int_{\zeta'}^1\frac{\der\xi}{\xi}\frac{\alpha_s}{2\pi}P_{qg}^{real}(\xi)D^{(0)}_{h/q}\left(\frac{\zeta'}{\xi}\right)\nonumber\\
    &\times\left[\frac{2}{\varepsilon}-\ln(e^{\gamma_E}\pi\mu^2/\mu_F^2)+...\right]\,.
\end{align}
The contribution from the integration of a quark can be obtained almost on identical lines. Combining these three contributions we find:

\begin{equation}
D_{h/g}(\zeta,\mu_F^2)
=
D_{h/g}^{(0)}(\zeta)
+
\frac{\alpha_s}{2\pi}
\left[
\frac{2}{\varepsilon}
-\ln\!\left(e^{\gamma_E}\pi\mu^2/\mu_F^2\right)
\right]
\int_{\zeta}^{1}\frac{\der \xi}{\xi}\,
\left[P_{qg}(\xi)\,
D_{h/\Sigma}^{(0)}\!\left(\frac{\zeta}{\xi} \right)   + P_{gg}(\xi)\,
D_{h/g}^{(0)}\!\left(\frac{\zeta}{\xi}\right) 
\right]\,,
\end{equation}
where we defined the singlet 
\begin{align}
    D_{h/\Sigma}^{(0)}\!\left(\zeta \right) = \sum_q \left[ D_{h/q}^{(0)}\!\left(\zeta \right) + D_{h/{\bar q}}^{(0)}\!\left(\zeta \right) \right] \,.
\end{align}
Ultimately, the hybrid ``collinear plus CGC" factorization formula for inclusive dijet production in pA that emerges from our calculation of the real NLO corrections is
\begin{align}
         \frac{\der\sigma^{pA\to h_1h_2+X}}{\der^2\boldsymbol{k}_{h_1\perp}\der\eta_{h_1} \der^2\boldsymbol{k}_{h_2\perp} \der\eta_{h_2}}=\int_0^1\der x_p &\int_0^1\frac{\der \zeta_1}{\zeta_1^2}\int_0^1\frac{\der \zeta_2}{\zeta_2^2} \ g(x_p,\mu_F^2)D_{h_1/g}(\zeta_1,\mu_F^2)D_{h_2/g}(\zeta_2,\mu_F^2)\nonumber\\
         &\times \left.\frac{\der\sigma^{gA\to gg+X}}{\der^2\boldsymbol{k}_{1}\der\eta_1 \der^2\boldsymbol{k}_{2\perp} \der\eta_{2}}\right|_{k_1^\mu= k_{h_1}^\mu/\zeta_1,k_{2}^\mu=k_{h_2}^\mu/\zeta_2,p^-=x_pq^-}\,,
\end{align}
for the $g\to gg$ partonic channel. A complete proof would require the calculation of the virtual corrections as well.

\subsection{DGLAP evolution of the initial state quark in the $q \to qg$ channel}

We now study the divergence when the final-state antiquark is collinear with the incoming gluon and is integrated out. This contribution should be absorbed into the gluon-to-quark DGLAP evolution (thus, to the renormalization of the \textit{quark} PDF) of the initial state quark in the quark-initiated channel \cite{Iancu:2020mos}.

This initial-state collinear divergence can be isolated along the same lines as the one described in the previous subsection. In coordinate space, this contribution arises from the limit where the transverse separation between the antiquark and the incoming gluon $r_{w_1 y}$ is much larger than that between the final state quark and gluon, $r_{xz}$. In this limit, the function $\Theta_{q,1}$ can be approximated to unity and the amplitude reduces to 
\begin{equation}
    \mathcal{M}^{\bar{\lambda}\lambda\sigma\sigma',ab}_{q,ij} \simeq \int \der^6 \boldsymbol{\Pi} \,(C_{R1,ij}^{ab} - C_{R2,ij}^{ab})\Ncal^{\bar{\lambda}\lambda\sigma\sigma'}_{q,reg}\,. 
\end{equation}
The perturbative factor can be factorized in terms of the $q\to qg$ LO perturbative factor as
\begin{equation}
    \Ncal^{\bar{\lambda}\lambda\sigma\sigma'}_{q,reg} = \frac{ig}{\pi} \sqrt{\xi (1-\xi)}\, \Gamma^{\sigma \bar{\lambda}}_{g\xrightarrow{}q\bar{q}}(1-\xi) \frac{\et^{\bar{\lambda}}\cdot \boldsymbol{r}_{w_1y}}{r_{w_1y}^2} \Ncal_{qg,LO}^{\sigma\sigma' \lambda }\left( \frac{z_q}{\xi}, \frac{z_g}{\xi}, \dipvec{xz} \right)\,,
\end{equation}
where $\xi = 1-z_{\bar{q}}$. The $q\to qg$ LO perturbative factor is given by
\begin{align}
    \Ncal^{\bar{\lambda} \sigma \sigma'}_{qg,LO}(z_q,z_{g},\boldsymbol{r}_{xz}) = -\frac{2igp^-}{\pi} \sqrt{z_q}\, \Gamma^{\sigma \bar{\lambda}}_{q\xrightarrow{}qg}(z_q) \delta^{\sigma,-\sigma'} \frac{\et^{\bar{\lambda}}\cdot \dipvec{xz} }{r_{xz}^2}\,.
\end{align}
The squared modulus of the color operator, summed over the partonic colors and after integrating over the transverse momentum of the antiquark $\Transv{k}{\bar{q}}$, which identifies the transverse coordinates $\Transv{y}{}'=\Transv{y}{}$ yields
\begin{align}
    &\Tr\bigg\{\left[ \WL{x}t^ct^b\WLadj{y} U_{ac}(\Perp{z}) - t^a V(\Transv{w}{1})t^b\WLadj{y} \right]\nonumber\\
    &\times \left[ V(\Transv{y}{})t^bt^cV^{\dagger}(\Transv{x}{}') U_{ca}(\Perp{z}') -  V(\Transv{y}{})t^b V(\Transv{w}{1}')t^a\right] \bigg\} =C_F^2 \N \Xi_{qg,LO}(\Transv{x}{}', \Transv{z}{}'; \Transv{x}{}, \Transv{z}{})\,,
\end{align}
where the color operator is given by
\begin{align}
    \Xi_{qg,LO}(\Transv{x}{}', \Transv{z}{}'; \Transv{x}{}, \Transv{z}{}) &= \frac{1}{2C_F\N} \Bigg[ \N^2\big( D_{zz.}Q_{xzz'x'} - D_{xz}D_{zw_1'} - D_{z'x'}D_{w_1z'} \big)\nonumber\\
    &+ D_{w_1x'} + D_{xw_1'} - D_{xx'} - D_{w_1 w_1'} \Bigg] \,.
\end{align}
Similarly to the previous subsection, we can obtain the $g\to q$ DGLAP splitting function by squaring the ``square root" of the $g\to q\bar{q}$ splitting function thanks to the identity
\begin{equation}\label{eq: square_q-qqbar_splitting_id}
    \sum_{\bar{\lambda}} \big[\Gamma^{\sigma \bar{\lambda}}_{g\xrightarrow{}q\bar{q}}(1-\xi)\big]^2 \et^{\bar{\lambda},i} \et^{\lambda*,j} = P_{qg}^{real}(\xi)\delta^{ij} - i\sigma\frac{(1-\xi)^2 - \xi^2}{2} \varepsilon^{ij}\,.
\end{equation}
The divergent piece of the transverse integral over $\Transv{y}{}$ is diagonal in the indices $i,j$, hence it will project onto the diagonal piece of Eq.\,\eqref{eq: square_q-qqbar_splitting_id}, which is precisely the term that contains the relevant splitting function. We then have:
\begin{align}
    &\int \frac{\der \zbar}{\zbar} \int \der^2 \Transv{y}{} \sum_{\sigma \sigma' \lambda \bar{\lambda}} \Ncal^{\bar{\lambda}\lambda\sigma\sigma', \dagger}_{q,reg} \Ncal^{\bar{\lambda}\lambda\sigma\sigma'}_{q,reg} = \nonumber\\
    &\int_{0}^{1} \der \xi \,\xi  \frac{g^2}{\pi} P_{qg}^{real}(\xi)\, \int \frac{\der^2 \Transv{y}{}}{\pi} \frac{\dipvec{w_1 y} \cdot \dipvec{w_1'y}}{r_{w_1y}^2 r_{w_1'y}^2} \sum_{\sigma,\sigma',\lambda} \Ncal_{qg,LO}^{\sigma\sigma' \lambda, \dagger }\left( \frac{z_q}{\xi}, \frac{z_g}{\xi}, \dipvec{xz} \right) \Ncal_{qg,LO}^{\sigma\sigma' \lambda }\left( \frac{z_q}{\xi}, \frac{z_g}{\xi}, \dipvec{xz} \right)\,.
\end{align}

The differential cross section, integrated over the antiquark kinematics, can be readily found to be
\begin{align}\label{eq: DGLAP_g_to_q_partonic}
    \frac{\der \sigma^{gA \to qg(\bar{q}) +X}}{\der \eta_{q} \der^2 \Transv{k}{q} \der \eta_{g} \der^2 \Transv{k}{g}} &= \frac{2\alpha_s C_F}{(2\pi)^6}\int_{0}^{1} d\xi \, \delta\left(1-\frac{z_q}{\xi} - \frac{z_g}{\xi}\right)  \frac{\alpha_s}{2\pi} P_{qg}^{real}(\xi)\, \int \der^4 \boldsymbol{\Pi}_{LO} \der^4\boldsymbol{\Pi}_{LO}'\nonumber\\
    &\mathcal{K}_{qg,LO}\left(\frac{z_q}{\xi}, \frac{z_g}{\xi}, \dipvec{xz}, \dipvec{x'z'} \right) \avg{\Xi_{qg,LO}(\Transv{x}{}, \Transv{z}{}'; \Transv{x}{}, \Transv{z}{})}{Y}\int \frac{\der^2 \Transv{y}{}}{\pi} \frac{\dipvec{w_1 y} \cdot \dipvec{w_1'y}}{r_{w_1y}^2 r_{w_1'y}^2}\,,
\end{align}
with the LO $q\to qg$ kernel defined as \cite{Iancu:2018hwa}
\begin{equation}
    \mathcal{K}_{qg,LO}\left(z_q, z_g, \dipvec{xz}, \dipvec{x'z'} \right) = z_q (1 + z_q^2) \frac{\dipvec{xz} \cdot \dipvec{x'z'}}{ r_{xz}^2 r_{x'z'}^2}\,.
\end{equation}
Eq.\,\eqref{eq: DGLAP_g_to_q_partonic} features the LO $q \to qg$ partonic cross section evaluated at $z_i\to z_i/\xi$. As in the previous case, we rescale the longitudinal momentum of the incoming gluon to $p^- = \xi x_pq^-$:
\begin{align}
    \frac{\der \sigma^{gA \to qg(\bar{q})+X}}{\der \eta_{q} \der^2 \Transv{k}{q} \der \eta_{g} \der^2 \Transv{k}{g}} = \int_{0}^{1} \der \xi \frac{\der \sigma^{qA \to qg +X}}{\der \eta_{q} \der^2 \Transv{k}{q} \der \eta_{g} \der^2 \Transv{k}{g}}\Bigg|_{p^- = \xi x_pq^-} \times\frac{\alpha_s}{2\pi} P_{qg}^{real}(\xi)\,\int \frac{\der^2 \Transv{y}{}}{\pi} \frac{\dipvec{w_1 y} \cdot \dipvec{w_1'y}}{r_{w_1y}^2 r_{w_1'y}^2}\,.
    \label{gtoqqbarg-antiquark-inregrated}
\end{align}
To obtain the \textit{dijet} cross section, one has to  convolve the partonic cross section with the ``bare" gluon PDF:
\begin{equation}
    \frac{\der \sigma^{pA \to qg(\bar{q}) + X}}{\der \eta_{q} \der^2 \Transv{k}{q} \der \eta_{g} \der^2 \Transv{k}{g}} = \int_{0}^{1} \der x_p \frac{\der \sigma^{gA \to qg(\bar{q})+ X}}{\der \eta_{q} \der^2 \Transv{k}{q} \der \eta_{g} \der^2 \Transv{k}{g}} \times g^{(0)}(x_p)\,.
    \label{eq:gtoqqbarg-convoluted}
\end{equation}
Substituting Eq.\,\eqref{gtoqqbarg-antiquark-inregrated} into Eq.\,\eqref{eq:gtoqqbarg-convoluted} and making the change of variables $x_p' = \xi x_p$ yields
\begin{align}
    \frac{\der\sigma^{pA\to qg (\bar{q})+X}}{\der^2\boldsymbol{k}_{q\perp}\der\eta_q \der^2\boldsymbol{k}_{{\bar{q}}\perp} \der\eta_{\bar{q}}}&=\int_0^1\der x_p' \ \left.\frac{\der\sigma^{qA\to qg +X}}{\der^2\boldsymbol{k}_{q\perp}\der\eta_q \der^2\boldsymbol{k}_{g\perp} \der\eta_{g}}\right|_{LO,p^-= x_p'q^-}\times \int_{x_p'}^1\frac{\der \xi}{\xi}\frac{\alpha_s}{2\pi}P_{qg}^{real}(\xi) g^{(0)}\left(\frac{x_p'}{\xi}\right)\nonumber\\
    &\times\left[\frac{2}{\varepsilon}-\ln(e^{\gamma_E}\pi\mu^2/\mu_F^2)+...\right]\,.\label{eq:pAtoqg-dglap-init}
\end{align}
The lower bound in the integral over $\xi$ comes from the restriction on the longitudinal momentum of the gluon relative to the proton $x_p\leq 1$. The collinear divergence isolated in Eq.\,\eqref{eq:pAtoqg-dglap-init} is absorbed into the renormalization of the quark PDF through the gluon-to-quark mixing term. In analogy with Eq.\,\eqref{eq:renorm_gluon-PDF}, we define the renormalized quark distribution as
\begin{equation}
q(x_p,\mu_F^2)
=
q^{(0)}(x_p)
+
\frac{\alpha_s}{2\pi}
\left[
\frac{2}{\varepsilon}
-\ln\!\left(e^{\gamma_E}\pi\mu^2/\mu_F^2\right)
\right]
\int_{x_p}^{1}\frac{\der \xi}{\xi}\,
P^{\rm real}_{qg}(\xi)\,
g^{(0)}\!\left(\frac{x_p}{\xi}\right)
+\cdots ,
\end{equation}
where the displayed term is the gluon-to-quark mixing contribution to the quark PDF renormalization. Here $P^{\rm real}_{qg}(\xi)$ is the real splitting kernel associated with the collinear splitting $g\to q\bar q$. The ellipsis denotes the usual quark-to-quark renormalization contribution as well as finite terms. This contribution can be obtained from the $q\to qgg$ channel in \cite{Iancu:2018hwa} in a similar fashion.

\section{Differential cross section for trijet production}\label{sec: Cross section trijets}
In this final section, we proceed to obtain the differential cross section for the trijet production in the gluon-initiated channel. The calculation is straightforward, but cumbersome due to the several subprocesses contributing to the amplitudes. Under the hybrid approximation, we can write the cross section for the trijet production initiated by a gluon as
\begin{multline}
    \frac{\der\sigma^{pA\rightarrow 3\textrm{jet} +X}}{\der^2\boldsymbol{k}_{1\perp}\der\eta_1 \der^2\boldsymbol{k}_{2\perp} \der\eta_{2} \der^2\boldsymbol{k}_{3\perp} \der\eta_3}\\
    = \int dx_p\,  g(x_p,\mu^2) \bigg(\frac{\der\sigma^{gA\rightarrow q\bar{q}g +X}}{\der^2\boldsymbol{k}_{1\perp}\der\eta_1 \der^2\boldsymbol{k}_{2\perp} \der\eta_{2} \der^2\boldsymbol{k}_{3\perp} \der\eta_3} + \frac{\der\sigma^{gA\rightarrow ggg +X}}{\der^2\boldsymbol{k}_{1\perp}\der\eta_1 \der^2\boldsymbol{k}_{2\perp} \der\eta_{2} \der^2\boldsymbol{k}_{3\perp} \der\eta_3}  \bigg) \,,
\end{multline}
where $x_p g(x_p,\mu^2)$ is the gluon PDF of the proton evaluated at the factorization scale $\mu^2$. The variable $x_p$ corresponds to the longitudinal momentum of the gluon relative to the proton. The partonic cross section is obtained by squaring the amplitude, averaging (summing) over initial (final) state quantum numbers, and multiplying by the flux factor and the phase space of the final-state particles:
\begin{align}\label{trijet cross section general}
    & \frac{\der\sigma^{gA\rightarrow 3\textrm{jet} +X}}{\der^2\boldsymbol{k}_{1\perp}\der\eta_1 \der^2\boldsymbol{k}_{2\perp} \der\eta_{2} \der^2\boldsymbol{k}_{3\perp} \der\eta_3} \nonumber \\
    & = \frac{1}{8(2\pi)^9}\frac{1}{2(p^-)^2}(2\pi)\delta(1-z_1- z_{2} - z_3)\times  \frac{1}{2(\N^2-1)} \sum_{\begin{smallmatrix} \mathrm{pol}, & \\ \mathrm{colors} \end{smallmatrix}} \left\langle  \big|\mathcal{M}[\rho_A]\big|^2\right\rangle_Y\,.
\end{align}
To keep this section relatively compact, as we have intended to do with the previous sections, for each contribution, we only present the main expression for the differential cross section. The definitions of the kernels and $S$-matrices involved on each channel are given in Appendix \ref{app: Kernels_and_Smatrices}.

\subsection{The $g \rightarrow q\bar{q}g$ channel}
At the cross section level for this channel, we have ``direct" contributions from the square of the gluon emission from quark, antiquark and gluon, and the corresponding interferences between these processes. The number of independent contributions from the different topologies is 6, which we can express schematically as
\begin{align}
    \der \sigma = \der \sigma \big|_q + \der \sigma \big|_{\bar q} + \der \sigma \big|_g + 2 \mathrm{Re}\left( \der \sigma \big|_{qg-int} + \der \sigma \big|_{\bar{q}g-int} + \der \sigma \big|_{q\bar{q}-int} \right)\,.
\end{align}
Let us take for example the contribution from the gluon emission by quark amplitudes. The cross section corresponding to this process will feature three kinds of kernels. The first one arises from the regular contributions of the diagrams. Since all diagrams contain a regular term, there will be nine contributions featuring this kernel. The second kernel comes from the interference between the regular and instantaneous amplitudes. There are twelve contributions coming from this interference. Finally, there will be four contributions coming from the instantaneous kernel. The color operators involved in the cross section can be simplified using Fierz identities to simplify the color matrices when we sum (average) over the color indices. The rest of the contributions to the total cross section can be obtained in a similar manner.\\

\subsubsection{Gluon emission by quark}
\begin{figure}
    \centering
    \includegraphics[width=1\linewidth]{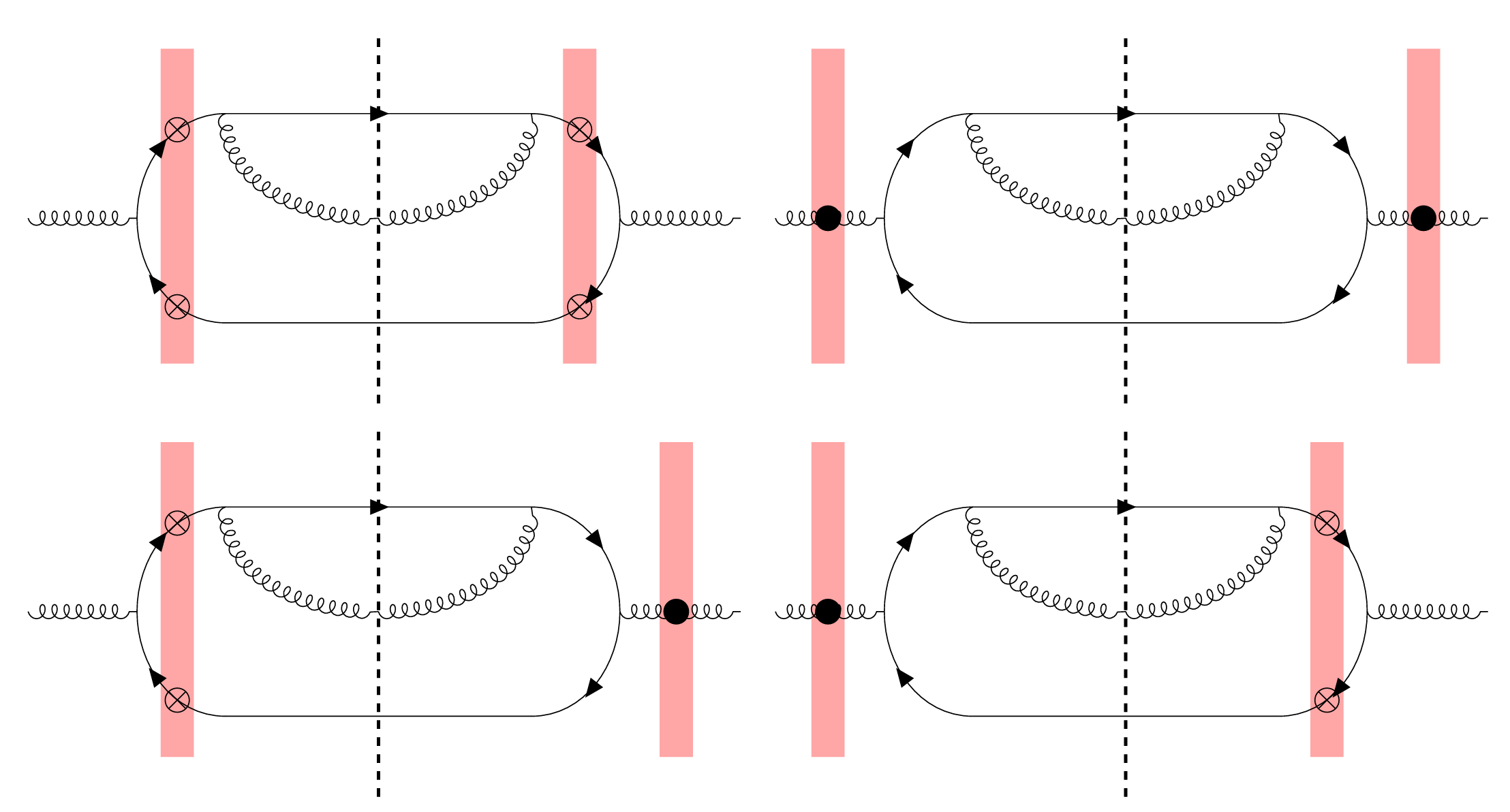}
    \caption{Diagrams contributing to the fourth term of the regular part of the cross  from the direct term of the quark contribution. The diagrams describe intermediate and initial state interactions.}
    \label{fig:Quark_intermediate_and_initial_interactions}
\end{figure}

Starting from the expression for the total amplitude from the gluon emission by a quark, Eq.\,\eqref{quark parent qbarqg amplitude}, and using Eq.\,\eqref{trijet cross section general} we obtain the following contribution to the cross section

\begin{multline}\label{cross_section_quark_contribution}
    \frac{\der\sigma^{gA\rightarrow q\bar{q}g +X}}{\der^2\boldsymbol{k}_{q\perp}\der\eta_q \der^2\boldsymbol{k}_{{\bar{q}}\perp} \der\eta_{\bar{q}} \der^2\boldsymbol{k}_{g\perp} \der\eta_g} \bigg|_{q} = \frac{2 C_F }{(2\pi)^{10}} \frac{\delta(1-z_q- z_{\bar{q}} - z_g)\pi^2}{4(p^-)^2\times2(2C_F)(\N^2-1)} \sum_{\begin{smallmatrix} \bar{\lambda}\lambda\sigma\sigma',  \\ ab,ij \end{smallmatrix}} \left\langle \big|\mathcal{M}_{q}^{\bar{\lambda}\lambda\sigma\sigma'}[\rho_A] \big|^2\right\rangle_Y \\
    =\frac{\alpha_s^2 \delta(1-z_q- z_{\bar{q}} - z_g)}{(2\pi)^{10}} \frac{C_F}{2} \int \der^6\boldsymbol{\Pi} \der^6\boldsymbol{\Pi}' \bigg\{ \mathcal{K}_1(\boldsymbol{r}_{x'z'},\boldsymbol{r}_{w_1'y'};\boldsymbol{r}_{xz},\boldsymbol{r}_{w_1y})\\
    \bigg[\Theta'_{q,1}\Theta_{q,1} \avg{\Xi_{1} (\Perp{x}', \Perp{y}', \Perp{z}';\Perp{x}, \Perp{y}, \Perp{z})}{Y} -(\{\Perp{x},\Perp{z}\} \to \Transv{w}{1}) - (\{\Perp{x}',\Perp{z}'\} \to \Transv{w}{1}')\\
    + (\{\Perp{x},\Perp{z}\} \to \Transv{w}{1}, \{\Perp{x}',\Perp{z}'\} \to \Transv{w}{1}') \bigg] + \mathcal{K}_2(\boldsymbol{r}_{x'z'},\boldsymbol{r}_{w_1'y'};X_R)\bigg[\Theta'_{q,1}\avg{\Xi_{1} (\Perp{x}', \Perp{y}', \Perp{z}';\Perp{x}, \Perp{y}, \Perp{z})}{Y}\\
    - (\{\Perp{x}',\Perp{z}'\} \to \Transv{w}{1}') \bigg] + \mathcal{K}_2(X_R';\boldsymbol{r}_{zx},\boldsymbol{r}_{w_1y})\bigg[\Theta_{q,1} \avg{\Xi_{1} (\Perp{x}', \Perp{y}', \Perp{z}';\Perp{x}, \Perp{y}, \Perp{z})}{Y}- (\{\Perp{x},\Perp{z}\} \to \Transv{w}{1}) \bigg]\\
    + \mathcal{K}_3(X_R';X_R) \avg{\Xi_{1} (\Perp{x}', \Perp{y}', \Perp{z}';\Perp{x}, \Perp{y}, \Perp{z})}{Y}\bigg\}\,.
\end{multline}
The terms in the cross section corresponding to the change of variables $\{\Perp{x},\Perp{z}\} \to \Transv{w}{1}$ replace the final state interaction terms in the direct amplitude with intermediate interactions (the primed coordinates replace the interactions in the conjugated amplitude). The diagrams contributing to the fourth term of the regular kernel, where we represent all final state interactions with intermediate interactions are shown in Figure \ref{fig:Quark_intermediate_and_initial_interactions}. As anticipated from the discussion at the beginning of this section, Eq.\,\eqref{cross_section_quark_contribution} features three different kernels. The first kernel corresponds to the square of the regular contribution. The second kernel refers to the interference between the regular and the instantaneous contribution. Finally, the third kernel encodes the square of the instantaneous contribution. 
The color operator appearing in the cross section is defined as
\begin{multline}
    \avg{\Xi_{1} (\Perp{x}', \Perp{y}', \Perp{z}';\Perp{x}, \Perp{y}, \Perp{z})}{Y}\\
    = S_{q\bar{q}g}^{q\bar{q}g}(\Perp{x}', \Perp{y}', \Perp{z}';\Perp{x}, \Perp{y}, \Perp{z}) - S_{q\bar{q}g}^{g}(\Perp{v}';\Perp{x}, \Perp{y}, \Perp{z}) - S^{q\bar{q}g}_{g}(\Perp{x}', \Perp{y}', \Perp{z}';\Perp{v}) + S_{gg}(\Perp{v}'; \Perp{v})\,.
\end{multline}

\begin{figure}
    \centering
    \includegraphics[width=1\linewidth]{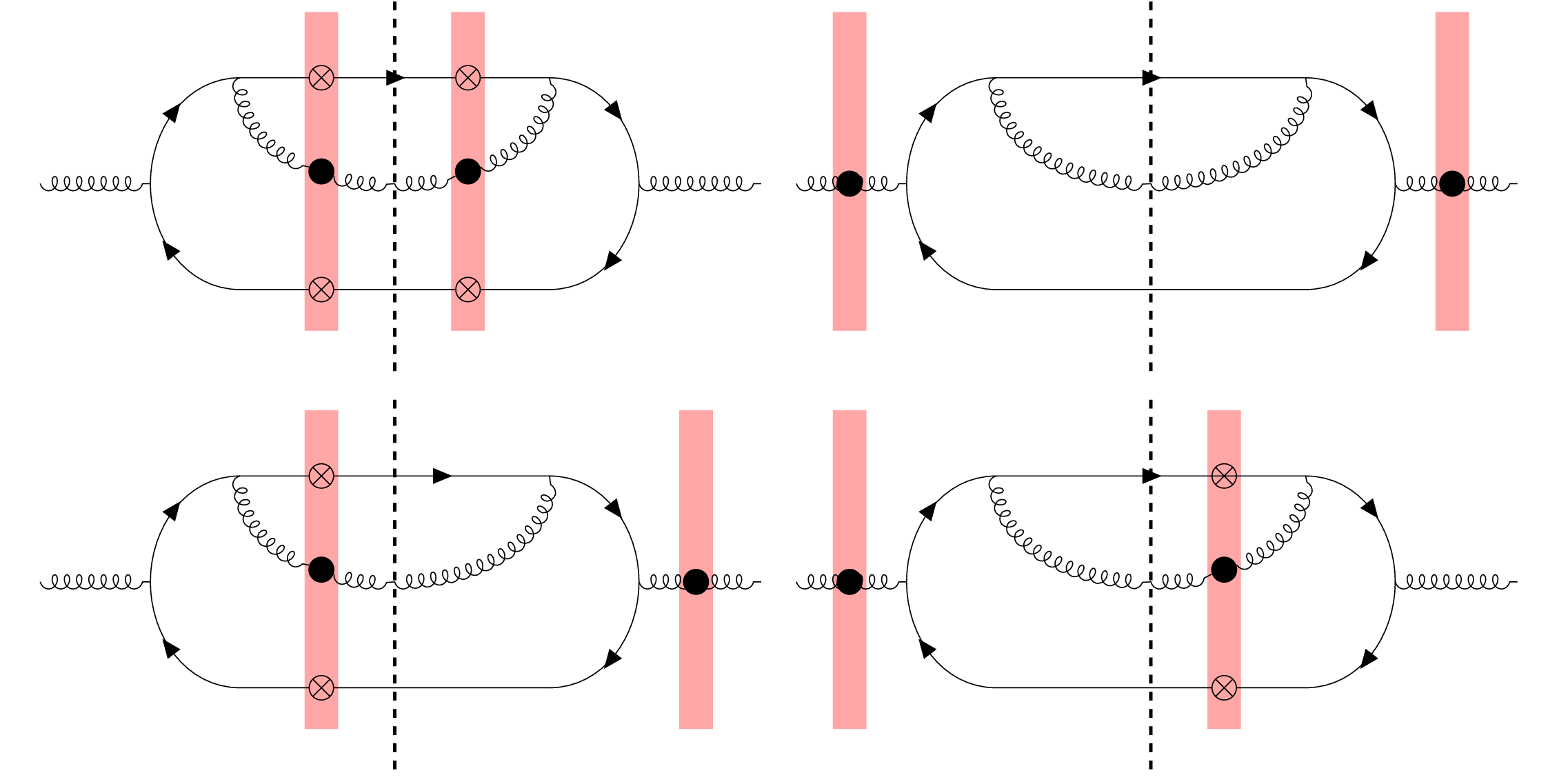}
    \caption{Diagrams involved in the color correlator $\Xi_{1} (\Perp{x}', \Perp{y}', \Perp{z}';\Perp{x}, \Perp{y}, \Perp{z})$. The correlator encodes final state and initial state interactions.}
    \label{fig:Quark_final_and_initial_interactions}
\end{figure}

This color correlator corresponds to the square of the difference between the color operators of diagrams R1 and R3. As in Section \ref{sec: Dijet prod LO} where the S-matrices are introduced in the LO dijet cross section, we intend to denote the partons participating in the multiple scattering with lower and upper indices for the $S$ symbol; the lower indices corresponding to the partons involved in the direct amplitude, and the upper indices to the partons involved in the conjugated amplitude. If the S-matrices are real, the correlator $S^{q\bar{q}g}_{g}(\Perp{x}', \Perp{y}', \Perp{z}';\Perp{v})$ can be expressed as 
$S_{q\bar{q}g}^{g}(\Perp{v};\Perp{x}', \Perp{y}', \Perp{z}')$. These correlators must be evaluated at the rapidity corresponding to the longitudinal momentum fraction transferred by the target
\begin{equation}
    x_g \equiv \frac{1}{x_p s}\left(\frac{\Transv{k}{q}^2}{z_q}+\frac{\Transv{k}{\bar{q}}^2}{z_{\bar{q}}}+\frac{\Transv{k}{g}^2}{z_g}\right)\,,
\end{equation}
 where $s$ is the squared center of mass energy of the proton-nucleus system. Naturally, by changing the corresponding labels of the momenta of the final state partons, an analogous value of $x_g$ is to be used in the $ggg$ case.

The normalization of the scattering matrices is such that in the non-interacting limit, they result in the identity operator. For a given S-matrix, the normalization factor is therefore the color factor of the ``bare" diagram that generates the matrix, i.e.~the diagram with no shockwave insertion.

\subsubsection{Gluon emission by gluon}

Similarly for the emission by the gluon, we start from Eq.\,\eqref{gluon parent qbarqg amplitude}. The resulting cross section will have contributions from three different kernels as well, corresponding to contributions from regular, instantaneous, and interferences between regular and instantaneous terms. The final expression for the cross section of this process reads
\begin{multline}
    \frac{\der\sigma^{gA\rightarrow q\bar{q}g +X}}{\der^2\boldsymbol{k}_{q\perp}\der\eta_q \der^2\boldsymbol{k}_{{\bar{q}}\perp} \der\eta_{\bar{q}} \der^2\boldsymbol{k}_{g\perp} \der\eta_g} \bigg|_{g} = \\
    =\frac{\alpha_s^2  \delta(1-z_q- z_{\bar{q}} - z_g)}{(2\pi)^{10}} \frac{\N}{2} \int \der^6\boldsymbol{\Pi} \der^6\boldsymbol{\Pi}' \bigg\{ \mathcal{K}_4(\boldsymbol{r}_{y'x'},\boldsymbol{r}_{w_2'z'};\boldsymbol{r}_{yx},\boldsymbol{r}_{w_2z})\\
    \bigg[\Theta'_{g,1}\Theta_{g,1} \avg{\Xi_{2} (\Perp{x}', \Perp{y}', \Perp{z}';\Perp{x}, \Perp{y}, \Perp{z})}{Y}-(\{\Perp{x},\Perp{y}\} \to \boldsymbol{w}_{2\perp}) - (\{\Perp{x}',\Perp{y}'\} \to \boldsymbol{w}_{2\perp}')\\
    + (\{\Perp{x},\Perp{y}\} \to \boldsymbol{w}_{2\perp}, \{\Perp{x}',\Perp{y}'\} \to \boldsymbol{w}_{2\perp}') \bigg]+ \mathcal{K}_5(\boldsymbol{r}_{y'x'},\boldsymbol{r}_{w_2'z'};X_R)\bigg[\Theta'_{g,1} \avg{\Xi_{2} (\Perp{x}', \Perp{y}', \Perp{z}';\Perp{x}, \Perp{y}, \Perp{z})}{Y}\\
    - (\{\Perp{x}',\Perp{y}'\} \to \boldsymbol{w}_{2\perp}') \bigg] + \mathcal{K}_5(X_R';\boldsymbol{r}_{yx},\boldsymbol{r}_{w_2z}) \bigg[\Theta_{g,1} \avg{\Xi_{2} (\Perp{x}', \Perp{y}', \Perp{z}';\Perp{x}, \Perp{y}, \Perp{z})}{Y}\\
    - (\{\Perp{x},\Perp{y}\} \to \boldsymbol{w}_{2\perp}) \bigg]+ \mathcal{K}_6(X_R';X_R) \avg{\Xi_{2} (\Perp{x}', \Perp{y}', \Perp{z}';\Perp{x}, \Perp{y}, \Perp{z})}{Y}\bigg\}\,.
\end{multline}
Analogous to the case of the gluon emission by a quark, the change of variables \((\{\boldsymbol{x}_\perp, \boldsymbol{y}_\perp\} \to \boldsymbol{w}_{2\perp})\) transforms the terms associated with final-state interactions in the direct amplitude into terms corresponding 
to intermediate state interactions --- in this case, two intermediate gluons. The same reasoning applies to the primed variables. 

Similarly, the color operator appearing in the color scattering matrices of the cross section contains the square of the sum of the diagrams with final state interactions (diagram R4) and initial state interactions (diagram R6). The expression reads
\begin{multline}
    \avg{\Xi_{2} (\Perp{x}', \Perp{y}', \Perp{z}';\Perp{x}, \Perp{y}, \Perp{z})}{Y}\\
    = S_{q\bar{q}g}^{q\bar{q}g, (2)}(\Perp{x}', \Perp{y}', \Perp{z}';\Perp{x}, \Perp{y}, \Perp{z}) - S_{q\bar{q}g}^{g, (2)}(\Perp{v}';\Perp{x}, \Perp{y}, \Perp{z}) - S_{q\bar{q}g}^{g, (2)}(\Perp{v};\Perp{x}', \Perp{y}', \Perp{z}') + S_{g}^{g}(\Perp{v}'; \Perp{v})\,.
\end{multline}

\subsubsection{Gluon-quark interference}

Next, we move to the first of the interference terms. The interference between the gluon emission by gluon diagrams and the gluon emission by quark diagrams features an extra kernel, giving a total of four of them in the cross section. We calculate the contribution from the gluon emission by gluon in the direct amplitude and the gluon emission by a quark in the conjugate amplitude, but the contribution from the mirror diagrams is identical. The cross section has a form similar to the previous contributions and reads

\begin{multline}
    \frac{\der\sigma^{gA\rightarrow q\bar{q}g +X}}{\der^2\boldsymbol{k}_{q\perp}\der\eta_q \der^2\boldsymbol{k}_{{\bar{q}}\perp} \der\eta_{\bar{q}} \der^2\boldsymbol{k}_{g\perp} \der\eta_g} \bigg|_{qg-int}= \\
    \frac{\alpha_s^2 \delta(1-z_q- z_{\bar{q}} - z_g)}{(2\pi)^{10}} \frac{\N}{4} \int \der^6\boldsymbol{\Pi} \der^6\boldsymbol{\Pi}' \bigg\{ \mathcal{K}_7(\boldsymbol{r}_{x'z'},\boldsymbol{r}_{w_1'y'};\boldsymbol{r}_{xy},\boldsymbol{r}_{w_2z})\\
    \bigg[\Theta'_{q,1}\Theta_{g,1} \avg{\Xi_{3} (\Perp{x}', \Perp{y}', \Perp{z}';\Perp{x}, \Perp{y}, \Perp{z})}{Y} -(\{\Perp{x},\Perp{y}\} \to \Transv{w}{2}) - (\{\Perp{x}',\Perp{z}'\} \to \Transv{w}{1}')\\
    + (\{\Perp{x},\Perp{y}\} \to \Transv{w}{2}, \{\Perp{x}',\Perp{z}'\} \to \Transv{w}{1}') \bigg]+ \mathcal{K}_8(\boldsymbol{r}_{x'z'},\boldsymbol{r}_{w_1'y'};X_R)\bigg[\Theta'_{q,1} \avg{\Xi_{3} (\Perp{x}', \Perp{y}', \Perp{z}';\Perp{x}, \Perp{y}, \Perp{z})}{Y}\\
    - (\{\Perp{x}',\Perp{z}'\} \to \Transv{w}{1}') \bigg] + \mathcal{K}_9(X_R';\boldsymbol{r}_{yx},\boldsymbol{r}_{w_2z}) \bigg[\Theta_{g,1} \avg{\Xi_{3} (\Perp{x}', \Perp{y}', \Perp{z}';\Perp{x}, \Perp{y}, \Perp{z})}{Y}\\
    - (\{\Perp{x},\Perp{y}\} \to \Transv{w}{2}) \bigg]+ \mathcal{K}_{10}(X_R';X_R) \avg{\Xi_{3} (\Perp{x}', \Perp{y}', \Perp{z}';\Perp{x}, \Perp{y}, \Perp{z})}{Y}\bigg\}\,.
\end{multline}
In the same way, the color operator in the cross section represents the square of the initial and final state interactions between the gluon emission by quark channel and the gluon emission by gluon channel. The expression reads
\begin{multline}
    \avg{\Xi_{3} (\Perp{x}', \Perp{y}', \Perp{z}';\Perp{x}, \Perp{y}, \Perp{z})}{Y}\\
    = S_{q\bar{q}g}^{q\bar{q}g, (3)}(\Perp{x}', \Perp{y}', \Perp{z}';\Perp{x}, \Perp{y}, \Perp{z}) - S_{q\bar{q}g}^{g, (3)}(\Perp{v}';\Perp{x}, \Perp{y}, \Perp{z}) - S_{q\bar{q}g}^{g, (3)}(\Perp{v};\Perp{x}', \Perp{y}', \Perp{z}') + S_{g}^{g}(\Perp{v}'; \Perp{v})\,.
\end{multline}

\subsubsection{Gluon-antiquark interference}
The cross section of the gluon-antiquark interference gives a similar expression as the previous contribution. The expression reads

\begin{multline}
    \frac{\der\sigma^{gA\rightarrow q\bar{q}g +X}}{\der^2\boldsymbol{k}_{q\perp}\der\eta_q \der^2\boldsymbol{k}_{{\bar{q}}\perp} \der\eta_{\bar{q}} \der^2\boldsymbol{k}_{g\perp} \der\eta_g} \bigg|_{\bar{q}g-int}= \\
    \frac{\alpha_s^2 \delta(1-z_q- z_{\bar{q}} - z_g)}{(2\pi)^{10}} \frac{\N}{4} \int \der^6\boldsymbol{\Pi} \der^6\boldsymbol{\Pi}' \bigg\{ \mathcal{K}_{11}(\boldsymbol{r}_{z'y'},\boldsymbol{r}_{w_3'z'};\boldsymbol{r}_{yx},\boldsymbol{r}_{w_2 z})\\
    \bigg[\Theta'_{\bar{q},1}\Theta_{g,1} \avg{\Xi_{4} (\Perp{x}', \Perp{y}', \Perp{z}';\Perp{x}, \Perp{y}, \Perp{z})}{Y}-(\{\Perp{x},\Perp{y}\} \to \Transv{w}{2}) - (\{\Perp{y}',\Perp{z}'\} \to \Transv{w}{3}')\\
    + (\{\Perp{x},\Perp{y}\} \to \Transv{w}{2}, \{\Perp{y}',\Perp{z}'\} \to \Transv{w}{3}') \bigg]+ \mathcal{K}_{12}(\boldsymbol{r}_{z'y'},\boldsymbol{r}_{w_3'x'}; X_R)\bigg[\Theta'_{\bar{q},1} \avg{\Xi_{4} (\Perp{x}', \Perp{y}', \Perp{z}';\Perp{x}, \Perp{y}, \Perp{z})}{Y}\\
    - (\{\Perp{y}',\Perp{z}'\} \to \Transv{w}{3}') \bigg] + \mathcal{K}_{13}(X_R' ;\boldsymbol{r}_{yx},\boldsymbol{r}_{w_2 z}) \bigg[\Theta_{g,1} \avg{\Xi_{4} (\Perp{x}', \Perp{y}', \Perp{z}';\Perp{x}, \Perp{y}, \Perp{z})}{Y}\\
    - (\{\Perp{x},\Perp{y}\} \to \Transv{w}{2}) \bigg]+ \mathcal{K}_{14}(X_R';X_R) \avg{\Xi_{4} (\Perp{x}', \Perp{y}', \Perp{z}';\Perp{x}, \Perp{y}, \Perp{z})}{Y}\bigg\}\,.
\end{multline}
The color operator appearing in the cross section is defined as
\begin{multline}
    \avg{\Xi_{4} (\Perp{x}', \Perp{y}', \Perp{z}';\Perp{x}, \Perp{y}, \Perp{z})}{Y}\\
    = S_{q\bar{q}g}^{q\bar{q}g, (4)}(\Perp{x}', \Perp{y}', \Perp{z}';\Perp{x}, \Perp{y}, \Perp{z}) - S_{q\bar{q}g}^{g, (4)}(\Perp{v}';\Perp{x}, \Perp{y}, \Perp{z}) - S_{q\bar{q}g}^{g, (4)}(\Perp{v};\Perp{x}', \Perp{y}', \Perp{z}') + S_{g}^{g}(\Perp{v}'; \Perp{v})\,.
\end{multline}
The S-matrices of the quark-gluon and the antiquark-gluon interferences are related to the S-matrices of the gluon contribution in the following way: 
\begin{align}
    \Xi_{2} (\Perp{x}', \Perp{y}', \Perp{z}';\Perp{x}, \Perp{y}, \Perp{z}) = (\Xi_{4} (\Perp{x}', \Perp{y}', \Perp{z}';\Perp{x}, \Perp{y}, \Perp{z}) - \Xi_{3} (\Perp{x}', \Perp{y}', \Perp{z}';\Perp{x}, \Perp{y}, \Perp{z}) )/2\,.
\end{align}
The intermediate state interaction matrices, however, do not satisfy this relation as they depend on the transverse coordinates of the parton that emits the final state gluon.

\subsubsection{Quark-antiquark interference}

The interference between the quark and the antiquark features four kernels at the level of the cross sections. However, the contribution from the product of the instantaneous terms will vanish after summing over the polarizations and helicities. This leaves us with only three kernels. The resulting cross section coming from this interference reads
\begin{multline}
    \frac{\der\sigma^{gA\rightarrow q\bar{q}g +X}}{\der^2\boldsymbol{k}_{q\perp}\der\eta_q \der^2\boldsymbol{k}_{{\bar{q}}\perp} \der\eta_{\bar{q}} \der^2\boldsymbol{k}_{g\perp} \der\eta_g} \bigg|_{q\bar{q}-int} = \\
    =\frac{\alpha_s^2 \delta(1-z_q- z_{\bar{q}} - z_g)}{(2\pi)^{10}} \frac{1}{8\N} \int \der^6\boldsymbol{\Pi} \der^6\boldsymbol{\Pi}' \bigg\{ \mathcal{K}_{15} (\boldsymbol{r}_{y'z'},\boldsymbol{r}_{w_3'x'};\boldsymbol{r}_{zx},\boldsymbol{r}_{w_1y})\\
    \bigg[\Theta'_{\bar{q},1}\Theta_{q,1} \avg{\Xi_{5} (\Perp{x}', \Perp{y}', \Perp{z}';\Perp{x}, \Perp{y}, \Perp{z})}{Y}-(\{\Perp{x},\Perp{z}\} \to \boldsymbol{w}_{1\perp}) - (\{\Perp{y}',\Perp{z}'\} \to \boldsymbol{w}_{3\perp}')\\
    + (\{\Perp{x},\Perp{z}\} \to \boldsymbol{w}_{1\perp}, \{\Perp{y}',\Perp{z}'\} \to \boldsymbol{w}_{3\perp}') \bigg]+ \mathcal{K}_{16} (\boldsymbol{r}_{y'z'},\boldsymbol{r}_{w_3'x'};X_{R})\bigg[\Theta'_{\bar{q},1} \avg{\Xi_{5} (\Perp{x}', \Perp{y}', \Perp{z}';\Perp{x}, \Perp{y}, \Perp{z})}{Y}\\
    - (\{\Perp{x}',\Perp{y}'\} \to \boldsymbol{w}_{3\perp}') \bigg] + \mathcal{K}_{17} (X_{R}';\boldsymbol{r}_{zx},\boldsymbol{r}_{w_1y}) \bigg[\Theta_{q,1} \avg{\Xi_{5} (\Perp{x}', \Perp{y}', \Perp{z}';\Perp{x}, \Perp{y}, \Perp{z})}{Y}\\
    - (\{\Perp{x},\Perp{z}\} \to \boldsymbol{w}_{1\perp}) \bigg]\bigg\}\,.
\end{multline}
Notice that this contribution is suppressed by a power of $\N^2$ compared to the rest of the contributions. The corresponding color operator is
\begin{multline}
    \avg{\Xi_{5} (\Perp{x}', \Perp{y}', \Perp{z}';\Perp{x}, \Perp{y}, \Perp{z})}{Y}\\
    = S_{q\bar{q}g}^{q\bar{q}g, (5)}(\Perp{x}', \Perp{y}', \Perp{z}';\Perp{x}, \Perp{y}, \Perp{z}) - S_{q\bar{q}g}^{g, (5)}(\Perp{v}';\Perp{x}, \Perp{y}, \Perp{z}) - S_{q\bar{q}g}^{g, (5)}(\Perp{v};\Perp{x}', \Perp{y}', \Perp{z}') + S_{g}^{g}(\Perp{v}'; \Perp{v})\,.
\end{multline}

\subsection{The $g\rightarrow ggg$ channel}
We now move to the cross section of the $ggg$ production. Analogous to the previous calculation, we start from the amplitude obtained in Section \ref{sec: Amplitudes trijet}, namely Eq.\,\eqref{gluon_to_ggg_total_amplitude}. We will split the cross section into three components due to the large amount of terms coming from the exchange of partons: the regular contribution, the ``total" instantaneous contribution, and the interference between the regular and the instantaneous contribution. As in the $q\bar{q}g$ case, the resulting cross section can be schematically written as
\begin{align}
    \der \sigma = \der \sigma \big|_{reg} + \der \sigma \big|_{inst} + 2 \mathrm{Re}\left( \der \sigma \big|_{reg-inst} \right)\,.
\end{align}

\subsubsection{Regular contribution}
\begin{figure}
    \centering
    \includegraphics[width=0.9\linewidth]{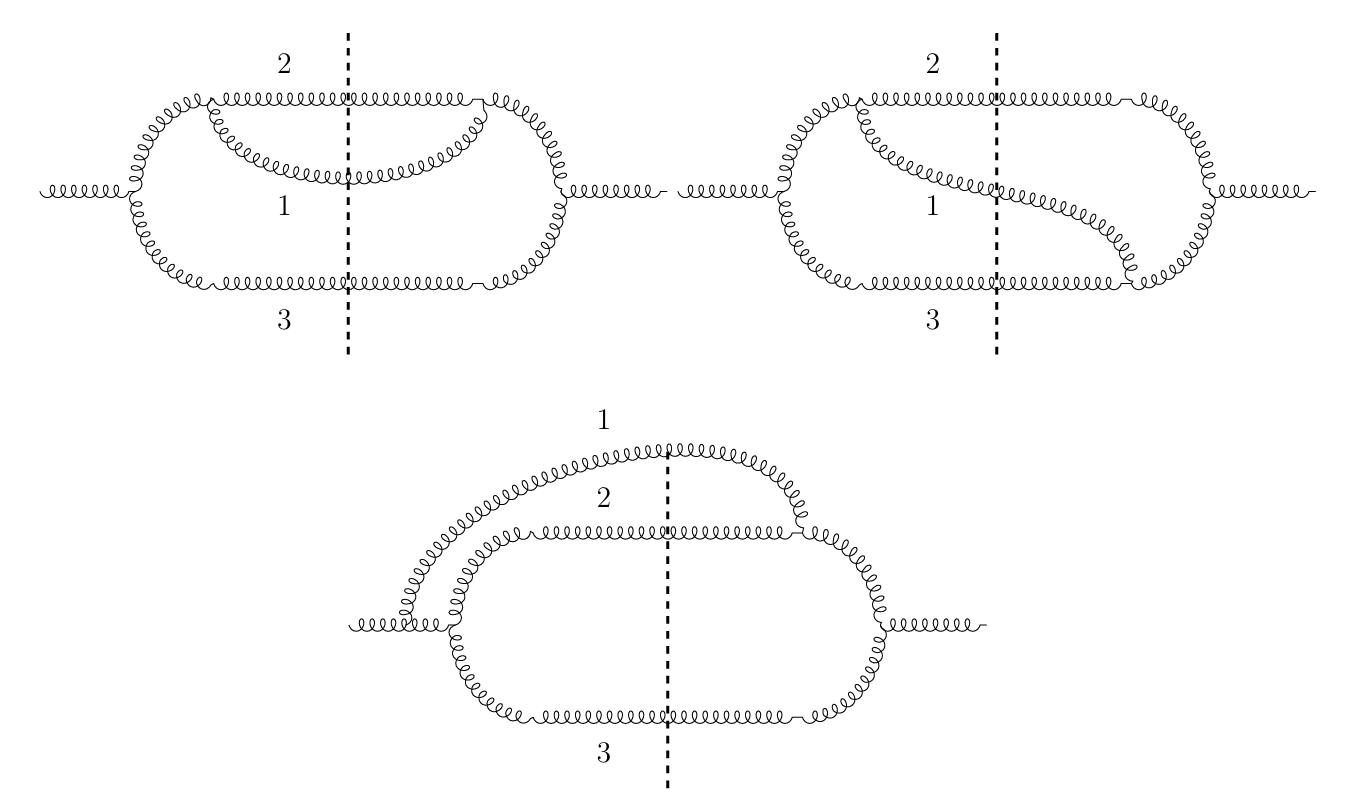}
    \caption{The three topologies appearing in the explicit terms of the regular cross section. The diagram in the upper left side corresponds to the square of configuration A. The topology to its right corresponds to the interference term AxC. Finally the topology at the bottom corresponds to the interference term AxB}
    \label{fig:Gluon squared amp topologies}
\end{figure}

The regular contributions consists of the square of three terms from the regular amplitude: the original configuration, the exchange of gluons one and three, and the exchange of gluons two and three. Because all final state particles are gluons, the interference between the exchange of gluons will not be suppressed by powers of $\N$, so we need to consider all the 9 possible terms. The expression for the cross section of this contribution reads

\begin{multline}
    \frac{\der\sigma^{gA\rightarrow ggg +X}}{\der^2\boldsymbol{k}_{1\perp}\der\eta_1 \der^2\boldsymbol{k}_{2\perp} \der\eta_{2} \der^2\boldsymbol{k}_{3\perp} \der\eta_3} \bigg|_{reg} = \\
    =\frac{\alpha_s^2 \delta(1-z_1- z_{2} - z_3)}{(2\pi)^{10}} \N^2  \int \der^6\boldsymbol{\Pi} \der^6\boldsymbol{\Pi}' \bigg\{ \mathcal{K}^{1}_{ggg}(\boldsymbol{r}_{y'x'},\boldsymbol{r}_{w_A'z'};\boldsymbol{r}_{yx},\boldsymbol{r}_{w_Az})\\
    \bigg[\Theta'_{ggg,1A}\Theta_{ggg,1A} \avg{\Xi_{ggg,1} (\Perp{x}', \Perp{y}', \Perp{z}';\Perp{x}, \Perp{y}, \Perp{z})}{Y}-(\{\Perp{x},\Perp{y}\} \to \boldsymbol{w}_{A\perp}) - (\{\Perp{x}',\Perp{y}'\} \to \boldsymbol{w}_{A\perp}')\\
    + (\{\Perp{x},\Perp{y}\} \to \boldsymbol{w}_{A\perp}, \{\Perp{x}',\Perp{y}'\} \to \boldsymbol{w}_{A\perp}') \bigg]+ \mathcal{K}^{2}_{ggg}(\boldsymbol{r}_{y'x'},\boldsymbol{r}_{w_A'z'};\boldsymbol{r}_{yz},\boldsymbol{r}_{w_Bx})\\
    \bigg[\Theta'_{ggg,1A}\Theta_{ggg,1B} \avg{\Xi_{ggg,2} (\Perp{x}', \Perp{y}', \Perp{z}';\Perp{x}, \Perp{y}, \Perp{z})}{Y}-(\{\Perp{x},\Perp{z}\} \to \boldsymbol{w}_{B\perp}) - (\{\Perp{x}',\Perp{y}'\} \to \boldsymbol{w}_{A\perp}')\\
    + (\{\Perp{x},\Perp{z}\} \to \boldsymbol{w}_{B\perp}, \{\Perp{x}',\Perp{y}'\} \to \boldsymbol{w}_{A\perp}') \bigg]+ \mathcal{K}^{3}_{ggg}(\boldsymbol{r}_{y'x'},\boldsymbol{r}_{w_A'z'};\boldsymbol{r}_{zx},\boldsymbol{r}_{w_Cy})\\
    \bigg[\Theta'_{ggg,1A}\Theta_{ggg,1C} \avg{\Xi_{ggg,3} (\Perp{x}', \Perp{y}', \Perp{z}';\Perp{x}, \Perp{y}, \Perp{z})}{Y} -(\{\Perp{z},\Perp{y}\} \to \boldsymbol{w}_{C\perp}) - (\{\Perp{x}',\Perp{y}'\} \to \boldsymbol{w}_{A\perp}')\\
    + (\{\Perp{z},\Perp{y}\} \to \boldsymbol{w}_{C\perp}, \{\Perp{x}',\Perp{y}'\} \to \boldsymbol{w}_{A\perp}') \bigg] \bigg\} + (k_1 \leftrightarrow k_3) + (k_2 \leftrightarrow k_3)\,.
\end{multline}
In the expression above, we have contributions from three kernels (see Figure \ref{fig:Gluon squared amp topologies}). The first kernel corresponds to the square of the regular term of the original configuration of gluons ($A\times A$). The second kernel corresponds to the interference between the original configuration and the exchange of gluons one and three ($A\times B$), the original configuration being in the conjugate amplitude. Finally, the third kernel encodes the contribution from the interference between the original configuration and the exchange of gluons two and three ($A\times C$), again, the original configuration being in the complex conjugated amplitude. The remaining terms are obtained by exchanging the momenta of the final state gluons of these contributions.

The three color correlators appearing in the regular contribution are defined as 
\begin{multline}
    \avg{\Xi_{ggg,i} (\Perp{x}', \Perp{y}', \Perp{z}';\Perp{x}, \Perp{y}, \Perp{z})}{Y}\\
    = S_{ggg}^{ggg, (i)}(\Perp{x}', \Perp{y}', \Perp{z}';\Perp{x}, \Perp{y}, \Perp{z}) - S_{ggg}^{g, (i)}(\Perp{v}';\Perp{x}, \Perp{y}, \Perp{z}) - S_{ggg}^{g, (i)}(\Perp{v};\Perp{x}', \Perp{y}', \Perp{z}') + S_{g}^{g}(\Perp{v}'; \Perp{v})\,,
\end{multline}
with $i=1,2,3.$. These correlators satisfy the relation
\begin{multline}
    \Xi_{ggg,1}(\Perp{x}', \Perp{y}', \Perp{z}';\Perp{x}, \Perp{y}, \Perp{z}) =\\
    (\Xi_{ggg,2}(\Perp{x}', \Perp{y}', \Perp{z}';\Perp{x}, \Perp{y}, \Perp{z}) + \Xi_{ggg,3}(\Perp{x}', \Perp{y}', \Perp{z}';\Perp{x}, \Perp{y}, \Perp{z}))/2\,.
\end{multline}
This is consistent with the discussion we had in the calculation of the amplitudes in Section~\ref{sec: Amplitudes trijet}, where we found that the color operators satisfied the relation Eq.\,\eqref{eq:Jacobi-color}. Notice however that, just as for the $q\bar{q}g$ case, the intermediate state interaction matrices do not satisfy this relation since they depend on the transverse coordinates of the intermediate gluon.

\subsubsection{Instantaneous contribution}

\begin{multline}
    \frac{\der\sigma^{gA\rightarrow ggg +X}}{\der^2\boldsymbol{k}_{1\perp}\der\eta_1 \der^2\boldsymbol{k}_{2\perp} \der\eta_{2} \der^2\boldsymbol{k}_{3\perp} \der\eta_3} \bigg|_{inst} = \\
    =\frac{\alpha_s^2 \delta(1-z_1- z_{2} - z_3)}{(2\pi)^{10}} \N^2 \int \der^6\boldsymbol{\Pi} \der^6\boldsymbol{\Pi}' \bigg\{ \mathcal{K}^{4}_{ggg}(X_{R}',X_{R}) \avg{\Xi_{ggg,1} (\Perp{x}', \Perp{y}', \Perp{z}';\Perp{x}, \Perp{y}, \Perp{z})}{Y}\\
    + \mathcal{K}^{5}_{ggg}(X_{R}',X_{R}) \avg{\Xi_{ggg,2} (\Perp{x}', \Perp{y}', \Perp{z}';\Perp{x}, \Perp{y}, \Perp{z})}{Y} \bigg\} + (k_1 \leftrightarrow k_3)\,.
\end{multline}
In this expression, we have contributions from the square of the instantaneous part of the original contribution (A), corresponding to the kernel 4, and the interference between A and B, kernel 5.

\subsubsection{Interference contribution}

The interference contribution between the regular and instantaneous amplitudes features six different kernels. However, we only write explicitly the contribution from the regular part in the direct amplitude and configuration A of the instantaneous amplitude in the complex conjugated amplitude. The remaining six terms can be obtained by exchanging the momentum labels $k_1 \leftrightarrow k_3$. The expression of the cross section reads
\begin{multline}\label{cross_section_ggg_interference_contr}
    \frac{\der\sigma^{gA\rightarrow ggg +X}}{\der^2\boldsymbol{k}_{1\perp}\der\eta_1 \der^2\boldsymbol{k}_{2\perp} \der\eta_{2} \der^2\boldsymbol{k}_{3\perp} \der\eta_3} \bigg|_{reg-inst} = \\
    =\frac{\alpha_s^2 \delta(1-z_1- z_{2} - z_3)}{(2\pi)^{10}} \N^2 \int \der^6\boldsymbol{\Pi} \der^6\boldsymbol{\Pi}' \bigg\{ \mathcal{K}^{6}_{ggg}(X_{R}';\boldsymbol{r}_{yx},\boldsymbol{r}_{w_Az})\\
    \bigg[\Theta_{ggg,1A} \avg{\Xi_{ggg,1} (\Perp{x}', \Perp{y}', \Perp{z}';\Perp{x}, \Perp{y}, \Perp{z})}{Y} -(\{\Perp{x},\Perp{y}\} \to \boldsymbol{w}_{A\perp}) \bigg]+ \mathcal{K}^{7}_{ggg}(X_{R}';\boldsymbol{r}_{yz},\boldsymbol{r}_{w_Bx})\\
    \bigg[\Theta_{ggg,1B} \avg{\Xi_{ggg,2} (\Perp{x}', \Perp{y}', \Perp{z}';\Perp{x}, \Perp{y}, \Perp{z})}{Y} -(\{\Perp{z},\Perp{y}\} \to \boldsymbol{w}_{B\perp}) \bigg]+ \mathcal{K}^{8}_{ggg}(X_{R}';\boldsymbol{r}_{zx},\boldsymbol{r}_{w_Cy})\\
    \bigg[\Theta_{ggg,1C} \avg{\Xi_{ggg,3} (\Perp{x}', \Perp{y}', \Perp{z}';\Perp{x}, \Perp{y}, \Perp{z})}{Y} -(\{\Perp{x},\Perp{z}\} \to \boldsymbol{w}_{C\perp}) \bigg]\bigg\}+ (k_1 \leftrightarrow k_3)\,.
\end{multline}
The kernels appearing in Eq.\,\eqref{cross_section_ggg_interference_contr} correspond to the interference between the regular part of configuration A, B and C and instantaneous part of configuration A, in that order.

\section{Conclusions and Outlook}\label{sec: Conclusions}

In this paper, we calculated for the first time the analytical results for forward trijet production in proton-nucleus collisions in the gluon-initiated channel, using the CGC effective theory. Combined with the real NLO correction to the dijet cross section in the quark-initiated channel computed in~\cite{Iancu:2020mos}, this work thus completes the calculation of the full trijet cross section in pA collisions. The total amplitude receives contributions from both the splitting of a gluon into a $q\bar{q}g$ state and the splitting of a gluon into a three-gluon state $ggg$. Particular attention was devoted to exploiting the unitarity relations between the amplitudes contributing to each channel to express the results in the most compact form. 

The $g \to q\bar{q}g$ amplitudes can be organized into three distinct topologies, corresponding to gluon emission by the quark, the gluon, and the antiquark. Our results for the amplitudes associated with each topology are presented in Eqs.\,\eqref{quark parent qbarqg amplitude}, \eqref{gluon parent qbarqg amplitude}, and \eqref{antiquark parent qbarqg amplitude}. This organization makes manifest a clear relation between the amplitudes corresponding to gluon emission by a quark or an antiquark and the amplitudes appearing in DIS trijet production of a $q\bar{q}g$ final state in the photoproduction limit. In DIS, only two topologies contribute to the total amplitude, namely gluon emission by a quark or an antiquark. While these topologies involve diagrams similar to those appearing in pA collisions, the colorless nature of the incoming parton in DIS implies the absence of diagrams with initial-state interactions with the shockwave. This difference leads to a modification of the color operator of the resulting amplitudes. The amplitudes of the quark and antiquark topologies, given in Eqs.\,\eqref{quark parent qbarqg amplitude} and \eqref{antiquark parent qbarqg amplitude}, can be related to the corresponding DIS amplitudes by making the replacements $g \to -e e_f, \,
C_{R3,ij}^{ab},\, C_{\overline{R3},ij}^{ab} \to [t^b]_{ij}$, and by removing the color matrix $t^a$ arising from the initial gluon splitting in the remaining color operators. 

Regarding the $g\to ggg$ channel, to our knowledge, this is the first calculation in the CGC literature to include the four-gluon vertex.
We found that this contribution exhibits a structure similar to that of the instantaneous pieces of the double-splitting vertex topology and its permutations. This observation allows us to absorb the corresponding impact factors into a single ``effective'' instantaneous amplitude for each permutation. Our final results for the total amplitude in this channel are presented in Eq. \eqref{gluon_to_ggg_total_amplitude}. A careful analysis of the symmetry properties of the amplitudes under gluon exchange enabled us to write these final formulas in a relatively compact form.

Our results in Section~\ref{sec:Slow gluon lim} focus on the slow-gluon limit of the trijet cross section which gives rise to the rapidity divergence of the dijet cross section after integrating over the slow gluon phase space. We verified that the diagrams contributing to the JIMWLK evolution for a given topology correspond precisely to configurations in which the slow gluon is emitted immediately before or after the shockwave, in agreement with the expectations of Refs.~\cite{Iancu:2018hwa, Iancu:2020mos}. This observation reduces the number of contributing diagrams to only two for each topology at the level of the amplitude. By expressing the color operators in terms of left and right generators, we were able to express the amplitudes, and subsequently the cross section, in a way that makes self-evident the connection to the JIMWLK evolution of the LO cross sections. In Appendix \ref{app: JIMWLK_cross_section} we further confirmed that the resulting cross section reproduces the JIMWLK evolution of the LO cross section, providing a nontrivial consistency check of our calculation. We also observed that operators involving products of two or more color traces acquire subleading-$N_c$ corrections in their evolution, originating from gluon exchange between distinct correlators.

In Section~\ref{sec:DGLAP}, we demonstrated that the collinear divergences, arising when the unobserved parton becomes collinear to one of the observed hadrons, can be factorized as a single real step of DGLAP evolution of the quark/gluon PDF or the quark/gluon fragmentation functions, convolved with the LO cross section for di-hadron production. This result not only provides a nontrivial cross-check of the amplitudes presented in the summary section, but also establishes the validity of the hybrid ‘dilute–dense’ factorization framework for dijet production in pA collisions at NLO.

The resulting \textit{trijet} cross section obtained in Section \ref{sec: Cross section trijets} involves a large number of terms containing nontrivial color correlators. The contributions from the $ggg$ channel are particularly involved, and for this reason we have presented only the leading large-$N_c$ contributions to the corresponding $S$-matrices. In this limit, the $S$-matrices are expressed solely in terms of dipole and quadrupole correlators, in agreement with the discussion in Section~\ref{sec: Dijet prod LO} (see also Ref. \cite{Dominguez:2013}). One can further verify by inspection that, in the large-$N_c$ limit, the $S$-matrices entering the $q\bar{q}g$ cross section are also fully described in terms of these two correlators. As briefly discussed in Section~\ref{sec: Dijet prod LO}, explicit expressions for the dipole and quadrupole operators (and, in principle, any higher-point correlator) can be obtained within the Gaussian approximation, where the color charges inside the hadron are assumed to be local in rapidity and distributed according to a Gaussian weight. In this approximation, the evolution of any $n$-point correlator --- whether single-traced or a product of traces --- can be reduced to a closed system of differential equations involving only the dipole operator.

Further simplifications are expected for the real corrections in this channel upon taking the back-to-back limit. In Ref.~\cite{Caucal:2025sea}, we studied the contribution of a $q\bar{q}g$ final state to dijet production in the case where one integrates out either the quark or the antiquark. We showed that these contributions can be expressed in terms of hard factors describing the $g(q)\to qg$ scattering process and sea-quark distributions at small $x$, expressed in terms of the dipole gluon TMD \footnote{Another way to obtain factorization in terms of quark TMDs for dilute-dense forward $pA$ collisions is to consider sub-eikonal corrections to the LO CGC result~\cite{Altinoluk:2024dba,Altinoluk:2024tyx}. In this approach, the quark TMD is formulated in terms of correlation functions of the fermionic background field.}. Our results were based on the amplitudes calculated in this paper, which serves as another robust cross-check for our results. In an upcoming work, we will extend this analysis to the quark-initiated channel in pA collisions. In order to obtain the complete real NLO corrections, one needs to also consider the contributions arising from integrating out a final-state gluon. These contributions require special care, as they involve large logarithms that must be systematically factorized into their corresponding evolution equations. In particular, gluon radiation gives rise to rapidity logarithms, Sudakov double and single logarithms, and collinear logarithms. The rapidity and Sudakov logarithms are associated with JIMWLK and CSS evolution, respectively, while the collinear logarithms must be absorbed into the DGLAP evolution of PDFs or fragmentation functions, depending on whether they originate from initial- or final-state radiation, as shown in section~\ref{sec:DGLAP} for the real part. A recent study addressing these issues in DIS dijet production can be found in Ref.~\cite{Caucal:2025mth}.

Another interesting kinematical limit occurs when the total momentum imbalance of three jets is small compared to the transverse momenta of each jet. In this limit, we expect it to be possible to recast our results in a TMD factorized form, as it has been shown for similar processes involving a real photon in the initial or final state \cite{Altinoluk:2018byz,Altinoluk:2020qet}. Trijet production has been also studied within Improved-TMD (ITMD) \cite{Bury:2020ndc} which relaxes the small momentum imbalance requirement, but neglects genuine saturation effects \cite{Altinoluk:2019wyu}. It would then be interesting to compare numerical results for trijet production in the full CGC, ITMD and TMD, and identify kinematical and genuine saturation effects \cite{Boussarie:2021ybe,Fujii:2020bkl}. The numerical computation of our results is nevertheless very challenging, as it will require the evaluation of multi-dimensional Fourier transforms of the product of perturbative factors with various multi-point correlators of Wilson lines \cite{Mantysaari:2019hkq}.

The results presented in this work also provide essential building blocks for the computation of real corrections to jet and hadron production at next-to-next-to-leading order (NNLO) accuracy within the CGC framework. Achieving NNLO precision is particularly important for observables sensitive to saturation effects, such as single-inclusive hadron production in proton--nucleus collisions, where experimental measurements have revealed significant nuclear suppression at forward rapidities~\cite{BRAHMS:2004xry, STAR:2008ixi}. While NLO calculations of hadron production in pA collisions have been achieved in Refs.~\cite{Chirilli:2011km, Chirilli:2012jd, Altinoluk:2014eka}, the computation of NNLO corrections would further increase the precision, stability and reduce the scale uncertainties, enabling precision comparisons with experimental data at RHIC and the LHC, as well as future measurements at the Electron–Ion Collider. The present calculation can be readily used to extract the ``double real" contributions to the two-loop evolution of the gluon dipole \cite{Balitsky:2013fea}, as well as the two-loop DGLAP $g \to g$ splitting function. The non-singlet DGLAP splitting function has been recently computed at two-loops within light-cone perturbation theory approach \cite{Lappi:2026xaa}.

\section*{Acknowledgements}
We thank Raju Venugopalan for enlightening discussions. P.C. is funded by the Agence Nationale de la Recherche under grant ANR-25-CE31-5230 (TMD-SAT). F.S. is supported by the Laboratory Directed Research and Development of Brookhaven National Laboratory and RIKEN-BNL Research Center, as well as the National Science Foundation (NSF)
within the framework of the JETSCAPE collaboration,
under grant number OAC-2514008 (CSSI:C-SCAPE). We are grateful for the support of the Saturated Glue (SURGE) Topical Theory Collaboration, funded by the U.S. Department of Energy, Office of Science, Office of Nuclear Physics. Finally, F.S. acknowledges support from the U.S. Department of Energy, Office of Science, Office of Nuclear Physics under the umbrella of the Quark-Gluon Tomography (QGT) Topical Collaboration with Award DE-SC0023646.

\appendix

\section{Conventions and general identities} \label{app:Conventions and general identities}
\subsection{Light-cone coordinates}
Our convention for light-cone coordinates is 
\begin{equation}
x^{+}=\frac{1}{\sqrt{2}}(x^{0}+x^{3})\,,\qquad
x^{-}=\frac{1}{\sqrt{2}}(x^{0}-x^{3})\,.
\end{equation}
A typical four-vector expressed in this basis is of the form $v^{\mu}=(v^{+}, v^{-}, \Transv{v}{})$, where $\Transv{v}{}$ is the transverse 2-component vector of $v$. The scalar product of two four vectors then is expressed as $v^{\mu}u_{\nu} = v^+u^- + v^-u^+ - \Transv{v}{}\cdot \Transv{u}{}$. The metric tensor can be deduced from this expression with components $g^{+-}=g^{-+} = 1$ and $g^{ij} = -\delta^{ij}$. The rest of the components are zero. We can define analogous Dirac matrices $\gamma^{+}, \gamma^{-}$ satisfying the usual commutation relation
\begin{equation}
    \left\{ \gamma^{\mu}, \gamma^{\nu} \right\} \equiv 2g^{\mu,\nu} \mathds{1}_{4}\,.
\end{equation}
\subsection{Feynman rules}\label{app: Feynman rules}

The free massless quark and gluon Feynman propagators are
\begin{align}
S^{0}_{\sigma\sigma',ij}(l) &= \frac{i\,\slashed{l}_{\sigma\sigma'}}{l^{2}+i\epsilon}\,\delta_{ij}, \\
G^{0}_{\mu\nu,ab}(l) &= \frac{i \Pi_{\mu\nu}(l)}{l^{2}+i\epsilon}\delta_{ab},
\end{align}
The gluon polarization tensor $\Pi_{\mu\nu}$ which appears in the free gluon propagator is defined as
\begin{equation}
\Pi_{\mu\nu}(l)=-g_{\mu\nu}+\frac{l_{\mu}n_{\nu}+n_{\mu}l_{\nu}}{n\cdot l}\,,
\end{equation}
where the light-cone vector $n$ is defined in the ``plus" direction, $n^{\mu}=(1,0,\boldsymbol{0}_{\perp})$.

\begin{figure}
    \centering
    \includegraphics[width=0.7\linewidth]{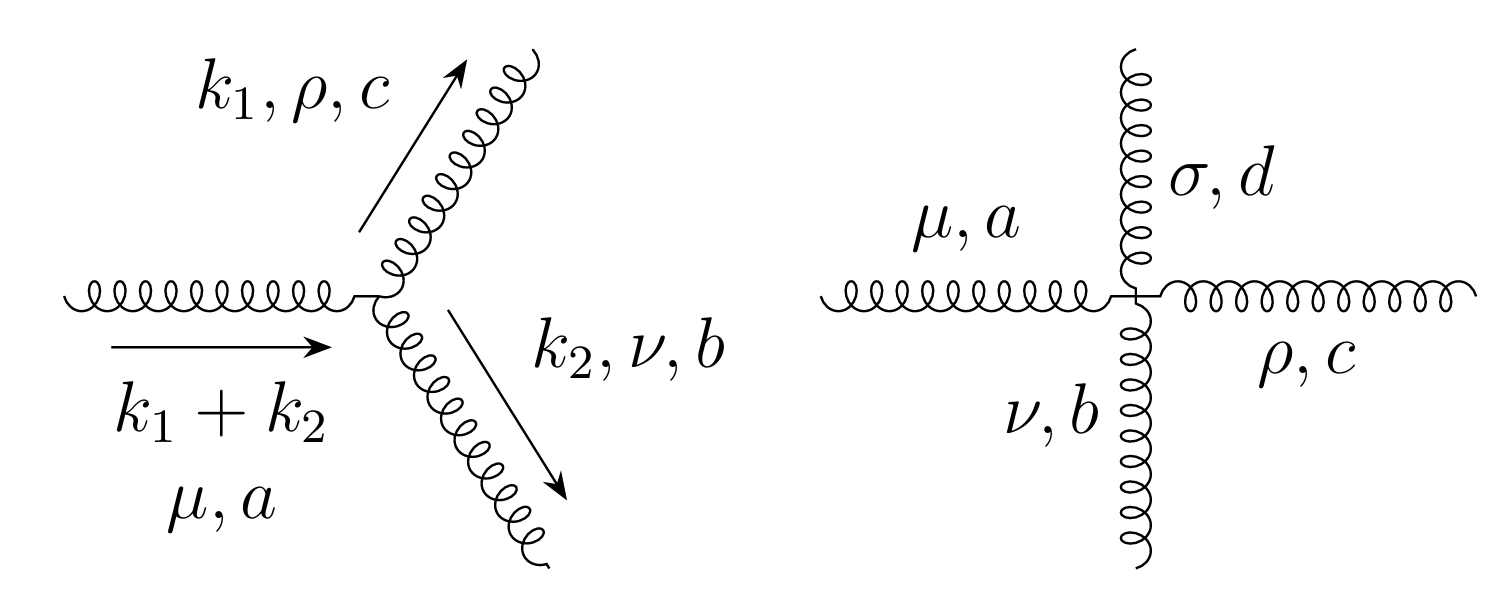}
    \caption{The triple and four gluon vertices of the QCD Feynman rules.}
    \label{fig:Gluon_splitting_vertices}
\end{figure}

The photon-quark and gluon-quark vertices are defined as
\begin{align}
    V_{q\gamma}^{\mu} = -ieq_f\gamma^{\mu}\,,\\
    V_{q\gamma,ij}^{a,\mu} = ig\gamma^{\mu} t^a_{ij}\,,
\end{align}
Similarly, the triple and four gluon vertices are defined as
\begin{align}
    V_{ggg}^{\mu\nu\rho}(k_1, k_2) &= gf^{abc}\mathcal{V}^{\mu\nu\rho}(k_1, k_2)\,, \label{triple_gluon_vertex}\\
    V_{gggg}^{\mu\nu\rho\sigma,abcd} &= -ig^2\bigg[f^{abe}f^{cde}(g^{\mu\rho}g^{\nu\sigma}-g^{\mu\sigma}g^{\nu\rho}) + f^{ace}f^{bde}(g^{\mu\nu}g^{\rho\sigma} - g^{\mu\sigma}g^{\nu\rho})\nonumber\\
    &+ f^{ade}f^{bce}(g^{\mu\nu}g^{\rho\sigma}-g^{\mu\rho}g^{\nu\sigma}) \bigg]\,. \label{four_gluon_vertex}
\end{align}
The tensor introduced in the triple gluon vertex is defined as 
\begin{equation}
    \mathcal{V}^{\mu\nu\rho} (k_1,k_2) = g^{\mu\nu}(2k_2 + k_1)^{\rho} + g^{\nu\rho}(k_1 - k_2)^{\mu} - g^{\rho\mu}(2k_1 + k_2)^{\nu}\,,
\end{equation}
where $k_1$ and $k_2$ are the momenta of the two outgoing gluons, with Lorentz indices $\rho$ and $\nu$ and colors $c$ and $b$, respectively, and the initial gluon has momentum $k_1 + k_2$, Lorentz index $\mu$ and color $a$ (see Figure \ref{fig:Gluon_splitting_vertices}). In the four gluon vertex, the gluons are arranged in alphabetic order with respect to their colors in the clockwise direction.

In the light-cone gauge $A^{-}=0$, the polarization vector for an on-shell gluon with non-zero transverse momentum $\Transv{k}{}$ is
\begin{equation}
    \epsilon(k,\lambda) \equiv \left( \frac{\PolVect{\lambda} \cdot \Transv{k}{}}{k^-}, \,0,\,\PolVect{\lambda} \right) \,,
\end{equation}
where $\lambda=\pm 1$ and $\et^{\lambda}= 1/\sqrt{2}\,(1,\,i\lambda)$. The transverse polarization vector satisfies the identity
\begin{equation}
    \varepsilon^{ij}\PolVect{\lambda,j}=i\lambda\,\PolVect{\lambda,i}\,.
\end{equation}

\subsection{Color identities}
In order to simplify the color operators we use the following relations for any $3\times3$ matrices $C$ and $D$:
\begin{align}
\mathrm{Tr}(C)\mathrm{Tr}(D) &= 2\,\mathrm{Tr}(C t^a D t^a)+\frac{1}{N_c}\mathrm{Tr}(CD)\,, \\
\mathrm{Tr}(CD) &= 2\,\mathrm{Tr}(C t^a)\mathrm{Tr}(D t^a)+\frac{1}{N_c}\mathrm{Tr}(C)\mathrm{Tr}(D)\,. 
\end{align}
We can also express any Wilson line in its adjoint representation in terms of a pair of Wilson lines in the fundamental representation:
\begin{equation}
V^\dagger(\Transv{x}{})t^aV(\Transv{x}{})=t^bU^{ba}(\Transv{x}{})\,.
\end{equation}

\section{Dirac algebra}\label{app: Spinor contractions}
In this appendix we provide a collection of spinor contractions and gluon tensor identities.

\subsection{Spinor contractions}
Following Ref.\,\cite{Caucal:2021lgf}, we work with Dirac spinors in the helicity basis which satisfy the Dirac equation
\begin{equation}
    \slashed{k} u(k, \sigma) =0 \,.
\end{equation}
The solutions to this equation are given by
\begin{equation}
    u(k, +)  = v(k, -) = \frac{1}{2^{1/4}}
    \begin{pmatrix}
    \sqrt{k^{+}}\, e^{-i\phi_k} \\
    \sqrt{k^{-}} \\
    \sqrt{k^{+}}\, e^{-i\phi_k} \\
    \sqrt{k^{-}}
    \end{pmatrix},
    \qquad
    u(k, -) = v(k, +)  = \frac{1}{2^{1/4}}
    \begin{pmatrix}
    \sqrt{k^{-}} \\
    - \sqrt{k^{+}}\, e^{i\phi_k} \\
    - \sqrt{k^{-}} \\
    \sqrt{k^{+}}\, e^{i\phi_k}
    \end{pmatrix}\,,
\end{equation}
where the sign $\pm$ denotes the helicity of the spinor, and $\phi_k$ is the azimuthal angle of $\Transv{k}{}$. The Dirac spinors satisfy the completeness relation
\begin{equation}
    \slashed{k} = \sum_{\sigma} u(k, \sigma) \, \bar{u}(k, \sigma)  \,.
\end{equation}
Then, the contraction of two Dirac spinors with $\gamma^{-}$ is given by \cite{Kovchegov:2012mbw}
\begin{equation}
     \bar{u}(k,\sigma) \gamma^- v(p,\sigma') = 2 \sqrt{k^- p^-}\, \delta^{\sigma, -\sigma'} \,.
\end{equation}
For the transverse gamma matrices we have:
\begin{equation}
    \bar{u}(k,\sigma) \gamma^i v(p,\sigma') = 2\sqrt{k^-p^-}\left[\frac{\boldsymbol{k}_\perp^i}{2k^-} + \frac{\boldsymbol{p}_\perp^i}{2p^-} +\frac{i\sigma}{2}\varepsilon^{ij}\left(\frac{\boldsymbol{k}_\perp^j}{k^-} - \frac{\boldsymbol{p}_\perp^j}{p^-} \right)  \right] \delta^{\sigma,-\sigma'} \,.
\end{equation}
Since the spinors $u$ and  $v$ are related by a helicity flip, the identities above under the exchange of a quark for an antiquark or vice versa can be obtained by flipping the helicity of the corresponding fermion in the right hand side of the equations. These identities allows us to obtain the spinor contractions for the ``fundamental" splitting vertices. The contraction of a $g\to q\bar{q}$ splitting vertex is given by

\begin{equation}\label{gluon->quark-antiquark wavefunction}
    \begin{split}
       \Bar{u}(k_1,\sigma)\slashed{\epsilon}(k_1+k_2,\lambda) v(k_2,\sigma')&= \frac{2\sqrt{k_1^- k_2^-}}{k_1^-} \et^{\lambda} \cdot \left(\Transv{k}{1} - \frac{k_1^-}{k_2^-}\Transv{k}{2}\right) \delta^{\sigma,-\sigma'} \left[\frac{k_1^- - k_2^-}{2(k_1^- + k_2^-)} - \frac{\sigma\lambda}{2}\right]\\
        &= -\frac{2\sqrt{k_1^- k_2^-}}{k_1^-} \et^{\lambda} \cdot \left(\Transv{k}{1} - \frac{k_1^-}{k_2^-}\Transv{k}{2}\right) \delta^{\sigma,-\sigma'} \Gamma^{\sigma \lambda}_{g\xrightarrow{}q\bar{q}}\left(\frac{k_1^-}{k_1^- + k_2^-}\right)\,. 
    \end{split}  
\end{equation}
Similarly, the $q \to qg$ vertex spinor contraction reads
\begin{equation}\label{quark->quark-gluon wavefunction}
    \begin{split}
        \Bar{u}(k_1-k_2,\sigma)\slashed{\epsilon}^*(k_2,\lambda) u(k_1,\sigma')= \frac{2k_1^-\sqrt{(k_1^- - k_2^-)k_1^-}}{k_2^-(k_1^- -  k_2^-)} \et^{\lambda} \cdot \left(\Transv{k}{2}- \frac{k_2^-}{k_1^-}\Transv{k}{1}\right) \delta^{\sigma,\sigma'} \Gamma^{\sigma \lambda}_{q\xrightarrow{}qg}\left(\frac{k_1^- - k_2^-}{k_1^-} \right)\,.
    \end{split}
\end{equation}
The contraction for a $\bar{q}\to \bar{q}g$ vertex can be obtained by flipping the helicity of the spinors in Eq. \eqref{quark->quark-gluon wavefunction}:
\begin{equation} \label{antiquark->antiquark-gluon wavefunction}
    \begin{split}
        \bar{v}(k_1,\sigma)\slashed{\epsilon}^*(k_2,\lambda) v(k_1-k_2,\sigma') &= \bar{u}(k_1-k_2,-\sigma) \slashed{\epsilon}^*(k_2,\lambda) u(k_1,-\sigma') \\
        &= \frac{2 k_1^-\sqrt{(k_1^- - k_2^-)k_1^-}}{k_2^-(k_1^- -  k_2^-)} \et^{\lambda} \cdot \left(\Transv{k}{2}- \frac{k_2^-}{k_1^-}\Transv{k}{1}\right) \delta^{\sigma,\sigma'} \Gamma^{-\sigma \lambda}_{q\xrightarrow{}qg}\left(\frac{k_1^- - k_2^-}{k_1^-} \right)\,.
    \end{split}
\end{equation}
In the case where one of the fermions is instantaneous, we need the following identities:  
\begin{equation}
    \bar{u}(k,\sigma) \gamma^i v(n,\sigma') = 2\sqrt{k^-}\et^{\sigma, i}\delta^{\sigma,-\sigma'} \,,
\end{equation}
\begin{equation} 
    \bar{u}(n,\sigma) \gamma^i v(k,\sigma') = 2\sqrt{k^-}\et^{\sigma', i} \delta^{\sigma,-\sigma'}\,.
\end{equation}
where $n$ is defined as in the previous section. These identities allow us to obtain the spinor contractions for the splitting vertices with an instantaneous fermion. For a $g\to q\bar{q}$ vertex:
\begin{equation} \label{gluon->quark-antiquark wavefunction inst}
    \begin{split}
       \Bar{u}(k_1,\sigma)\slashed{\epsilon}(k_1+k_2,\lambda) v(n,\sigma') &= -2\sqrt{k_1^-} \et^{\lambda}\cdot \et^{\sigma} \delta^{\sigma,-\sigma'} \\
        &= -2\sqrt{k_1^-} \delta^{\sigma,-\lambda} \delta^{\sigma,-\sigma'}\,. \\
    \end{split}  
\end{equation}
For a $q\to qg$ vertex in the case where the incoming quark is instantaneous we have:
\begin{equation} \label{quark->quark-gluon wavefunction inst}
    \begin{split}
        \bar{u}(k_1,\sigma)\slashed{\epsilon}^*(k_1+k_2,\lambda) u(n,\sigma')&= -2\sqrt{k_1^-} \et^{\lambda*} \cdot \et^{\sigma} \delta^{\sigma,\sigma'} \\
        &= -2\sqrt{k_1^-} \delta^{\sigma,\lambda} \delta^{\sigma,\sigma'} \,. \\
    \end{split}  
\end{equation}
For a $q\to qg$ vertex in the case where the outgoing quark is instantaneous:
\begin{equation} \label{quark->quark-gluon wavefunction inst 2}
    \begin{split}
        \bar{u}(n,\sigma)\slashed{\epsilon}^*(k_1+k_2,\lambda) u(k_2,\sigma') &= -2\sqrt{k_2^-} \et^{\lambda*} \cdot \et^{\sigma} \delta^{\sigma,\sigma'} \\
        &= -2\sqrt{k_1^-} \delta^{\sigma,\lambda} \delta^{\sigma,\sigma'} \,. \\
    \end{split}  
\end{equation}

\subsection{Gluon tensor identities}
Now let us introduce some useful identities related to the gluon propagator and polarization vector. The polarization vectors can be decomposed into a symmetric and an antisymmetric component as
\begin{equation}
    \et^{\lambda, i}\et^{-\lambda, j} = \frac{1}{2}(\delta^{ij} - i\lambda\varepsilon^{ij})\,.
\end{equation}
This identity is useful to perform the sum over the polarization of the gluons. Another useful identity to convert external gluons into internal ones when scattering off the shockwave is
\begin{equation}\label{external_to_internal_gluon_id}
    \epsilon^*_{\rho}(k_g, \bar{\lambda}) g^{\rho \beta} \Pi_{\beta \alpha}(l) = -\epsilon^*_{\alpha}(l) \,.
\end{equation}
The product of two gluon propagators is given by
\begin{equation}
    \Pi_{\alpha\beta}(l) \Pi^{\beta\delta}(l') = -\sum_{\eta} \epsilon^{*}_{\alpha}(l,\eta) \epsilon^{\delta}(l',\eta) \,.
\end{equation}
Notice the absence of the instantaneous contributions in the expression above. This is due to the fact that the polarization vectors are orthogonal to the light-like vectors $n$ and $\bar{n}$: $\epsilon(k,\lambda)\cdot n =0$.

Finally, let us define the contraction for the fundamental $g\to gg$ vertex
\begin{multline}\label{g-> gg wavefunction}
    \epsilon^*_{\gamma} (k_1,\eta) \left[g^{\alpha\beta} (2k_2 + k_1)^{\gamma} + g^{\beta\gamma}(k_1 - k_2)^{\alpha} - g^{\gamma\alpha}(2k_1 + k_2)^{\beta} \right] \epsilon^*_{\beta}(k_2,\lambda) \epsilon_{\alpha}(k_1 + k_2,\bar{\lambda})\\ 
        = -2\frac{(k_1^- + k_2^-)}{k_1^-} \Gt^{\bar{\lambda}\lambda\eta}\left(\frac{k_1^{-}}{k_1^{-} + k_2^{-}}\right)\cdot \left(\Transv{k}{1} - \frac{k_1^-}{k_2^-} \Transv{k}{2}\right)\,,
\end{multline}
where the $g\to gg$ splitting function is defined in Eq. \eqref{g_to_gg_splitting_function}.

\section{Useful integrals}\label{app: Useful integrals}
In this section we present the integrals required to perform the ``plus" and the transverse integration over the internal momentum integrals. We start by presenting the ``plus" contour integrals, then we introduce some useful identities to tackle the transverse integrals.

\subsection{Contour integrals}
Let us start with an integral containing two poles
\begin{equation}
    \mathcal{I}^{+}_{2}(l_1,l_2) = \int \frac{dl_2^+}{2\pi} \frac{1}{((l_1- l_2)^2 + i\varepsilon)(l_2^2+ i\varepsilon)}\,.
\end{equation}
The details of the calculation of this integral can be found in Appendix D of Ref. \cite{Caucal:2021lgf}. The resulting expression is
\begin{equation}\label{plus_integral_two_poles_general}
    \mathcal{I}^{+}_{2}(l_1,l_2) =  \frac{i \Theta\left(l_2^- \right) \Theta\left(l_1^- -l_2^- \right)}{-4l_2^- (l_1^- - l_2^-)\left[\frac{2l_1^+(l_1^- - l_2^-) - (\Transv{l}{1} - \Transv{l}{2})^2}{2(l_1^- - l_2^-)} + \frac{i\varepsilon}{2(l_1^- - l_2^-)} - \frac{\Transv{l}{2}^2}{2l_2^-} + \frac{i\varepsilon}{2l_2^-}\right]} \,.
\end{equation}
In the case where the momentum vector is on-shell, $l_1^2 =0$, and the transverse component of $l_1$ is zero, the integral simplifies to
\begin{equation}\label{plus_integral_two_poles_on_shell}
    \mathcal{I}^{+}_{2}(l_1,l_2)\bigg|_{l_1^2 =0,\Transv{l}{1}=0} =  \frac{i \Theta\left(l_2^- \right) \Theta\left(l_1^- -l_2^- \right)}{2l_1^- \left[\Transv{l}{2}^2 + i\varepsilon\right]} \,.
\end{equation}
This is the integral needed for the LO perturbative factors calculated in Section~\ref{sec: Dijet prod LO}.

In the trijet production amplitudes, specifically in diagrams with the shockwave interacting with the three partons, we have the appearance of double ``plus" integrals with four poles. Let
\begin{equation}
    \mathcal{I}_{4,reg}^{+} = \int \frac{dl_1^+}{2\pi} \frac{dl_2^+}{2\pi} \frac{1}{((l_1- l_2)^2 + i\varepsilon)(l_2^2+ i\varepsilon)(l_1^2 + i\varepsilon)((p-l_1)^2+i\varepsilon)} \,.
\end{equation}
Making use of the residue theorem and simplifying the expression further gives
\begin{align}
    \mathcal{I}_{4,reg}^{+} = \frac{i^2}{2l_1^- \left[\Transv{L}{}^2 + \frac{p^-l_2^-(l_1^- - l_2^-)}{l_1^{-2}(p^- - l_1^-)}\Transv{l}{1}^2 \right] 2p^- \Transv{l}{1}^2},
\end{align}
with $\Transv{L}{} = \Transv{l}{2} - \frac{l_2^-}{l_1^-}\Transv{l}{1}$.

The instantaneous pieces of the final state diagrams will modify the integral above by canceling out one of the poles. Define
\begin{equation}
    \mathcal{I}_{3,inst}^{+} (l_1, l_2) = \int \frac{dl_1^+}{2\pi} \frac{dl_2^+}{2\pi} \frac{1}{((l_1- l_2)^2 + i\varepsilon)(l_2^2+ i\varepsilon)((p-l_1)^2+i\varepsilon)} \,.
\end{equation}
The details of the calculation of this integral can also be found in Appendix D of Ref. \cite{Caucal:2021lgf}. The resulting expression is
\begin{align}
    \mathcal{I}_{3,inst}^{+} (l_1, l_2) = \frac{1}{4l_1^- (p^- - l_1^-) \left[\Transv{L}{}^2 + \Delta^2 \Transv{l}{1}^2 \right]} \,,
\end{align}
with $\Transv{L}{}$ defined as for the regular integral case.

\subsection{Transverse integrals}

Let us now perform the transverse integrals appearing in the calculations. The details of these calculations can be found in section 4 and Appendix E of Ref. \cite{Caucal:2021lgf}. The simplest transverse integral appears in the LO calculations and reads
\begin{equation}\label{LO_transverse_integral}
    \frac{\et^{\bar{\lambda}*}\cdot \Transv{L}{}}{\Transv{L}{}^2} = \frac{i}{2\pi} \int \der^2 \Transv{R}{} e^{-i \Transv{R}{}\cdot\Transv{L}{}}\frac{\et^{\bar{\lambda}*}\cdot \Transv{R}{}}{\Transv{R}{}^2}\,.
\end{equation}
For the trijet calculations, we require the integral
\begin{equation}
     \int \frac{\der^2 \Transv{l}{1}}{2\pi} \frac{\der^2 \Transv{l}{2}}{2\pi} e^{i\Transv{l}{1}\cdot \dipvec{xy}} e^{i\Transv{l}{2}\cdot \dipvec{zx}} \frac{\Transv{l}{1}^i\Transv{L}{}^j}{\Transv{l}{1}^2\left[\Transv{L}{}^2 + \Delta^2 \Transv{l}{1}^2 \right]},
\end{equation}
with $\Transv{L}{} = \Transv{l}{2} - \alpha \Transv{l}{1}$. The integral gives
\begin{align}\label{Trijet_Transverse_integral_final_state_reg}
    \int \frac{\der^2\Transv{l}{1}}{2\pi} \frac{\der^2 \Transv{l}{2}}{2\pi} e^{i\Transv{l}{1}\cdot \dipvec{xy}} e^{i\Transv{l}{2}\cdot \dipvec{zx}} \frac{\Transv{l}{1}^i\Transv{L}{}^j}{l_{1\perp}^2\left[\Transv{L}{}^2 + \Delta^2 \Transv{l}{1}^2 \right]} = -\frac{\dipvec{zx}^j}{\dipvec{zx}^2} \frac{\Transv{R}{}^i}{\Transv{R}{}^2 + \Delta^2 \dipvec{zx}^2}\,,
\end{align}
where $\Transv{R}{} = \dipvec{xy} + \frac{1}{\alpha}\dipvec{zx}$.

For the instantaneous contributions, we need the following integral
\begin{align}\label{Trijet_Transverse_integral_final_state_inst}
    \int \frac{\der^2\Transv{l}{1}}{2\pi} \frac{\der^2\Transv{l}{2}}{2\pi} e^{i\Transv{l}{1}\cdot \dipvec{xy}} e^{i\Transv{l}{2}\cdot \dipvec{zx}} \frac{1}{L_{\perp}^2 + \Delta^2 l_{1\perp}^2 } = \frac{1}{\Transv{R}{}^2 + \Delta^2 \dipvec{zx}^2}\,,
\end{align}
where $\Transv{R}{}$ is defined as above.

\section{Detailed calculation of diagram R7}\label{app: Detailed of calculation of diagram R7}
\begin{figure}[H]
    \centering
    \includegraphics[width=0.5\linewidth]{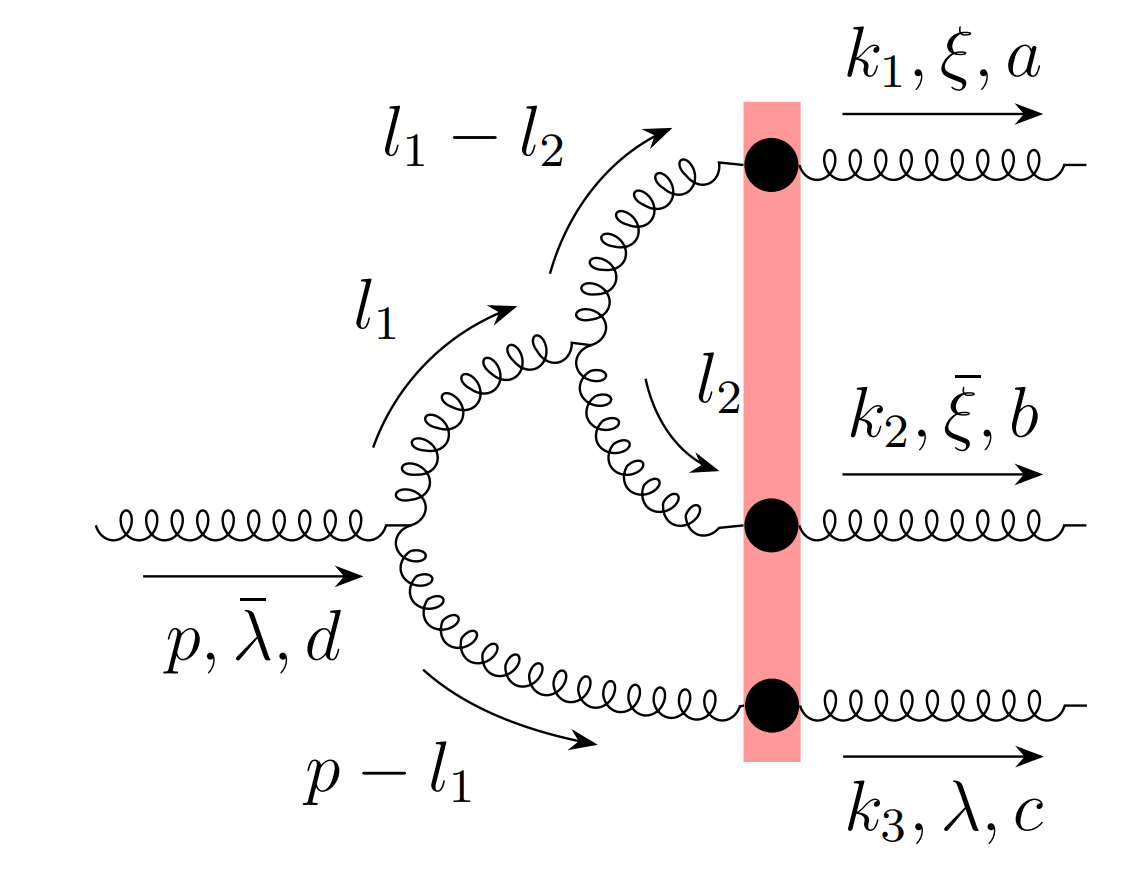}
    \caption{Diagram R7: Shockwave after second $gg$ splitting.}
    \label{fig: gqqNLOR7}
\end{figure}

The purpose of this appendix is to present the detailed calculation of the diagram R7, corresponding to the $g\to ggg$ channel diagram where all final state partons scatter off the shockwave. In Figure \ref{fig: gqqNLOR7}, we illustrate the corresponding Feynman diagram with the inner momentum labels, polarizations and helicities of the participating partons.
The scattering amplitude of the diagram then reads
\begin{multline}
    S = \int \frac{\der^4 l_1}{(2\pi)^4} \frac{\der^4 l_2}{(2\pi)^4} \epsilon^{*}_{\alpha}(k_1,\xi) T^{g,\alpha\beta}_{ae}(k_1,l_1-l_2)\frac{i\Pi_{\beta\epsilon}(l_1 - l_2)}{(l_1 - l_2)^2 + i\epsilon}   \left\{gf^{fge} \mathcal{V}^{\gamma\delta\epsilon}(l_1-l_2,l_2) \right\} \frac{i\Pi_{\gamma\rho}(l_1)}{l_1^2 + i\epsilon}\\
    \times\left\{gf^{dhf} \mathcal{V}^{\mu\nu\rho}(l_1, p - l_1) \right\} \epsilon_{\mu}(p,\bar{\lambda})\epsilon^{*}_{\bar{\mu}}(k_2,\bar{\xi}) T^{g,\bar{\mu}\sigma}_{bg}(k_2,l_2) \frac{i\Pi_{\sigma\delta}(l_2)}{l_2^2 + i\epsilon} \epsilon^{*}_{\tau}(k_3,\lambda) T^{g,\tau\theta}_{ch}(k_3,p-l_1) \frac{i\Pi_{\theta\nu}(p-l_1)}{(p-l_1)^2 + i\epsilon} \,.
\end{multline}
The physical amplitude, after factoring out a delta function and a factor of $2\pi$, is
\begin{equation}
    \mathcal{M}_{R7}^{\bar{\lambda}\lambda\bar{\xi}\xi} =  \int \der^6 \boldsymbol{\Pi} \left[f^{efg}f^{dhf} U_{ae}(\boldsymbol{x}_{\perp})U_{bg}(\boldsymbol{y}_{\perp})U_{ch}(\boldsymbol{z}_{\perp}) \right] \Ncal^{\bar{\lambda}\lambda\xi \bar{\xi}}_{R7}(\boldsymbol{x}_{\perp}, \boldsymbol{y}_{\perp}, \boldsymbol{z}_{\perp})\,,
\end{equation}
where
\begin{multline}
    \Ncal^{\bar{\lambda}\lambda\xi \bar{\xi}}_{R7}(\Transv{x}{}, \Transv{y}{}, \Transv{z}{}) = \frac{g^2}{4\pi^2} \int \frac{\der^4 l_1}{(2\pi)^2} \frac{\der^4 l_2}{(2\pi)^2} e^{-\Transv{l}{1}\cdot \dipvec{xz}} e^{i\Transv{l}{2}\cdot \dipvec{yx}} \\
    \times \frac{-\delta(l_2^- - k_2^-) \delta(l_1^- - k_1^- - k_2^-)(2k_1^-)(2k_2^-)(2k_3^-)\mathcal{G}_{A}^{\gamma}\Pi_{\gamma\rho}(l_1)\mathcal{G}_{B}^{\rho} }{D_1}\,.
\end{multline}
Here we have defined the gluon structures as 
\begin{align}
    \mathcal{G}_{A}^{\gamma} &= \epsilon^{*}_{\alpha}(k_1,\xi) g^{\alpha\beta} \Pi_{\beta\epsilon}(l_1 - l_2) \mathcal{V}^{\gamma\delta\epsilon}(l_1-l_2,l_2) \epsilon^{*}_{\xi}(k_2,\bar{\xi}) g^{\xi\sigma} \Pi_{\sigma\delta}(l_2)\,,\\
    \mathcal{G}_{B}^{\rho} &= \epsilon^{*}_{\tau}(k_3,\lambda) g^{\tau\theta} \Pi_{\theta\nu}(p-l_1)\mathcal{V}^{\mu\nu\rho}(l_1, p - l_1)\epsilon_{\mu}(p,\bar{\lambda})\,.
\end{align}
and $D_1 = ((l_1- l_2)^2 + i\varepsilon)(l_2^2+ i\varepsilon)(l_1^2 + i\varepsilon)((p-l_1)^2+i\varepsilon)$. By decomposing $\Pi_{\gamma\rho}(l_1)$ in terms of the polarization vectors using Eq.\,\eqref{gluon propagator decomposition} we can split the contributions into regular and instantaneous terms:
\begin{align}
    \mathcal{G}_{A,reg} = \mathcal{G}_{A}^{\gamma} \epsilon_{\gamma}(l_1,\eta)\,, &\qquad \mathcal{G}_{B,reg} = \mathcal{G}_{B}^{\rho} \epsilon^{*}_{\rho}(l_1,\eta)\,,\\
    \mathcal{G}_{A,inst} = \mathcal{G}_{A}^{\gamma} n_{\gamma}\,, &\qquad \mathcal{G}_{B,inst} = \mathcal{G}_{B}^{\rho} n_{\rho}\,.
\end{align}

\vspace{3mm}

\textit{Regular contribution}

\vspace{3mm}

Using Eqs.\, \eqref{external_to_internal_gluon_id} and \eqref{g-> gg wavefunction} we obtain for the regular gluon structure
\begin{equation}
    \begin{split}
        \mathcal{G}_{A,reg} &= \mathcal{G}_{A}^{\gamma} \epsilon_{\gamma}(l_1,\eta) \\
        &= \epsilon^{*}_{\epsilon}(l_1-l_2,\xi) \mathcal{V}^{\gamma\delta\epsilon}(l_1-l_2,l_2) \epsilon^{*}_{\delta}(l_2,\bar{\xi}) \\
        &= \frac{2(z_1+z_2)^2}{z_1 z_2}\Gt^{\eta\bar{\xi}\xi}\left( \frac{z_1}{z_1 + z_2}\right) \cdot \Transv{L}{} \,,
    \end{split}
\end{equation}
and
\begin{equation}
    \begin{split}
        \mathcal{G}_{B,reg} &= \mathcal{G}_{B}^{\rho} \epsilon^{*}_{\rho}(l_1,\eta) \\
        &= -\epsilon^{*}_{\nu}(p-l_1,\lambda) \mathcal{V}^{\gamma\delta\epsilon}(l_1-l_2,l_2)\epsilon_{\mu}(p,\bar{\lambda})\epsilon^{*}_{\rho}(l_1,\eta) \\
        &= \frac{2}{(z_1+z_2)(1-z_1-z_2)} \Gt^{\bar{\lambda} \lambda\eta}(z_1+z_2) \cdot \Transv{l}{1}\,.
    \end{split}
\end{equation}
The remaining task is to perform the integrals over the inner momenta:
\begin{equation}
    \int \frac{\der^4 l_1}{(2\pi)^2} \frac{\der^4 l_2}{(2\pi)^2} e^{-\Transv{l}{1}\cdot \dipvec{xz}} e^{i\Transv{l}{2}\cdot \dipvec{yx}} \delta(l_2^- - k_2^-) \delta(l_1^- - k_1^- - k_2^-) \frac{\Transv{l}{1}^i \Transv{L}{}^j}{D_1}\,.
\end{equation}
Notice that the structure of the integrals over $l_1$ and $l_2$ is the same as for diagram R1. Using the identity Eq.\,\eqref{reg integral 4 poles} for the plus integrals and Eq.\,\eqref{transverse integral diag R1}
\begin{multline}
    \int \frac{\der^4 l_1}{(2\pi)^2} \frac{\der^4 l_2}{(2\pi)^2} e^{-\Transv{l}{1}\cdot \dipvec{xz}} e^{i\Transv{l}{2}\cdot \dipvec{yx}} \delta(l_2^- - k_2^-) \delta(l_1^- - k_1^- - k_2^-) \frac{\Transv{l}{1}^i \Transv{L}{}^j}{D_1}\\
    = -\frac{1}{4p^-l_1^-} \int \frac{\der^2 \Transv{l}{1}}{2\pi} \frac{\der^2 \Transv{l}{2}}{2\pi} e^{-\Transv{l}{1}\cdot \dipvec{xz}} e^{i\Transv{l}{2}\cdot \dipvec{yx}} \frac{\Transv{l}{1}^i \Transv{L}{}^j}{l_{1\perp}^2[L_{\perp}^2 + \Delta^2l_{1\perp}^2]}\\
    = -\frac{1}{4p^-l_1^-} \left(-\frac{z_3(z_1 + z_2)\dipvec{yx}^j}{\dipvec{yx}^2} \frac{\dipvec{wz}^i}{X_R^2} \right)= \frac{z_3}{4(p^-)^2} \frac{\dipvec{yx}^j}{\dipvec{yx}^2} \frac{\dipvec{wz}^i}{X_R^2}\,.
\end{multline}
Therefore, the regular part of the impact factor becomes
\begin{equation}
    N_{R7,reg}^{\bar{\lambda}\lambda\bar{\xi}\xi}(\Transv{x}{}, \Transv{y}{}, \Transv{z}{}) = - \frac{2g^2p^-}{\pi^2} \frac{\Gt^{\eta\bar{\xi}\xi}\left( \frac{z_1}{z_1 + z_2}\right) \cdot \dipvec{yx}}{r_{yx}^2} \frac{\Gt^{\bar{\lambda} \lambda\eta}(z_1+z_2)\cdot \dipvec{w_Az}}{r_{w_Az}^2 } \Theta_{ggg,1}\,,
\end{equation}
with $\Theta_{ggg,1} = r_{w_Az}^2/(r_{w_Az}^2 + \frac{z_1 z_{2}}{z_3(z_1 + z_{2})^2} r_{yx}^2)$.

\vspace{3mm}

\textit{Instantaneous contribution}

\vspace{3mm}

The instantaneous term of the gluon structure simplifies as follows. Since the instantaneous vector $n_{\rho}$ satisfies the identity $\epsilon(k,\lambda)\cdot n = 0$, we get, after contraction of the indices,
\begin{equation}
    \begin{split}
        \mathcal{G}_{A,inst} &= \mathcal{G}_{A}^{\gamma} n_{\gamma}\\
        &= -\epsilon^{*}_{\epsilon}(l_1-l_2,\xi) g^{\epsilon\delta}(l_1^- - 2l_2^-) \epsilon^{*}_{\delta}(l_2,\bar{\xi}) \\
        &= \frac{z_2 - z_1}{p^-} \delta^{\xi,-\bar{\xi}}\,,
    \end{split}
\end{equation}
and
\begin{equation}
    \begin{split}
        \mathcal{G}_{B,inst} &= \mathcal{G}_{B}^{\rho} n_{\rho}\\
        &= -\epsilon^{*}_{\nu}(p-l_1,\lambda) g^{\mu\nu} (2p^- - l_1^-)\epsilon_{\mu}(p,\bar{\lambda})\\
        &= p^-(1+z_3) \delta^{\lambda\bar{\lambda}}\,.
    \end{split}
\end{equation}
The integrals on $l_1$ and $l_2$ are identical to those of the instantaneous part of diagram R1. Using equations \ref{inst integral diagram R1} and \ref{transverse integral diag R1}, with the respective change of variables, and after doing some algebra we obtain,
\begin{equation}
    N_{R7,inst}^{\bar{\lambda}\lambda\bar{\xi}\xi}(x_{\perp}, y_{\perp}, z_{\perp}) = -\frac{g^2p^- z_1z_2z_3(z_2-z_1)(1 + z_3)}{2\pi^2(z_1 + z_2)^2} \frac{\delta^{\xi,-\bar{\xi}} \delta^{\lambda \bar{\lambda}}}{X_R^2}.
\end{equation}
The total impact factor then reads 
\begin{multline}
    N_{R7}^{\bar{\lambda}\lambda\bar{\xi}\xi}(\Transv{x}{}, \Transv{y}{}, \Transv{z}{}) = - \frac{2g^2p^-}{\pi^2} \frac{\Gt^{\eta\bar{\xi}\xi}\left( \frac{z_1}{z_1 + z_2}\right) \cdot \dipvec{yx}}{r_{yx}^2} \frac{\Gt^{\bar{\lambda} \lambda\eta}(z_1+z_2)\cdot \dipvec{w_Az}}{r_{w_Az}^2} \Theta_{ggg,1}\\
    -\frac{g^2p^- z_1z_2z_3(z_2-z_1)(1 + z_3)}{2\pi^2(z_1 + z_2)^2} \frac{\delta^{\xi,-\bar{\xi}} \delta^{\lambda \bar{\lambda}}}{X_R^2}\,.
\end{multline}

\section{Definitions of the kernels and $S$-matrices featured in the trijet cross section} \label{app: Kernels_and_Smatrices}
In this Appendix we provide the definitions and explicit expressions for the kernels and $S$-matrices involved in our results from Section \ref{sec: Cross section trijets}. Throughout this section, we assume that repeated color indices (e.g. $a,c,d,d$) are summed over.

\subsection{The $g \to q\bar{q}g$ channel}
Let us start by presenting the kernels involved in thr direct contribution of the quark topology. We have:
\begin{multline}
    \mathcal{K}_1(\boldsymbol{r}_{z'x'},\boldsymbol{r}_{w_1'y'};\boldsymbol{r}_{zx},\boldsymbol{r}_{w_1y}) = \frac{2\pi^4}{g^4 (p^-)^2} \sum_{\sigma \sigma' \lambda \bar{\lambda}} \left(\Ncal^{\bar{\lambda}\lambda\sigma\sigma'}_{q,reg}\right)^{\dagger}\Ncal^{\bar{\lambda}\lambda\sigma\sigma'}_{q,reg}\\
    =\frac{ 4z_{\bar{q}} z_q}{(z_g + z_q)^2}
    \left[ (z_q^2+ (z_q + z_g)^2) (z_{\bar{q}}^2 + (z_g + z_q)^2) \delta^{k',k} \delta^{l',l}\right.\\
    \left.+ z_g (z_g + 2z_q) ( (z_g + z_q)^2 - z_{\bar{q}}^2 )\epsilon^{k',k} \epsilon^{l',l} \right] \times \frac{\dipvec{z'x'}^{k'}\dipvec{zx}^{k}}{r_{z'x'}^2 r_{zx}^2} \frac{\dipvec{w_1'y'}^{l'} \dipvec{w_1y}^l}{r_{w_1'y'}^2 r_{w_1y}^2}\,,\label{eq:K1-def}
\end{multline}
\begin{align}
    \mathcal{K}_2(\boldsymbol{r}_{z'x'},\boldsymbol{r}_{w_1'y'};X_R) &= \frac{2\pi^4}{g^4 (p^-)^2} \sum_{\sigma \sigma' \lambda \bar{\lambda}} \left(\Ncal^{\bar{\lambda}\lambda\sigma\sigma'}_{q,reg}\right)^{\dagger}\Ncal^{\bar{\lambda}\lambda\sigma\sigma'}_{q,inst} = -\frac{ 4z_{\bar{q}}^3 z_g z_q^2}{z_q + z_g} \times \frac{\dipvec{z'x'}\cdot \dipvec{w_1'y'}}{r_{z'x'}^2r_{w_1'y'}^2} \frac{1}{X_R^2}\,,
\end{align}
\begin{align}
    \mathcal{K}_3(X_R';X_R) = \frac{2\pi^4}{g^4 (p^-)^2} \sum_{\sigma \sigma' \lambda \bar{\lambda}} \left(\Ncal^{\bar{\lambda}\lambda\sigma\sigma'}_{q,inst}\right)^{\dagger}\Ncal^{\bar{\lambda}\lambda\sigma\sigma'}_{q,inst} = \frac{4z_q^3 z_{\bar{q}}^3 z_g^2 }{(z_q + z_g)^2} \times \frac{1}{X_R^{2'}} \frac{1}{X_R^2}\,.
\end{align}
The S-matrices of this contribution are given by:
\begin{multline}
    S_{q\bar{q}g}^{q\bar{q}g}(\Perp{x}', \Perp{y}', \Perp{z}';\Perp{x}, \Perp{y}, \Perp{z}) = \frac{1}{NC_F^2} \left \langle \Tr \left[C_{R1}^{ab,\dagger}(\Perp{x}', \Perp{y}', \Perp{z}') C_{R1}^{ab}(\Perp{x}, \Perp{y}, \Perp{z}) \right] \right\rangle_Y\\
    =\frac{\N^2}{4C_F^2} \bigg\langle D_{y'y}D_{zz'}Q_{z'x'xz} - \frac{1}{\N^2}\bigg(D_{xx'}D_{y'y} + Q_{z'x'xz}Q_{y'z'zy} \bigg) + \frac{1}{\N^4} Q_{y'x'xy} \bigg\rangle_{Y}\,,
\end{multline}
\begin{multline}
    S_{q\bar{q}g}^{g}(\Perp{v}';\Perp{x}, \Perp{y}, \Perp{z})  = \frac{1}{NC_F^2} \left \langle\Tr \left[C_{R3}^{ab,\dagger}(\Perp{x}', \Perp{y}', \Perp{z}') C_{R1}^{ab}(\Perp{x}, \Perp{y}, \Perp{z}) \right] \right\rangle_Y\\
    = \frac{\N^2}{4C_F^2} \bigg\langle D_{xz}D_{v'y}D_{zv'} - \frac{1}{\N^2}\bigg(D_{xv'}D_{v'y} + D_{xz}D_{zy} \bigg) + \frac{1}{\N^4} D_{xy} \bigg\rangle_{Y}\,,
\end{multline}
\begin{multline}
    S^{q\bar{q}g}_{g}(\Perp{x}', \Perp{y}', \Perp{z}';\Perp{v})  = \frac{1}{NC_F^2} \left\langle\Tr \left[C_{R1}^{ab,\dagger}(\Perp{x}', \Perp{y}', \Perp{z}') C_{R3}^{ab}(\Perp{x}, \Perp{y}, \Perp{z}) \right] \right\rangle_Y\\
    = \frac{\N^2}{4C_F^2} \bigg\langle D_{y'v}D_{vz'}D_{z'x'} - \frac{1}{\N^2}\bigg(D_{vx'}D_{y'v} + D_{z'x'}D_{y'z'} \bigg) + \frac{1}{\N^4} D_{y'x'} \bigg \rangle_{Y} \,,
\end{multline}
where we have made use of Fierz identities to arrive at the second expression for each correlator. The color trace comes from the sum over the fermionic color indices, these being in the fundamental representation.

Let us move to the direct contribution of the gluon topology. The three kernels appearing in the cross section are
\begin{align}
    &\mathcal{K}_4(\boldsymbol{r}_{y'x'},\boldsymbol{r}_{w_2'z'};\boldsymbol{r}_{yx},\boldsymbol{r}_{w_2z}) =\frac{2\pi^4}{g^4 (p^-)^2} \sum_{\sigma \sigma' \lambda \bar{\lambda}} \left(\Ncal^{\bar{\lambda}\lambda\sigma\sigma'}_{g,reg}\right)^{\dagger}\Ncal^{\bar{\lambda}\lambda\sigma\sigma'}_{g,reg}\\
    &=\frac{8 z_{\bar{q}} z_q }{(z_{\bar{q}} + z_q)^2}  \bigg[((z_{q}^2 + z_{\bar{q}}^2)(1+z_g^2) + z_g^2) \delta^{k',k} \delta^{l',l} - z_g\left( (z_q + z_{\bar{q}})^2 + (z_q - z_{\bar{q}})^2 +z_g \right) \delta^{k,l} \delta^{k',l'}\nonumber\\
    &+  z_g\left( (z_q + z_{\bar{q}})^2 + (z_q - z_{\bar{q}})^2 +z_g \frac{(z_q - z_{\bar{q}})^2}{(z_q + z_{\bar{q}})^2}\right)\delta^{k',l} \delta^{l',k}  \bigg]\times \frac{\dipvec{y'x'}^{k'}\dipvec{yx}^k}{r_{y'x'}^2 r_{yx}^2} \frac{\dipvec{w_2'z'}^{l'} \dipvec{w_2z}^l}{r_{w_2'z'}^2 r_{w_2z}^2}\label{eq:K4-def}\,,
\end{align}
\vspace{-5mm}
\begin{align}
    \mathcal{K}_5(X_R';\boldsymbol{r}_{yx},\boldsymbol{r}_{w_2z}) &= \frac{2\pi^4}{g^4 (p^-)^2} \sum_{\sigma \sigma' \lambda \bar{\lambda}} \left(\Ncal^{\bar{\lambda}\lambda\sigma\sigma'}_{g,inst}\right)^{\dagger}\Ncal^{\bar{\lambda}\lambda\sigma\sigma'}_{g,reg}\nonumber\\  &=\frac{4z_{\bar{q}}^2z_q^2 \, z_g \, (z_g + 1) \,  \, (z_{q} - z_{\bar{q}}) \left( 1+z_g^2 \right)}{ (z_{\bar{q}} + z_q)^4 } \frac{1}{X_R^{'2}} \frac{\dipvec{yx} \cdot \dipvec{w_2 z}}{r_{yx}^2 r_{w_2 z}^2}\,, \\
    \mathcal{K}_6(X_R';X_R) &= \frac{2\pi^4}{g^4 (p^-)^2} \sum_{\sigma \sigma' \lambda \bar{\lambda}} \left(\Ncal^{\bar{\lambda}\lambda\sigma\sigma'}_{g,inst}\right)^{\dagger}\Ncal^{\bar{\lambda}\lambda\sigma\sigma'}_{g,inst}=  \frac{8 \, z_{\bar{q}}^3 \, z_g^2 \, (z_g + 1)^2 \, z_q^3}{(z_{\bar{q}} + z_q)^4} \frac{1}{X_R^{'2}} \frac{1}{X_R^2}\,.
\end{align}
The multiple scattering matrices of this contribution are defined as 
\begin{align}
    S_{q\bar{q}g}^{q\bar{q}g, (2)}(\Perp{x}', \Perp{y}', \Perp{z}';\Perp{x}, \Perp{y}, \Perp{z}) &= \frac{1}{\N^2 C_F} \left\langle\Tr \left[C_{R4}^{ab,\dagger}(\Perp{x}', \Perp{y}', \Perp{z}') C_{R4}^{ab}(\Perp{x}, \Perp{y}, \Perp{z}) \right] \right\rangle_Y \nonumber\\
    &=\frac{\N}{4C_F} \bigg\langle D_{xx'}D_{z'z}Q_{y'z'zy} + D_{y'y}D_{zz'}Q_{z'x'xz}\\ \nonumber
    &\qquad \qquad \qquad- \frac{1}{\N^2}\bigg(O_{xz'zyy'zz'x'} + O_{y'z'zx'xzz'y} \bigg) \bigg\rangle_{Y}\,,\\
    S_{q\bar{q}g}^{g,(2)}(\Perp{v}';\Perp{x}, \Perp{y}, \Perp{z}) &= \frac{1}{\N^2 C_F} \left\langle\Tr \left[C_{R6}^{ab,\dagger}(\Perp{x}', \Perp{y}', \Perp{z}') C_{R4}^{ab}(\Perp{x}, \Perp{y}, \Perp{z}) \right] \right\rangle_Y\nonumber\\
    &= \frac{\N}{4C_F} \bigg\langle D_{xz}D_{v'y}D_{zv'} + D_{zy}D_{xv'}D_{v'z}\nonumber \\ 
    &\qquad \qquad \qquad \qquad \qquad - \frac{1}{\N^2}\bigg(S_{zv'xzv'y} + S_{xv'zyv'z} \bigg)\bigg\rangle_{Y}\,.
\end{align}
For the quark-gluon interference topology, the four kernels featured in the cross section are
\begin{align}
    &\mathcal{K}_7(\boldsymbol{r}_{z'x'},\boldsymbol{r}_{w_1'y'};\boldsymbol{r}_{yx},\boldsymbol{r}_{w_2z}) =\frac{2\pi^4}{g^4 (p^-)^2} \sum_{\sigma \sigma' \lambda \bar{\lambda}} \left(\Ncal^{\bar{\lambda}\lambda\sigma\sigma'}_{q,reg}\right)^{\dagger}\Ncal^{\bar{\lambda}\lambda\sigma\sigma'}_{g,reg} =\frac{ 4z_q z_{\bar{q}} }{(z_q + z_{\bar{q} })^2 (z_q + z_g)}\nonumber\\
    &\times \bigg[z_g\left \{ z_{\bar{q}}^2(1 + z_q(1+ z_q + z_{\bar{q}}) + z_q(z_q + z_g)(z_q - z_{\bar{q}}(z_q + z_{\bar{q}})) \right\} \, (\delta^{k',k} \delta^{l',l} + \delta^{k,l} \delta^{k',l'})\nonumber\\ 
     &- \left\{
    z_{\bar{q}}^2 (z_q + z_g + z_q(z_q + z_{\bar{q}})^2)  + z_q(z_q + z_g)\left[(z_q + z_{\bar{q}})(z_q + (z_q + z_g)(z_q + z_{\bar{q}})) + z_qz_g \right]
    \right\} \delta^{k,l'} \delta^{l,k'} \nonumber\\
     & + 2z_qz_{\bar{q}}z_g^2(z_q + z_g)( \epsilon^{k'l}\epsilon^{l'k} - \delta^{k,l'} \delta^{l,k'})\bigg] \times \frac{\dipvec{z'x'}^{k'}\dipvec{yx}^k}{r_{z'x'}^2 r_{yx}^2} \frac{\dipvec{w_1'y'}^{l'} \dipvec{w_2z}^l}{r_{w_1'y'}^2 r_{w_2z}^2}\,,
\end{align}
\begin{align}
    \mathcal{K}_8(X_R';\boldsymbol{r}_{yx},\boldsymbol{r}_{w_2z}) &= \frac{2\pi^4}{g^4 (p^-)^2} \sum_{\sigma \sigma' \lambda \bar{\lambda}} \left(\Ncal^{\bar{\lambda}\lambda\sigma\sigma'}_{q,inst}\right)^{\dagger}\Ncal^{\bar{\lambda}\lambda\sigma\sigma'}_{g,reg}\nonumber\\
    &= \frac{4
    z_{\bar{q}}^2 \, z_g \, z_q^2 
    \left( z_qz_g^2 - z_{\bar{q}} \right)
    }{
    (z_{\bar{q}} + z_q)^2 (z_g + z_q)
    }  \frac{1}{X_R^{'2}} \frac{\dipvec{yx} \cdot \dipvec{w_2 z}}{r_{yx}^2 r_{w_2 z}^2}\,, \\
    \mathcal{K}_9(\dipvec{z'x'}, \dipvec{w_1'y'}; X_R) &= \frac{2\pi^4}{g^4 (p^-)^2} \sum_{\sigma \sigma' \lambda \bar{\lambda}} \left(\Ncal^{\bar{\lambda}\lambda\sigma\sigma'}_{q,reg}\right)^{\dagger}\Ncal^{\bar{\lambda}\lambda\sigma\sigma'}_{g,inst}\nonumber\\
    =&-\frac{
    16 \,z_g z_q^2 z_{\bar{q}}^2  \, (z_g + 1) \, (z_{\bar{q}} - z_q)
    }{
    (z_{\bar{q}} + z_q)^2
    } \frac{1}{X_R^{2}}  \frac{\dipvec{z'x'} \cdot \dipvec{w_1' y'}}{r_{z'x'}^2 r_{w_1' y'}^2}\,, \\
    \mathcal{K}_{10}(X_R';X_R) &=  \frac{2\pi^4}{g^4 (p^-)^2} \sum_{\sigma \sigma' \lambda \bar{\lambda}} \left(\Ncal^{\bar{\lambda}\lambda\sigma\sigma'}_{q,inst}\right)^{\dagger}\Ncal^{\bar{\lambda}\lambda\sigma\sigma'}_{q,inst}=\frac{
    16 \,  z_q^3 z_{\bar{q}}^3 \, z_g^2 \, (z_g + 1)
    }{
    (z_{\bar{q}} + z_q)^2 \, (z_g + z_q)
    } \frac{1}{X_R^{'2}} \frac{1}{X_R^2} \,,
\end{align}
and the multiple scattering matrices involved in the process are
\begin{align}
    S_{q\bar{q}g}^{q\bar{q}g, (3)}(\Perp{x}', \Perp{y}', \Perp{z}';\Perp{x}, \Perp{y}, \Perp{z}) &= \frac{2}{\N^2 C_F} \left\langle\Tr \left[C_{R1}^{ab,\dagger}(\Perp{x}', \Perp{y}', \Perp{z}') C_{R4}^{ab}(\Perp{x}, \Perp{y}, \Perp{z}) \right] \right\rangle_Y \nonumber\\
    &= -\frac{\N}{2C_F} \bigg\langle D_{y'y}D_{zz'}Q_{z'x'xz} - \frac{1}{\N^2} O_{xz'zyy'zz'x'} \bigg\rangle_{Y}\,,\\
    S_{q\bar{q}g}^{g,(3)}(\Perp{v}';\Perp{x}, \Perp{y}, \Perp{z}) &= \frac{2}{\N^2 C_F} \left\langle\Tr \left[C_{R3}^{ab,\dagger}(\Perp{x}', \Perp{y}', \Perp{z}') C_{R4}^{ab}(\Perp{x}, \Perp{y}, \Perp{z}) \right] \right\rangle_Y \nonumber\\
    &= -\frac{\N}{2C_F} \bigg\langle D_{xz}D_{v'y}D_{zv'} - \frac{1}{\N^2}S_{xv'zyv'zx}\bigg\rangle_{Y}\,.
\end{align}
For the gluon-antiquark interference, the four kernels in the cross section have the following expressions:
\begin{align}
    &\mathcal{K}_{11}(\boldsymbol{r}_{z'y'},\boldsymbol{r}_{w_3'z'};\boldsymbol{r}_{yx},\boldsymbol{r}_{w_2 z}) = \frac{2\pi^4}{g^4 (p^-)^2} \sum_{\sigma \sigma' \lambda \bar{\lambda}} \left(\Ncal^{\bar{\lambda}\lambda\sigma\sigma'}_{\bar{q},reg}\right)^{\dagger}\Ncal^{\bar{\lambda}\lambda\sigma\sigma'}_{g,reg} = \frac{4 z_{\bar{q}} z_q}{(z_{\bar{q}} + z_g) (z_q + z_{\bar{q}})^2}\nonumber\\ 
    &\times\bigg[ (z_q + z_{\bar{q}})\big\{(z_{\bar{q}} + z_g)(2\zbar z_g z_q -( z_q^2 + \zbar^2)) -(1-z_g)(z_q^2\zbar - z_g(\zbar + z_g)) \big\}\epsilon^{kl} \epsilon^{k'l'}\nonumber\\
    &-\big\{(z_q + \zbar)[ (\zbar + z_g) (2\zbar z_g z_q + \zbar^2 + z_q^2) - \zbar z_q^2(1-z_g) ] - z_g(\zbar - z_q)^2(\zbar + z_g) \big\} \delta^{kl} \delta^{k'l'}\nonumber\\
    &-\zbar (z_q + \zbar)^2(\zbar + z_g)^2(\delta^{kk'}\delta^{ll'} + \epsilon^{kk'} \epsilon^{ll'}) \bigg]
    \times \frac{\dipvec{z'y'}^{k'}\dipvec{yx}^k}{r_{z'y'}^2 r_{yx}^2} \frac{\dipvec{w_3'y'}^{l'} \dipvec{w_2z}^l}{r_{w_3'y'}^2 r_{w_2z}^2}\,,
\end{align}
\begin{align}
    \mathcal{K}_{12}(\boldsymbol{r}_{z'y'},\boldsymbol{r}_{w_3'x'}; X_R) &= \frac{2\pi^4}{g^4 (p^-)^2} \sum_{\sigma \sigma' \lambda \bar{\lambda}} \left(\Ncal^{\bar{\lambda}\lambda\sigma\sigma'}_{\bar{q},reg}\right)^{\dagger}\Ncal^{\bar{\lambda}\lambda\sigma\sigma'}_{g,inst}\nonumber\\
    &= -\frac{
    16 z_{\bar{q}}^2 z_g \left(z_g + 1\right) z_q^2 \left(z_{\bar{q}} - z_q\right)}{
    \left(z_q + z_{\bar{q}}\right)^2} \frac{1}{X_R^{2}}  \frac{\dipvec{z'y'} \cdot \dipvec{w_3' x'}}{r_{z'y'}^2 r_{w_3' x'}^2}\,, \\
    \mathcal{K}_{13}(X_R';\boldsymbol{r}_{yx},\boldsymbol{r}_{w_2z}) &= \frac{2\pi^4}{g^4 (p^-)^2} \sum_{\sigma \sigma' \lambda \bar{\lambda}} \left(\Ncal^{\bar{\lambda}\lambda\sigma\sigma'}_{\bar{q},inst}\right)^{\dagger}\Ncal^{\bar{\lambda}\lambda\sigma\sigma'}_{g,reg}= \frac{
    4 z_{\bar{q}}^2 z_g z_q^2 ( z_q - \zbar z_g^2 )
    }{
    \left( z_q + z_{\bar{q}} \right)^2 \left( z_{\bar{q}} + z_g \right)
    }  \frac{1}{X_R^{'2}} \frac{\dipvec{yx} \cdot \dipvec{w_2 z}}{r_{yx}^2 r_{w_2 z}^2}\,, \\
    \mathcal{K}_{14}(X_R';X_R) &= \frac{2\pi^4}{g^4 (p^-)^2} \sum_{\sigma \sigma' \lambda \bar{\lambda}} \left(\Ncal^{\bar{\lambda}\lambda\sigma\sigma'}_{q,inst}\right)^{\dagger}\Ncal^{\bar{\lambda}\lambda\sigma\sigma'}_{g,inst}=  -\frac{
    16 z_q^3 z_{\bar{q}}^3 z_g^2 \left(z_g + 1\right) }{\left(z_q + z_{\bar{q}}\right)^2 \left(z_{\bar{q}} + z_g\right)} \frac{1}{X_R^{'2}} \frac{1}{X_R^2}\,,
\end{align}
and the multiple scattering matrices have the following definitions:
\begin{align}
    S_{q\bar{q}g}^{q\bar{q}g, (4)}(\Perp{x}', \Perp{y}', \Perp{z}';\Perp{x}, \Perp{y}, \Perp{z}) &= \frac{2}{\N^2 C_F} \left\langle\Tr \left[C_{\overline{R1}}^{ab,\dagger}(\Perp{x}', \Perp{y}', \Perp{z}') C_{R4}^{ab}(\Perp{x}, \Perp{y}, \Perp{z}) \right] \right\rangle_Y \nonumber\\
    &= \frac{\N}{2C_F} \bigg\langle D_{xx'}D_{zz'}Q_{zyy'z'} - \frac{1}{\N^2} O_{xzz'yy'z'zx'} \bigg\rangle_{Y}\,,\\
    S_{q\bar{q}g}^{g,(4)}(\Perp{v}';\Perp{x}, \Perp{y}, \Perp{z}) &= \frac{2}{\N^2 C_F} \left\langle\Tr \left[C_{\overline{R3}}^{ab,\dagger}(\Perp{x}', \Perp{y}', \Perp{z}') C_{R4}^{ab}(\Perp{x}, \Perp{y}, \Perp{z}) \right] \right\rangle_Y \nonumber\\
    &= \frac{\N}{2C_F} \bigg\langle D_{zy}D_{xv'}D_{v'z} - \frac{1}{\N^2}S_{zv'xzv'y}\bigg\rangle_{Y}\,.\\
\end{align}
Finally, for the quark-anitquark interference, the kernels of this contribution are defined as 
\begin{multline}
    \mathcal{K}_{15} (\boldsymbol{r}_{y'z'},\boldsymbol{r}_{w_3'x'};\boldsymbol{r}_{zx},\boldsymbol{r}_{w_1y}) = \frac{2\pi^4}{g^4 (p^-)^2} \sum_{\sigma \sigma' \lambda \bar{\lambda}} \left(\Ncal^{\bar{\lambda}\lambda\sigma\sigma'}_{\bar{q},reg}\right)^{\dagger}\Ncal^{\bar{\lambda}\lambda\sigma\sigma'}_{q,reg}\\
   =\frac{
    4 z_{\bar{q}} z_q}{
    \left(z_{\bar{q}} + z_g\right) \left(z_g + z_q\right)
    } \bigg[
    \left( z_q (z_q + z_g)  + \zbar (\zbar + z_g) \right) \left( z_q (\zbar + z_g)  + \zbar (z_q + z_g) \right) \delta_{k,k'} \delta_{l,l'}\\
    + z_g \left(z_{\bar{q}} - z_q\right)^2 \epsilon_{k,k'} \epsilon_{l,l'}
    \bigg] \times \frac{\dipvec{z'y'}^{k'}\dipvec{zx}^k}{r_{z'y'}^2 r_{zx}^2} \frac{\dipvec{w_3'y'}^{l'} \dipvec{w_1y}^l}{r_{w_3'y'}^2 r_{w_1y}^2}\,,
\end{multline}
\begin{align}
    \mathcal{K}_{16} (\boldsymbol{r}_{y'z'},\boldsymbol{r}_{w_3'x'};X_{R}) &= \frac{2\pi^4}{g^4 (p^-)^2} \sum_{\sigma \sigma' \lambda \bar{\lambda}} \left(\Ncal^{\bar{\lambda}\lambda\sigma\sigma'}_{\bar{q},reg}\right)^{\dagger}\Ncal^{\bar{\lambda}\lambda\sigma\sigma'}_{q,inst}= -\frac{
    4 z_g z_{\bar{q}}^3 z_q^2}{ z_g + z_q} \frac{1}{X_R^{2}}  \frac{\dipvec{z'y'} \cdot \dipvec{w_3' x'}}{r_{z'y'}^2 r_{w_3' x'}^2}\,, \\
    \mathcal{K}_{17} (X_{R}';\boldsymbol{r}_{zx},\boldsymbol{r}_{w_1y}) &= \frac{2\pi^4}{g^4 (p^-)^2} \sum_{\sigma \sigma' \lambda \bar{\lambda}} \left(\Ncal^{\bar{\lambda}\lambda\sigma\sigma'}_{\bar{q},inst}\right)^{\dagger}\Ncal^{\bar{\lambda}\lambda\sigma\sigma'}_{q,reg}= -\frac{
    4 z_g z_{q}^3 z_{\bar{q}}^2}{ z_g + z_{\bar{q}} }  \frac{1}{X_R^{'2}} \frac{\dipvec{zx} \cdot \dipvec{w_1y}}{r_{zx}^2 r_{w_1y}^2}\,,
\end{align}
and the $S$-matrices:
\begin{align}
    S_{q\bar{q}g}^{q\bar{q}g, (5)}(\Perp{x}', \Perp{y}', \Perp{z}';\Perp{x}, \Perp{y}, \Perp{z}) &= \frac{2}{\N^2 C_F} \left\langle\Tr \left[C_{\overline{R1}}^{ab,\dagger}(\Perp{x}', \Perp{y}', \Perp{z}') C_{R1}^{ab}(\Perp{x}, \Perp{y}, \Perp{z}) \right] \right\rangle_Y \nonumber\\
    &= -\frac{\N}{2C_F} \bigg\langle D_{xx'}D_{y'y} + O_{y'z'zx'xzz'y} - Q_{z'x'xz}Q_{y'z'zy}\\
    &\qquad \qquad+ \frac{1}{\N^2} Q_{y'x'xy} \bigg\rangle_{Y}\,,
\end{align}
\begin{align}
    S_{q\bar{q}g}^{g,(5)}(\Perp{v}';\Perp{x}, \Perp{y}, \Perp{z}) &= \frac{1}{\N^2 C_F} \left\langle\Tr \left[C_{\overline{R3}}^{ab,\dagger}(\Perp{x}', \Perp{y}', \Perp{z}') C_{R1}^{ab}(\Perp{x}, \Perp{y}, \Perp{z}) \right] \right\rangle_Y \nonumber\\
    &= -\frac{\N}{2C_F} \bigg\langle D_{xv'}D_{v'y} + D_{xz}D_{zy} - S_{zv'xzv'y} - \frac{1}{\N^2}D_{xy}\bigg\rangle_{Y}\,.
\end{align}

\subsection{The $g \to ggg$ channel}
We now present the definition of the objects featured in the $ggg$ cross section. Since the tensor structure and the color operators appearing in this channel are more complicated we will only present the definition of the kernels without showing the explicit sum over the helicities and polarizations. For the color operators we only present the large $\N$ approximation and assume that the correlators are real to simplify the structures further.

Let us start with the definitions of the objects appearing in the regular contribution. The expressions of the kernels appearing in the cross section are
\begin{multline}
    \mathcal{K}^{1}_{ggg}(\boldsymbol{r}_{y'x'},\boldsymbol{r}_{w_A'z'};\boldsymbol{r}_{yx},\boldsymbol{r}_{w_Az}) = \frac{2\pi^4}{g^4 (p^-)^2} \sum_{\lambda \bar{\lambda} \xi \bar{\xi}} \left(\Ncal^{\bar{\lambda}\lambda\bar{\xi}\xi}_{ggg,reg\,A}\right)^{\dagger}\Ncal^{\bar{\lambda}\lambda\bar{\xi}\xi}_{ggg,reg\,A}\\ 
    =8 \sum_{\xi\bar{\xi}\lambda\bar{\lambda}} \left(\Gt^{\eta'\bar{\xi}\xi}\left( \frac{z_1}{z_1 + z_2}\right)\right)^{k'*} \left(\Gt^{\bar{\lambda} \lambda\eta'}(z_1+z_2)\right)^{l'*} \left(\Gt^{\eta\bar{\xi}\xi}\left( \frac{z_1}{z_1 + z_2}\right)\right)^{k} \left(\Gt^{\bar{\lambda} \lambda\eta}(z_1+z_2)\right)^{l}\\
    \times \frac{\dipvec{y'x'}^{k'}\dipvec{yx}^k}{r_{y'x'}^2 r_{yx}^2} \frac{\dipvec{w_A'z'}^{l'} \dipvec{w_Az}^l}{r_{w_A'z'}^2 r_{w_Az}^2}\,,
\end{multline}
\vspace{-5mm}
\begin{multline}
    \mathcal{K}^{2}_{ggg}(\boldsymbol{r}_{y'x'},\boldsymbol{r}_{w_A'z'};\boldsymbol{r}_{yz},\boldsymbol{r}_{w_Bx}) =\frac{\pi^4}{g^4 (p^-)^2} \sum_{\lambda \bar{\lambda} \xi \bar{\xi}} \left(\Ncal^{\bar{\lambda}\lambda\bar{\xi}\xi}_{ggg,reg\,A}\right)^{\dagger}\Ncal^{\bar{\lambda}\lambda\bar{\xi}\xi}_{ggg,reg\,B}\\ 
     =4\sum_{\xi\bar{\xi}\lambda\bar{\lambda}} \left(\Gt^{\eta'\bar{\xi}\xi}\left( \frac{z_1}{z_1 + z_2}\right)\right)^{k'*} \left(\Gt^{\bar{\lambda} \lambda\eta'}(z_1+z_2)\right)^{l'*} \left(\Gt^{\eta\bar{\xi}\lambda}\left( \frac{z_3}{z_3 + z_2}\right)\right)^{k} \left(\Gt^{\bar{\lambda} \xi\eta}(z_3+z_2)\right)^{l}\\
     \times \frac{\dipvec{y'x'}^{k'}\dipvec{yz}^k}{r_{y'x'}^2 r_{yz}^2} \frac{\dipvec{w_A'z'}^{l'} \dipvec{w_Bx}^l}{r_{w_A'z'}^2 r_{w_Bx}^2}\,,
\end{multline}
\vspace{-5mm}
\begin{multline}
    \mathcal{K}^{3}_{ggg}(\boldsymbol{r}_{y'x'},\boldsymbol{r}_{w_A'z'};\boldsymbol{r}_{zx},\boldsymbol{r}_{w_C y}) = \frac{\pi^4}{g^4 (p^-)^2} \sum_{\lambda \bar{\lambda} \xi \bar{\xi}} \left(\Ncal^{\bar{\lambda}\lambda\bar{\xi}\xi}_{ggg,reg\,A}\right)^{\dagger}\Ncal^{\bar{\lambda}\lambda\bar{\xi}\xi}_{ggg,reg\,C}\\ 
    =4 \sum_{\xi\bar{\xi}\lambda\bar{\lambda}} \left(\Gt^{\eta'\bar{\xi}\xi}\left( \frac{z_1}{z_1 + z_2}\right)\right)^{k'*} \left(\Gt^{\bar{\lambda} \lambda\eta'}(z_1+z_2)\right)^{l'*} \left(\Gt^{\eta \lambda \xi}\left( \frac{z_1}{z_1 + z_3}\right)\right)^{k} \left(\Gt^{\bar{\lambda} \bar{\xi} \eta}(z_1+z_3)\right)^{l}\\
    \times \frac{\dipvec{y'x'}^{k'}\dipvec{zx}^k}{r_{y'x'}^2 r_{zx}^2} \frac{\dipvec{w_A'z'}^{l'} \dipvec{w_Cy}^l}{r_{w_A'z'}^2 r_{w_Cy}^2}\,.
\end{multline}
In the interference kernels we multiplied by an additional factor of $1/2$ due to the normalization of their color correlators (which are normalized as in the $q\bar{q}g$ case).

The multiple scattering matrices have the following expressions in the large $\N$ limit (the sum over colors in the first line of the right hand side is implied). For the direct term A$\times$A:
\begin{multline}
    S_{ggg}^{ggg, (1)}(\Perp{x}', \Perp{y}', \Perp{z}';\Perp{x}, \Perp{y}, \Perp{z}) = \frac{1}{2N^3C_F} \avg{C_{R7,A}^{abcd,\dagger}(\Perp{x}', \Perp{y}', \Perp{z}') C_{R7,A}^{abcd} (\Perp{x}, \Perp{y}, \Perp{z})}{Y}\\
    \approx \frac{1}{4}  \bigg\{ \Dcal(\Perp{x}, \Perp{x}')\Dcal(\Perp{z}', \Perp{z})\Qcal(\Perp{x}',\Perp{y}',\Perp{y},\Perp{x}) \Qcal(\Perp{y}',\Perp{z}',\Perp{z},\Perp{y})\\
    + \Dcal(\Perp{y}, \Perp{y}')\Dcal(\Perp{z}', \Perp{z})\Qcal(\Perp{y}',\Perp{x}',\Perp{x},\Perp{y}) \Qcal(\Perp{x}',\Perp{z}',\Perp{z},\Perp{x}) +  \mbox{c.c.}\bigg\}\\
    = \frac{1}{2} \Dcal(\Perp{z}', \Perp{z}) \bigg\{ \Dcal(\Perp{x}, \Perp{x}')\Qcal(\Perp{x}',\Perp{y}',\Perp{y},\Perp{x}) \Qcal(\Perp{y}',\Perp{z}',\Perp{z},\Perp{y})\\
    + \Dcal(\Perp{y}, \Perp{y}')\Qcal(\Perp{y}',\Perp{x}',\Perp{x},\Perp{y}) \Qcal(\Perp{x}',\Perp{z}',\Perp{z},\Perp{x}) \bigg\}\,,
\end{multline}
\begin{multline}
    S_{ggg}^{g,(1)}(\Perp{x}', \Perp{y}', \Perp{z}';\Perp{v}) = \frac{1}{2N^3C_F} \left\langle C_{R7,A}^{abcd,\dagger}(\Perp{x}', \Perp{y}', \Perp{z}') C_{R9,A}^{abcd} (\Perp{x}, \Perp{y}, \Perp{z}) \right\rangle_Y\\
    \approx \frac{1}{4} \bigg\{\Dcal(\Perp{v}, \Perp{y}')\Dcal(\Perp{z}', \Perp{v})\Dcal(\Perp{y}', \Perp{x}')\Dcal(\Perp{x}', \Perp{z}')\\
    + \Dcal(\Perp{v}, \Perp{x}')\Dcal(\Perp{z}', \Perp{v})\Dcal(\Perp{x}', \Perp{y}')\Dcal(\Perp{y}', \Perp{z}') + c.c.\bigg\}\\
    = \frac{1}{2} \bigg\{\Dcal(\Perp{v}, \Perp{y}')\Dcal(\Perp{z}', \Perp{v})\Dcal(\Perp{y}', \Perp{x}')\Dcal(\Perp{x}', \Perp{z}')\\
    + \Dcal(\Perp{v}, \Perp{x}')\Dcal(\Perp{z}', \Perp{v})\Dcal(\Perp{x}', \Perp{y}')\Dcal(\Perp{y}', \Perp{z}')\bigg\}\,.
\end{multline}
For the interference between A and B:
\begin{multline}
    S_{ggg}^{ggg, (2)}(\Perp{x}', \Perp{y}', \Perp{z}';\Perp{x}, \Perp{y}, \Perp{z}) = \frac{1}{N^3C_F} \left \langle C_{R7,A}^{abcd,\dagger}(\Perp{x}', \Perp{y}', \Perp{z}') C_{R7,B}^{abcd} (\Perp{x}, \Perp{y}, \Perp{z})\right\rangle_Y\\
    \approx \frac{1}{2}  \bigg\{ \Dcal(\Perp{x}, \Perp{x}')S(\Perp{z}', \Perp{z})\Qcal(\Perp{x}',\Perp{y}',\Perp{y},\Perp{x}) \Qcal(\Perp{y}',\Perp{z}',\Perp{z},\Perp{y}) + \mbox{c.c.}\bigg\}\\
    = \Dcal(\Perp{x}, \Perp{x}')\Dcal(\Perp{z}', \Perp{z})\Qcal(\Perp{x}',\Perp{y}',\Perp{y},\Perp{x}) \Qcal(\Perp{y}',\Perp{z}',\Perp{z},\Perp{y})\,,
\end{multline}
\begin{multline}
    S_{ggg}^{g,(2)}(\Perp{x}', \Perp{y}', \Perp{z}';\Perp{v}) = \frac{1}{N^3C_F} \left \langle C_{R7,A}^{abcd,\dagger}(\Perp{x}', \Perp{y}', \Perp{z}') C_{R9,B}^{abcd} (\Perp{x}, \Perp{y}, \Perp{z})\right\rangle_Y\\
    \approx \frac{1}{2}  \bigg\{ \Dcal(\Perp{v}, \Perp{x}')\Dcal(\Perp{z}', \Perp{v})\Dcal(\Perp{x}', \Perp{y}')\Dcal(\Perp{y}', \Perp{z}') + \mbox{c.c.}\bigg\}\\
    = \Dcal(\Perp{v}, \Perp{x}')\Dcal(\Perp{z}', \Perp{v})S(\Perp{x}', \Perp{y}')\Dcal(\Perp{y}', \Perp{z}')\,.
\end{multline}
And finally for the interference between A and C
\begin{multline}
    S_{ggg}^{ggg, (3)}(\Perp{x}', \Perp{y}', \Perp{z}';\Perp{x}, \Perp{y}, \Perp{z}) = \frac{1}{N^3C_F} \left \langle C_{R7,A}^{abcd, \dagger}(\Perp{x}', \Perp{y}', \Perp{z}') C_{R7,C}^{abcd} (\Perp{x}, \Perp{y}, \Perp{z}) \right\rangle_Y\\
    \approx \frac{1}{2}  \bigg\{ \Dcal(\Perp{y}, \Perp{y}')\Dcal(\Perp{z}', \Perp{z})\Qcal(\Perp{y}',\Perp{x}',\Perp{x},\Perp{y}) \Qcal(\Perp{x}',\Perp{z}',\Perp{z},\Perp{x}) + \mbox{c.c.}\bigg\}\\
    = \Dcal(\Perp{y}, \Perp{y}')\Dcal(\Perp{z}', \Perp{z})\Qcal(\Perp{y}',\Perp{x}',\Perp{x},\Perp{y}) \Qcal(\Perp{x}',\Perp{z}',\Perp{z},\Perp{x})\,,
\end{multline}
\begin{multline}
    S_{ggg}^{g,(3)}(\Perp{x}', \Perp{y}', \Perp{z}';\Perp{v}) = \frac{1}{N^3C_F} \left \langle C_{R7,A}^{abcd, \dagger}(\Perp{x}', \Perp{y}', \Perp{z}') C_{R9,C}^{abcd} (\Perp{x}, \Perp{y}, \Perp{z}) \right\rangle_Y\\
    \approx \frac{1}{2}  \bigg\{ \Dcal(\Perp{v}, \Perp{y}')\Dcal(\Perp{z}', \Perp{v})\Dcal(\Perp{y}', \Perp{x}') \Dcal(\Perp{x}', \Perp{z}') + \mbox{c.c.}\bigg\}\\
    = \Dcal(\Perp{v}, \Perp{y}')\Dcal(\Perp{z}', \Perp{v})\Dcal(\Perp{y}', \Perp{x}')\Dcal(\Perp{x}', \Perp{z}')\,.
\end{multline}
For the direct term of the instantaneous contributions the expressions of the kernels read
\begin{multline}
    \mathcal{K}^{4}_{ggg}(X_{R}',X_{R}) =\frac{2\pi^4}{g^4 (p^-)^2} \sum_{\lambda \bar{\lambda} \xi \bar{\xi}} \left(\Ncal^{\bar{\lambda}\lambda\bar{\xi}\xi}_{ggg,inst\,A}\right)^{\dagger}\Ncal^{\bar{\lambda}\lambda\bar{\xi}\xi}_{ggg,inst\,A}\\
    = (z_1 z_2 z_3)^2 \left[
\left( g(z_1,z_3,z_2) + 1 \right)^2
+
\left( g(z_1,z_2,z_3) + 1 \right)^2\right.\\
\left.
+
\left(
g(z_1,z_3,z_2)
+
g(z_1,z_2,z_3)
- 2
\right)^2
\right] \times \frac{1}{X_{R}^{'2} X_{R}^{2}} \,,
\end{multline}
\begin{multline}
    \mathcal{K}^{5}_{ggg}(X_{R}',X_{R}) =\frac{\pi^4}{g^4 (p^-)^2} \sum_{\lambda \bar{\lambda} \xi \bar{\xi}} \left(\Ncal^{\bar{\lambda}\lambda\bar{\xi}\xi}_{ggg,inst\,A}\right)^{\dagger}\Ncal^{\bar{\lambda}\lambda\bar{\xi}\xi}_{ggg,inst\,B}\\
    =\frac{ (z_1 z_2 z_3)^2}{4} \Bigg[
\left(
g(z_1,z_2,z_3) - 3
\right)
\left(
g(z_1,z_3,z_2)
+
g(z_1,z_2,z_3)
- 2
\right)\\
+
\left(
g(z_1,z_3,z_2) + 1
\right)
\left(
g(z_2,z_1,z_3) - 1
\right)\\
+
\left(
g(z_1,z_2,z_3) + 1
\right)
\left(
g(z_2,z_1,z_3)
-
g(z_1,z_2,z_3)
\right)
\Bigg]\times \frac{1}{X_{R}^{'2} X_{R}^{2}}\,,
\end{multline}
where we have defined the function
\begin{align}
    g(z_1,z_2,z_3) \equiv \frac{(z_1 - z_3)(1+z_2)}{(z_1 + z_3)^2}\,.
\end{align}
Again, due to the difference in the color operators of the direct and the interference terms, the kernel 5 contains an additional factor of $1/2$.

Finally, we define the kernels for the interference between the regular and the instantaneous contributions. Their expressions read
\begin{align}
    \mathcal{K}^{6}_{ggg}(X_{R}';\boldsymbol{r}_{yx},\boldsymbol{r}_{w_Az}) &= \frac{2\pi^4}{g^4 (p^-)^2} \sum_{\lambda \bar{\lambda} \xi \bar{\xi}} \left(\Ncal^{\bar{\lambda}\lambda\bar{\xi}\xi}_{ggg,inst\,A}\right)^{\dagger}\Ncal^{\bar{\lambda}\lambda\bar{\xi}\xi}_{ggg,reg\,A}\nonumber\\
    &= 2(z_1 z_2 z_3)\sum_{\xi \bar{\xi} \lambda \bar{\lambda}} \Pi^{\xi \bar{\xi} \lambda \bar{\lambda}}\left(\Gt^{\eta\bar{\xi}\xi}\left( \frac{z_1}{z_1 + z_2}\right)\right)^{k} \left(\Gt^{\bar{\lambda} \lambda\eta}(z_1+z_2)\right)^{k} \frac{1}{X_{R}^{'2}} \frac{\dipvec{yx} \cdot \dipvec{w_Az}}{r_{yx}^2 r_{w_A z}^2}\,,
\end{align}
\begin{align}
    \mathcal{K}^{7}_{ggg}(X_{R}';\boldsymbol{r}_{yz},\boldsymbol{r}_{w_Bx}) &= \frac{\pi^4}{g^4 (p^-)^2} \sum_{\lambda \bar{\lambda} \xi \bar{\xi}} \left(\Ncal^{\bar{\lambda}\lambda\bar{\xi}\xi}_{ggg,inst\,A}\right)^{\dagger}\Ncal^{\bar{\lambda}\lambda\bar{\xi}\xi}_{ggg,reg\,B}\nonumber\\
    &= (z_1 z_2 z_3) \sum_{\xi \bar{\xi} \lambda \bar{\lambda}} \Pi^{\xi \bar{\xi} \lambda \bar{\lambda}} \left(\Gt^{\eta\bar{\xi}\lambda}\left( \frac{z_3}{z_3 + z_2}\right)\right)^{k} \left(\Gt^{\bar{\lambda} \xi\eta}(z_3+z_2)\right)^{k} \frac{1}{X_{R}^{'2}} \frac{\dipvec{yz} \cdot \dipvec{w_Bx}}{r_{yz}^2 r_{w_B x}^2}\,,
\end{align}
\begin{align}
     \mathcal{K}^{8}_{ggg}(X_{R}';\boldsymbol{r}_{zx},\boldsymbol{r}_{w_Cy}) &= \frac{\pi^4}{g^4 (p^-)^2} \sum_{\lambda \bar{\lambda} \xi \bar{\xi}} \left(\Ncal^{\bar{\lambda}\lambda\bar{\xi}\xi}_{ggg,inst\,A}\right)^{\dagger}\Ncal^{\bar{\lambda}\lambda\bar{\xi}\xi}_{ggg,reg\,C}\nonumber\\
     &= (z_1 z_2 z_3) \sum_{\xi \bar{\xi} \lambda \bar{\lambda}} \Pi^{\xi \bar{\xi} \lambda \bar{\lambda}} \left(\Gt^{\eta \lambda \xi}\left( \frac{z_1}{z_1 + z_3}\right)\right)^{k} \left(\Gt^{\bar{\lambda} \bar{\xi} \eta}(z_1+z_3)\right)^{k} \frac{1}{X_{R}^{'2}} \frac{\dipvec{zx} \cdot \dipvec{w_C y}}{r_{zx}^2 r_{w_C y}^2}\,,
\end{align}
with the tensor $\Pi^{\xi \bar{\xi} \lambda \bar{\lambda}}$ defined as 
\begin{multline}
    \Pi^{\xi \bar{\xi} \lambda \bar{\lambda}} = \left( \frac{(z_2 - z_1)(1+z_3)}{(z_1+z_2)^2} +1 \right)\delta^{\bar{\lambda}\lambda}\delta^{\xi,-\bar{\xi}}  + \left( \frac{(z_3 - z_1)(1+z_2)}{(z_1+z_3)^2} + 1 \right) \delta^{\bar{\lambda}\bar{\xi}}\delta^{\xi,-\lambda} -2\delta^{\bar{\lambda}\xi}\delta^{\lambda, -\bar{\xi}}\,.
\end{multline}

\section{Recovering the JIMWLK evolution from the cross section}\label{app: JIMWLK_cross_section}
In this appendix we provide explicit expressions for the JIMWLK evolution of the LO $g \to q\bar{q}$ and $g \to gg$ cross sections. These results were obtained from the differential cross section expressions obtained in Section \ref{sec: Cross section trijets}, so they provide an important cross-check for our results. This Appendix is meant to be an extension of our results obtained in section \ref{sec:Slow gluon lim}, were we only expressed the JIMWLK evolution in terms of the JIMWLK Hamiltonian.

\subsection{The $g \to q\bar{q}g$ channel}
The combined contribution from the nine different channels contributing to the $q\bar{q}g$ trijet cross section, integrated over the gluon kinematics $(z_g, \Transv{k}{g})$ provide the following expression for the differential cross section in the slow gluon approximation:
\begin{align}\label{JIMWLK real contr qqbarg}
    &\frac{\der\sigma^{gA\rightarrow q\bar{q}g +X}}{\der^2\boldsymbol{k}_{q\perp}\der\eta_q \der^2\boldsymbol{k}_{{\bar{q}}\perp} \der\eta_{\bar{q}} } \bigg|_{slow} = \nonumber\\
    &\frac{\alpha_s \delta(1-z_q- z_{\bar{q}} )}{8(2\pi)^{6}} \frac{\N}{2 C_F} \int_{\begin{smallmatrix}
        \Perp{x}, \Perp{y}\\ \Perp{x}', \Perp{y}'
    \end{smallmatrix}} e^{-i \Transv{k}{q}\cdot(\Perp{x} - \Perp{x}')} e^{-i \Transv{k}{\bar{q}}\cdot (\Perp{y} - \Perp{y}')} \mathcal{K}_{q\bar{q},LO}(\boldsymbol{r}_{y'x'}, \boldsymbol{r}_{yx})\frac{\alpha_s \N}{(2\pi)^2}\int_{\Perp{z}} \int \der\eta_g \nonumber\\
    &\Bigg\langle \frac{2 \, \boldsymbol{r}_{xz} \cdot \boldsymbol{r}_{zx'}}{r_{zx'}^2 r_{zx}^2}\, \bigg\{ -2D_{y'y} \left(D_{xx'} - D_{xz}D_{zx'} \right) + \frac{1}{\N^2} \left(Q_{y'x'xy} -  Q_{y'x'zy}D_{xz} - Q_{y'zxy}D_{zx'} + D_{xx'}D_{y'y} \right) \bigg\} \nonumber\\
    &+ \frac{2\boldsymbol{r}_{yz} \cdot \boldsymbol{r}_{zy'}}{r_{zy'}^2 r_{zy}^2}\,\bigg\{ -2D_{xx'} \left(D_{y'y} - D_{y'z}D_{zy} \right) + \frac{1}{\N^2} \bigg( Q_{y'x'xy} -  Q_{zx'xy}D_{y'z} - Q_{y'x'xz}D_{zy} + D_{xx'}D_{y'y} \bigg) \bigg\} \nonumber\\
    &+ \frac{4\,\boldsymbol{r}_{vz} \cdot \boldsymbol{r}_{zv'}}{r_{v'z}^2 r_{vz}^2}\, \bigg\{ -D_{vv'} \left( D_{v'v} - D_{v'z}D_{zv} \right) - D_{v'v} \left( D_{vv'} - D_{vz}D_{zv'} \right) - \frac{1}{\N^2}\left( S_{zvv'zvv'} + S_{zv'vzv'v} -2 \right) \bigg\} \nonumber\\
    &+ \Bigg(\frac{ 2\boldsymbol{r}_{vz} \cdot \boldsymbol{r}_{zx'}}{r_{zx'}^2 r_{vz}^2}\, \bigg\{ 2D_{y'v} \left(D_{vx'} - D_{vz}D_{zx'} \right) + \frac{1}{\N^2} \bigg( S_{zvy'zvx'} + S_{zx'vzy'v} - 2D_{y'x'} \bigg) \bigg\} \nonumber\\
    &+ \frac{2 \,\boldsymbol{r}_{vz} \cdot \boldsymbol{r}_{zy'}}{r_{zy'}^2 r_{vz}^2}\, \bigg\{ 2D_{vx'} \left(D_{y'v} - D_{y'z}D_{zv} \right) + \frac{1}{\N^2} \bigg( S_{zvy'zvx'} + S_{zx'vzy'v} - 2D_{y'x'} \bigg) \bigg\} \nonumber\\
    &+ \frac{2\boldsymbol{r}_{xz} \cdot \boldsymbol{r}_{zy'}}{r_{zy'}^2 r_{zx}^2}\,\bigg\{  \frac{1}{2\N^2} \bigg( - Q_{y'zxy}D_{zx'} - Q_{y'x'xz}D_{zy} + D_{xx'}D_{y'y} + D_{xy}D_{y'x'} \nonumber\\
     &+ S_{xx'zyy'z} + S_{zx'xzy'y} - Q_{y'x'xy} -Q_{xx'y'y}  \bigg) \bigg\}  + (\Transv{x}{}, \Transv{y}{} \leftrightarrow \Transv{x}{}', \Transv{y}{}' )\Bigg) 
    \Bigg\rangle_{Y}\,,
\end{align}
where the terms $\Transv{x}{}, \Transv{y}{} \leftrightarrow \Transv{x}{}', \Transv{y}{}'$ come from the mirror diagrams of the interference terms. As discussed in Section~\ref{sec:Slow gluon lim}, the lower bound in the rapidity integral in Eq.\,\eqref{JIMWLK real contr qqbarg} provides the real contribution of one step in the JIMWLK evolution of the LO cross section of $q\bar{q}$ dijet production, Eq.\,\eqref{LO cross section qqbar}.

\subsection{The $g \to ggg$ channel}
We now present the results for the LO $g\to gg$ cross section JIMWLK evolution. In this case, just as for the $ggg$ cross section, we present the results in the large $\N$ limit and assuming that all correlators are real. Analogous to the previous case, the sum of all the contributions at the level of the cross section integrated over the kinematics of the slow gluon $(z_3, \Transv{k}{3})$ gives
\begin{align}
    &\frac{\der\sigma^{gA\rightarrow gg +X}}{ \der^2\boldsymbol{k}_{1\perp} \der\eta_{1} \der^2\boldsymbol{k}_{2\perp} \der\eta_2} \bigg|_{slow} = \nonumber\\
    &=\frac{\alpha_s \delta(1- z_{1} - z_2)}{(2\pi)^{6}} \N \int_{\begin{smallmatrix}
        \Perp{x}, \Perp{y}\\ \Perp{x}', \Perp{y}'
    \end{smallmatrix}}  e^{-i \Transv{k}{1}\cdot (\Perp{x} - \Perp{x}')} e^{-i \Transv{k}{2}\cdot (\Perp{y} - \Perp{y}')} \mathcal{K}_{gg,LO}(\boldsymbol{r}_{xy}, \boldsymbol{r}_{x'y'}) \nonumber\\
    &\frac{\alpha_s \N}{(2\pi)^2} \int \der\eta_{3} \int_{\Perp{z}} \Bigg\langle  \frac{2\boldsymbol{r}_{x z} \cdot \boldsymbol{r}_{z x'}}{r_{x' z}^2 r_{x z}^2} \bigg\{ -2 D_{y' y} Q_{x' y' y x} (D_{x x'} - D_{x z} D_{z x'}) \nonumber\\
    &- D_{x x'} D_{y' y} \big( Q_{x' y' y x} - Q_{x' y' y z} D_{z x} - Q_{z y' y x} D_{x' z} + D_{x' x} D_{y y'} \big) \bigg\} + \frac{2\boldsymbol{r}_{v z} \cdot \boldsymbol{r}_{zv'}}{r_{v' z}^2 r_{v z}^2} \bigg\{ -4 D_{v v'} \big( D_{v' v} - D_{v' z} D_{z v} \big) \bigg\} \nonumber\\
    &+ \frac{2\,\boldsymbol{r}_{y z} \cdot \boldsymbol{r}_{z y'}}{r_{y' z}^2 r_{y z}^2} \bigg\{ -2 D_{x'x} Q_{y' x' x y} \big( D_{y y'} - D_{y z} D_{z y'} \big) \nonumber\\
    &- D_{x'x} D_{y y'} \big( Q_{y' x' x y} - Q_{z x' x y} D_{y' z} - Q_{y' x' x z} D_{z y} + D_{x x'} D_{y' y} \big) \bigg\} \nonumber\\
    &+ \Bigg( \frac{2\,\boldsymbol{r}_{v z} \cdot \boldsymbol{r}_{zx'}}{r_{x' z}^2 r_{x z}^2} \bigg\{  2 D_{y' v} D_{x' y'} \,(D_{v x'} - D_{v z} D_{z x'}) \bigg\} + \frac{2\,\boldsymbol{r}_{y z} \cdot \boldsymbol{r}_{z v'}}{r_{v' z}^2 r_{y z}^2} \bigg\{  -2 D_{y x} D_{x v'} \big( D_{v' y} - D_{v' z} D_{z y} \big)  \bigg\} \nonumber\\
    &+ \frac{2\,\boldsymbol{r}_{y z} \cdot \boldsymbol{r}_{zx'}}{r_{x' z}^2 r_{y z}^2} \bigg\{  D_{x x'} D_{y' y} \big( -\,Q_{x' y' y z}\, D_{z x} - Q_{x' z y x}\, D_{z y'} + D_{x' x} D_{y y'} + D_{y x} D_{x' y'} \big) \bigg\} \nonumber\\
    &+ (\Transv{x}{}, \Transv{y}{} \leftrightarrow \Transv{x}{}', \Transv{y}{}' ) \Bigg)
    \Bigg\rangle_{Y}\,.
\end{align}
The lower bound of the rapidity integral provides the real contributions to the JIMWLK evolution of the LO cross section of $gg$ dijet production, Eq.\,\eqref{LO gg cross section} in the large $\N$ limit.

\bibliographystyle{utcaps}
\bibliography{refs}

\end{document}